\documentclass[onecolumn,sort&compress,numbers]{els-mrw} 

\usepackage{amsmath,amssymb,amsfonts,amsthm,makeidx,graphicx,xcolor}
\usepackage{txfonts}
\usepackage{helvet}
\usepackage{bigints}
\newcommand{\hmu}{\hat{\mu}}
\newcommand{\calD}{\mathcal{D}}
\newcommand{\calO}{\mathcal{O}}
\newcommand{\calZ}{\mathcal{Z}}

\begin{document}


\chapter{The QCD phase diagram}\label{chap1}

\author[1]{Szabolcs Bors\'anyi}%
\author[2]{Paolo Parotto}%


\address[1]{\orgname{University of Wuppertal}, \orgdiv{Theoretical physics}, \orgaddress{42119 Gausstr 20, Wuppertal, Germany}}
\address[2]{\orgname{Universit\`a di Torino}, \orgdiv{Dipartimento di Fisica} \orgname{and INFN Torino}, \orgaddress{via P. Giuria 1, I-10125 Torino, Italy}}

\articletag{Chapter Article tagline: update of previous edition, reprint.}

\maketitle

\begin{abstract}[Abstract]
Strongly interacting matter exhibits new phases under extreme conditions.
Matter was exposed to such extremes not only in the Early Universe, but
also today in the cores of neutron stars, as well as in laboratory experiments
at a much smaller scale. We study the underlying theory, 
Quantum Chromodynamics (QCD) with the methods of statistical physics
and explore the various phases we may encounter in experiment, such as
the Quark Gluon Plasma. We briefly summarize the experimental evidence
for the new forms of matter and review the theoretical efforts to embed
these findings in the broader context of quantum field theory,
with special attention to exact and broken symmetries and critical behaviour.
\end{abstract}

\begin{keywords}
Quark gluon plasma \sep 
Quark matter \sep
Color superconductivity \sep
Chiral symmetry \sep
Lattice QCD \sep
Heavy ion experiments
\end{keywords}


\begin{glossary}[Nomenclature]
	\begin{tabular}{@{}lp{34pc}@{}}
		QCD & Quantum Chromodynamics, the theory of strongly
interacting matter based on quarks and gluons\\
		Lattice QCD & The discipline where 
theorists use large computers as experimental devices to solve the QCD with stochastic methods. \\
RHIC & Relativistic Heavy Ion Collider, a leading facility at Brookhaven National Lab to study the phase diagram\\
		HRG & Hadron Resonance Gas, a successful model that approximates
QCD thermodynamics at low temperature\\
		QGP & Quark Gluon Plasma, a phase of QCD at large energy density where quarks are liberated from hadrons\\
		CFL & Color Flavor Locking, a phase of QCD at large baryon density\\
physical mass&Theorists can study QCD with whatever quark masses. 'Physical mass' means that Nature's choice is taken\\
chiral limit&The up and down quarks are very light. It is instructive
to study what would happen if they were massless.\\
	\end{tabular}
\end{glossary}

\section*{Objectives}
\begin{itemize}
	\item The reader is first introduced to the most studied representation
of the phase diagram in the temperature -- chemical potential plane with an
emphasis on existing evidence. Our focus will be what we have learned on the phase diagram from
heavy ion experiments, lattice QCD and other theoretical approaches.
	\item The subsequent section is more theoretical: we review the predictions based on the symmetry breaking pattern of QCD and study indirect evidences from lattice QCD. We discuss the famous Columbia plot and the role of imaginary chemical potentials in the uncovering of finite density QCD.
	\item We then extend the scope to include the effects of an external
magnetic field, isospin density or a theta parameter that couples to topology.
	\item The final section looks into the more elusive phases, with
a special emphasis of large density QCD. We relate the theoretical expectation to future astrophysical observations.
\end{itemize}

\section{Introduction}\label{sec:intro} 

The strong force is manifest in our Universe as the short-range interaction
that holds together the massive hadrons, most importantly protons and neutrons.
We are fortunate to possess a powerful theory, Quantum Chromodynamics (QCD),
which is able to explain and to predict the masses, decay properties
and internal structure of the hundreds of hadrons and resonance states
Nature presents us with.
QCD is a quantum field theory with six quark species, playing the role 
of matter, and an eight-component gauge field, the gluons,
responsible for interactions. As opposed to other gauge theories, like
quantum electrodynamics (QED), self interactions between gluons play
a crucial role. Their most important consequence is asymptotic freedom:
the theory is weakly coupled at very short distances or at high energy
scales. Thus, at sufficiently high temperatures, 
gluon fields lose the strength to bind hadrons together.

The extreme energy density required to unbind hadrons was indeed realized in
nature immediately after the Big Bang. The hot early Universe cooled below this 
threshold when it was approximately 10 $\mu$s old, and the temperature was around
$ 2\cdot 10^{12}$~K. Whether still today the extreme densities inside the cold cores of
neutron stars could induce a similar transition from hadronic matter to quarks is the
subject of intensive research.  Thanks to recent and future gravitational wave 
observatories, we can observe simultaneously hot and dense matter in the mergers 
of binary neutron stars. The first observation of such an event, GW170817, showed the 
potential of multi-messenger astronomy, where gravitational and neutrino 
measurements complement the photon-based 
observations~\cite{LIGOScientific:2017vwq,LIGOScientific:2017ync}.

The transition between the quark and hadronic phases is, however, best 
studied in the laboratory with heavy ion collisions. After compelling hints from the 
SPS at CERN, the Relativistic Heavy Ion Collider (RHIC) facility at BNL has uncovered 
the features of a hot new phase, the quark gluon plasma. Extreme energy collisions 
at the LHC at CERN have subsequently pushed the experimental study of the 
formation, flow and freeze-out of the plasma into the realm of precision science.
The precious synergy between theory and experiment has allowed a new field to 
emerge, focused on the mapping of the phase diagram of strongly interacting matter 
through the prediction of new phases and the identification of observable signatures.
This chapter seeks to summarize what has been learned in the process.

\begin{figure}[t]
\centering
\includegraphics[width=8cm]{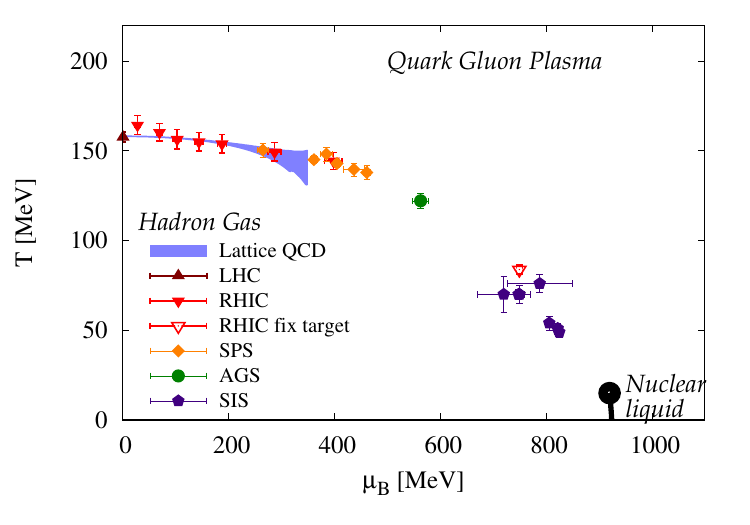}
\includegraphics[width=7cm]{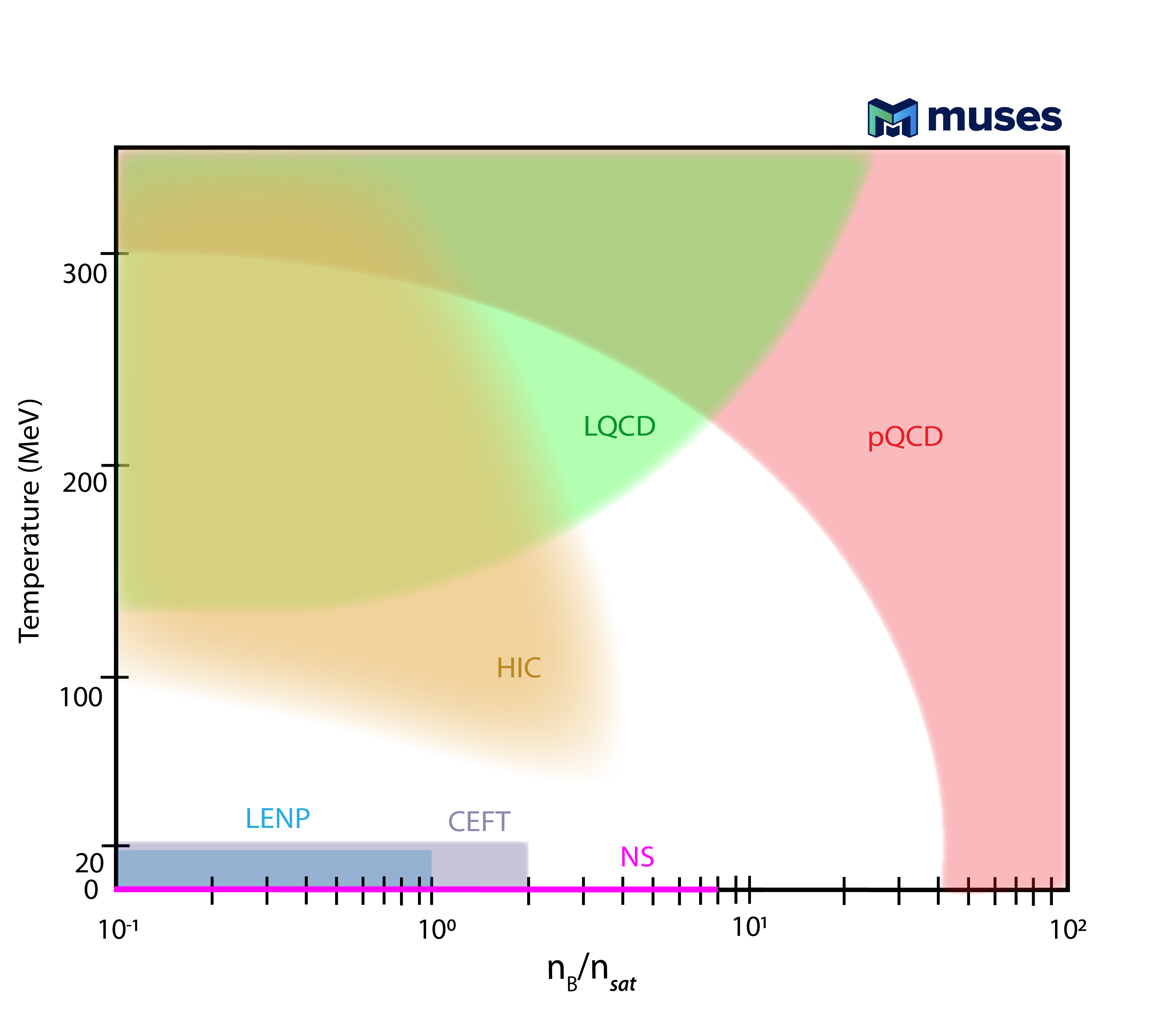}
\caption{\label{fig:muses}
Data on the QCD phase diagram. The left panel shows
the diagram in the temperature ($T$) --
baryon chemical potential ($\mu_B$) plane.
We know from lattice simulations \cite{Aoki:2006we} that
on the temperature axis the transition is a cross-over. A
cross-over line starts at $T\approx 157$~MeV
\cite{Borsanyi:2010bp,Bazavov:2011nk} and stretches into
the bulk of the phase diagram \cite{Borsanyi:2020fev}. The width
of the band refers to the theoretical uncertainties not to the width
of the transition.
The data points on and beyond the cross-over line show the
thermodynamic parameters of the chemical freeze-out
\cite{Cleymans:1998fq,Vovchenko:2015idt,Becattini:2016xct,Vovchenko:2018fmh,STAR:2017sal,Andronic:2017pug,Lysenko:2024hqp}.
The small feature at $T\approx18$~MeV is the liquid gas transition
and the corresponding end-point, measured in low energy heavy ion
collision experiments \cite{Elliott:2013pna}.
The right panel uses the density, normalized to the nuclear
saturation density $n_0=0.17 \, {\rm fm}^{-3}$ as the first axis. It shows
not the data, but where we can expect information from.
(Picture presented by the Muses collaboration \cite{MUSES:2023hyz})
}
\end{figure}

Although there are numerous different sketches of possible structures
for the QCD phase diagram, our actual knowledge is rather limited. This is
summarized in the left panel of Fig.~\ref{fig:muses} in the most common picture 
of temperature vs baryo-chemical potential. Since in QCD the baryon number is integer
valued and conserved, a grand canonical description with a well defined chemical 
potential can be employed. The baryo-chemical potential expresses the 
thermodynamic force associated with the creation of a single baryon. 

We have good reason to believe that nature is described in the energy range of
0.1\dots10~GeV by quantum chromodynamics~\cite{Gross:2022hyw}. 
Simulations of QCD on the lattice have calculated the masses of most low lying 
hadrons \cite{Durr:2008zz,BMW:2014pzb}, and nuclear lattice simulations 
have reproduced the radii and binding energies of the lightest 
nuclei~\cite{Lahde:2013uqa}. 

Also at finite temperature -- and, though with severe limitations, finite density --
lattice QCD simulations can provide rigorous evidence, and thus constrain the 
structure of the phase diagram. Continuum extrapolated simulations have shown 
that a broad transition around $155-160$~MeV \cite{Borsanyi:2010bp,Bazavov:2011nk} 
separates a low temperature phase where the hadronic description is successful from
a high temperature one, where observables are still non-perturbative but dominated 
by light degrees of freedom. 
The transition for vanishing chemical potential is a crossover \cite{Aoki:2006we} 
marked by the spontaneous breaking of the chiral symmetry. By reducing the quark
masses below their physical value, the crossover was connected to a second-order chiral
transition at $T_c=132^{+3}_{-6}$~MeV in the chiral limit \cite{Ding:2019prx}. At finite
chemical potential, the crossover is believed~\cite{Stephanov:2006dn} to become stronger
and eventually turn into a first-order transition, thus implying the existence of a critical 
point. The transition temperature in the chiral limit is expected to be an upper bound for
the temperature at this critical end-point.

Lattice simulations are severely limited by a sign problem at finite density, which originates
from the gluon-based approach of lattice QCD. In order to avoid the impossible task of
representing Grassmann variables, the quark fields are integrated out and the resulting
bosonic effective action is simulated. The resulting euclidean path integrals are computed
via importance sampling, which is possible if the effective action is real and positive, but
for a generic chemical potential it is not. Only in a few special cases is the effective action
strictly positive, i.e. if i) the chemical potential is purely imaginary; ii) the chemical potential
is real but has opposite sign for two degenerate quarks (e.g. with finite isospin chemical
potential). Additionally, non-vanishing electric fields introduce a sign problem, while 
magnetic fields don't. We will discuss these extra directions in a separate section below.

The phase boundary at finite density has thus been explored by determining the curvature
of the crossover line, which has been computed in the continuum limit by several lattice
groups \cite{Bonati:2015bha,Bellwied:2015rza,Bazavov:2018mes}. The band in the left
panel of Fig.~\ref{fig:muses} shows the result from Ref.~\cite{Borsanyi:2020fev}.
In the same work the broadness of the transition $\Delta T\approx 15~\mathrm{MeV}$ 
was also determined, defined as the width of the chiral susceptibility peak. While the
transition line unambiguously shows a negative curvature, the width does not 
significantly change for $\mu_B<300~\mathrm{MeV}$. 
We deliberately omitted from the left panel of Fig.~\ref{fig:muses} possible locations
of the chiral end-point, as no reliable extrapolations exist, let alone continuum
extrapolated results. Still, exlusion regions can be set up: a recent continuum
extrapolated result claims $\mu_{B,c}>450$~MeV with a $2\sigma$ 
confidence~\cite{Borsanyi:2025dyp}.

Experiments, too, seek to chart the QCD phase diagram and possibly locate the 
chiral critical endpoint, which indeed has been a major motivation behind several 
generations of collider experiments. Among the wealth of data collected in 
two decades of the Beam Energy Scan program at RHIC, let us highlight two classes 
of observables particularly relevant for QCD thermodynamics.
First, it was observed that the global abundance of hadrons can be modeled by a
thermal ensemble. The associated grand canonical parameters $T$ and $\mu_B$ 
provide a snapshot of the chemical freeze-out, the moment when inelastic scatterings
cease. The data points in the left panel of Fig.~\ref{fig:muses} show such conditions for
different collision energies at different facilities. For small through intermediate
chemical potentials, where lattice data are available, the freeze-out line closely follows
the chiral transition line. 

The other class of observables is based on event-by-event fluctuations of conserved
charges. These were advocated for their sensitivity to criticality
if freeze-out occurs at near-critical parameters \cite{Stephanov:1999zu}.
In fact, high order fluctuations of the net proton number have the advantage to
be accessible to experiment and to diverge with a high inverse power of the
correlation length \cite{Stephanov:2008qz}. If the crossover line Fig.~\ref{fig:muses}
ends in a chiral critical point, a non-monotonic pattern in the fourth-to-second
cumulant (or factorial cumulant) is expected, based on computations in effective
models \cite{Stephanov:2011pb}. The STAR experiment at RHIC has measured the 
energy dependence of  high order proton fluctuations
\cite{Adamczyk:2013dal,STAR:2021rls,STAR:2021iop}. Even in the latest update
\cite{STAR:2025zdq} no unambiguous critical signal was found.

While the search for the chiral critical point of QCD intensifies, another, well 
established critical endpoint exists at low temperature: it corresponds to the 
liquid-gas transition of nuclear matter.
The lightest hadrons that carry baryon charge are the nucleons with a mass 
$m_N \approx940$~MeV, which up to a small binding energy sets the threshold
for the baryo-chemical potential where nucleons can form a condensate. 
This happens in a first order transition at $\mu^{LG}\approx 924$~MeV, which for
increasing temperature weakens and eventually terminates in an endpoint at
$T_c^{LG}=17.9(4)$~MeV, as found by low energy heavy ion collision 
experiments~\cite{Elliott:2013pna}. 
This transition and the end-point have also been located by lattice simulations 
of the low energy effective theory of baryons \cite{Lu:2019nbg}. 
Additionally, functional methods in QCD can also provide quantitative predictions
in this region of the phase diagram \cite{Fukushima:2023wnl}.

One may notice that the freeze-out line in Fig.~\ref{fig:muses} appears to approach 
the cross-over line emerging from the liquid-gas critical point. Since chiral symmetry 
is broken on both sides of this crossover, it must be distinct from the chiral transition 
line. The latter -- be it first order or crossover -- has to separate from the liquid-gas 
crossover line somewhere between $T \approx 145 - 20$~MeV. 
It is possible that the chiral endpoint exists but is far from the conditions accessible 
at chemical freeze-out. In this case, new means to detect it need to be found, otherwise
it would be entirely out of the reach of heavy ion collision experiments.

While the well established facts on the QCD phase diagram are not many, several 
different approaches provide insight into the behavior of the theory in different 
regimes. This is summarized in the right panel of Fig.~\ref{fig:muses}, from the 
MUSES collaboration~\cite{MUSES:2023hyz}, where the chemical potential axis 
is replaced by the net baryon density, expressed in units of the nuclear saturation
density $\sim 0.17$~fm$^{-3}$. While this phase diagram does not reveal any
features, it displays the validity range of several complementary approaches:
i) perturbative methods (red) are reliably applicable at very high 
temperature or density, thanks to asymptotic freedom; ii) lattice QCD simulations 
(green) are most successful at zero density, in the transition region or above, and 
finite-density extrapolations rely on expansions in $\mu_B/T$, hence the range 
in $\mu_B$ increases at high temperature; iii) heavy ion collisions (orange) 
explore intermediate densities near and above the transition, but can hardly go 
beyond saturation density; iv) chiral effective theories and lattice simulations 
where gluons are not resolved are limited to very low temperatures (blue), but 
cover nuclear physics up to twice saturation density; v) neutron stars probe the 
equation of state of nuclear matter to even higher densities; vi) the upcoming 
generation of gravitational wave observations promises abundant
merger signals, which will give us experimental access to the white region in the 
center of the diagram~\cite{MUSES:2023hyz}.

We start this chapter with a phenomenological review of available evidence from
experiments, lattice results and other theoretical computations on the phase 
diagram in Section \ref{sec:pheno}.
Much of the research on the phase diagram is centered on the hypothetical chiral 
critical endpoint. We discuss its theoretical motivation and the expected 
structure of the phase diagram in general in Section \ref{sec:theory}.
While the phase diagram is most frequently presented in the temperature vs 
baryochemical plane, more external variables can be considered, such as other 
chemical potentials (e.g. isospin) or the strength of an external magnetic field.
The phase diagram in these alternative representations will be addressed
in Section \ref{sec:multi}.
The densities that we can study in an Earth-bound laboratory are limited
below the nuclear saturation density. Beyond that, astrophysical observation
will let us explore the remote parts of the phase diagram. 
In Section \ref{sec:dense} we briefly summarize the theoretical knowledge toward 
the high density limit, where exotic phases are expected.


\section{Phenomenology of the phase diagram}\label{sec:pheno} 

\subsection{Evidence based on collision experiments}

\begin{figure*}
    \centering
    \includegraphics[width=0.48\linewidth]{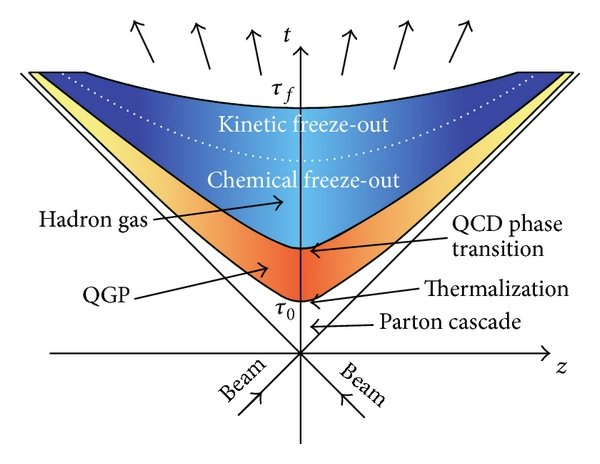}
    \includegraphics[width=0.48\linewidth]{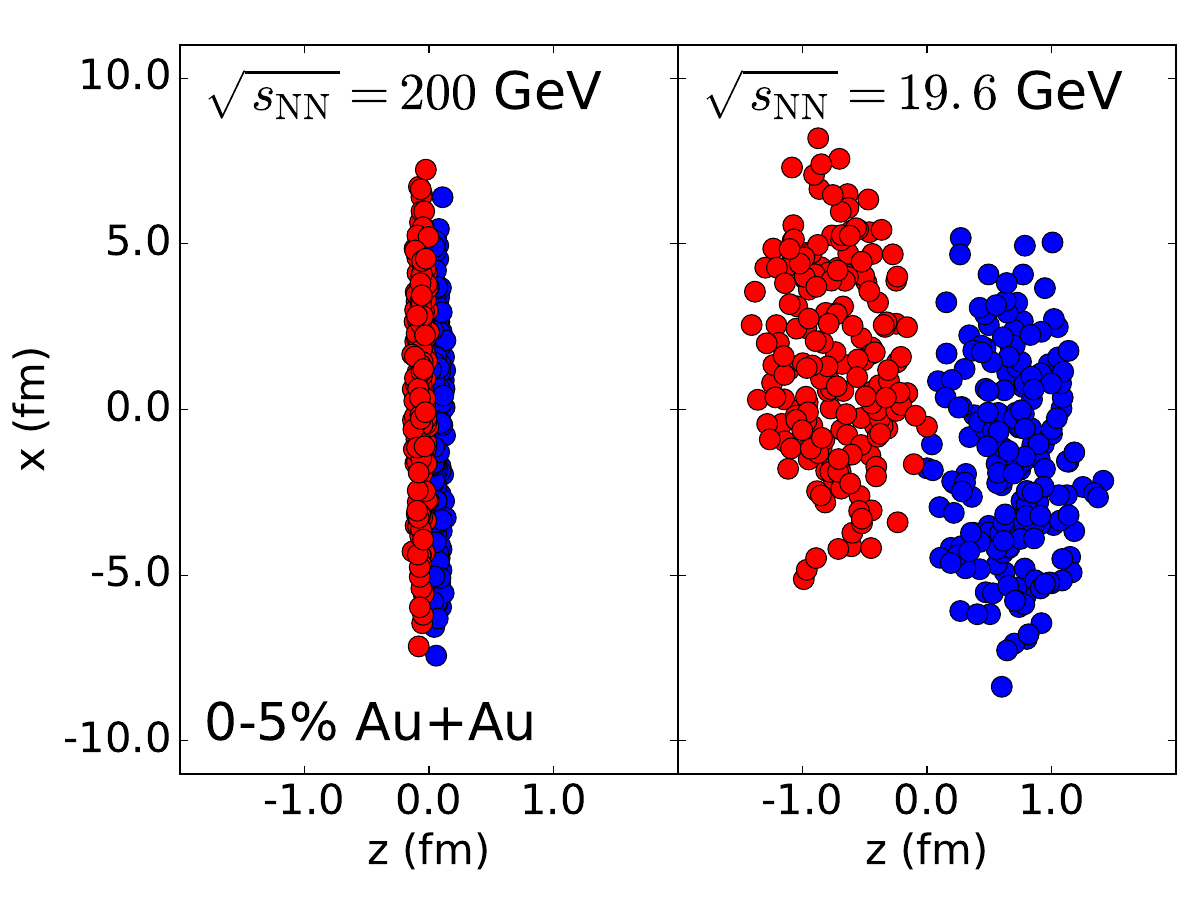}
    \caption{Left: schematic spacetime evolution of the system created in a heavy ion collision. Right: initial position of the nucleons in the laboratory frame for different collision energies~\cite{Du:2024wjm}.}
    \label{fig:HIC}
\end{figure*}

Heavy ion collisions are the tool of choice for experimentally investigating the 
thermodynamics of QCD for densities up to around nuclear saturation density, and 
temperatures up to a few times larger than the QCD transition temperature. They are 
carried out in the largest particle accelerators in the world, like the Large Hadron 
Collider (LHC) and the Super Proton Synchrotron (SPS) 
at CERN and the Relativistic Heavy Ion Collider (RHIC) at the 
Brookhaven National Laboratory, as well as future 
facilities such as the Facility for Antiproton and Ion Research (FAIR) at GSI and the 
Japan Proton Accelerator Research Complex (J-PARC). 

Heavy atomic nuclei, typically lead or gold, are collided at ultrarelativistic 
speeds, with beam energies ranging from a few GeV to a few TeV. Immediately 
after the collision, a large amount of energy is deposited in a very small 
volume in the form of color fields. The subsequent evolution of the medium 
created in heavy ion collisions is commonly divided in a few steps, 
schematically pictured with a spacetime diagram in Fig.~\ref{fig:HIC}. 
Within a very short time, typically less than 1 fm/c, the system reaches an 
approximate thermal equilibrium, and -- if the energy density is sufficient -- a 
quark gluon plasma state is formed. At top LHC energies the initial temperature 
can easily exceed 400 MeV~\cite{Harris:2023tti,Plumberg:2024leb}. This very hot 
and dense medium then expands very quickly, driven by the internal pressure 
gradients, and accordingly cools down. The quark gluon plasma phase of this 
evolution is now successfully described by means of relativistic hydrodynamic 
simulations (see e.g. Refs.~\cite{Gale:2013da,An:2021wof} for a review). 
Comparison of hydrodynamic simulations and experimental results has led to the 
discovery that the quark gluon plasma is the most perfect fluid currently known, 
as its specific shear and bulk viscosities (namely, normalized by the entropy 
density, $\eta/s$, $\zeta/s$) are extremely small, more than an order of magnitude 
smaller than in superfluid helium. This stage of the evolution ends around 
$6-10$~fm/c ($\sim 2-3\cdot10^{-23}$~s)~\cite{Lisa:2016buz}, when the temperature eventually drops below the 
QCD transition temperature $T_c$ and quarks and gluons become confined, at the 
so-called \textit{hadronization}. The still very hot and dense medium is at this 
point characterized by both elastic and inelastic collisions, which can alter 
the system's chemical composition. The latter cease at a stage named chemical 
freeze-out, after which the relative abundance of different hadron species is 
fixed. Finally, also elastic collisions cease as the system becomes very dilute, 
and thus particle spectra, too, are fixed at the kinetic freeze-out, after which 
only strong or electromagnetic decay processes may take place before the 
particles reach the detector. The characteristic time of weak decay processes is 
too long for them to take place during the system's evolution, of the order of about 
$\sim 10^{-22}$~seconds. It is remarkable that most of current experimental 
knowledge on the structure of the QCD phase diagram was inferred 
from the measurement of final state hadrons in such events.

\begin{figure}
    \centering
    \includegraphics[height=0.38\linewidth]{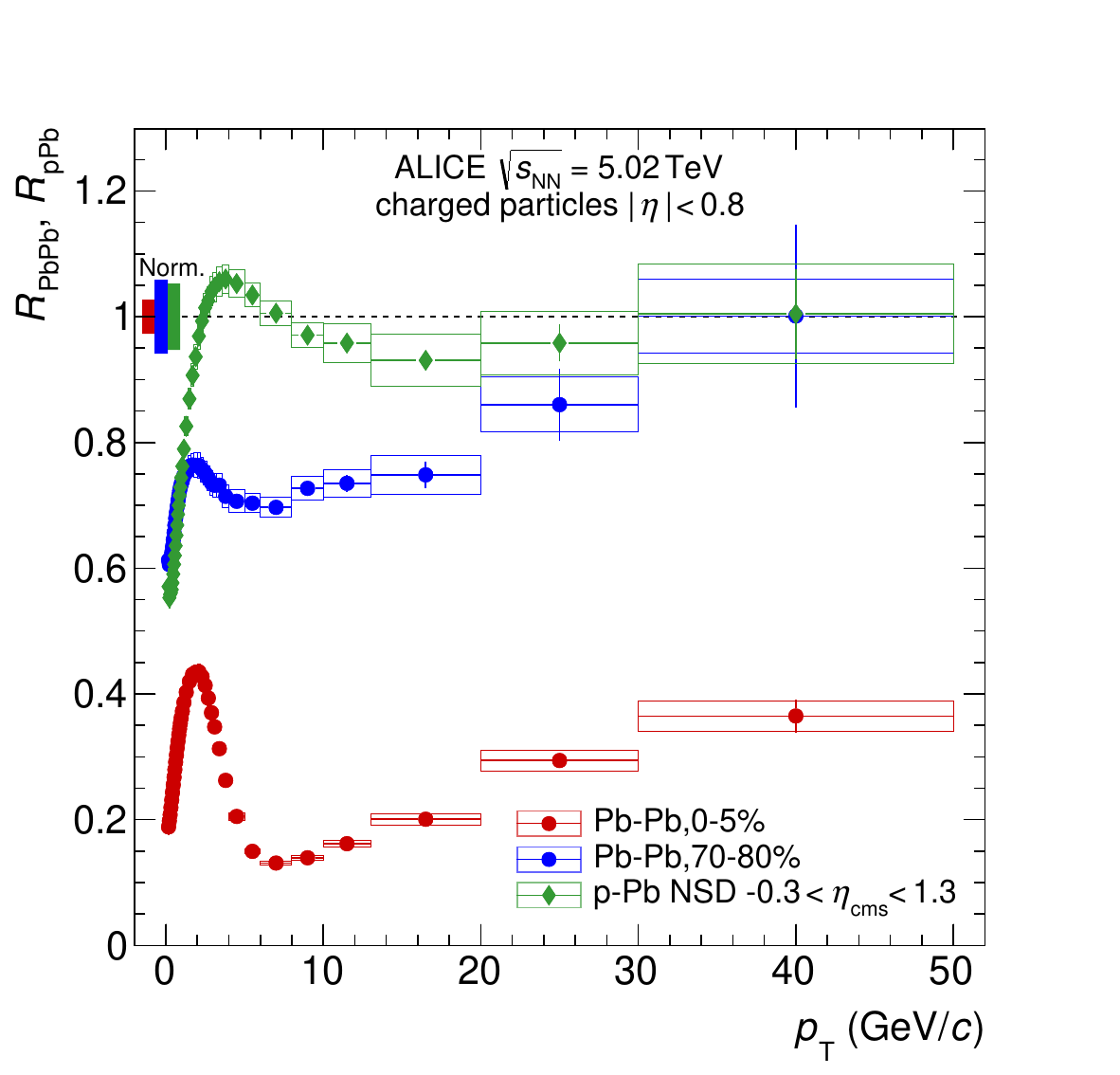} \hfill
    \includegraphics[height=0.34\linewidth]{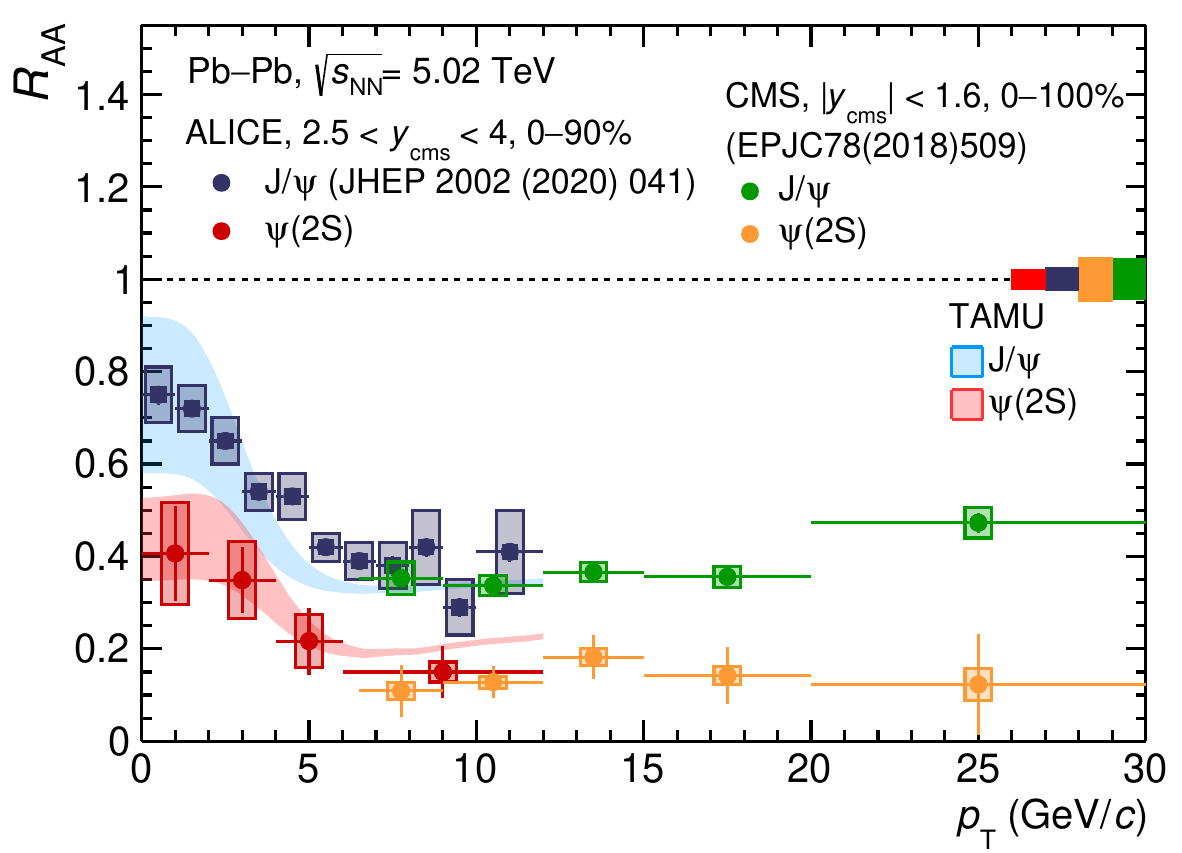}
    \caption{Left: jet nuclear modification factor in p-Pb as well as peripheral and central Pb-Pb collisions at the LHC~\cite{ALICE:2022wpn}.
    Right: charmonium nuclear modification factor in Pb-Pb collisions for $1S$ and $2S$ states at CMS~\cite{CMS:2017uuv} and ALICE~\cite{ALICE:2018rtz}. 
    }
    \label{fig:QGPsig1}
\end{figure}

By varying the system's colliding energy, the net density of the medium 
created can be modified. While at very large energy the nuclei are essentially 
transparent due to the extreme Lorentz contraction (see Fig.~\ref{fig:HIC}, right), and only color fields are left behind at the collision site, when 
the energy is lowered Lorentz contraction is smaller, and the overlap time 
significantly increases. This results in 
baryon stopping, i.e. a deposition of baryon number at mid-rapidity, increasing with decreasing collision energy, which drives the net baryon density to positive values. This allows to probe 
the phase diagram over a broad range of chemical potential.

\subsubsection{QGP formation in experiments}

Evidence of the production in the lab of a deconfined medium comes from a number 
of different signatures~\cite{Harris:1996zx,Harris:2023tti} and the strong 
synergy with a comprehensive theoretical description which was built in the past
decades. Among the most popular effects attributed to QGP formation are 
jet-quenching, the suppression (and re-generation) of quarkonium and flow.

\begin{figure}
    \centering
    \includegraphics[height=0.35\linewidth]{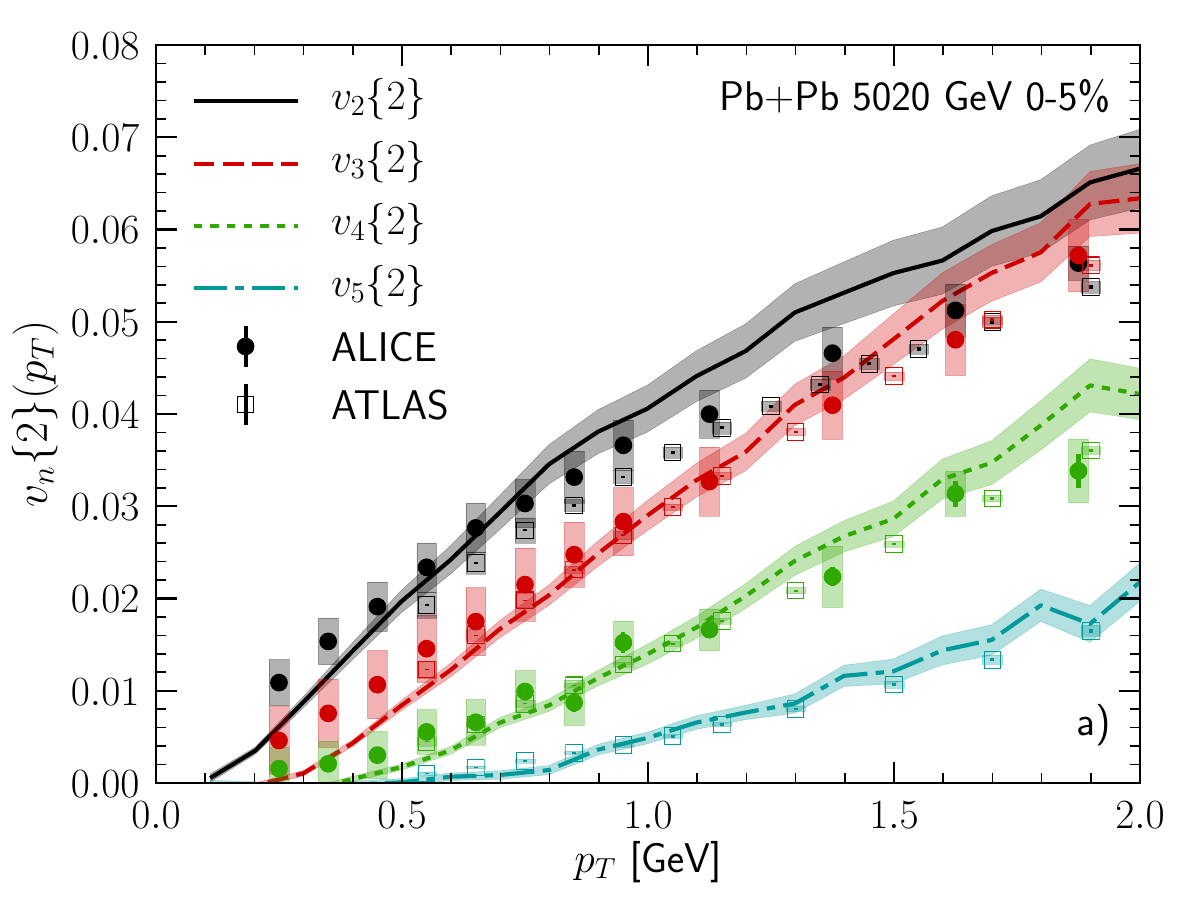} \hfill
    \includegraphics[height=0.35\linewidth]{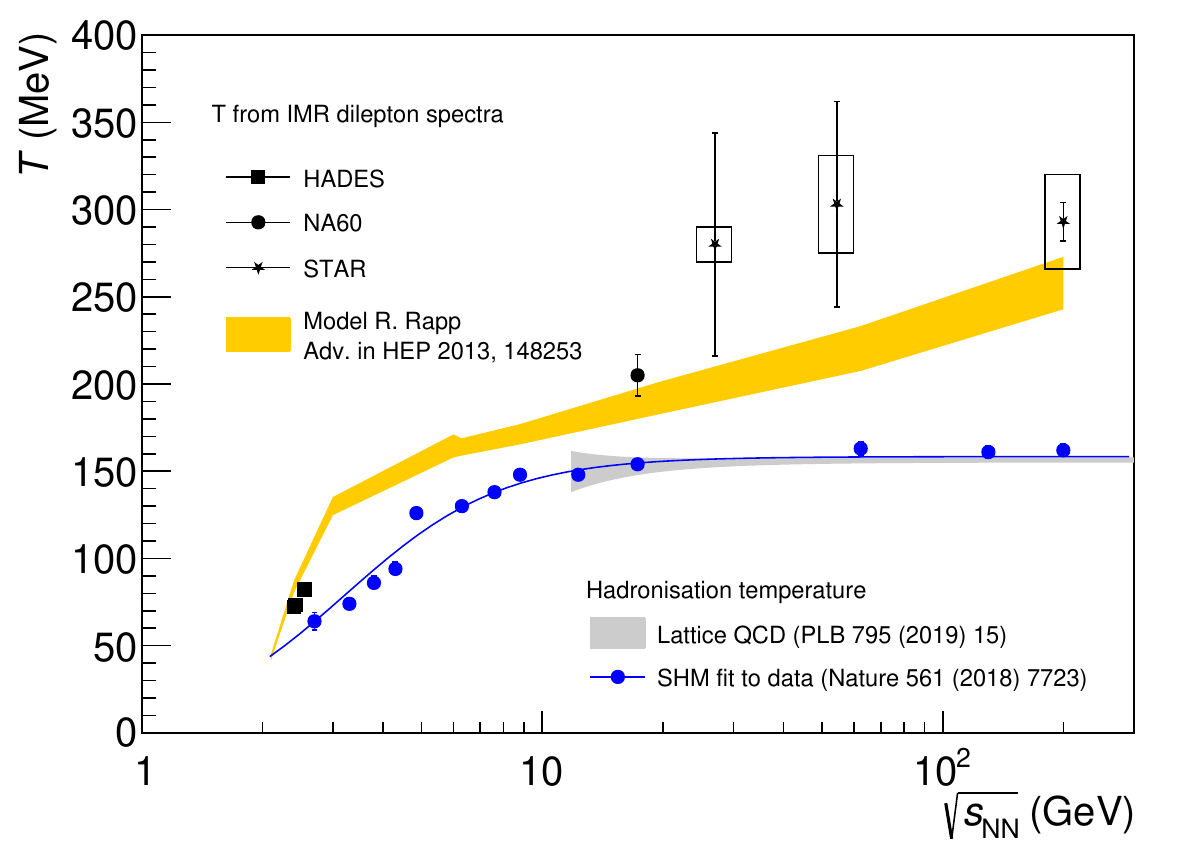}
    \caption{ 
    Left: flow coefficients in semi-central Pb-Pb collisions at the LHC, compared to results from a hybrid formalism including a viscous hydrodynamics description of the QGP phase~\cite{Schenke:2020mbo}.
    Right: temperatures extracted from dilepton invariant mass spectra from NA60, STAR and HADES, compared to chemical freezeout temperatures and model expectations. Figure from Ref.~\cite{Ahdida:2932302}.
    }
    \label{fig:QGPsig2}
\end{figure}

Jet quenching is the reduction in the energy of high momentum partons (jets) as 
they travel through the deconfined medium, mostly due to collisions and gluon 
emission. This results in a reduced jet yield, and can be quantified by the 
relative number of high momentum hadrons compared to properly scaled p-p 
collisions, usually referred as the nuclear modification factor $R_{AA}$. A 
value below one indicates that, differently from p-p collisions, the produced 
medium has caused a yield reduction. This is shown in Fig.~\ref{fig:QGPsig1} (left) for 
different colliding systems at the LHC, highlighting how the effect increases 
when a larger -- and longer-lasting -- QGP phase is 
produced~\cite{ALICE:2022wpn}.

A similar argument is the basis of $J/\psi$ suppression. Matsui and 
Satz~\cite{Matsui:1986dk} suggested that in a deconfined medium quarkonium would 
melt because of the screening of the color force. Based on this argument, a 
smaller yield of quarkonium states would be expected in heavy-ion collisions, 
compared to p-p ones, which again could be quantified through a nuclear 
modification factor. Additionally, the effect would be larger for more weakly 
bound states, a phenomenon known as sequential suppression. This can be seen in 
Fig.~\ref{fig:QGPsig1} (right), where $R_{AA}$ at LHC is shown for $J/\psi$ (a 
$1S$ state) and $\psi(2S)$, in both cases smaller than one, and displaying a more 
pronounced suppression for the 2S state. In Fig.~\ref{fig:QGPsig1} (right) a 
complementary effect is also visible at low-$p_T$: quarkonium regeneration. Bound 
states can form by a recombination mechanism between the c and $\bar{c}$ quarks, 
during the deconfined phase and/or at the hadronisation of the system. This 
effect is sizable for charmonium at high LHC energies due to the large density of 
$c\bar{c}$ pairs, while for bottomonium states it's negligible, because the 
density of $b\bar{b}$ pairs is much lower~\cite{Andronic:2025jbp}.  
The original view on quarkonium suppression has been revisited in recent years to
include the effect of dissociation due to collisions with the medium on top (or 
instead~\cite{Bazavov:2023dci}) of  color screening, which can be encoded in an 
imaginary potential~\cite{Laine:2006ns,Beraudo:2007ky} and described within the 
formalism of open quantum 
systems~\cite{Akamatsu:2014qsa,Brambilla:2016wgg,Akamatsu:2020ypb,Brambilla:2023hkw}.

While these observables were proposed decades ago, new opportunities for QGP 
signaling were provided more recently by flow observables. Collectively, they 
relate to the fact that the strongly coupled nature of the deconfined medium 
results in the appearance of collective behavior; in short, the QGP ``flows''.
A quantitative description is made possible by the flow coefficients $v_n$, 
namely the Fourier coefficients of the azimuthal distribution of final state 
particles. Non-zero values for these coefficients reflect the anisotropy of 
particle emission driven by the initial spatial anisotropy of the medium. If the 
QGP were not strongly coupled, or not present, free streaming would result in
a washing away of these anisotropies. In Fig.~\ref{fig:QGPsig2} (left) results from 
ALICE~\cite{ALICE:2018rtz} and ATLAS~\cite{ATLAS:2018ezv} for the first flow 
coefficients of charged hadrons are shown, and compared to results from a hybrid 
model including viscous hydrodynamic description of the plasma phase, with 
excellent agreement~\cite{Schenke:2020mbo}.

Evidence on the formation of a thermalized medium at temperatures well
exceeding the transition temperature predicted by lattice QCD can be provided by photon and dilepton spectra~\cite{Shuryak:1978ij,PhysRevLett.97.102301,Rapp:2014hha,Seck:2020qbx,Salabura:2020tou,Geurts:2022xmk,Xu:2022mqn,Savchuk:2022aev}. Photons and dileptons are radiated by the medium throughout the system's evolution, and their thermal distributions can be reconstructed from momentum and invariant mass spectra, respectively. However, while photons spectra are heavily affected by the ``blue shift'' effect due to the rapid expansion of the system, dilepton measurements allow for a clean determination of the temperature of the medium. In recent years, such measurements have been carried out by the NA60~\cite{NA60:2008dcb,Specht:2010xu}, STAR~\cite{STAR:2015zal,PhysRevC.107.L061901,STAR:2024bpc} and HADES~\cite{HADES:2019auv} 
collaborations, yielding results consistently above the chemical freeze-out and 
QCD transition lines, see Fig.~\ref{fig:QGPsig2} (right). 

\subsubsection{Phase separation and critical point}

From an experimental standpoint, it is not possible to directly locate the QCD phase 
transition on the phase diagram. However, by comparing results to theoretical 
models for the production of hadrons, it is possible to estimate the location
of the chemical freeze-out in different experimental setups. Thermal fits, i.e. 
fits to hadron yields based on thermal models such as the hadron resonance gas 
(HRG) model (see Sec.~\ref{sec:hrg}) have been widely employed to chart the 
location of chemical freeze-out in high energy collisions.
A collection of these from different collision energies and experimental 
facilities is shown in Fig.~\ref{fig:muses} (left), returning a consistent picture 
that is also in quite good agreement with current results for the QCD transition 
temperature at finite density. At large collision energies, the chemical freeze-out 
occurs shortly after hadronization, at temperatures around $T = 150-160$~MeV, due 
to the explosive expansion of the fireball that cause a quick dilution of the 
system. At larger chemical potentials, the line of freeze-out points approaches the 
location of ordinary, cold and dense nuclear matter. While results for the 
transition temperature above $\mu_B\simeq 400$~MeV are not available, it is expected
that the two curves detach at some point, due to the fact that lower collision 
energies create a less explosive system which ultimately freezes out at a lower 
temperature. 

Chemical freeze-out has also been investigated by studying low order net-particle 
fluctuations~\cite{Alba:2014eba,Bellwied:2018tkc,Bellwied:2019pxh}, namely proton, 
pion, kaon, lambda serving as proxies for conserved $B,Q,S$ charges as functions of 
the collision energy, and yielding comparable results as thermal fits to yields. 
Often, separate fits to strange and non-strange degrees of freedom result in 
improved fit quality and a consistent temperature gap between the freeze-out of 
strange and light particles, in a picture dubbed flavour hierarchy, which appears 
present at different collision energies in yields and fluctuations 
alike~\cite{Bluhm:2018aei,Bellwied:2018tkc,Bellwied:2019pxh,Flor:2020fdw}.

\begin{figure}
    \centering
    \includegraphics[width=\linewidth]{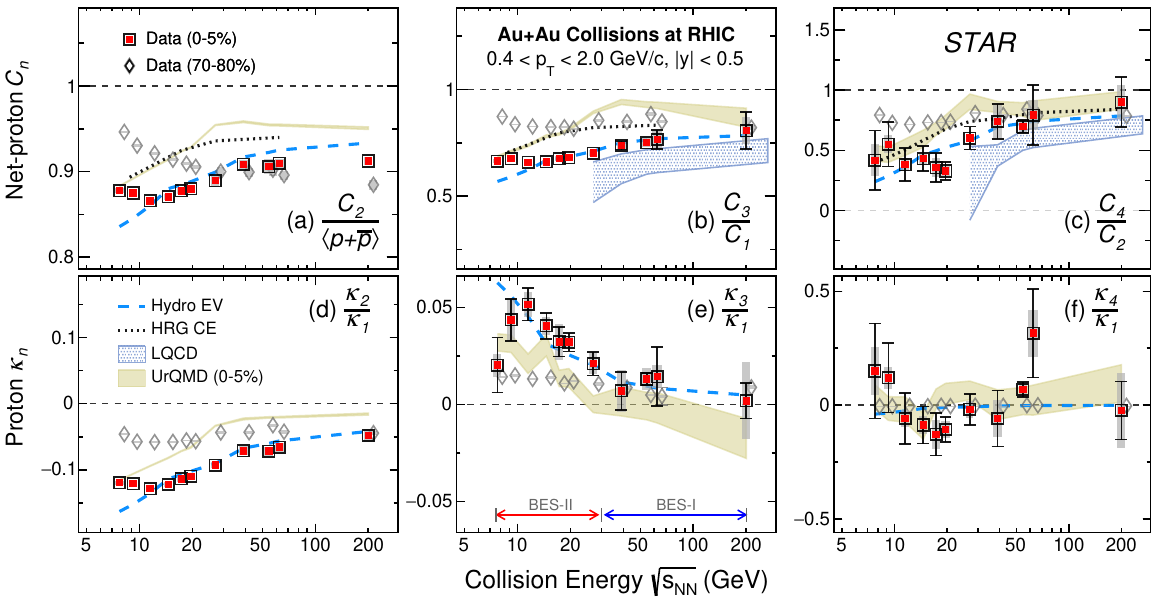}
    \caption{Net proton cumulant ratios (top) and proton factorial cumulant ratios (bottom) as functions of the collision energies from the STAR collaboration~\cite{STAR:2025zdq}, together with non-critical baselines.}
    \label{fig:STARnetp}
\end{figure}

The rather complete picture of where collision systems freeze out in the phase 
diagram clashes with the fact that it is not known whether the QCD transition in 
its close vicinity is a true phase transition or not. While it is well established 
that at small chemical potential QCD undergoes a smooth transition, at larger 
density the transition might be of the first order. Should there be a critical 
point separating the two regimes, as many models of QCD suggest, at some 
intermediate collision energy the system would be bound to evolve in its vicinity. 
Identifying signals of criticality in experimental measurements is largely based on 
the corresponding divergence in the correlation length, which should result in 
large fluctuations across the system. In particular, event-by-event fluctuations in 
the baryon number have been identified as the most promising observables, with 
higher order fluctuations diverging with increasing powers of the correlation 
length~\cite{Stephanov:2008qz}. Arguments based on the universality of critical 
behavior made it possible to identify a non-monotonic dependence on the collision 
energy as the typical smoking gun signature of criticality~\cite{Stephanov:2011pb}.
State-of-the-art results for net-proton fluctuation ratios from the STAR 
collaboration~\cite{STAR:2025zdq} are shown in Fig.~\ref{fig:STARnetp} together with
existing estimates of a non-critical baseline. It has been pointed out that 
factorial cumulants are likely to be more sensitive to criticality, as they isolate 
irreducible particle correlations of order $n$. Results from the HADES 
collaboration at $\sqrt{s} = 2.4$~GeV appear to follow a similar trend. Although the 
quality of the data has sensibly increased thanks to the new fixed target program 
at RHIC, identifying a clear pattern for criticality is not straightforward. Though 
deviations from the baseline below $\sqrt{s} \simeq 10$~GeV are certainly 
suggestive, better precision from the data, and more importantly a firmer hold on 
the correct non-critical baseline will be needed to draw more precise conclusions. 
Additional data from the HADES and CBM experiments are expected in the future, as
well as increased statistics on the STAR data.

\subsection{Evidence based on lattice QCD\label{sec:lat}}

The theoretical investigation of the QCD phase diagram can count on a number of
different approaches, each with its advantages and limitations. In 
Section~\ref{sec:approx} we will describe a number of methods to 
study the thermodynamics of QCD in different regimes, retaining more or less of 
the rich physics the theory entails. In this Section, we discuss the large body 
of evidence gathered in the past decades from the most robust method to study 
it, namely lattice simulations.  In fact, lattice QCD translates the problem of
solving the theory of strong interactions into a numerical experiment.
Most of the established features of QCD thermodynamics are indeed due to lattice 
simulations. 

In short, the formulation of QCD on a lattice amounts to calculating
\textit{euclidean} path integrals like:
\begin{equation} \label{eq:calZ1}
    \calZ = \bigintssss \calD U \calD \bar{\psi} \calD \psi \, e^{-S_G[U]} e^{-S_F[U,\bar{\psi},\psi]}
\end{equation}
where $U$ are the gauge fields and $\bar{\psi},\psi$ the fermion fields, and 
$S_G$, $S_F$ are the gauge and fermion actions. The fermion part is bilinear in the fermion fields, and can thus be integrated analytically, yielding:
\begin{equation}
     \calZ = \bigintssss \calD U \, e^{-S_G[U]} \det M[U]
\end{equation} 
where $M$ is the Dirac matrix involving all fermion fields on the lattice, and 
its determinant depends only on the gauge configuration $\{U\}$. 
The path integral in Eq.~\eqref{eq:calZ1} is formally analogous
to the partition function of a statistical system~\cite{Wilson:1974sk}, thus
expectation values can be evaluated as:
\begin{equation}
    \left\langle \hat{O} \right\rangle = \frac{1}{\calZ} \bigintssss \calD U \, \hat{O} \, e^{-S_G[U]} \det M[U] \, \, .
\end{equation}  
By discretizing spacetime, the integral over the field configurations becomes
numerically tractable and can be computed by means of Monte Carlo methods by
identifying $e^{-S_G[U]} \det M[U]$ with a statistical weight assigned to
configuration $U$. In practice, importance sampling methods are employed to
estimate the integral by summing over only a small portion of statistically
relevant configurations. This approach breaks down when a real chemical
potential is included because then $\det M[U]$ becomes complex making the would-be
probability weight of a gauge configuration complex. This is the complex action problem in lattice QCD.
Alternatives have been developed over the years to circumvent this problem and
somehow bridge the gap between the theory accessible to simulations ($\mu_B =
0$) and the theory of interest ($\mu_B > 0$). These include Taylor expansion
around $\mu_B=0$
\cite{Allton:2003vx,Allton:2005gk,Kaczmarek:2011zz,Endrodi:2011gv,Borsanyi:2012cr,Bazavov:2017dus,Bonati:2018nut,Bazavov:2018mes,HotQCD:2018pds,Bollweg:2022rps}
and analytic continuation from imaginary chemical 
potential~\cite{DElia:2002tig,deForcrand:2003bz,DElia:2007bkz,Cea:2009ba,Bonati:2015bha,Cea:2015cya,Bellwied:2015rza,Vovchenko:2017xad,Vovchenko:2017gkg,Bonati:2018nut,Borsanyi:2020fev,Borsanyi:2021sxv,Borsanyi:2022qlh}. Reweighting methods have also been widely employed~\cite{Barbour:1997ej,Fodor:2001au,Fodor:2001pe,Fodor:2004nz,deForcrand:2002pa,Alexandru:2005ix,Fodor:2007vv}, and have recently enjoyed increased interest~\cite{Endrodi:2018zda,Giordano:2020uvk, Giordano:2020roi, Borsanyi:2021hbk,Borsanyi:2022soo}. 

Most notably for the thermodynamics of QCD, lattice results include the 
determination of the transition line, the equation of state, and fluctuations of 
conserved charges, which are the main topics of this Section.

\subsubsection{QCD transition}

The fate of the hadronic phase at increasing temperature was initially proposed 
by Hagedorn to be a limiting temperature~\cite{Hagedorn:1965st}, known as 
Hagedorn temperature. Cabibbo and Parisi later showed that a more likely 
scenario would be the existence of a phase transition to a deconfined state of 
matter~\cite{Cabibbo:1975ig}. The nature and location of the transition remained 
uncertain for about 30 more years, until lattice simulations became powerful 
enough to provide definitive answers.

\begin{figure}
    \centering
    \includegraphics[height=0.36\linewidth]{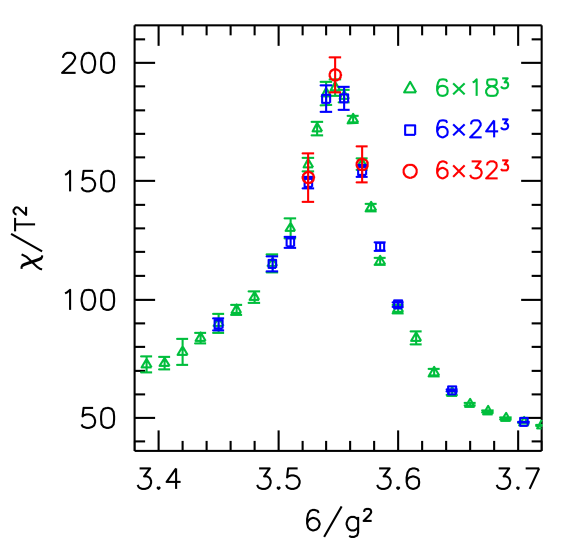} \hfill
         \includegraphics[height=0.35\linewidth]{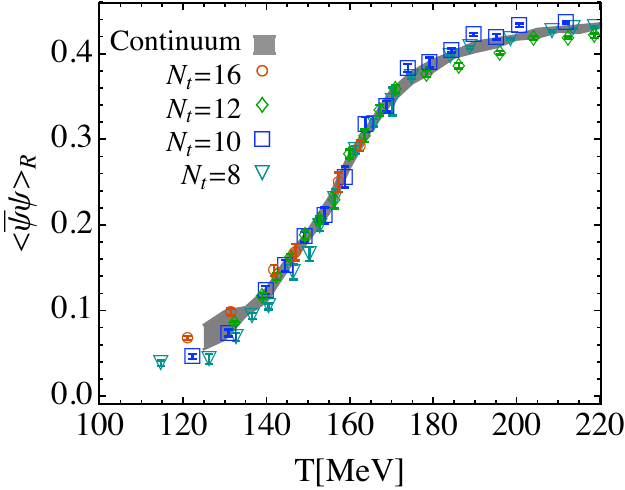}
    \caption{Left: chiral susceptibility as a function of the gauge coupling (equivalently, of the temperature) for different volumes. Figure from Ref.~\cite{Aoki:2006we}. Right: chiral condensate as functions of the temperature, at finite lattice spacing and in the continuum. Figure from Ref.~\cite{Borsanyi:2010bp}.} \label{fig:crossover}
    \label{fig:enter-label}
\end{figure}

Simulations of the pure Yang-Mills theory have quickly shown that the QCD
transition at vanishing baryon density is of the first order~\cite{Brown:1988qe,Fukugita:1989yb}, 
yet it was a much harder task to apply lattice methods to full QCD with physical quark masses.
By means of a finite size scaling of continuum extrapolated results it was shown in Ref.~\cite{Aoki:2006we},
that in full QCD the transition is a smooth crossover. In general, the susceptibility peak of
the order parameter of a first-order transition is expected to grow in height
and shrink proportionally to the system's volume, while in a crossover
the susceptibilities converge. The left panel of Fig.~\ref{fig:crossover} shows this for the chiral
susceptibility with no visible dependence on the volume,
indicating that the transition with Nature's choice of quark masses is a crossover.

The transition separating the hadronic phase from the quark gluon plasma is, in 
fact, related to two distinct, albeit related, phenomena, namely the 
restoration of chiral symmetry and deconfinement, see Section~\ref{sec:theory}.
In the right panel of Fig.~\ref{fig:crossover} we show the temperature dependence of the pseudo-order 
parameter of the chiral transition, the renormalized chiral condensate.
One observes a smooth rise around $T\simeq 155$~MeV, as expected in a crossover.
Determinations of the pseudocritical temperature have been based on different observables, 
yielding slightly different values~\cite{Aoki:2006br,Aoki:2009sc,Borsanyi:2010bp,Bazavov:2011nk}. More
recent results based on chiral observables yield $T\sim 157$~MeV with
uncertainties around 1 MeV~\cite{Bazavov:2018mes,Borsanyi:2020fev}. 
The deconfinement aspect of the transition is more subtle, as will be discussed in Section~\ref{sec:columbia}.

\begin{figure}
    \centering
    \includegraphics[width=0.5\linewidth]{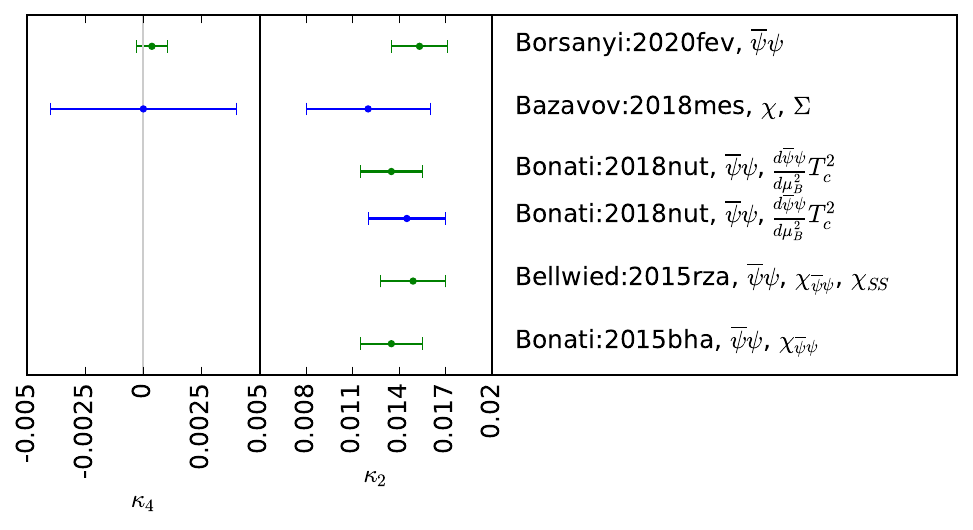}
    \includegraphics[width=0.45\linewidth]{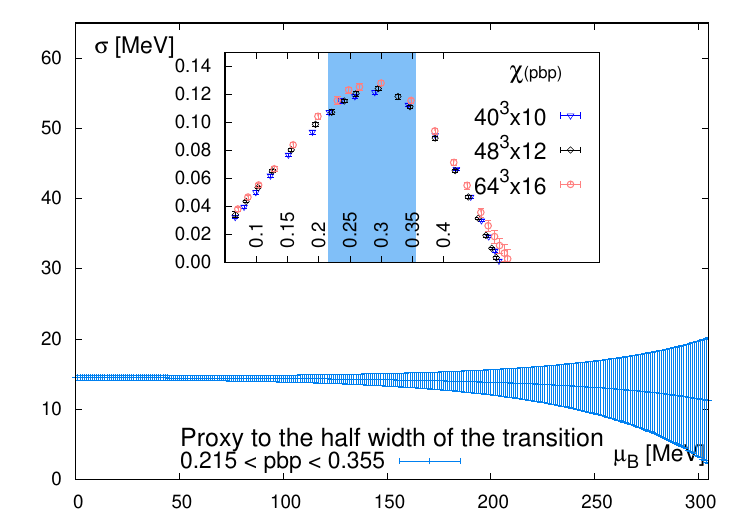}
    \caption{Left: world plot of the coefficients $\kappa_2, \kappa_4$ of Eq.~\ref{eq:tcline} (Taylor method in blue, imaginary $\mu_B$ in green). Right: width of the transition at real chemical potential. Figures from Ref.~\cite{Borsanyi:2020fev}.}
    \label{fig:kappaN}
\end{figure}

At finite chemical potential, it is customary to express the transition 
temperature as an expansion:
\begin{equation} \label{eq:tcline}
    T_c (\hmu_B) = T_c(0) \left( 1 - \kappa_2 \hmu_B^2 - \kappa_4 \hmu_B^4 + \calO (\mu_B^6) \right) \, \, ,
\end{equation}
where $\hmu_B = \mu_B/T$, and the coefficients $\kappa_n$ encode the chemical 
potential dependence of $T_c$. Results have been obtained making use of 
simulations at zero chemical potential, and directly evaluating the 
coefficients $\kappa_n$, or by determining the transition temperature for 
imaginary chemical potentials and analytically continuing to $\mu_B^2>0$~\cite{Bonati:2015bha,Bellwied:2015rza,Bonati:2018nut,Bazavov:2018mes,Ding:2024sux}. 
Results based on chiral observables from several groups and both
methods are shown in Fig.~\ref{fig:kappaN} (left), displaying good agreement. 
The NLO coefficient $\kappa_4$ appears to be sensibly smaller than $\kappa_2$, 
and compatible with zero at the current precision. The transition line shown in 
Fig.~\ref{fig:muses} was obtained from analytical continuation in 
Ref.~\cite{Borsanyi:2020fev}. While estimates for the transition temperature 
have become extremely precise, this does not imply that the transition takes 
place over such a narrow range in temperature. The crossover transition is 
rather broad, and its width was estimated in Ref.~\cite{Borsanyi:2020fev} as a 
function of the chemical potential, by studying the chiral susceptibility peak. 
As shown in Fig.~\ref{fig:kappaN} (right), the half-width of the crossover stays 
constant around $\Delta T \simeq 15$~MeV for $\mu_B < 300$~MeV, showing no hint 
of strengthening of the transition. Similarly, the chiral transition peak height 
is shown to remain constant in the same range. The extrapolation to higher
chemical potentials is extremely difficult, but a proxy to the transition temperature
could be computed in a smaller volume ($16^3 \times 8$) up to 400~MeV in $\mu_B$ \cite{Borsanyi:2024xrx},
The chosen proxy was the peak of the static quark entropy $S_Q \sim
\partial \ln L/\partial T$ as suggested in Ref.~\cite{Bazavov:2016uvm},
because its value is more stable under the reduction of volume than the direct chiral observables
\cite{Borsanyi:2025lim}.
The peak of $S_Q(T)$ is found to roughly follow the chiral transition line of Ref.~\cite{Borsanyi:2020fev}.

\subsubsection{Equation of state}

\begin{figure}
    \centering
    \includegraphics[height=0.3\linewidth]{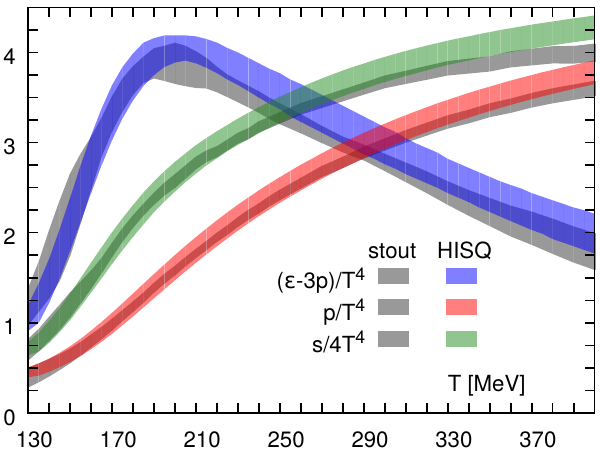} \hfill
    \includegraphics[height=0.33\linewidth]{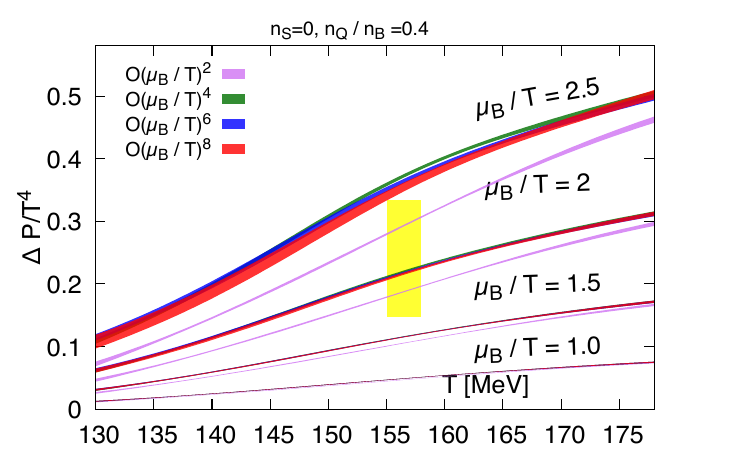}
    \caption{Left: equation of state at $\mu_B=0$ from the hotQCD and Wuppertal-Budapest collaborations, showing excellent agreement. Figure from~\cite{Bazavov:2014pvz}. Right: reduced pressure as a function of the temperature for different chemical potentials and orders of the Taylor expansion~\cite{Bollweg:2022fqq}.}
    \label{fig:eos}
\end{figure}

The equation of state is the fundamental quantity to describe the thermodynamic
behavior of a system in different conditions. Its qualitative features can 
provide information on the effective degrees of freedom in a system, and help 
locate where phase transitions occur. For the modeling of heavy ion collisions, 
the equation of state of QCD is a fundamental quantity, which allows to take 
into account the thermodynamics of the theory and to close the set of equations
to solve. Given the values of the volume $V$ and of the control parameters 
$T,\mu_B$, the equation of state is uniquely determined by the partition 
function $Z(V,T,\mu_B)$, from which the free energy $f = - T \ln Z(V,T,\mu_B)$ 
is defined. The pressure is (minus) the derivative of $f$ with respect to the 
volume, which under the assumption of a homogeneous system simply amounts to 
$p = -f/V = T/V \ln Z$. All thermodynamic quantities are then defined as 
derivatives of $p$.

On the lattice, it is not possible to calculate the pressure directly. Most 
commonly, the integral method~\cite{Engels:1990vr} is employed, writing the 
pressure as:
\begin{equation}
    \frac{p (T)}{T^4} = \frac{p (T_0)}{T^4} + \bigintssss_{T_0}^T \frac{dT}{T} \frac{I(T)}{T^4} \, \, ,
\end{equation}
where the trace anomaly $I(T)$ is determined by the renormalized values of the 
chiral condensates and the gauge action, and the integration constant 
$p(T_0)/T^4$ is determined e.g., as an integral of the chiral condensates in the 
quark masses.

For about a decade now there has been substantial agreement over the equation of 
state of QCD at vanishing density between different collaborations, with 
physical values of the quark masses and in the continuum limit, see e.g. 
Fig.~\ref{fig:eos} (left)~\cite{Borsanyi:2013bia,Bazavov:2014pvz}. A new 
determination was presented in a recent work~\cite{Borsanyi:2025dyp}, with 
smaller uncertainties but compatible with the previous ones.

At finite density the equation of state can be determined as a Taylor expansion~\cite{Allton:2002zi,Allton:2005gk}:
\begin{equation}
    \frac{p(T,\mu_B)}{T^4} = \sum_n \frac{1}{n!} \chi_n^B (T) \left( \frac{\mu_B}{T} \right)^n \, \, ,  \qquad \qquad    
    {\rm with} \qquad \chi_n^B (T) = T^n \frac{\partial^n \left( p/T^4 \right)}{\partial \mu_B^n} \, \, ,
\end{equation}
where only even coefficients contribute due to charge conjugation symmetry. The 
determination of the susceptibilities $\chi_n^B$ is very costly when the order 
is increased, and convergence of the series is usually checked by comparing 
subsequent orders. Results for coefficients up to order 8 have been 
published~\cite{Kaczmarek:2011zz,Endrodi:2011gv,Borsanyi:2012cr,Bazavov:2017dus,Bonati:2018nut,Bazavov:2018mes,HotQCD:2018pds,Bollweg:2022rps}, but only up to 
order 4 were continuum extrapolated (and order 6 in a smaller 
volume~\cite{Borsanyi:2023wno}).
Fig.~\ref{fig:eos} shows the baryon density as a function of the temperature, 
for different chemical potentials and different expansion orders, obtained from 
a Taylor expansion in Ref.~\cite{Bollweg:2022fqq}. A multidimensional Taylor 
expansion of the QCD pressure was constructed in 
Refs.~\cite{Monnai:2019hkn,Noronha-Hostler:2019ayj} to cover the full 3D space 
of chemical potentials $\mu_B,\mu_Q,\mu_S$ needed in the hydrodynamic 
description of heavy ion collisions with multiple conserved charges.

\begin{figure}
    \centering
    \includegraphics[width=0.465\linewidth]{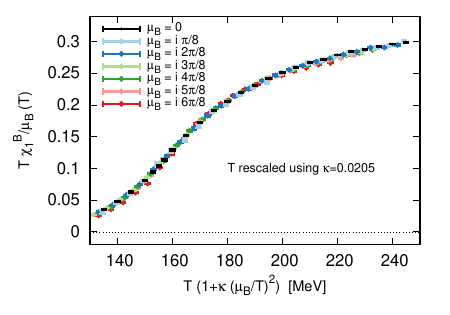}
    \includegraphics[width=0.45\linewidth]{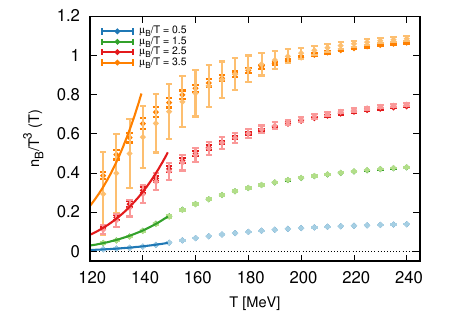}
    \caption{Left: collapse of curves of $n_B/\hmu_B$ for different imaginary valued $\mu_B$ when $T$ is replaced by $T^\prime = T (1 + 0.0205 \hmu_B^2)$. Right: baryon density in the continuum limit, for different chemical potentials, at two subsequent orders in the expansion (darker shade is $\calO (\hmu_B^2)$, lighter shade is $\calO (\hmu_B^4)$). The perfect agreement up to $\hmu_B = 3.5$ shows the good convergence of the series. Figures from Ref.~\cite{Borsanyi:2021sxv}.}
    \label{fig:eos2}
\end{figure}

Imaginary $\mu_B$ simulations have been used in the past to determine 
fluctuation observables~\cite{DElia:2016jqh,Borsanyi:2018grb}. In 
Ref.~\cite{Borsanyi:2021sxv} an alternative approach -- sometimes dubbed 
$T^\prime$-expansion and generalized in 
Refs.~\cite{Borsanyi:2022qlh,Abuali:2025tbd}-- was developed to extrapolate the 
equation of state to finite chemical potential. This approach is based on the 
ansatz of a simplified $\mu_B$ dependence of the baryon density in the vicinity 
of $\mu_B=0$:
\begin{equation}
    n_B (T,\mu_B) = \hmu_B \, \chi_2^B (T^\prime, 0) \, \, \, \qquad \qquad
    {\rm with} \qquad T^\prime = T (1 + \kappa_2(T) \hmu_B^2 + \kappa_4(T) \hmu_B^4 + \calO (\hmu_B^6)) \, \, ,
\end{equation}
where the coefficients $\kappa_n(T)$, similarly to Eq.~\eqref{eq:tcline}, encode
the $\mu_B$ dependence, but here are crucially $T$-dependent themselves to 
define the expansion rigorously. This expansion amounts to a reorganization of 
the Taylor series, carried out along lines of constant $n_B/\hmu_B$. This 
results in better convergence, and smaller uncertainties, as shown in 
Fig.~\ref{fig:eos2} (right).

\subsection{Diagrammatic approaches to QCD\label{sec:approx}}

\subsubsection{Weak coupling expansion\label{sec:htl}}
\begin{figure}
\centering
\includegraphics[height=0.35\textwidth]{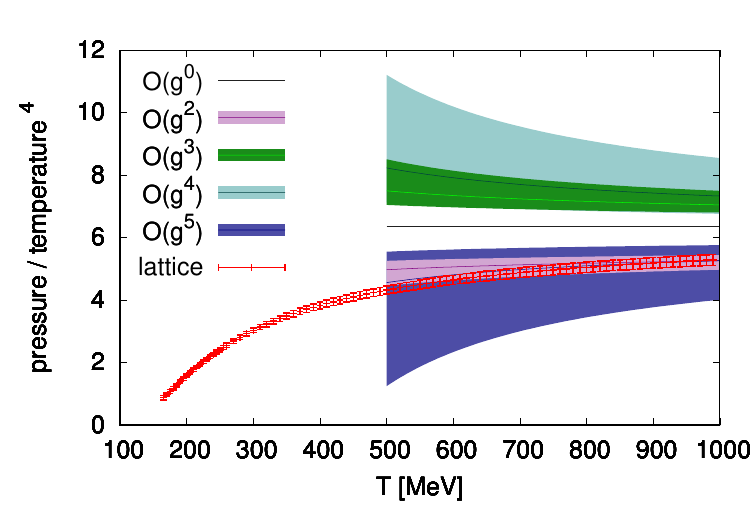}
\includegraphics[height=0.35\textwidth]{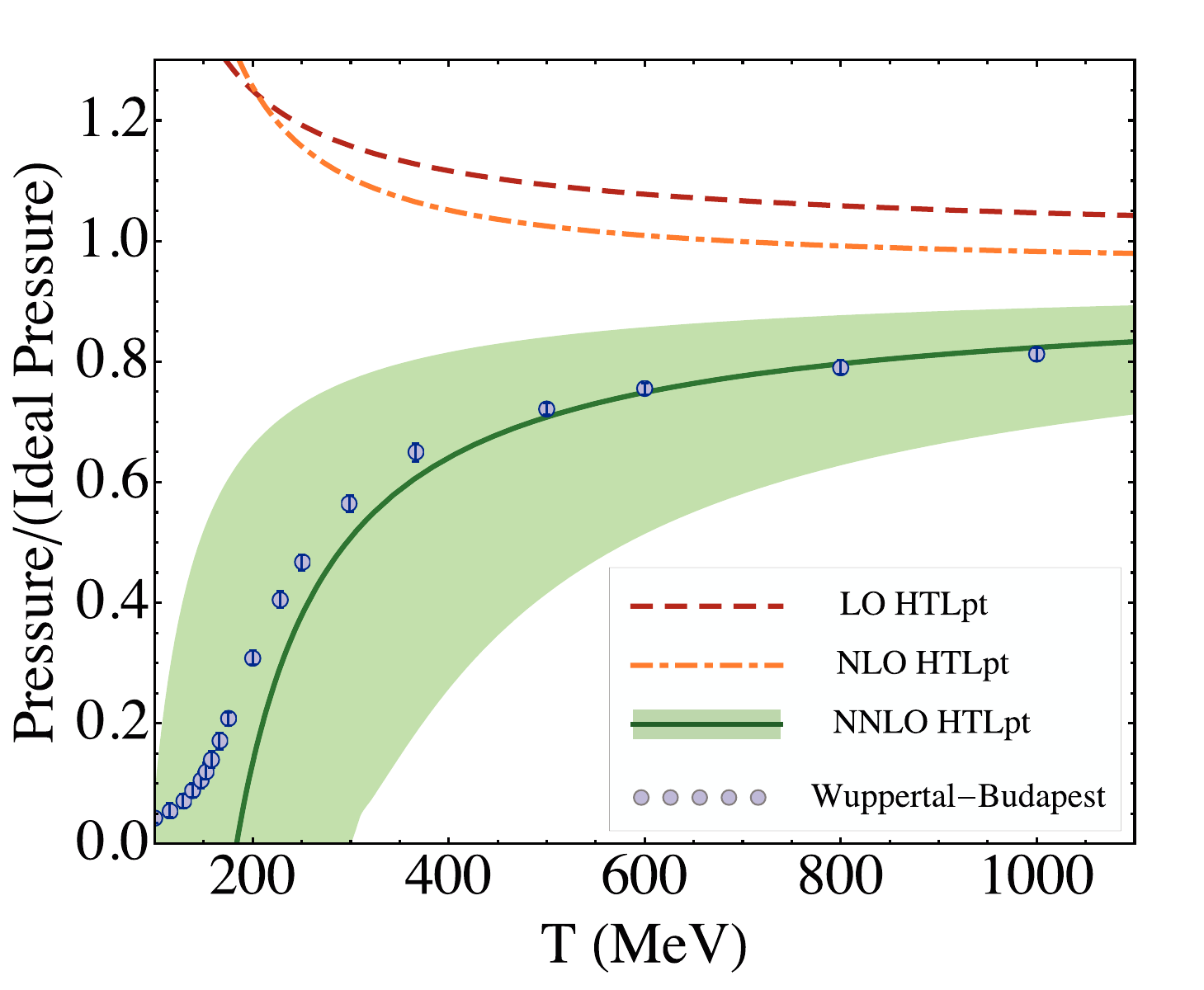}
\caption{\label{fig:htl}
The normalized QCD pressure ($p/T^4$) as a function of temperature (4-flavor case).
The pressure approaches the Stefan-Boltzmann limit at high $T$. The pace at
which this happens should be controlled by perturbation theory, but its orders
show an oscillatory pattern (left panel). Three-loop hard thermal loop
perturbation theory cures this and agrees with lattice data, though with large
theoretical uncertainties \cite{Andersen:2011sf}. (Lattice data taken from
Ref.~\cite{Borsanyi:2016ksw}). }
\end{figure}

Asymptotic freedom relates large temperatures to small values of the running coupling.
However, at phenomenological temperatures of the high temperature phase the couplings are
of $\mathcal{O}(1)$, and it is not natural to expect a well converging perturbative series.
Subsequent orders of the expansion of the thermodynamic potential in the coupling constant
$\alpha_s$ exhibit a wildly oscillating pattern (see Fig.~\ref{fig:htl} left).
We know this to order
$\alpha_s^3\log\alpha_s$ \cite{Kajantie:2002wa} and there is no hope for higher orders
with perturbative methods. Their computation is prevented
by the magnetic mass problem \cite{Linde:1980ts,Gross:1980br}, which limits the
calculable order in this case to $\alpha_s^3$ due to infrared divergences in the Feynman diagrams:
an additional power of the effective infrared cutoff ($\sim g^2T$) contributes to 
the denominator for each new loop order.

Hard thermal loop perturbation theory (HTLpt) has brought much improvement to the 
poor prospects of ordinary perturbation theory. The idea is to consider massive
quasi-particles instead of massless gluons and quarks as the basis of the expansion, 
thus, moving the expansion point. While such shifting by a mass term can be easily 
introduced in a scalar theory,
the gauge invariant equivalent is more elaborate \cite{Braaten:1991gm}, but
still depends on a mass parameter $m_D$. The expansion scheme has been implemented to
two-loop \cite{Andersen:2002ey} and three-loop \cite{Andersen:2011sf} order. The
loop order does not completely determine the procedure, because there are multiple
schemes for selecting the optimal $m_D$. One option is to use the mass parameter of the
effective theory obtained by dimensional reduction \cite{Braaten:1995cm}.

The improved behaviour of the HTLpt is shown in the right panel of Fig.~\ref{fig:htl}.
Note, that these results are phenomenologically relevant only if calculated to three-loop order.
Higher loop results are not expected since HTLpt suffers from the 
magnetic mass problem just as ordinary perturbation theory does \cite{Andersen:2011sf,Ghiglieri:2020dpq}.
Also, the results shown in Fig.~\ref{fig:htl} exhibit a large renormalization scale dependence.
Renormalization is typically performed at the energy scale of $\mu=2\pi T$, but variations by a factor
two on both sides are also shown, to estimate the systematic errors of the truncation.

The formalism was extended to finite quark-chemical potential and quark number
susceptibilities \cite{Blaizot:2001vr,Haque:2013sja,Andersen:2012wr,Haque:2013qta,Andersen:2015eoa}.
Especially the fourth-order baryon and isospin fluctuations exhibit very small scale
dependence. Dedicated lattice studies computed these at high temperatures in the
continuum limit and found good agreement with HTLpt down to 250 MeV temperature
\cite{Bellwied:2015lba,Ding:2015fca}.

\begin{figure}
\centering
\includegraphics[width=0.45\textwidth]{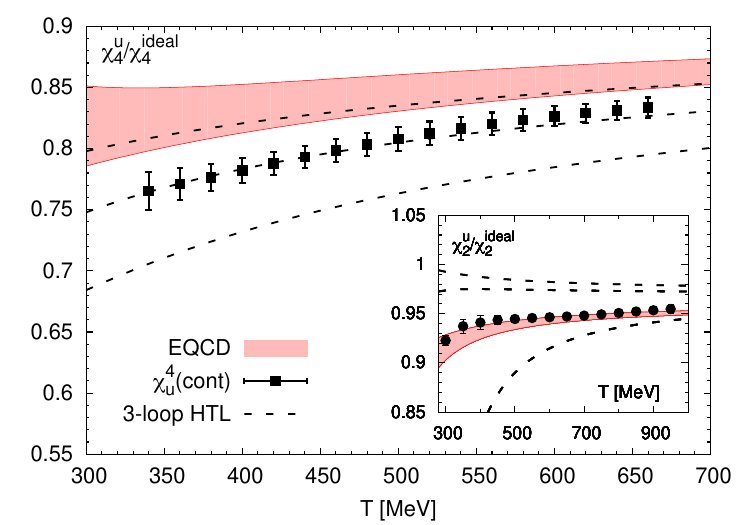}
\caption{\label{fig:chi_eqcd}
Fourth and second order (in the inset) quark number susceptibilities found with two approaches to
weak coupling expansion:  3-loop HTLpt \cite{Haque:2014rua} and EQCD
\cite{Mogliacci:2013mca,Bazavov:2013uja}, compared with lattice data
\cite{Bazavov:2013uja,Ding:2015fca} The uncertainties of EQCD are smaller than
in the case of HTLpt (the latter shown with dashed lines).
}
\end{figure}

The framework of electrostatic QCD (EQCD)
\cite{Appelquist:1981vg,Kajantie:1995dw,Braaten:1995jr} offers a systematic approach to weak
coupling expansion. It is based on the reduction of the four dimensional full theory
to a three dimensional bosonic gauge theory with a scalar field in the adjoint representation.
Even the order $\alpha^3_s$ can be included by simulating the reduced theory on
the lattice. Among other applications, such results are available for the
pressure \cite{Hietanen:2008tv,Mogliacci:2013mca} and  quark number
susceptibilities \cite{Hietanen:2008xb,Rummukainen:2021qyl}. The use of non-zero
quark masses was discussed in Ref.~\cite{Laine:2006cp}. In
Fig.~\ref{fig:chi_eqcd} we show a comparison of results on light quark number
susceptibilities  with lattice data \cite{Bazavov:2013uja,Ding:2015fca}.

\subsubsection{Functional QCD \label{sec:fun}}

Hard thermal loop thermodynamics is just one example for the resummation schemes
that approximate the QCD thermodynamical functional. In this section, we discuss
two generic approximation schemes that run under the common label of functional
methods for QCD. In these schemes, equations for the $n$-point functions of the
underlying quantum field theory are constructed; this often starts with $n=2$,
that is, the propagators. However, these equations depend on higher-point
functions, such that an infinite hierarchy of equations arises, that requires a
truncation for practical use. The various approaches clearly differ in technical
terms, but the resummation they implement is often guided by the same principle.

The functional renormalization group (FRG) offers a self-consistent framework
for solving a quantum field theory. It integrates the path integral starting
with the UV degrees of freedom and sequentially progressing towards the
infrared. This is done by solving a flow equation, e.g. the Wetterich equation
\cite{Wetterich:1992yh} (see also \cite{Ellwanger:1993mw,Morris:1993qb}), 
that describes the response of the
effective action or its derivatives to the inclusion of a new momentum scale as
the infrared cutoff of the theory is reduced towards zero. This principle has
already been used successfully applied in several field theories, including those 
with phase transitions \cite{Dupuis:2020fhh}. For a detailed discussion of the
systematics of this approach in the QCD context, see Ref.~\cite{Ihssen:2024miv}.

An initial success in the application to strong interactions was the computation of the
gauge and ghost propagators in the vacuum of the Yang-Mills theory \cite{Fischer:2008uz}. 
These are, by now, consistent among different functional methods, agree with lattice and satisfy the
criteria for color confinement \cite{Cyrol:2016tym,Huber:2020keu}. Quark confinement and the
deconfinement transition at finite temperature were shown to follow from the
features of these propagators by solving the flow equation for the effective
potential of the temporal gauge background, strongly related to the Polyakov
loop \cite{Braun:2007bx}.

A key feature in the FRG approach to full QCD is dynamical hadronization~\cite{Gies:2001nw,Gies:2002hq}
(see Ref.~\cite{Pawlowski:2005xe} for a detailed discussion).
This means that the flow equations for the gluon, ghost and quark correlators are
extended with fields representing composite (meson) operators (see
Ref.~\cite{Fukushima:2021ctq} for the generalization to other channels). 
As the infrared cut-off is reduced, these degrees of freedom are automatically
activated and account for the dynamical breaking of chiral symmetry and the
emergence of hadrons \cite{Braun:2014ata}. For this to happen, one formulates coupled flow equations
for the propagators and vertices of the fundamental theory (QCD) and 
of the composite operators in some parametrized form \cite{Braun:2014ata,Fu:2019hdw}.
This is unlike effective models (to be discussed in section \ref{sec:eff}) where
the gauge fields are already integrated, leaving a handful of pre-set hadrons as
degrees of freedom. It is also in contrast with the brute force approach of
lattice QCD, that computes the hadronic phase solely in terms of gluonic and
quark variables.

The state-of-the-art formulation of the truncated flow equations is the result
of decades of research \cite{Braun:2006jd,Mitter:2014wpa,Braun:2014ata,Fu:2019hdw,Fukushima:2021ctq}.
The transition temperature in Yang-Mills theories \cite{Braun:2007bx} and in the two-flavor theory \cite{Braun:2009gm}
were important milestones. By today, functional results for the QCD transition
with 2+1 flavors include the chiral condensate as a function of temperature, which 
agrees with lattice results \cite{Fu:2019hdw,Borsanyi:2010bp}. 
The chemical potential dependence can also be studied, from which the curvature of the transition
line emerges ($\kappa=0.0142(2)$ in the $\mu_S=0$ scheme, 
close to lattice results \cite{Cea:2015cya,Ding:2024sux}). The location of the critical end-point was
computed to be ($T_{\rm CEP},\mu_{B,\rm CEP}=(107,635)~\mathrm{MeV}$) \cite{Fu:2019hdw}.
Note, that to demonstrate the critical behaviour itself would require further ingredients.

Another successful guiding principle for resummation schemes is offered by the Dyson-Schwinger
approach \cite{Fischer:2018sdj}. Dyson-Schwinger equations are path integrals of
total derivatives that reveal non-trivial identities between the full,
non-perturbative $n$-point functions. These can be conveniently set up in the Landau gauge.
A notable feature of these equations is that they can be solved in the the infrared limit
and the solution is consistent with certain confinement criteria \cite{Fischer:2008uz}.
The momentum scales achievable by this scaling solution is, however, below what
is relevant for the phase diagram ($\Lambda_{\rm QCD}$). 
Unlike the $n$PI effective actions, a truncation is introduced through the use
of an approximate full vertex \cite{Eichmann:2016yit}. One example is the rainbow
ladder truncation, where the $qq$ interaction is reduced to an effective gluon
exchange. Due to the complexity of the gluon self-interaction, in one possible
approach the gluon propagator is taken from finite-temperature Yang-Mills lattice simulations
\cite{Fischer:2010fx}. The gluon propagators (electric and magnetic) are unquenched
by solving a coupled propagator equation for gluons and quarks. This approximation
misses implicit, in $1/N_c$ subleading quark loop effects in the Yang-Mills self-energies.
Nevertheless, the first order transition of the Yang-Mills theory as well as the existence
of a heavy critical mass (see section \ref{sec:columbia}) including its chemical
potential dependence could be calculated with these equations \cite{Fischer:2014vxa}.
Addressing the chiral limit with the Dyson-Schwinger equations is more challenging.
The rainbow-ladder truncation correctly predicts the second order transition in the chiral
limit, although with mean field exponents \cite{Roberts:2000aa}. It is crucial for the study
of the chiral transition to include the pion and sigma propagators in the quark equations \cite{Fischer:2011pk}.

The Dyson-Schwinger equations were first solved for the 2+1 flavor case with physical quark masses
in Refs.~\cite{Fischer:2012vc,Fischer:2013eca}, and later extended with the charm quarks in Ref.~\cite{Fischer:2014ata}.
While resulting transition temperature agrees with lattice by construction, 
the quantitative agreement of the chiral condensate as a function of temperature
with lattice simulations is non-trivial. The solution features a chiral
crossover and a critical end-point in the phase diagram at $\mu_B\approx
500$~MeV. Later, to refine the quantitative analysis, baryon effects were
considered in the quark-gluon vertex \cite{Eichmann:2015kfa}. In
Ref.~\cite{Gunkel:2021oya} the quark self-energy was extended with mesonic
backcoupling terms, simplifying the vertex at the same time. In this approximation
the critical end-point is predicted to be at  ($T_{\rm CEP},\mu_{B,\rm CEP}=(117,600)~\mathrm{MeV}$)

Apart from the location of the critical point, many observables have been computed using
the Dyson-Schwinger equations. The equation of state was extracted using a 2PI effective action
of the quark sector in the rainbow-ladder truncation \cite{Gao:2015kea}. The fluctuations
of conserved charges were computed in detail in a more elaborate approximation of the quark-gluon
vertex in Ref.~\cite{Isserstedt:2019pgx}.

\begin{figure}
    \centering
    \includegraphics[height=0.30\textwidth]{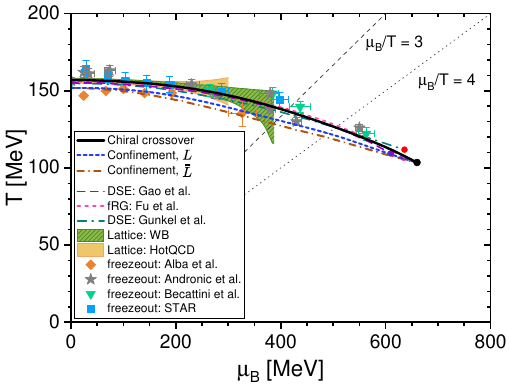}
    \includegraphics[height=0.30\textwidth]{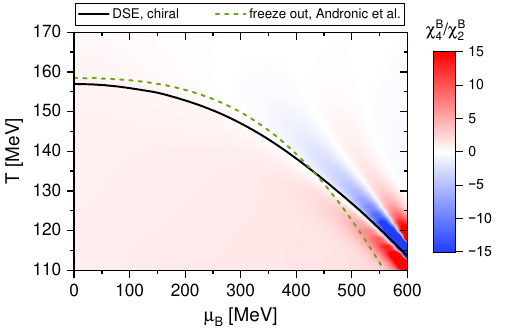}
    \caption{\label{fig:funphase}Left: Summary plot of the QCD phase diagram based on functional approaches.
    The additional data points refer to the empirical chemical freeze-out parameters.
    Right: baryon cumulant ratio $\chi^B_4/\chi^B_2$ on the phase diagram ($\mu_S=0$ setup).
    (Plots taken from Ref.~\cite{Lu:2025cls})
    }
\end{figure}

Thanks to new solutions to the functional renormalization group equations
the vacuum quark-gluon vertex and the unquenched gluon propagator are available for use
in the Dyson-Schwinger equations  \cite{Cyrol:2017ewj}. The latter can be matched with the propagator from
lattice \cite{Zafeiropoulos:2019flq}. This way new forms of the Dyson-Schwinger equations
have been introduced and solved \cite{Gao:2021wun}.
With the new truncation the equation of state \cite{Lu:2023mkn} and fluctuations of net-baryon number
could be computed at finite temperature and density \cite{Lu:2025cls}.
Most importantly a new result on the transition line (with the curvature parameter $\kappa=0.0147(5)$)
and the critical end-point was found in \cite{Gao:2020fbl}
at ($T_{\rm CEP},\mu_{B,\rm CEP}=(109,610)~\mathrm{MeV}$) (see Fig.~\ref{fig:funphase}).
Remarkably, in the works here cited, the crossover temperatures from
the chiral observables and the Polyakov loop are always close and are equal at the critical endpoint.

Functional approaches to QCD have presented a consistent view on the QCD phase diagram.
This is summarized in the left panel of Fig.~\ref{fig:funphase}. All approximations feature
a chiral critical endpoint near ($T,\mu_B)\approx (105-115~\mathrm{MeV},600-650~\mathrm{MeV}$).
The right panel shows the net-baryon cumulant ratio $\chi^B_4/\chi^B_2$ from the most recent DSE computation,
This is the central observable for the experimental search (then replacing baryons by protons) \cite{Lu:2025cls},
it was previously mapped by low energy models embedded in functional QCD~\cite{Fu:2021oaw,Fu:2023lcm}.

\subsection{Models that approximate QCD\label{sec:models}}

\subsubsection{Hadron resonance gas model\label{sec:hrg}}

The better known limits of the thermodynamic structure of QCD matter correspond to 
large or small temperature and chemical potential. 

When both are small, QCD matter is in a hadron gas phase, in which quarks and 
gluons are \textit{confined} inside composite particles with neutral color charge 
called hadrons, like protons and neutrons. At very low temperature and density, this
phase is very dilute, interactions are negligible and the thermodynamics is dominated
by the lightest degrees of freedom, namely pions. When either $T$ or $\mu_B$ is 
increased, heavier states are thermally excited and interactions become more 
prominent. Before the advent of QCD, Hagedorn proposed \cite{Hansen:1990yg} that a
thermally excited strongly interacting medium would be made up of composite objects 
populating an exponentially growing spectrum for increasing masses. While in 
Hagedorn's view this was a continuous ``effective'' spectrum modeling the formation 
of strongly interacting objects, it predicted a maximum possible temperature for 
strongly interacting matter of the order of $T\simeq 200$~MeV, which now can be seen 
as a surprisingly accurate estimate for the transition to the \textit{deconfined} 
phase~\cite{Cabibbo:1975ig}.

In the modern view, this spectrum is replaced by the actually observed hadrons, 
which indeed populate an exponentially growing spectrum. The modern version of 
Hagedorn's model is the hadron resonance gas (HRG) model, which in its simplest 
(ideal) version describes an interacting hadron gas in terms of a non-interacting 
gas of hadrons and all their resonant states. 
The system's partition function factorizes into single-particle contributions as $\ln Z (T,\mu_B,V) = \sum_i \ln Z_i (T,\mu_B,V)$
with:
\begin{equation}
    \ln Z_i (T,\mu_B,V) = \pm \frac{d_i V}{2 \pi^2} \bigintssss_0^\infty \!\!\!\! dp p^2 \ln \left( 1 \pm e^{\frac{\sqrt{p^2+m_i^2}-\mu_i}{T}} \right) \, \, ,
\end{equation}
where the sum runs over all hadron states and $d_i$, $m_i$, $\mu_i$ are the spin 
degeneracy, mass and single-particle chemical potential of species $i$.
Even with its rather crude setup, the HRG model has proven extremely successful 
in reproducing experimental results as well as theoretical predictions in the low 
temperature phase. This is quite impressive, 
considering that the model has virtually no free parameter: given a value for the 
temperature, the baryon chemical potential and the system volume, all thermodynamic 
quantities are determined.

Perhaps the most striking success of the model is its ability to 
\textit{simultaneously} describe the abundances of detected hadrons in heavy ion 
collisions over several orders of magnitude, from pions to omega baryons and even 
light nuclei, with a single temperature, baryon chemical potential and 
volume~\cite{Alba:2014eba,Andronic:2017pug}.  Among the perks of this model, 
especially useful when comparing to experimental measurements, is the possibility 
to take into account genuine experimental effects, such as the contributions from 
decay feed-down and the implementation of acceptance cuts. These effects cannot 
easily be included in other theoretical approaches to the study of QCD 
thermodynamics. Although initially the model has been applied to large colliding 
systems in the grand canonical formulation, it has been later shown that canonical 
corrections can -- and should -- be included for smaller colliding systems, such 
as heavy ion collisions at lower energy and proton-proton or proton-nucleus 
collisions at high energies~\cite{Beutler:2010cha,Vovchenko:2019kes}. With these 
corrections, a good description is achieved for a huge variety of colliding setups.
Furthermore, the comparison with other theoretical calculations is another great 
success of the HRG model, which can precisely reproduce most thermodynamics 
results from lattice 
QCD~\cite{Borsanyi:2011sw,Bazavov:2012jq,HotQCD:2014kol,Bellwied:2015lba,Borsanyi:2021hbk}. 

Although the HRG has proven extremely successful even in its ideal form, a more 
realistic treatment of interactions is necessary to describe observables
which are very sensitive to the details of the hadron spectrum or the strength of 
interactions in the 
medium~\cite{Vovchenko:2017xad,Vovchenko:2020lju,Bollweg:2022fqq,Bollweg:2022rps}, 
as for example high order fluctuations of conserved 
charges~\cite{Huovinen:2017ogf}.
In the S-matrix approach to relativistic statistical mechanics based on the virial 
expansion, first introduced by Dashen, Ma and Bernstein~\cite{Dashen:1969ep}, 
information on the phase shifts due to hadron interactions can be used to 
construct the second virial coefficient. The ideal HRG model thus corresponds to 
the leading order of the virial expansion, in which it also includes contributions 
by all known resonant states. However, all interactions not 
related to resonance formation are neglected. It was later shown that a large 
cancellation appears in the pressure between non resonant attractive and repulsive
contributions~\cite{Venugopalan:1992hy}. The virial expansion in the S-matrix 
formalism was employed e.g. to study the effect of interactions on baryon 
fluctuations~\cite{Huovinen:2017ogf}, and to investigate the existence 
of unknown strange baryon states~\cite{Fernandez-Ramirez:2018vzu}. 
In Ref.~\cite{Andronic:2018qqt} it was argued that the quality of thermal fits to 
particle yields can be dramatically improved by including the phase shift 
information on pion-nucleon scattering. The main limitation of the S-matrix  
approach is the limited available information on hadronic scattering phase shifts, 
which have to be modeled when not experimentally available.

For this reason, a number of modifications have been considered to further improve 
the accuracy of HRG model predictions. The hadron spectrum employed in the model 
can be adjusted by explicitly including additional 
states~\cite{Bazavov:2014xya,Alba:2017mqu,Alba:2020jir,Bollweg:2021vqf}, for 
example those predicted by theoretical 
models~\cite{Capstick:1986bm,Ebert:2009ub,Ferraris:1995ui}. Repulsive mean field 
interactions were included in Ref.~\cite{Huovinen:2017ogf}, showing that the effect
on baryon fluctuations is the same as in the relativistic virial 
expansion, and later employed to study the equation of state~\cite{Pal:2021qav}. 
The same setup was applied in Ref.~\cite{Biswas:2024xxh} -- also including 
additional hadronic states -- to study the equation of state and the chiral 
transition line at finite baryon chemical potential, finding good agreement with 
lattice results.
In a similar fashion, repulsive excluded-volume and attractive van der Waals-like 
interactions have also been extensively studied in recent years (see 
Ref.~\cite{Vovchenko:2020lju} for a review), which also provide an improved 
description of lattice QCD fluctuations data over the ideal formulation. An 
advantage of the van der Waals approach is the natural emergence of the liquid-gas
transition and associated critical point, which can be reproduced by tuning the 
values of the repulsive and attractive terms~\cite{Vovchenko:2017xad}.

\subsubsection{Effective models \label{sec:eff}}

Due to the difficulty in solving the full theory of strong interactions, one has often resorted
to low-energy effective theories to predict the features of the QCD phase diagram. 
Such models are sharing the symmetries of the full theory, and go beyond the claim
of being in the universality class of QCD near some critical point. Effective models
identify the relevant degrees of freedom and formulate a field theory in terms of these.
The resulting field theory is not always renormalizable, a cut-off scale is needed,
above which the model is not defined. Typical cut-off scales are below 1~GeV.

Effective models can be classified by their principal ingredients. One of the earliest
is the linear sigma model where the pions and an isospin singlet sigma are combined to model
the chiral symmetry breaking as in a scalar O(4) model \cite{Gell-Mann:1960mvl}.
Integrating out the sigma degree of freedom defines the non-linear sigma model.
Chiral perturbation theory, too, expresses QCD in terms of its lightest degree of freedom, 
the pions, or light pseudoscalar mesons, in general~\cite{Meissner:2024ona}. This leading Lagrangian coincides with
the non-linear sigma model \cite{Gasser:1983yg}. A chiral transition was
predicted to happen at 190~MeV in a low temperature expansion of the two-flavor
theory to three-loop order \cite{Gerber:1988tt}. In the chiral limit the transition temperature
would be 20~MeV lower. 

The Nambu-Jona--Lasino model is also an effective theory for chiral symmetry breaking
\cite{Nambu:1961tp,Nambu:1961fr}. The principal degrees of freedom are fermions that form,
in analogy to superconducting, quark-antiquark pairs as composite particles, the light mesons.
(see Ref.~\cite{Klevansky:1992qe} for a historic review). Effective models are often
solved in the mean field approximation, where the operator valued equations of motions are
linearized around a nontrivial (mean field) value of composite operators, facilitating an
order parameter for chiral symmetry breaking. A shortcoming of the NJL model is the lack
of confinement: in the chirally broken phase the fermions describe the constituent quarks.
In this framework the QCD phase diagram was sketched in the $T-\mu_B$ plane for
the two flavor theory \cite{Asakawa:1989bq}. This diagram featured a first order
transition at finite $\mu_B$ and $T=0$ and a smooth transition at finite $T$ and
$\mu_B=0$, indicating the necessity of an end-point on the $T-\mu_B$ plane. The
position of the end-point can be tuned with the parameters, and it could be
removed entirely if we consider a sufficient repulsive vector interaction among
the possible four-fermion terms \cite{Buballa:2003qv}.

If we explicitly include the pion and sigma fields together with the quark fields
in the effective Lagrangian, with the gluon fields already integrated, we get the quark meson model.
Here the phase structure has been computed in Ref.~\cite{Schaefer:2004en} using a renormalization
group method. They find a second order transition in the chiral limit that smoothens into a crossover
for physical pion mass. This crossover also turns first order in an end-point, however, the structure
at low temperature, high density is more complicated.

Since neither the Nambu-Jona-Lasino, nor the quark meson model can account for the deconfinement,
newer models include the order parameter of the deconfinement, the Polyakov loop as a dynamical
variable in the effective Lagrangian. At the heart of these models there is a potential for
the Polyakov loop with a temperature dependent parametrization such that the
center symmetry is respected and the lattice results for the quarkless gauge
theory are reproduced. The Polyakov loop is then brought into direct connection with the temporal
gauge field, which is fed into the Dirac operator of the quarks degrees of freedom.
These extended models, dubbed PNJL and PQM, respectively,
have been used extensively in the literature to explore the phase diagram (see \cite{Fukushima:2008wg,Ratti:2006wg,Ratti:2006gh} for the PNJL
model and \cite{Schaefer:2007pw,Haas:2013qwp,Skokov:2012kw} for the PQM).

\begin{figure}
    \centering
    \includegraphics[width=2.8in]{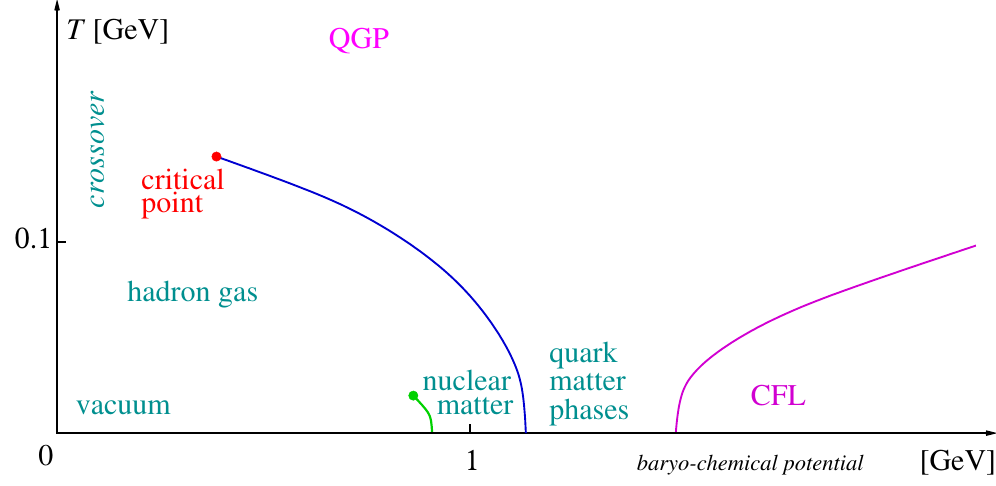}
    \caption{The standard view on the QCD phase diagram as suggested by several effective theories
    \cite{Stephanov:2004wx}.
    }\label{fig:pdstephanov}
\end{figure}

A common feature in these results is the existence of a critical end-point that separates
the crossover at high temperature from the first order line at high density.
Although the position of the end-point varies, one can conclude with the semi-quantitative
picture of Fig.~\ref{fig:pdstephanov}. The most distinctive feature is the blue transition
line that separates nuclear matter from quark matter. The labels suggest deconfinement,
but the definition is based on chiral symmetry breaking. The phase in the right
end of the diagaram (Color Flavor Locking) is a superconducting phase \cite{Alford:1999pb,Alford:2007xm},
see Section \ref{sec:cfl}.

The predictive power of such effective models has been greatly enhanced by their
embedding into the functional renormalization group framework. The two-flavor PQM
model was revisited in this sense in Ref.~\cite{Herbst:2013ail} and found
a chiral crossover and a critical end-point near the expected position of the liquid-gas
transition.  Several improvements have been introduced to the truncation of the
effective action: the inclusion of quark backreaction to the Polyakov loop potential \cite{Haas:2013qwp}, 
the use of the `t Hooft determinant to consider the $U(1)_A$ breaking \cite{Herbst:2013ufa}. 
Thanks to the fluctuations included by the FRG formalism a smooth transition is obtained
in accordance with lattice simulations. The slopes of the remnant order parameters were
larger in the mean field approach. Using a simple approximation for the strange degree
of freedom this effective model was used to derive very high order fluctuations, showing promising
smoking-gun signal for the baryon cumulant ratio $\chi^B_4/\chi^B_2$ on the phase diagram
in a large region around the critical end-point \cite{Fu:2021oaw}. 
These results were further refined by directly evolving the RG flow of quark-meson scattering processes
with a new result for the critical endpoint at $T_{CEP}=98$~MeV and $\mu_B=643$~MeV \cite{Fu:2023lcm},
which is fairly close to what was found in full functional QCD \cite{Fu:2019hdw} of Section \ref{sec:fun}.

A different class of models of QCD are those based on the gauge-gravity duality, 
in particular the Einstein-Maxwell-Dilaton (EMD) type 
models~\cite{Gubser:2008yx,Gubser:2008ny,Rougemont:2023gfz}. In these models the QCD 
phase diagram is mapped onto asymptotically Anti-de Sitter (AdS) charged black hole 
geometries in five dimensions. They are constructed with the explicit intent to 
reproduce the observed features of the QCD equation of state at zero chemical 
potential. The minimal ingredients they contain are the bulk metric field, a Maxwell 
field whose value at the boundary sets the chemical potential, and a dilaton field 
to break conformal invariance. A number of free parameters are fixed by requiring
agreement with lattice QCD results at zero chemical potential. These models 
inherently include the almost perfect fluidity of the QGP, and in all different 
realizations predict the existence of a critical 
point~\cite{DeWolfe:2010he,DeWolfe:2011ts}. They have been employed to 
determine transport coefficients of the 
QGP~\cite{Gubser:2008sz,Finazzo:2014cna,Rougemont:2015ona,Grefa:2022sav}, the 
location of the critical point~\cite{Critelli:2017oub,Hippert:2023bel} and 
the equation of state of QCD at finite density~\cite{Grefa:2021qvt}, showing 
agreement with lattice simulations where results from the latter are available.

\section{Theory of the phase diagram}\label{sec:theory} 

\subsection{Quark masses and symmetries: the Columbia plot\label{sec:columbia}}

\begin{figure*}[t]
\begin{center}
\includegraphics[width=0.32\textwidth]{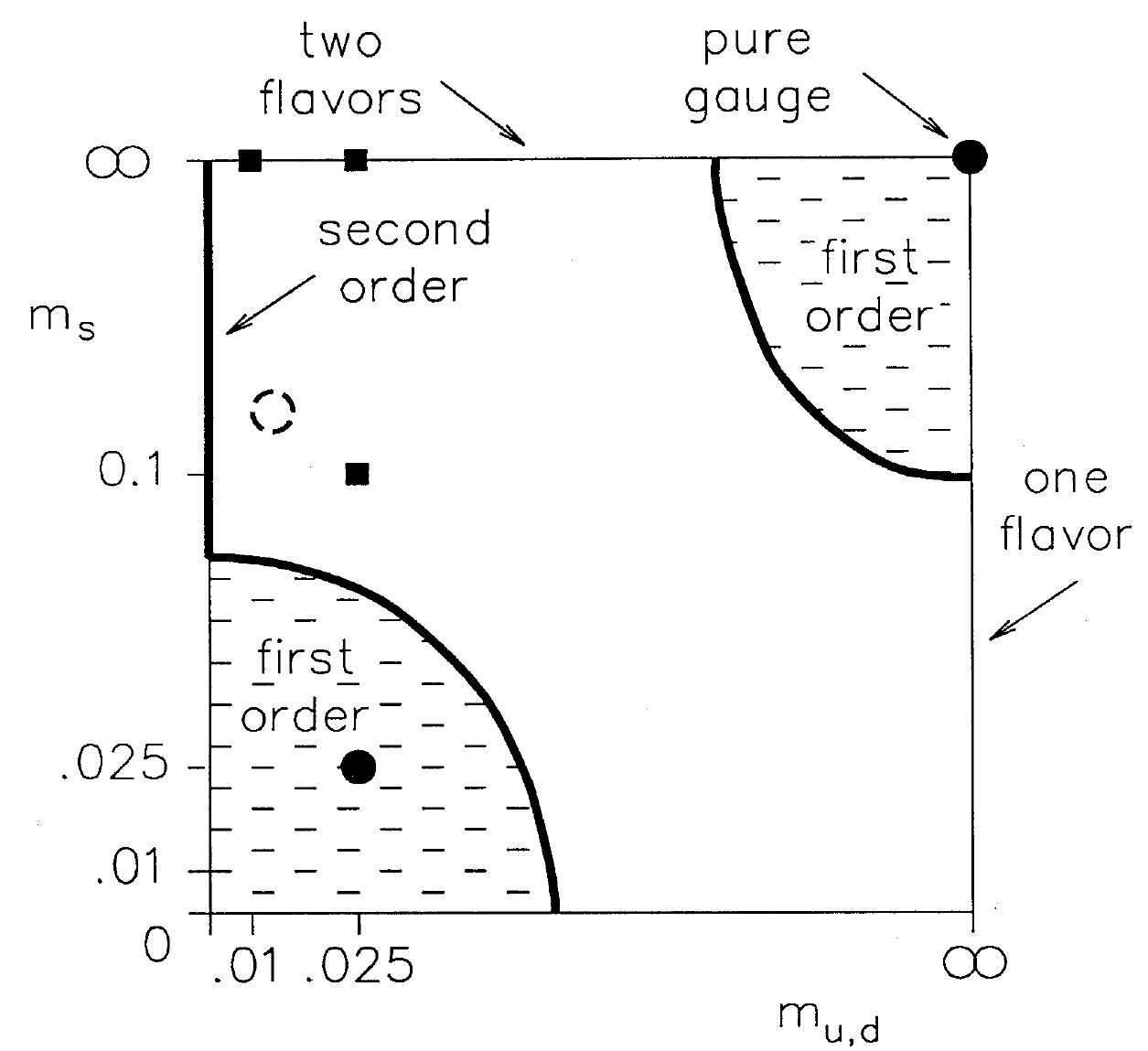}
\raisebox{3mm}{\includegraphics[width=0.28\textwidth]{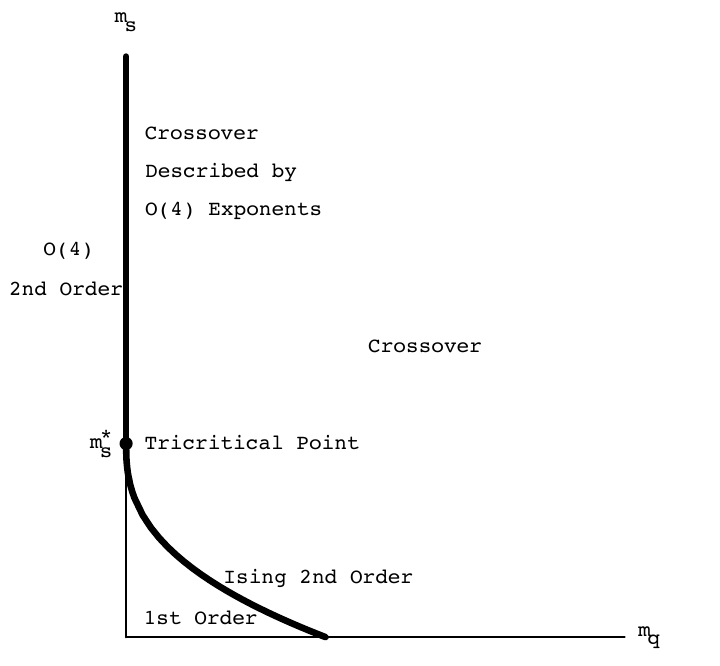}}
\raisebox{3mm}{
\includegraphics[width=0.28\textwidth]{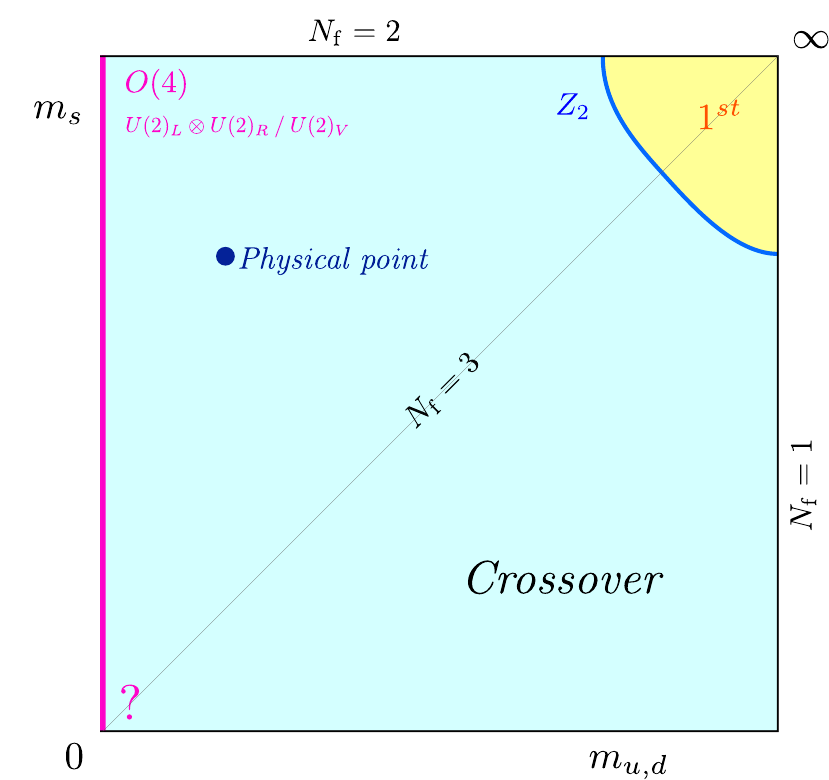}
}
\caption{\label{fig:columbia}
The Columbia plot in its original version (left) with a handful
of simulation points (black dots) \cite{Brown:1990ev}, in
agreement with the theoretical arguments of \cite{Pisarski:1983ms}.
The middle panel shows the detailed theoretical expectations for the lower left corner
\cite{Rajagopal:1995bc}.
New lattice results challenge the original view and propose a new version (right
)
\cite{Cuteri:2021ikv}. Recent effective model studies find room in the parameter
space for this scenario, too (see e.g.~\cite{Giacosa:2024orp}).
}
\end{center}
\end{figure*}

In the following discussion, we review the phase diagram in a specific 
representation: instead of temperature and chemical potential, the axes indicate
the values of the strange and light quark masses, thus positioning QCD in theory 
space.
This is the famous Columbia plot \cite{Brown:1990ev}, see
Fig.~\ref{fig:columbia}. The physics of the transition can be vastly different in different
corners of this diagram: in the top right corner all quark masses are infinite, and purely the dynamics
of gluons determine how confinement stops as temperature increases. However, in the lower left corner, light quarks take the dominant role and restore chiral symmetry at high temperature.
QCD with physical quark masses resides in the bulk of the diagram where the transition is a
crossover driven by an interplay of deconfinement and chiral dynamics.
In the Columbia plot the order of the transition is marked for each possible
pair of light and strange masses. The left panel shows the first attempt to
explore this vast range of parameters~\cite{Brown:1990ev}. Even after 35
years there are still unsettled issues, mostly centered around its lower left corner.

\subsubsection{The deconfinement transition}

A key feature of QCD is the running of the coupling constant, that is,
the effective coupling depends on the energy scale. A generic feature of 
non-abelian quantum field theories (below a threshold in the number of flavours)
is asymptotic freedom, which means that the coupling goes to zero
in the ultraviolet. Thus, at extreme temperatures, the theory is weakly coupled 
where the fundamental degrees of freedom (quarks and gluons) are also the relevant ones.
By their nature, these are color-nonsinglets.
This is in contrast to color confinement at low temperatures with color singlet
effective degrees of freedom. Whether or not the two temperature regions
are separated by a phase transition depends on the presence of matter fields. 

Specifically for the quark-less case (Yang-Mills theory), the path integral
has an exact symmetry related to the center of the gauge group (Z$(N_c)$). This means,
that for every gauge configuration, there are two others with the same weight.
The expectation value of the temporal Wilson line (the Polyakov loop) \cite{McLerran:1981pb}
\begin{equation}
    L(\vec x) = \frac{1}{N_c} \mathrm{Tr}\,\mathcal{P} \exp\left[\int_0^{1/T} A_0(t,\vec x)dt \right]
\end{equation}
transforms non-trivially, a non-vanishing expectation value signals the spontaneous breaking
of the center symmetry. ($N_c$ stands for the number of colors and $\mathcal{P}$ prescribes path ordering).
At the same time, the expectation value of the Polyakov loop is related to the free energy of a
static color source $F_q$ \cite{McLerran:1981pb}, with $\langle L\rangle=e^{-F_q/T}$. 
A zero value can be identified with a divergent
free energy associated with a color-nonsinglet static (infinitely heavy) quark.
This idea connects the center symmetry breaking with quark deconfinement.
The perturbative vacuum is clearly not center symmetric. A highly non-trivial
feature of SU($N_c$) theories is the restoration of this symmetry at low
temperature. The determination of the temperature dependence of $\langle L\rangle$, and thus the numerical observation of
the deconfinement transition, was the first successful application of lattice simulations to QCD thermodynamics
\cite{McLerran:1980pk,Kuti:1980gh}. The order of the deconfinement transition depends
on the number of colors. Monte Carlo simulations with finite size scaling have established that
the transition is second order for $N_c=2$ \cite{Engels:1989fz}
and first order for $N_c=3$ \cite{Brown:1988qe,Fukugita:1989yb}, which persists for
larger $N_c$ \cite{Lucini:2002ku}. The first order nature was predicted by
Svetitsky and Yaffe based on generic arguments on the effective potential of the Polyakov loop \cite{Yaffe:1982qf}. In fact, the cubic term in $L$, as required by the Z(3) symmetry,
is not compatible with a continuous transition and $Z(3)$ is not among the known universality classes.
Detailed studies of the latent heat \cite{Lucini:2005vg,Borsanyi:2022xml} have showed that among the SU($N$) theories the
transition in SU(3) is weak, with a latent heat 3-4 times smaller than in
the large-$N$ limit $L_h/T_c^4\approx  0.42(7) N_c^2$ \cite{Lucini:2005vg}. It was 
recently determined to be $L_h/T_c^4=1.025(21)_{\rm (stat)} (27)_{\rm
(sys)}$ \cite{Borsanyi:2022xml} (see also \cite{Shirogane:2020muc,Giusti:2025fxu}). 
The pressure function $p/T^4$ versus $T$ has been a subject of several large-scale lattice studies
up to perturbative temperatures \cite{Boyd:1996bx,Borsanyi:2012ve,Caselle:2018kap}.

The temperature of this first-order transition is sometimes quoted
in MeV units. Since lattice simulations can only produce dimensionless
ratios or products, this assumes the universality of a reference scale, that can be also defined
in full QCD with 2+1+1 flavors, where it is then given in MeV.
By choosing the reference scale
(spatial string tension, Sommer scale ($r_0$), or the flow-based scales $w_0$ and $\sqrt{t_0}$)
one gets 275~MeV, 322~MeV, 344~MeV and 285~MeV, respectively. This ambiguity
is inevitable when the running of the coupling differs between the simulated theory and Nature.

Matter fields break the center symmetry explicitly. To be precise, a center 
transformation can be re-interpreted as a shift in the quark chemical potential by 
the imaginary value $\mp i 2\pi T /N_c$. Due to the explicit breaking of this 
symmetry, the Polyakov loop
ceases to be an exact order parameter for the transition. In the
language of an effective potential this means a linear tilt in the effective
potential towards positive real values to leading order in the hopping parameter
expansion~\cite{Kiyohara:2021smr}. Allowing for a continuous dependence of the
latent heat as a function of control parameters, such as the inverse mass, one
can expect a finite first-order region in the top-right corner of the Columbia
plot (Fig.~\ref{fig:columbia}). This region is delimited by a line of critical
masses stretching between the two-flavor (top) and single-flavor (right) edges
of the diagram. Along this critical line the effective potential of the real
part of the Polyakov loop is akin to that of the Ising model, where the quark
masses provide the fine-tuning to make the positive and negative valued minima
of equal statistical weight.  Without quarks the negative minimum has double
weight at high temperatures, reflecting an intact Z(3) symmetry. 
In this context QCD can be mapped to a three state Potts model in an external field
that also features an Ising end-point \cite{Karsch:2001ya}.
Lattice studies have confirmed this, by finding (for two flavors) a critical
point in the Ising class at the pseudo-scalar (``pion'') mass $m_\pi/T_c\approx
18.1$~\cite{Saito:2011fs,Cuteri:2020yke}, which is compatible with earlier
estimates based on random matrix theory \cite{Kashiwa:2012wa}.

The meaning of the Polyakov loop, which is the order parameter in the heavy quark
limit, is less clear in the case of physical quark masses. Its renormalized
expectation value is related to the free energy ($F_Q)$ of a static quark as 
$L_{\rm ren} = \exp( -F_Q/T)$, which in turn, admits the definition of the
entropy of the static quark as $S_Q = -\partial F_Q/\partial T$. While $F_Q(T)$
is very smooth and monotonic across the cross-over, and its value is subject to
an arbitrary constant shift, $S_Q(T)$ is well defined and features a peak at $T_c$~\cite{Bazavov:2016uvm}.
Actually, Ref.~\cite{Bazavov:2016uvm} observes, that the peak position is consistent with
the chiral susceptibility's peak for various quark masses, while the more obvious choices
(e.g. the susceptibility of the Polyakov loop) did not give a conclusive result.
It was pointed out in Ref.~\cite{Clarke:2020htu} that $F_Q$ behaves as an energy-like observable 
in the vicinity of the chiral phase transition temperature, and for small quark masses
the corresponding O(4) scaling applies to its derivatives.

\subsubsection{QCD in the chiral limit}

As we have already discussed in Section~\ref{sec:lat}, QCD with physical quark masses
exhibits a crossover transition. It was not a simple achievement to establish this
fact from lattice simulations, but it also turns out to be a robust statement, as the transition
remains a crossover for a very broad range of parameters, as shown in Fig.~\ref{fig:columbia}.

While the upper right corner of the Columbia plot refers to the Yang-Mills theory, the
lower left corner corresponds to the chiral limit with three flavors. However,
let us consider the two-flavor chiral limit (left edge) first. As long as the third
(strange) quark is much heavier than the light flavors it will have no impact
on symmetry considerations. In the low
temperature phase the massless QCD Lagrangian has a
$U(N_f)\times U(N_f)$ symmetry, where the left and right-handed
(or, equivalently, the axial and vector combinations) can be rotated in flavor
space independently.  The $U(1)_V$ symmetry (charge conservation) is a true
symmetry. We can also regard the isospin symmetry as unbroken, as long
as the light flavors are degenerate ($m_u = m_d$) and QED effects are neglected.
Note that also a spontaneous symmetry breaking of the vector symmetries is ruled
out~\cite{Vafa:1983tf}. On the other hand, the chiral symmetry $SU(2)_A$ is
spontaneously broken, with the pseudo-scalar pions playing the role of Goldstone
bosons, as they are the lightest hadrons even in the presence of an explicit breaking
due to finite quark masses \cite{Weingarten:1983uj}. The last element of the classical
symmetry group is the flavor singlet chiral symmetry $U(1)_A$. This is broken by
a quantum anomaly, resulting in a large mass for the $\eta'$ meson. The mechanism
of this anomaly is the existence of topologically non-trivial configurations in
the path integral. The localized finite-action solutions of the Euclidean
gauge theory with non-trivial topology are called instantons -- the reader is
referred to the in-depth review on the subject in this volume or in Ref.~\cite{Shuryak:2021vnj}.

\begin{figure}
    \centering
    \includegraphics[width=0.45\textwidth]{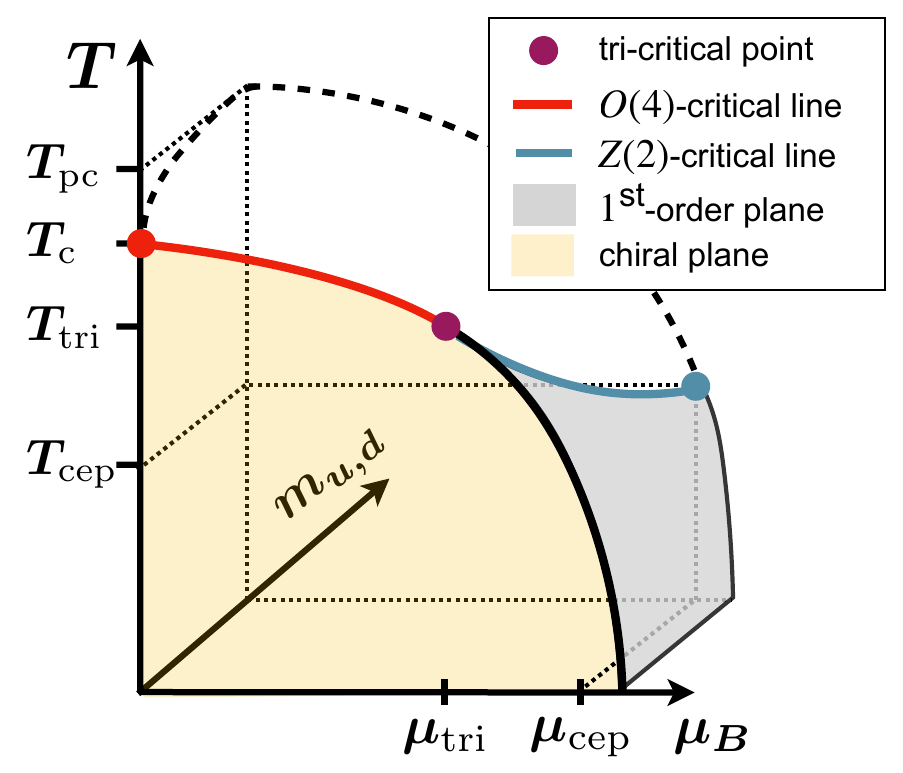}
    \caption{\label{fig:3dpdiag}
    The phase diagram extended to three dimensions: temperature ($T$), light quark mass (normalized to the strange mass $h=m_l/m_s$) and baryo-chemical potential ($\mu_B$). The front plane at $h=0$ shows the 2nd order chiral transition
    line in red, starting at the $O(4)$ critical temperature $T_c$. Two dashed lines indicate the chiral crossover surface: i) the quark mass is varied keeping $\mu_B=0$,
    ii) the chemical potential is varied with physical quark masses, starting at the pseudocritical temperature $T_{\rm pc}$ at $\mu_B=0$ and ending at the chiral critical endpoint at $T_{\rm cep}$. The universality class of the second order line where these two examples of a crossover end is different, but can be connected in a tri-critical point at $T_{\rm tri}$.
    (Plot from Ref.~\cite{Ding:2024sux})
    }
\end{figure}

The fate of the flavor-singlet and non-singlet chiral symmetries determines the order
of the transition in the QCD phase diagram. At perturbative temperatures the $SU(N_f)_A$ symmetry
is completely restored if the explicit breaking is strictly zero. This requires
at least a continuous transition between the chirally broken and unbroken phases at $T_\chi$.
If the $U(1)_A$ symmetry is also restored at the chiral transition $T_\chi$,
such transition may be either first or second order \cite{Pelissetto:2013hqa,Azcoiti:2021gst}.
If, however, the breaking of the $U(1)_A$ axial symmetry persists for $T>T_\chi$, the transition
is of second order in the $O(4)$ universality class. Recent lattice simulations agree
with the O(4) hypothesis and determine $T_\chi$ in the chiral limit to
be $T_\chi\approx 132$~MeV \cite{HotQCD:2019xnw,Kotov:2021rah}.
Truncated functional approximations to QCD have also determined the critical
temperature in the chiral limit, with results varying between 141 and 147 
MeV~\cite{Braun:2020ada,Gao:2021vsf,Bernhardt:2023hpr}, however, the critical region
was found to be much smaller than the pion masses used by any lattice group \cite{Braun:2023qak}.
A likely scenario for the embedding of this critical point in the broader picture of 
the phase diagram is shown in Fig.~\ref{fig:3dpdiag}.

\subsubsection{Instantons and the fate of the $U(1)_A$ symmetry}

Several lattice groups have studied the effective restoration of the $U(1)_A$
symmetry. The qualifier ``effective'' refers to the fact, that anomalous
fluctuations break the $U(1)_A$ at any temperature. 
For physical quark masses the instanton density, observed through the topological susceptibility,
does not completely vanish at perturbative temperatures, but drops with
temperature according to a power law with an exponent that depends on the number of 
flavors as given by the theory of the dilute instanton gas~\cite{Gross:1980br},
and was also computed on the lattice \cite{Petreczky:2016vrs,Borsanyi:2016ksw}.

The question of effective restoration of the $U(1)_A$ symmetry is usually
posed in the massless limit. This is understood as a limit of QCD theories with light quarks,
where first the infinite volume limit is taken, then the chiral limit ($m_u=m_d\to0$).
Performing the limits in reverse order the zero quark masses would suppress all instantons.
The question to be answered is whether the $U(1)_A$ (chiral singlet) and $SU(2)_A$ (chiral nonsinglet)
symmetries are restored at the same temperature in this particular limit, or not.

If present, the effective restoration of the $U(1)_A$ symmetry can be observed in the mesonic correlators
as the degeneracy of the pion and the scalar iso-triplet $a_0$ meson  \cite{Shuryak:1993ee}.
To find a direct evidence for the restoration from lattice QCD is extremely difficult, since
this requires simulations at very small quark masses and large volumes, while the effect of
$U_A(1)$ breaking may be very small (but non-zero) in the chiral limit.
Several lattice groups have made significant progress, but there is no conclusive answer found yet
\cite{Chiu:2011dz,Cossu:2013uua,Brandt:2016daq,Tomiya:2016jwr,Aoki:2020noz,Ding:2020xlj}. 
Initially, works with chiral fermions saw a restoration, and those working with an improved staggered
discretization did not. It is also not clear how an effective restoration can be established
at a given temperature based on finite-error data points.

The quantitative study of the anomalous breaking of $U(1)_A$ requires good
control on the low lying eigenvalues of the Dirac operator, since these
determine the statistical weight of configurations with instantons.
Staggered fermions are known to be prone to severe discretization effects at the low end of
the Dirac spectrum. A recent work with domain wall fermions at finite lattice
spacing~\cite{Gavai:2024mcj} has presented new results that are likely to be
unaffected from such errors, yet in agreement with the staggered results
\cite{Ding:2020xlj}, supporting on the non-restoration scenario. Let us remark,
though, that the actual temperature where the $U(1)_A$ symmetry is approximately
restored is very similar in the pro-restoration~\cite{Tomiya:2016jwr} and
non-restoration ~\cite{Ding:2020xlj} papers, but the pro-restoration papers
consistently use a 2 flavor setup (top left corner) while the non-restoration
papers work with physical strange mass (middle left edge). In the latter case
the chiral transition temperature can be about 30~MeV lower. 

In view of the available evidence a picture with interacting quasi-instantons
emerges~\cite{Kanazawa:2014cua}. In the deconfined phase the field configurations
in the pure gluonic theory are characterized with a non-interacting 
gas~\cite{Vig:2021oyt} of topological defects, instantons and anti-instantons. Their
density approximately follows a power law in the temperature~\cite{Gross:1980br}. 
Instantons (anti-instantons) contribute with a positive
(negative) integer to the topological charge of the configuration. By means of
the index theorem~\cite{Atiyah:1968mp} the number of zero eigenvalues of the
Dirac operator $D$ is given by the difference between the number of instantons and
anti-instantons, i.e. the topological charge. In addition to the exact zero modes, a
peak appears in the infrared end of the Dirac spectral 
density~\cite{Edwards:1999zm}. This peak can be explained by the non-interacting
instanton gas picture~\cite{Dick:2015twa,Azcoiti:2023xvu,Kovacs:2023vzi}. 

In dynamical QCD the statistical weight of a gauge configuration is given by
the fermion determinant $\det(D+m)^{N_f}$, where $m$ is the bare quark mass and $N_f$
is the number of light flavors. In a sufficiently large volume the zero mode
zone of the spectral density is dominated by eigenvalues below the quark mass,
numbering $\sim V\chi_0$ where $V$ is the four-volume and $\chi_0$ is the topological
susceptibility of the quark-less theory. Configurations with $n_i$ instantons and $n_a$
anti-instantons are, thus, suppressed by $n_i+n_a$ factors of the bare quark
mass. If this mass is small (near the chiral limit), configurations with many instantons
are suppressed by a higher factor, and those configurations will be favored where
instantons are sparse \cite{Gross:1980br}.

The Banks-Casher relation is known to connect the spectral density $\rho(\lambda)$ 
at zero eigenvalue with the value of the chiral condensate \cite{Banks:1979yr}. Indeed, in the low temperature
phase, chiral symmetry breaking is manifest through a finite limit of $\rho(\lambda)$
at $\lambda\to0$. Above the actual chiral transition temperature we are now confronted
with a power law with negative exponent, thus, a divergent limit at
$\lambda\to0$. It is tempting to think, that what we are facing is a new
''anomalous'' phase \cite{Alexandru:2019gdm}. A closer look, however, reveals
that a re-derivation of the Banks-Casher relation in the presence of an instanton-gas yields
a chiral condensate proportional to $\sim m^{N_f-1}$ \cite{Kovacs:2023vzi},
admitting a zero value in the chiral limit, corresponding to the
complete restoration of the chiral symmetry. At the same time, specific mesonic correlators,
often used to characterize the $U(1)_A$ breaking, are predicted to have a finite value for
$N_f=2$ \cite{Kovacs:2023vzi}.

It is, thus, likely, that the $U(1)_A$ restoration happens gradually as temperature is
increased. Its significance may depend on innocuous parameters, like the strange quark mass
that sets the temperature of interest, that, in turn, controls the instanton density.
In this scenario, QCD transition belongs to O(4) universality class for two massless flavors.

\subsubsection{Lower left corner of the Columbia plot}

The order of the transition with three light flavors is also much debated.
The standard picture in the three flavor chiral limit of QCD is shown in the middle
panel of Fig.~\ref{fig:columbia}. The detailed discussion in Ref.~\cite{Rajagopal:1995bc} 
is partly based on the seminal work of Pisarski and Wilczek \cite{Pisarski:1983ms}.
The latter solves an effective theory in $\epsilon$-expansion, that is, working in
$d=4-\epsilon$ dimensions, expanding in $\epsilon$ and setting $\epsilon=1$. Modeling
the axial anomaly and the chiral symmetry breaking and their relation to actual
hadron masses a first order transition was suggested for the chiral limit. The first
order nature persists for small perturbations in the mass. The middle panel of
Fig.~\ref{fig:columbia} shows a second order boundary that ends in a tri-critical point
where the boundary line touches the chiral edge of the Columbia plot. 

The standard picture with the first order transition was challenged by lattice studies
that did not find the predicted first order region \cite{Cuteri:2021ikv},
suggesting the right panel of Fig.~\ref{fig:columbia}.
This results, as well as earlier unsuccessful searches for the first order region 
using improved quarks \cite{Ding:2011du,Varnhorst:2015lea} are likely to be 
influenced by severe cut-off effects. The difficulty that lattice QCD faces
in the lower left corner of the Columbia plot is similar to those hindering the
conclusive study of the effective $U(1)_A$ restoration.

The doubt in the existence of the first order corner in the Columbia plot sparked a new
wave of interest in effective models and advanced solution techniques. For example,
in the case of restored $U(1)_A$ symmetry the existence of an infrared fixed
point is suggested, admitting the possibility of a 2nd order transition
\cite{Fejos:2022mso}. In the non-restoration scenario, however, the first order
prediction can be confirmed with renormalization group
equations (without relying on the $\epsilon$ expansion) \cite{Fejos:2024bgl}. 
The first order region also appears in model computations, see Ref.~\cite{Resch:2017vjs}
for an FRG based solution of the quark-meson model. The full computation suggests the
a small critical pseudo-scalar mass of 17~MeV, which is currently out of reach for lattice.
An effective model with a more generic $U(1)_A$ breaking finds parameters for both
first and second order \cite{Giacosa:2024orp}. If, however, effective models
find a sensitivity to the tunable parameters, lattice artefacts may also
shift the order of the transition, as it has already happened when unimproved
quarks were used with coarse lattices in the past.

We conclude this section admitting that the lower left corner of the Columbia plot is one
of the remaining white spots on our map of the QCD phase diagram. Lattice studies, though they
are not hindered by a sign problem, struggle to make predictions for small quarks and
extrapolate to the continuum. Firm knowledge in this region would tightly constrain  
effective models, and concrete predictions could be inferred for the hypothetical
chiral critical endpoint in the $T-\mu_B$ phase diagram.

\subsection{Imaginary valued chemical potentials \label{sec:immu}}

We do not study imaginary valued chemical potentials of pure academic interest,
but to gain valuable insight for the analytic structure of the QCD partition function,
which is immediately related the QCD phase diagram in the phenomenological domain.
The significance of imaginary $\mu_B$ is further raised by the applicability of
Monte Carlo simulations on the lattice: purely imaginary chemical potentials do not
invoke the sign problem. Their use in practical simulations open the possibility
of gaining information in the phenomenological domain via analytical continuation.
In this section we will mainly address an imaginary
baryo-chemical potential ($\mu_B$) defined along with the electric charge ($\mu_Q$) and
strangeness chemical potentials ($\mu_S$) as
\begin{eqnarray}
\mu_u&=&\frac{1}{3}\mu_B+\frac{2}{3}\mu_Q\,,\nonumber\\
\mu_d&=&\frac{1}{3}\mu_B-\frac{1}{3}\mu_Q\,,
\label{eq:BQSbasis}\\
\mu_s&=&\frac{1}{3}\mu_B-\frac{1}{3}\mu_Q-\mu_S\,,\nonumber
\end{eqnarray}

\begin{figure}[t]
    \centering
    \includegraphics[width=0.95\linewidth]{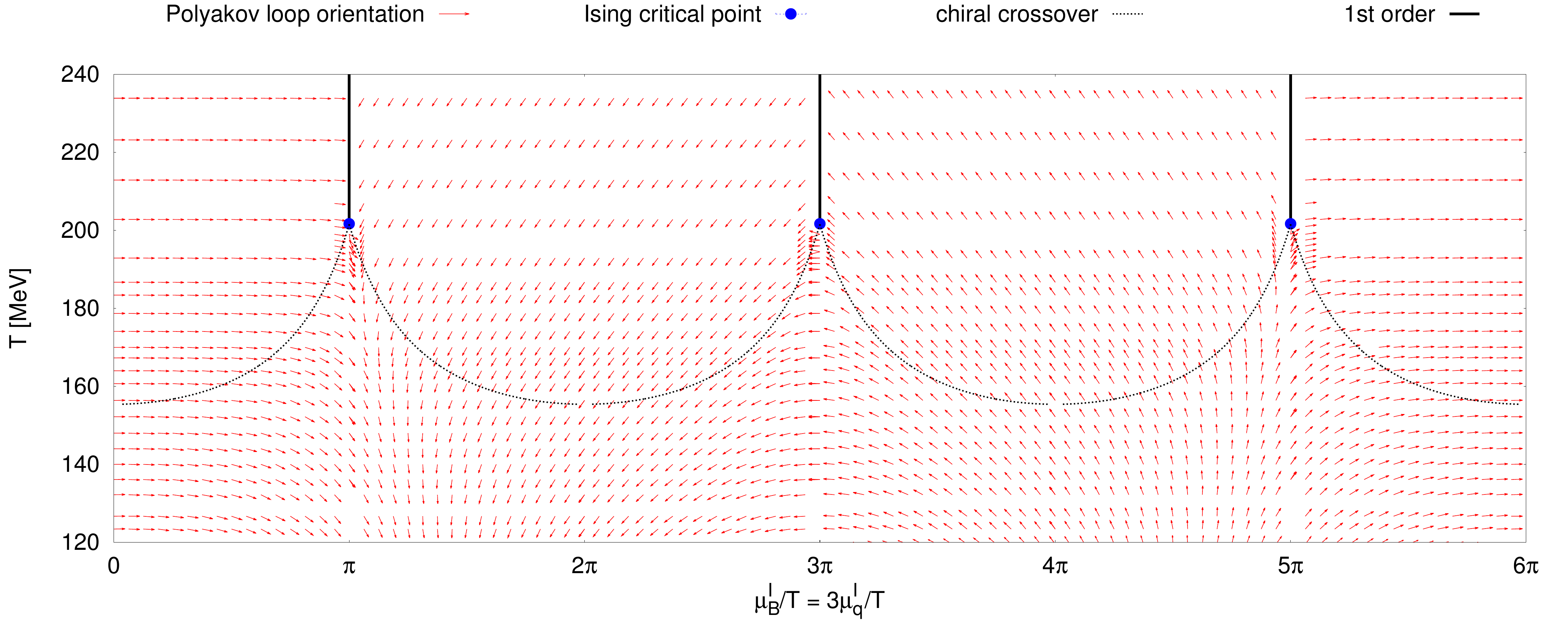}
    \caption{\label{fig:ploopdiagram}
    The phase diagram at imaginary chemical potential with physical quark masses in the temperature -- imaginary $\mu_B$ plane.
    The thermodynamic potential is periodic, $\mu_B \to \mu_B+i 2\pi T$, though after such a shift 
    in imaginary $\mu_B$ the Polyakov loop is rotated by a center element. The phase diagram features
    a first order line at $\mu_B = i \pi T + i 2n\pi T$ for any integer $n$ above a critical temperature
    $T_{RW}=208\pm 5$~MeV \cite{Bonati:2016pwz}.  The small red arrows show the orientation of the
    Polyakov loop.
    }
\end{figure}

The basic expectations for the phase diagram in the $T - \mathrm{Im}~(\mu_B/T)$ plane
were formulated in the seminal work of Roberge and Weiss \cite{Roberge:1986mm}. The quark chemical
potential $\mu_q$ appears as the pre-factor of a simple density term $\sim \psi^+\psi$ in the Lagrangian.
This $\mu_q$ plays the role of a homogeneous imaginary $A_0$ field of a U(1) gauge theory,
and can be transformed such that it only affects the boundary condition of the quark field as
\begin{equation}
    \psi(x,0) = -\exp(i\mu_q/T) \psi(x,1/T)
    \label{eq:immubc}
\end{equation}
This formula has a trivial symmetry for each quark flavor, independently: $\mu_q \to \mu_q + i 2\pi T$.
More importantly, though, the simultaneous shift of all quark chemical potential $\mu_q \to \mu_q + i 2\pi T/N_c$
is also an exact symmetry even in the presence of quarks, as it can be compensated
by a center transformation of the gauge field.
The same transformation can be simpler expressed as $\mu_B\to \mu_B+i2\pi T$, 
the corresponding periodicity is visible in Fig.~\ref{fig:ploopdiagram}.
Only color-non-singlet observables, such as the Polyakov loop are affected by this transformation.
A look at the definitions in Eq.~(\ref{eq:BQSbasis}) reveals that not just $\mu_B$, but 
$\mu_S$ and $\mu_Q$, too, are periodic with a period of $i2\pi T$.

\begin{figure}
    \centering
    \includegraphics[width=0.47\textwidth]{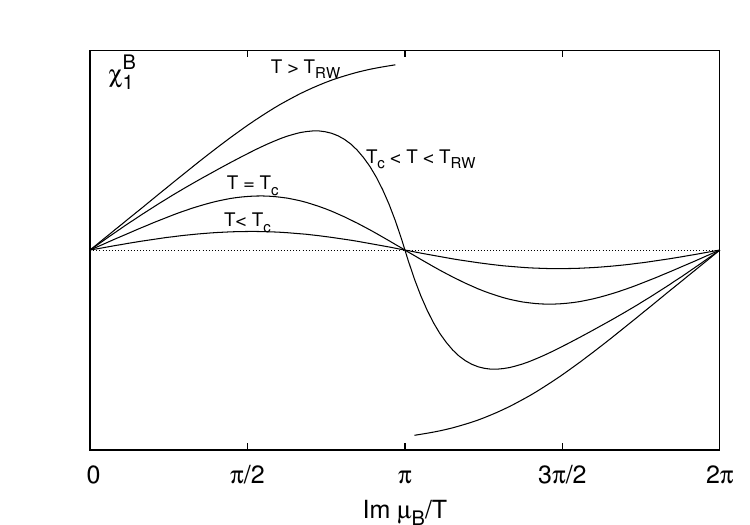}
    \includegraphics[width=0.47\textwidth]{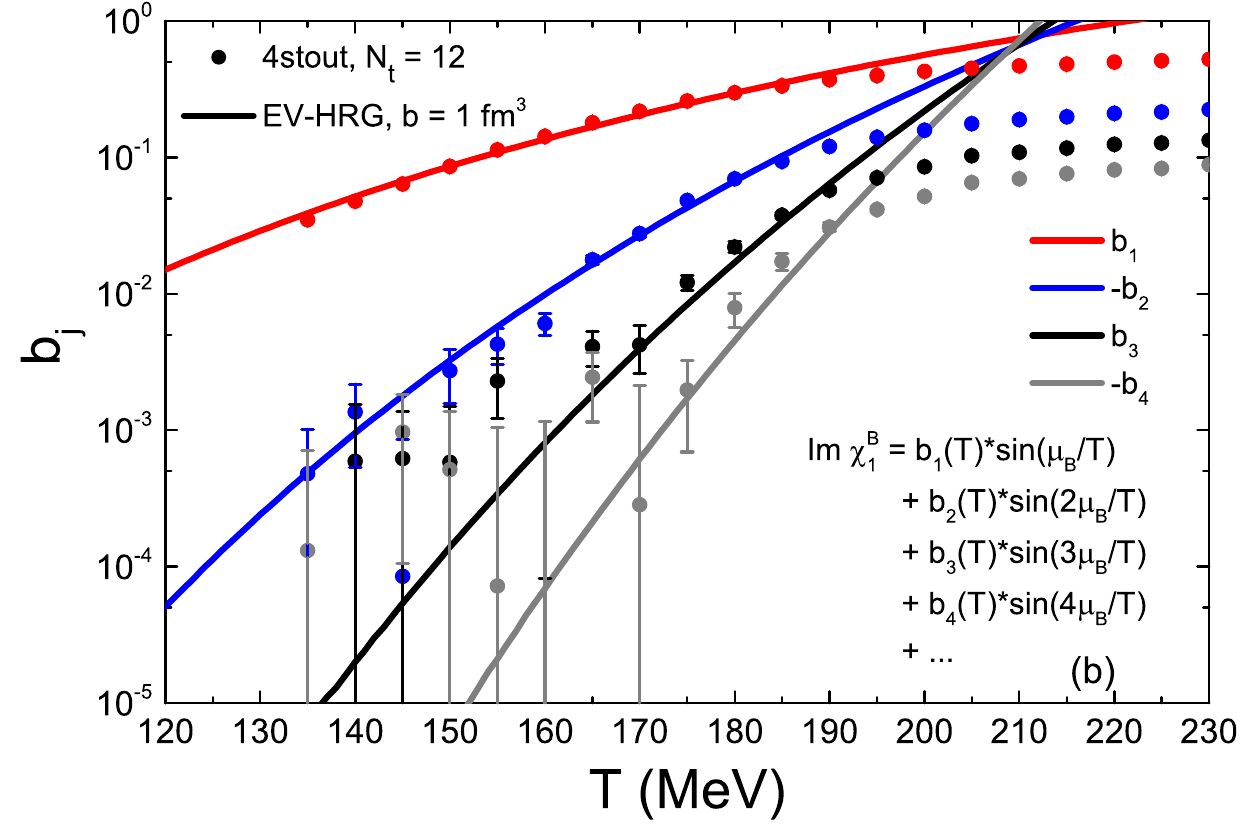}
    \caption{\label{fig:imnq}
    Left: The imaginary valued baryon density as a function of the imaginary chemical potential.
    Above the critical temperature $T_{RW}$ the baryon number shows a discontinuity, and indicates
    a first order transition. Below the chiral chiral transition $T_c$ this is a sine function
    to very good approximation.
    Right: The function in the left panel is periodic, its Fourier components on logarithmic scale
    are shown as a function of temperature \cite{Vovchenko:2017xad}. (In the latter the performance of a
    HRG-motivated excluded volume model is also shown.)
    }
\end{figure}

A further noteworthy feature of the partition sum $Z(\mu_B,\mu_Q,\mu_S)$ is charge conjugation symmetry, that
leaves $Z$ invariant to a \textit{simultaneous} sign flip in \textit{all} chemical potentials.
For example, with $\mu_S=\mu_Q=0$, the $\mu_B$ dependence of $Z$ is entirely encoded in the range
$\mu_B=0\to i\pi/T$. We know the grand canonical potential $\sim\log Z(\mu_B)$ 
at asymptotic temperatures analytically \cite{kapusta:book}. For QCD ($N_c=3$) we have
\begin{equation}
    \frac{1}{VT^3} \log Z(\mu_B) = \frac{8\pi^2}{45} + \frac{7\pi^2 N_f}{60}
    + \frac{1}{2} \sum_f \left( \frac{\mu_f^2}{T^2} + \frac{\mu_f^4}{2\pi^2T^4}\right)
\end{equation}
which is a polynomial in $\mu_B$ and is valid between $-\pi < \mathrm{Im}\mu_B/T < \pi$.
The imaginary baryon density $n_B= \mu_B/3 + \mu_B^3/27\pi^2$ is monotonic and periodic
at the same time, thus, it must be discontinuous at $\mu_B/T=i\pi+2\pi n$, for all integer $n$.
For this reason the phase diagram in the $T-\mathrm{Im}~\mu_B$ plane features a vertical
first order line, that is repeated periodically (see Fig.~\ref{fig:ploopdiagram}). This figure
is a simulation result on a $32^3\times8$ lattice with staggered fermions in the physical point.
The high and low temperature phases are separated by a crossover. The red arrows
show the orientation of the Polyakov loop. Notice, that below and above the chiral transition
temperature the Polyakov loop follows a different pattern. This reflects the different 
behavior in the imaginary $n_B(\mu_B)$ function (see Fig.~\ref{fig:imnq}). 
The Polyakov loop has a discontinuity where $n_B$ has, and both disappear in the
same critical point $T_{RW}=208\pm 5$~MeV \cite{Bonati:2016pwz}. The same study
has investigated the universality class of the Roberge-Weiss critical point and
found compatible results with an Ising-like critical endpoint.

In a cluster or fugacity expansion -- closely related to the virial expansion 
discussed in section~\ref{sec:hrg}, the (imaginary) baryon density at imaginary 
valued baryo-chemical potential reads 
\begin{equation}
\mathrm{Im} \frac{n_B}{T^3} = \sum_{k=1}^\infty b_k \sin\left(k \mathrm{Im} \mu_B\right) 
\end{equation}
ignoring the other chemical potentials. In the HRG model
$b_{k+1}\ll b_k$, thus in practice only $b_1$ contributes to baryon-related 
observables. In Fig.~\ref{fig:imnq} we show the function $n_B(\mu_B)$ for
imaginary arguments and its Fourier coefficients. The breakdown of the HRG model 
can be observed at $T_c$ where $b_2(T)$ and then at $T_{RW}$ all higher 
coefficients become relevant.

In the context of the Columbia plot (see section \ref{sec:columbia}) the fate of the Roberge-Weiss critical
point is often discussed near the chiral limit \cite{Philipsen:2014rpa}. In one possible
scenario, the Roberge-Weiss point turns into a first order triple point, while two critical end-points
progress along the crossover line towards the $\mu_B=0$ as all quark masses are simultaneously reduced,
eventually turning the chiral transition entirely first order. This picture is consistent with the
hypothetical existence of a first order region on the lower left corner of the Columbia plot, and
also predicts a first order triple-point already for not-yet-chiral quark masses. 
As of today most lattice studies addressing it use coarse lattices with
non-chiral (Wilson or staggered) discretizations, thus, the conclusions are not
final. A first order triple-point was observed only with un-improved actions on coarse lattices.
In more modern setups the Roberge-Weiss transition was always 2nd order, even for very small quark masses
\cite{Cuteri:2022vwk}.
Nevertheless, it was suggested that at $\mu_B=i\pi T$ the Roberge-Weiss critical point
is also a chiral critical point \cite{Bonati:2018fvg,Cuteri:2022vwk}.

\subsection{Yang-Lee edge singularities}

In recent years, the search for the QCD critical point saw the advent of a new method
based on the study of the analytic structure of the theory's partition function. In a
couple of seminal papers \cite{Yang:1952be,Lee:1952ig} Lee and Yang showed the 
profound connection between the analytic structure of the partition function and the 
presence and location of phase transitions. They showed that, although in finite 
systems a true phase transition cannot take place, the partition function of a 
physical system always has zeroes (the Lee-Yang zeroes) for complex values of the 
control parameter (in our case, the chemical potential $\mu_B$). The zeroes of the 
partition function correspond to singularities of the free energy, and thus to phase 
transitions. In the thermodynamic limit, an infinite number of Lee-Yang zeroes 
appear, which accumulate onto a branch cut in the complex $\mu_B$ plane, that 
terminates at branch points called Yang-Lee edge (YLE) singularities. In the 
presence of a critical point, these singularities fall onto the real axis at the 
critical value of $\mu_B$. Crucially, even away from a critical point, the YLE 
singularities are present in the complex $\mu_B$ plane, and are continuously 
connected to the critical point itself. 

The location of the YLE singularities is universal when expressed in terms of
the scaling variable $z = h/t^{\beta\delta}$, at: 
\begin{equation}
    z_c = \left| z_c \right|  \exp \left( \frac{i \pi}{2 \beta \delta} \right) \, \, ,
\end{equation}
where $\left| z_c \right|$ is a non-universal quantity which has only recently
been estimated with functional methods and on the
lattice~\cite{Connelly:2020gwa,Johnson:2022cqv,Rennecke:2022ohx,Karsch:2023rfb}.
Not just the branch point, but the other zeros on the branch cut follow 
a universal pattern, and the ratios of subsequent Lee-Yang zeros can be used
to locate the critical point \cite{Wada:2024qsk}.

The presence and universal behavior of the branching point allows new strategies
to locate critical points in the QCD phase diagram \cite{Skokov:2024fac}. Near
the critical point, one can create a map between the QCD phase diagram and the
Ising model one:
\begin{align}
    t &= a \Delta T + b \Delta \mu_B \\ \nonumber
    h &= c \Delta T + d \Delta \mu_B 
\end{align}
where $t$ and $h$ are the reduced temperature and magnetic field in the Ising
model, while $\Delta T$ and $\Delta \mu_B$ are the coordinates in the QCD phase
diagram relative to the critical point. Once the YLE singularity is located on
the complex plane,  the expected behavior for the approach towards the critical
point is known.  
Exploiting the relation between $z,h,t$ and then $T,\mu_B$, it follows that (see e.g., \cite{Skokov:2024fac}):
\begin{equation}
    \left( a \Delta T + b \Delta \mu_B \right) = i \left| z_c \right|^{\beta \delta}
    \left( c \Delta T + d \Delta \mu_B \right)^{\beta \delta} \, \, ,
\end{equation}
from which one can show that, as a function of the temperature, the real and imaginary parts of the critical chemical potential follow:
\begin{align} \label{eq:YLElaw}
    {\rm Re} \Delta \mu_B = \mu_{B,c} + c_1 \Delta T \, \, , \\ \nonumber
    {\rm Im} \Delta \mu_B = c_2 \Delta T^{\beta\delta} \, \, , 
\end{align}
where $\mu_{B,c}$ is the chemical potential at the critical point. 

It is not necessarily the chiral critical endpoint that dominates the parameter
space where observations are made. For example, above 160~MeV Ref.~\cite{Dimopoulos:2021vrk}
finds evidence for the Yang-Lee edge corresponding to the Roberge-Weiss critical point
from lattice simulations at imaginary $\mu_B$.
In a finite volume, the closest singularity of the free energy is located on the complex
$\mu_B$ plane and interpreted as the branching point. This implies an extrapolation
from imaginary to complex valued $\mu_B$, performed by fitting some ansatz that can account
for poles to the imaginary density. The systematics of this \textit{first extrapolation}
is less controlled in the vicinity of the chiral critical endpoint, where the magnitude
of $|\mu_B|$ is challenging. Once the leading singularity is found for the temperature
where we have a crossover, Eq.~(\ref{eq:YLElaw}) is applied as a \textit{second extrapolation}.
As of this writing, lattice studies have not yet been able to perform the first extrapolations
at low enough temperatures to guarantee the success of the second extrapolation.
In practice, additional terms can be included when fitting this temperature
dependence, and a potentially infinite number of functional forms ought to yield
the same estimates for $T_c, \mu_{B,c}$.

\begin{figure*}
    \centering
    \includegraphics[height=0.33\linewidth]{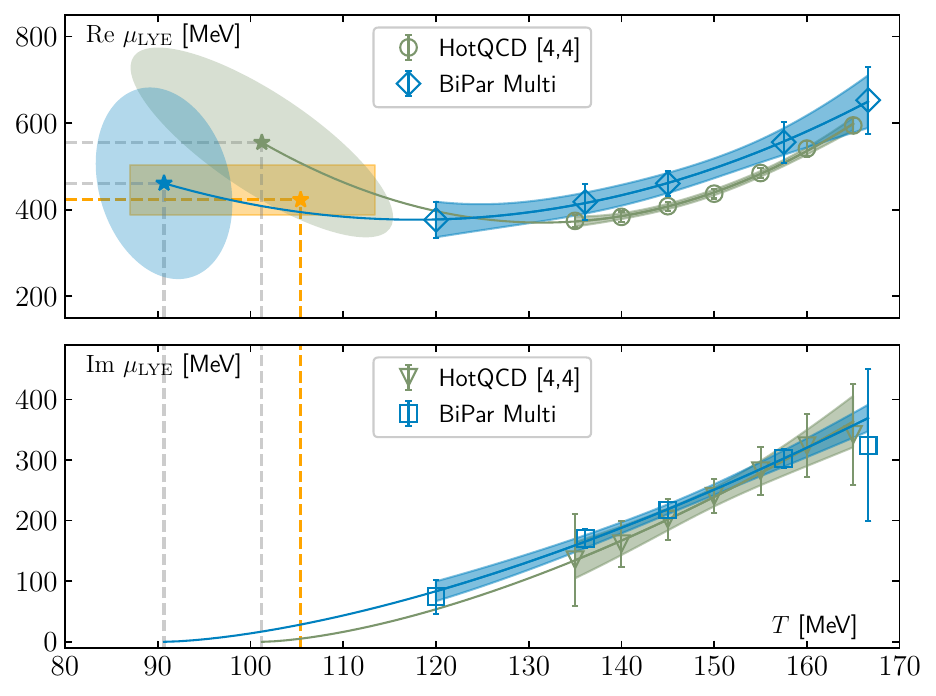}
    \includegraphics[height=0.37\linewidth]{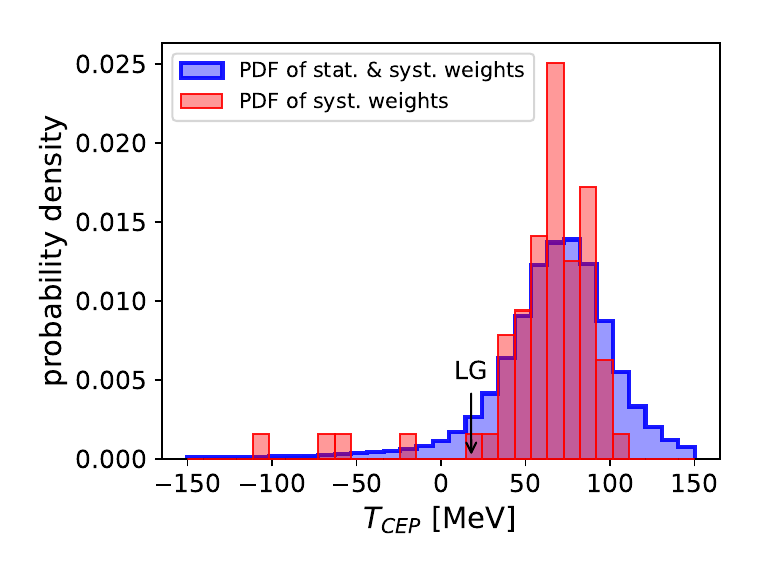}
    \caption{Left: estimates for the YLE singularities from Refs.~\cite{Bollweg:2022rps,Clarke:2024ugt}, and their $T$-extrapolations. Right: systematic analysis of the critical temperature from Ref.~\cite{Adam:2025phc}.}
    \label{fig:YLE}
\end{figure*}

Most commonly, a rational ansatz is applied to the free energy or its derivatives to 
estimate the YLE singularities locations. In the left panel of Fig.~\ref{fig:YLE} 
results for the real and imaginary parts are shown for different temperatures, 
together with their extrapolation in $T$ based on Eq.~(\ref{eq:YLElaw})~\cite{Bollweg:2022rps, Clarke:2024ugt}. In Ref.~\cite{Basar:2023nkp}, the procedure 
was modified by first applying a uniformizing map to the complex $\mu_B$ plane, then 
carrying out the YLE singularities estimation, before mapping back to $\mu_B$; this 
alternative yielded comparable results.

The authors of \cite{Clarke:2024ugt} estimated the location of the critical
point based on large volume simulations on a coarse lattice. Their main analysis
gives the estimate $(T_c,\mu_{B,c})=(105^{+8}_{-18},422^{80}_{-35})$~MeV, while
further systematic effects are expected that could push the critical chemical potential
towards $\mu_{B,c}\approx 650$~MeV.
Uncertainty estimation is all but obvious, given that 
both the YLE singularities estimation, as well as the temperature extrapolation can
in principle be carried out in and infinite number of ways. The authors of 
Ref.~\cite{Adam:2025pii} attempted to probe the predictive power of such a strategy, 
based on high statistics, $N_\tau = 8$ lattice results. In particular, different 
sources of systematics were considered: i) the rational ansatz was applied either to 
the free energy, to its first or second $\mu_B$-derivative; ii) the $T$ dependence 
was modeled by different functional forms, and iii) the fits carried out over 
different $T$-ranges. Moreover, differently from previous determinations, the 
rational ans\"atze for locating the YLE singularities were chosen to explicitly 
reflect CP symmetry, i.e. to be even in $\mu_B$. The results for the estimated 
value of the critical temperature are summarized in the histogram in the right panel 
of Fig.~(\ref{fig:YLE}), while no results were reported for the critical chemical 
potential, as the corresponding uncertainty would be very large. The main result is 
an upper bound for the location of the critical endpoint, which at the 1$\sigma$ 
level is either below $T = 103$~MeV or it does not exist. It should be reminded 
that, although quite promising, at present the strategy of locating the critical 
point by means of the YLE singularities hinges on substantial caveats. First, the 
continuum and infinite volume limits ought to be taken. Second, the systematics due 
to a truncated expansion and, most importantly, on the temperature range from which
the extrapolation is carried out, ought to be quantified.


\section{The multidimensional phase diagram\label{sec:multi}} 

\subsection{External magnetic fields\label{sec:magnetic}}

Let us now consider a new axis in the QCD phase diagram, the strength of an external
homogeneous magnetic field.
The strength is of the magnetic field is often expressed as the rooted
 product with the elementary charge in MeV units:
typical values in the interior of
magnetars are $\sqrt{eB}\sim 1~\mathrm{MeV}$ ($\sim 10^{15}~\mathrm{G}$),
but in the early Universe much larger fields were realized, such as
$\sqrt{eB}\gtrsim 1~\mathrm{GeV}$.
Transient magnetic fields are present in non-central collision experiments
\cite{Kharzeev:2007jp} are expected to be of order $\sqrt{eB}\sim 0.1-0.5~\mathrm{GeV}$. These 
have impact on the early plasma phase, the hydrodynamic
evolution or possibly even the chemical freeze-out. 

The magnetic field is time dependent in the examples above. In contrast,
in theoretical models we assume a constant background field in equilibrium.
The reader will find a summary of the theoretical work in Ref.~\cite{Andersen:2014xxa}
or with a focus at lattice QCD in Ref.~\cite{Endrodi:2024cqn}. As opposed
to electric fields or a baryo-chemical potential, magnetic fields can be simulated on the lattice
without encountering a sing problem. One important restriction is that in a finite volume
the total flux, thus, the $B$ field, too, is quantized.

Near the chiral transition, we have to study its order parameter, the chiral condensate,
as a function of the magnetic field. In the QCD vacuum the condensate is enhanced, this
is the \textit{magnetic catalysis} \cite{Shovkovy:2012zn}. At finite temperature the situation
is more complicated: the magnetic catalysis is still active for valence quarks, however,
sea quarks behave differently \cite{DElia:2011koc}.  While the former
is a monotonic function of $B$ at all temperatures, the sea quark
contribution exhibits a reversal of the trend in the transition region and gives the
dominant effect \cite{Bali:2012zg}. This is the \textit{inverse magnetic catalysis} and works
towards chiral symmetry restoration. In the case of inhomogeneous magnetic fields
the spatial response of the sea quarks follows a broadening pattern. Together with the more
localized valence quarks a non-uniform response emerges \cite{Brandt:2023dir}.

The magnetic fields have an impact not just on chiral symmetry,
but an analogous effect can be observed on the order parameter of the center symmetry, the Polyakov loop, too.
Small values of the Polyakov loop, characteristic to
confinement, correlate with an abundance of small eigenvalues ($\lambda_{\rm
min}$) of the Dirac operator. The statistical weight of a configuration can be
approximated as $\lambda_{\rm min}^{q|B|}$, thus, suppressing small values of
the Polyakov loop in the ensemble \cite{Bruckmann:2013oba}. 

\begin{figure}[t] \begin{center}
\includegraphics[width=0.4\textwidth]{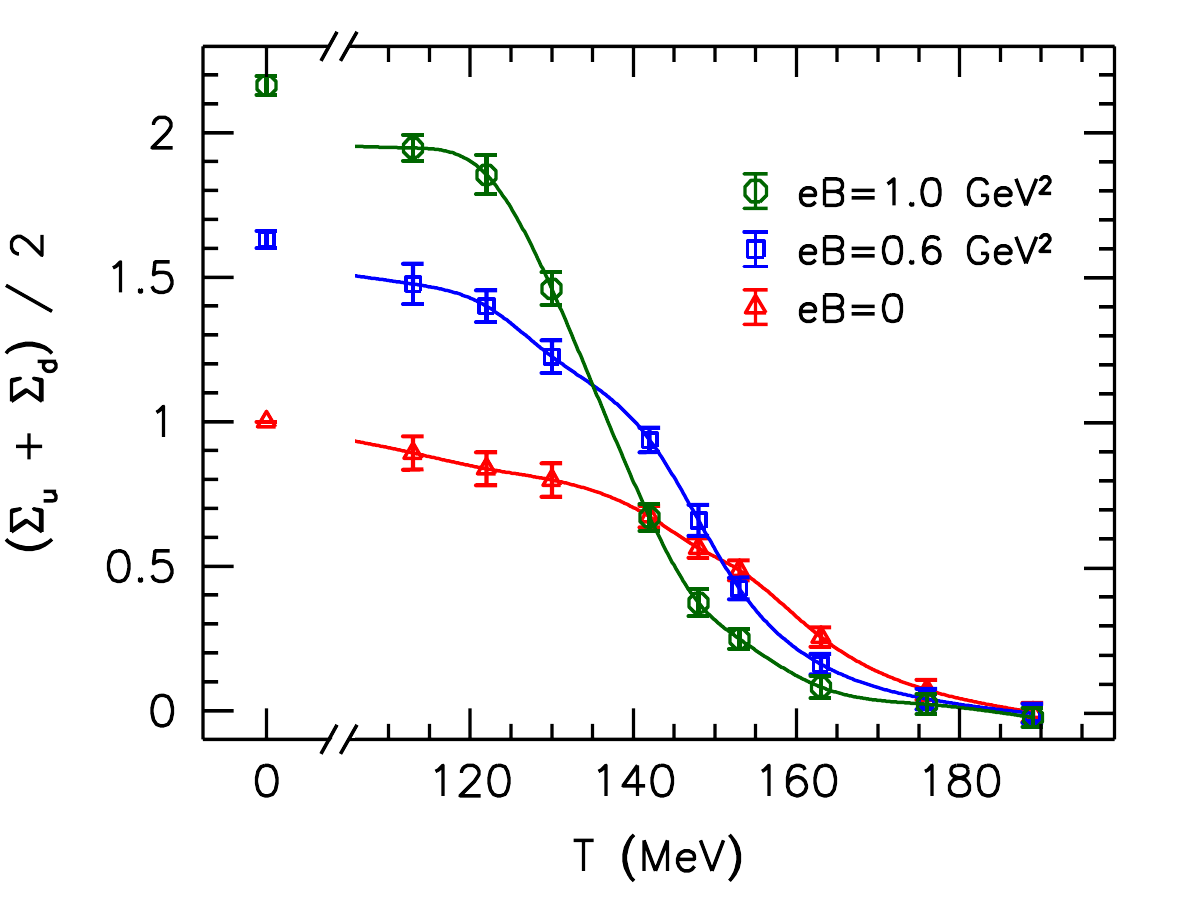}
\includegraphics[width=0.4\textwidth]{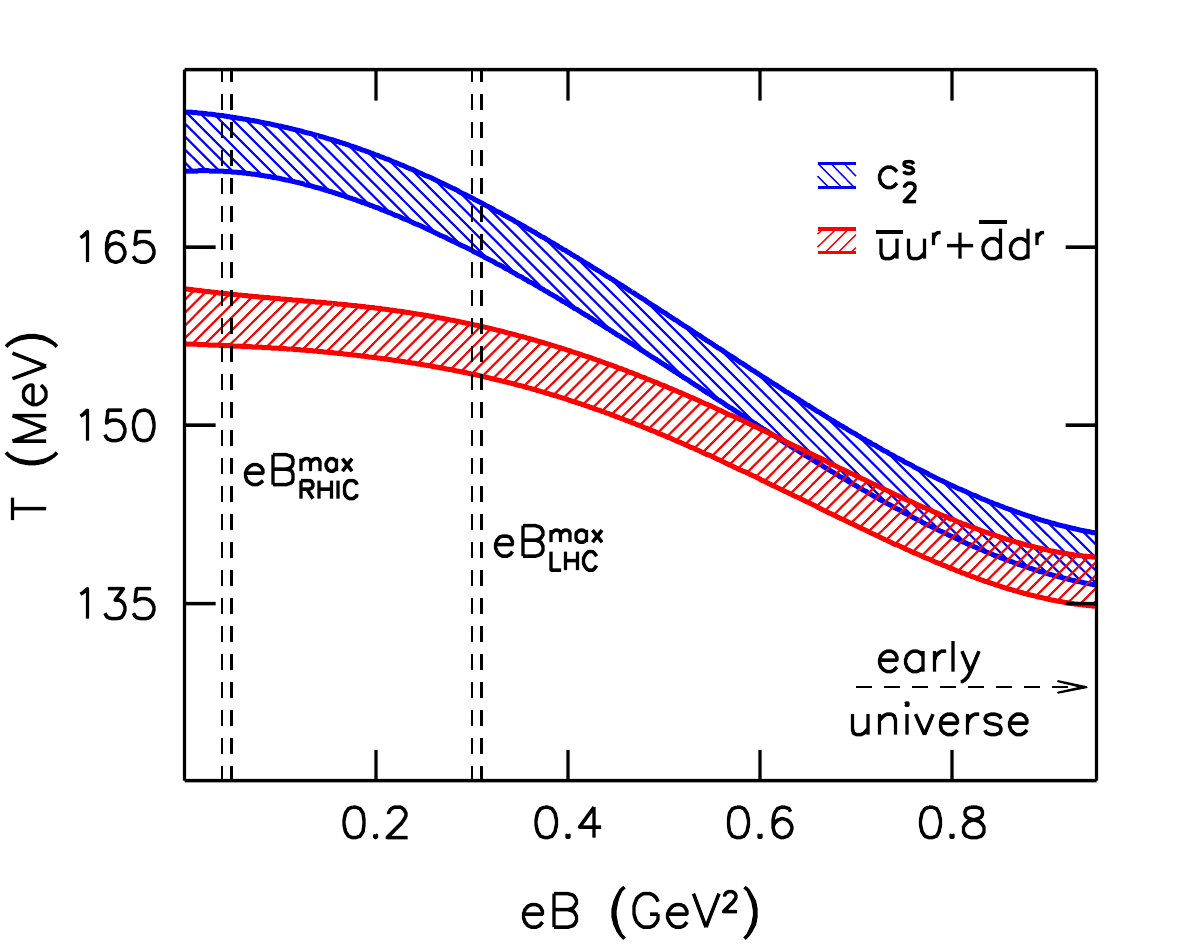}
\caption{\label{lattice:fig:magnetic} Left: The renormalized chiral condensate
(averaged over the two light flavors) as a function of the magnetic field for
temperatures below and above the crossover. One observes the magnetic catalysis
at low temperatures. This trend reverses near $T_c$
\cite{Endrodi:2014vza}.\\
Right: The QCD phase diagram in the $T-B$ plane from continuum extrapolated lattice simulations
\cite{Bali:2011qj}.
}
\end{center}
\end{figure}

These effects lead to a monotonic decrease of the transition temperature with
$B^2$ as first demonstrated on the lattice in Ref.~\cite{Bali:2011qj}. The main
result of Ref.~\cite{Bali:2011qj} is shown in Fig.~\ref{lattice:fig:magnetic}.
Due to their different electric charge, the up and down chiral condensates show
a slightly different pattern, the extracted transition temperatures are still
compatible \cite{Endrodi:2015oba}.  In the same work the outlined behaviour of
the Polyakov loop was confirmed.

\begin{figure}[t]
\begin{center}
\includegraphics[width=0.48\textwidth]{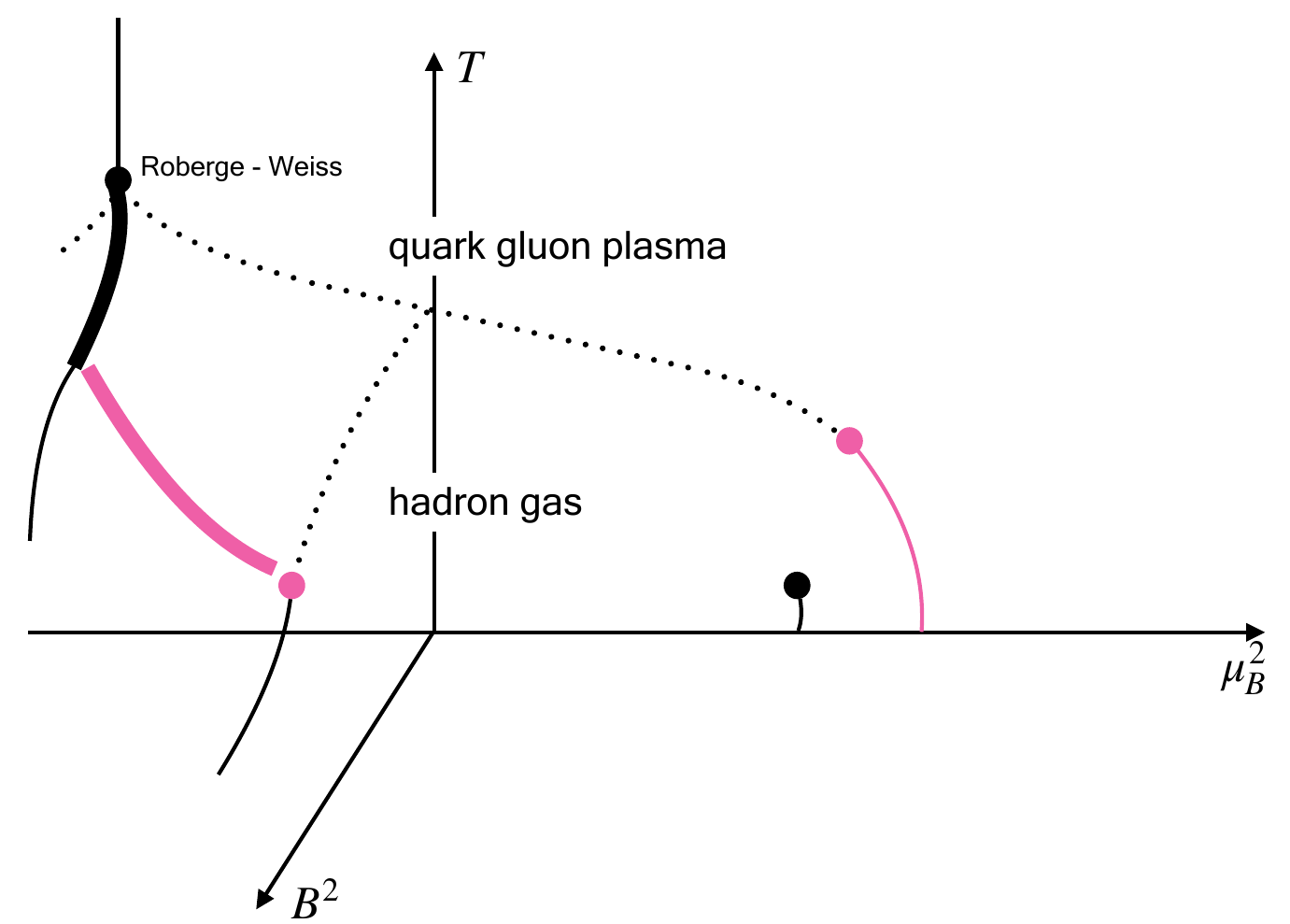}
\caption{\label{lattice:fig:muandBpd}
The three dimensional phase diagram with temperature, magnetic field
and chemical potential (left imaginary, right real) on the axis.
The $B$ and $\mu_B$ variables are squared to reflect symmetries.
The magenta features are hypothetical.
}
\end{center}
\end{figure}

Whether the crossover line in Fig.~\ref{lattice:fig:magnetic} ends
in a new critical point has been subject to various speculations
\cite{Cohen:2013zja}.  The phase diagram was later extended towards stronger
magnetic fields.  With increasing $B$ the crossover temperature further
decreases, and so does its width \cite{Endrodi:2015oba}
Simulating at isolated choices of $B$ Ref.~\cite{DElia:2021yvk} has found evidence
for a first-order transition at $eB= 9~\mathrm{GeV}^2$ and
$T_c =  63\pm 5 $~MeV. Thus, a second-order endpoint is expected to be
between 4 and $9~\mathrm{GeV}^2$ in $B$.
It is an intriguing question, whether this new critical point is
analytically connected to the expected chiral endpoint on the $T-\mu_B$
phase diagram, or perhaps it is related to the Roberge-Weiss transition
at $\mu_B=\pi i\,T$, see illustration in Fig.~\ref{lattice:fig:muandBpd}.
If it was connected with the chiral critical point at $\mu_B>0$ one might
attempt an analytical continuation of the crossover line (with question marks)
using imaginary $\mu_B$ simulations. Very recently, an argument was
presented for this being unlikely. Ref.~\cite{DElia:2025ybj} follows
the critical point at $\mu_B=\pi i\,T$ in a finite magnetic field,  observes
the drop in the critical temperature, and finds that it turns first order at
$B<2.5~\mathrm{GeV}^2$. In Fig. ~\ref{lattice:fig:muandBpd} we show
a possible scenario where the end-point in the $T-B$ plane is connected
to the Roberge-Weiss point with a critical line. Direct evidence
for this connection has not yet been found.

\subsection{Isospin chemical potential}

In the space of the quark chemical potentials ($\mu_u,\mu_d,\mu_s$) the baryon-chemical potential
($\mu_B$) represents a direction with $\mu_u=\mu_d$. The orthogonal combination ($\mu_u=-\mu_d$)
corresponds to the third component of the isospin. If there is no other source of isospin breaking
(non-degenerate up and down quarks, QED effects) any other component can be considered.
The third component is particularly interesting, e.g. in neutron stars with $\mu_I<0$ due to
the weak interaction. In that case $\mu_B\gg\mu_I$, however, we consider the (unstable) $\mu_B=0$ setup here.

The motivation behind the phase diagram with an isospin chemical potential ($\mu_I)$ axis with fixed
$\mu_B=0$ comes from theory: \\
i) There is no sign problem in this setup; thus, lattice simulations are possible.
In general, a real quark chemical potential introduces a complex fermion determinant
after quark degrees of freedom are integrated. However, as long as the up and down quarks are degenerate in mass,
the complex phase of these two determinants exactly cancel, leaving us with a positive semidefinite
combined quark matrix.\\
ii) The grand canonical pressure function $p_I(\mu)$ at
$\mu=\mu_u=-\mu_d=\mu_I/2$ is always greater than or equal to the pressure $P_B(\mu)$
with $\mu=\mu_u=\mu_d=\mu_B/3$ at the same temperature. This is a simple path integral relation:
we can rewrite the partition sum of baryon-dense matter to isospin-dense matter simply by
replacing the complex determinant of the $u,d$ quark pair $(\det M)^2$ by its modulus $|\det M|^2$.
In the process, both the partition sum $Z$ and its logarithm, the pressure, increase.

\begin{figure}[t]
\begin{center}
\includegraphics[width=0.48\textwidth]{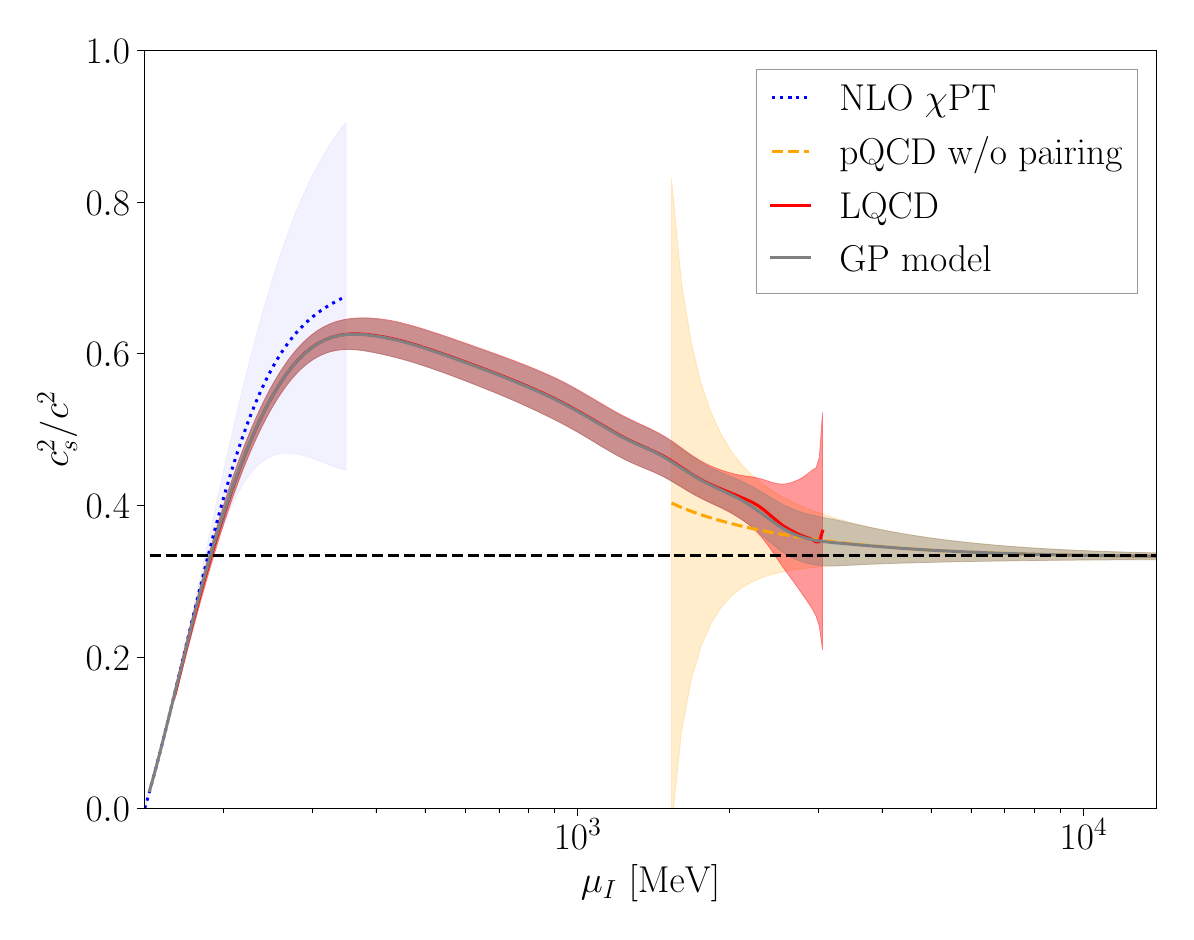}
\includegraphics[width=0.48\textwidth]{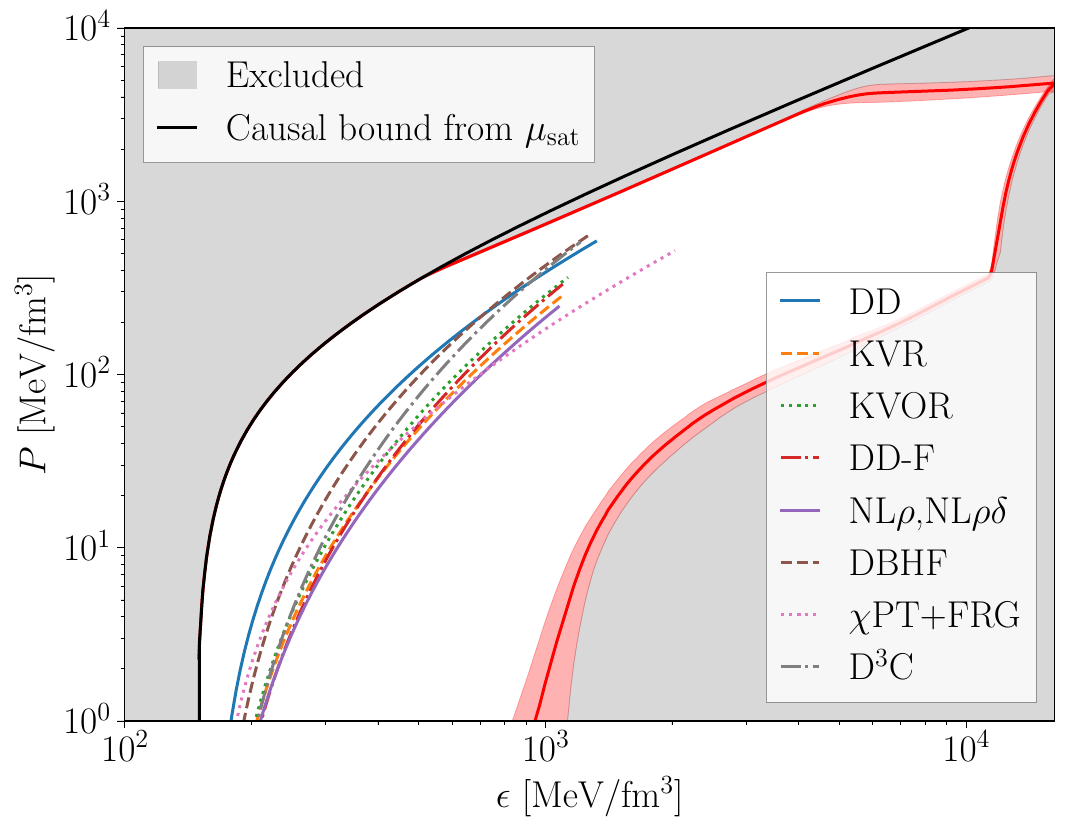}
\caption{\label{fig:abbott}
Left: The squared speed of sound ($c_s^2$) as a function of the isospin chemical
potential ($\mu_I$). The lattice QCD determination (red), pQCD determination
(orange), and $\chi$PT determination (blue) are combined into a model (GP for
Gaussian Process).
Right: Constrain to the equation of state for baryon-dense matter (the border of the possible
range is marked by the red bands). Several phenomenological equations of states are shown in the non-excluded region \cite{Abbott:2024vhj}.
}
\end{center}
\end{figure}

These points were used to constrain the QCD pressure at
zero temperature, for finite $\mu_B$ through the inequality $p_B\lesssim p_I$
using continuum extrapolated lattice results in $p_I$
\cite{Abbott:2023coj,Abbott:2024vhj}. The resulting equation of state (see
Fig.~\ref{fig:abbott}) shows a speed of sound that exceeds the conformal limit
$c_s^2>1/3$ over a large range of $\mu_I$. The right panel indicates the
exclusion region from the isospin constraint.

Besides these results at zero temperature, the phase diagram in the $T-\mu_I$ plane also received
attention from theorists. The expected structures were discussed in Ref.~\cite{Son:2000xc}. For
this one considers three regions in $\mu_I$. In the first region near $\mu_I<m_\pi$ an ordinary hadron
gas is expected. At the threshold value of $m_\pi$ pions condensate, and the $SU(2)$ isospin symmetry
spontaneously breaks to the $U(1)$ subgroup generated by $I_3$. A Goldstone boson appears, while two
other pions remain massive, their masses will approach $|\mu_I|$ for large densities.
The order parameter of this phase is $\langle \bar u \gamma_5 d\rangle\sim\Delta$. The non-zero gap $\Delta$
persists for large values of $\mu_I$. Eventually, the phase will exhibit Cooper pairs of color singlet
combinations of the $\bar u$ and $d$ quarks. The order parameter of this phase is not different from the pion
condensation, thus, no real transition is expected to separate the large-$\mu_I$ superconducting phase from the pion condensation at $\mu_I\gtrsim m_\pi$. The gap $\Delta$ is much larger in comparison to the
analogous phase of baryon-dense matter \cite{Son:2000xc}.

\begin{figure}
    \centering
    \includegraphics[width=0.48\textwidth]{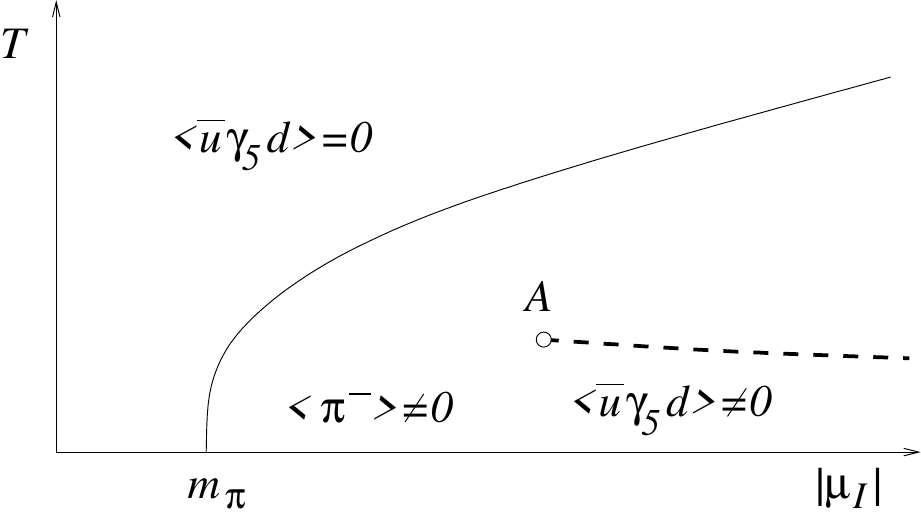}
    \includegraphics[width=0.48\textwidth]{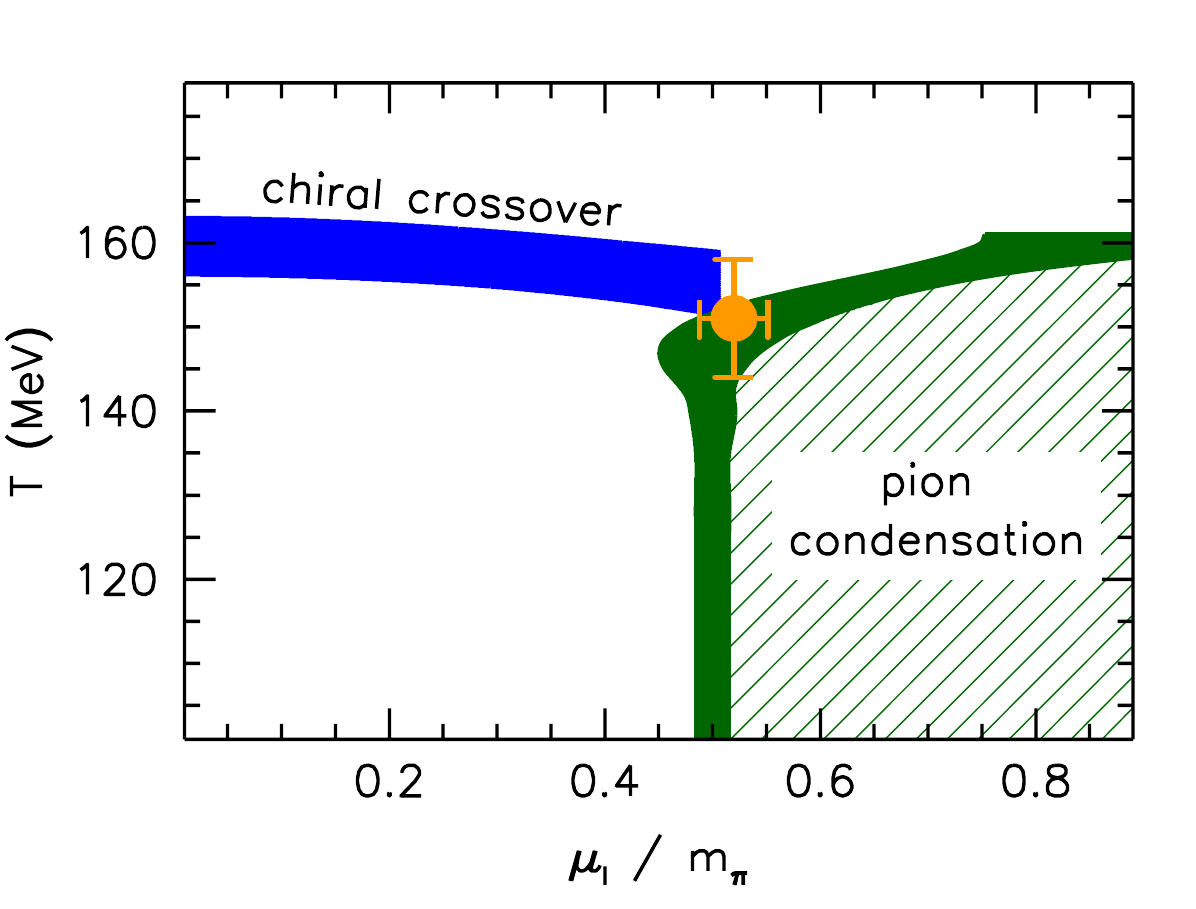}
    \caption{\label{fig:isospin}
    The QCD phase diagram in the temperature -- isospin chemical potential plane. The left panel
    shows the theoretical expectation based on effective field theory in Ref.~\cite{Son:2000xc}.
    The gross picture was confirmed on the lattice \cite{Brandt:2017oyy} (right panel, this plot
    introduces an extra factor two in the definition of $\mu_I$, thus, the onset of the condensation appears
    to be at $m_\pi/2$).
    }
\end{figure}

The progress in lattice QCD at finite temperature and isospin density was, indeed, quick exploring
the phase diagram. Polynomial extrapolation from $\mu=0$ as well as direct simulations are unproblematic
up to onset of the condensation phase. The Taylor coefficients indicate a divergence of the series
for $\mu_I>m_\pi$ \cite{Brandt:2018omg,Borsanyi:2023tdp}. In the pion
condensation phase non-trivial technical issues have to be addressed: the
presence of a Goldstone boson hinders the convergence of the iterative solver of
the Dirac equation: the number of iterations is proportional to the reciprocal
of the (square root of the) smallest eigenvalue of the fermion matrix, and the latter is zero in the
condensation phase \cite{Brandt:2017oyy}. The authors of Ref.~\cite{Brandt:2017oyy} offered
a solution in form of an infrared regulator term in the action that is removed in a second, reweighting-based
step. The resulting phase diagram together is shown together with the prediction of \cite{Son:2000xc} in Fig.~\ref{fig:isospin}. On the phase boundary between the hadron gas and pion condensation phases the
Polyakov loop indicates no deconfinement, as expected. The order of the transition is second order in
the O(2) universality class due to the pattern of symmetry breaking, and this, too was confirmed 
on lattice \cite{Brandt:2017oyy}. At higher temperatures, in the absence of pions, the condensation does not occur.

\section{Phase diagram of dense matter\label{sec:dense}} 

\subsection{Dense matter equation of state} 

\begin{figure}
    \centering
    \hspace{4mm}
    \includegraphics[height=0.36\linewidth]{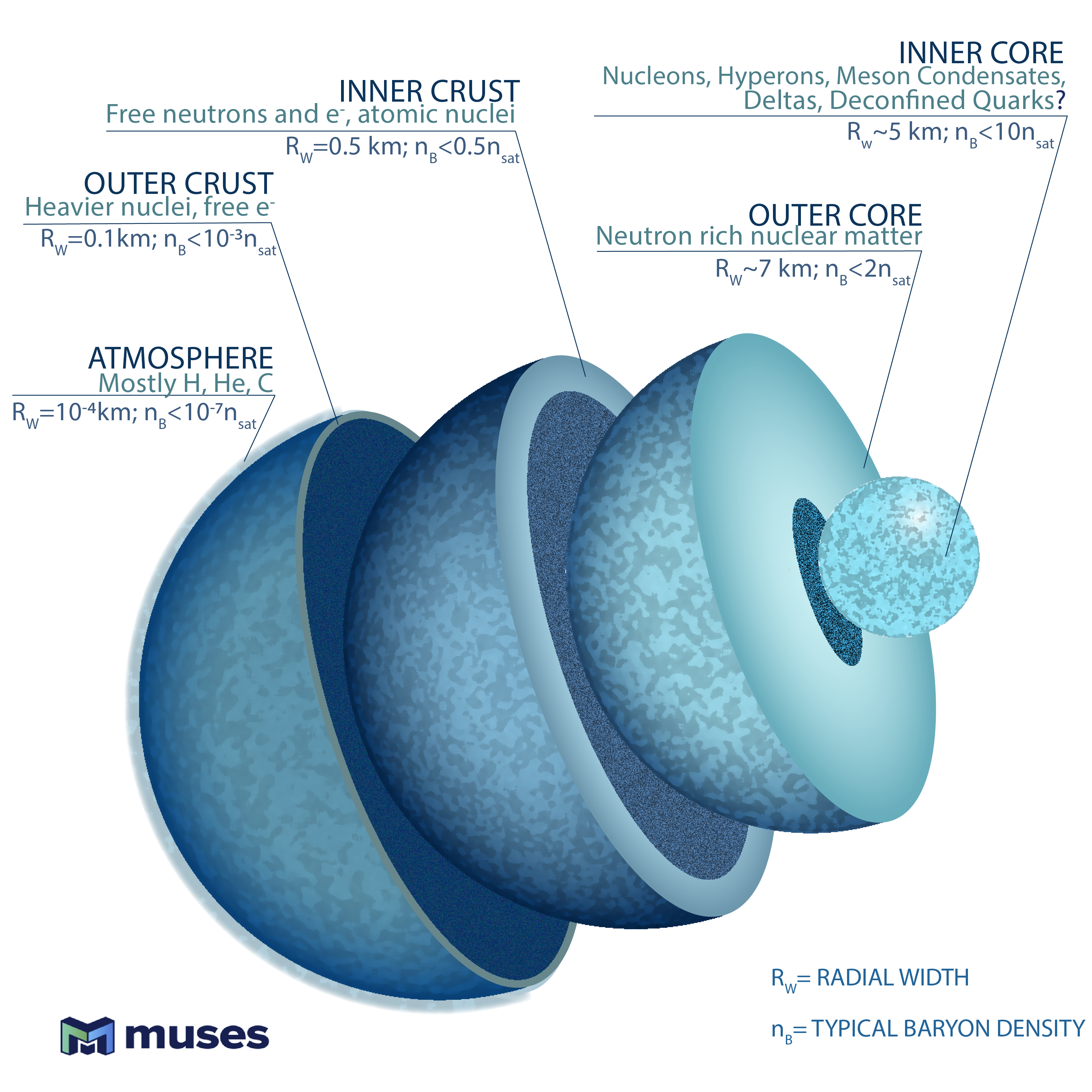} \hfill
    \includegraphics[height=0.34\linewidth]{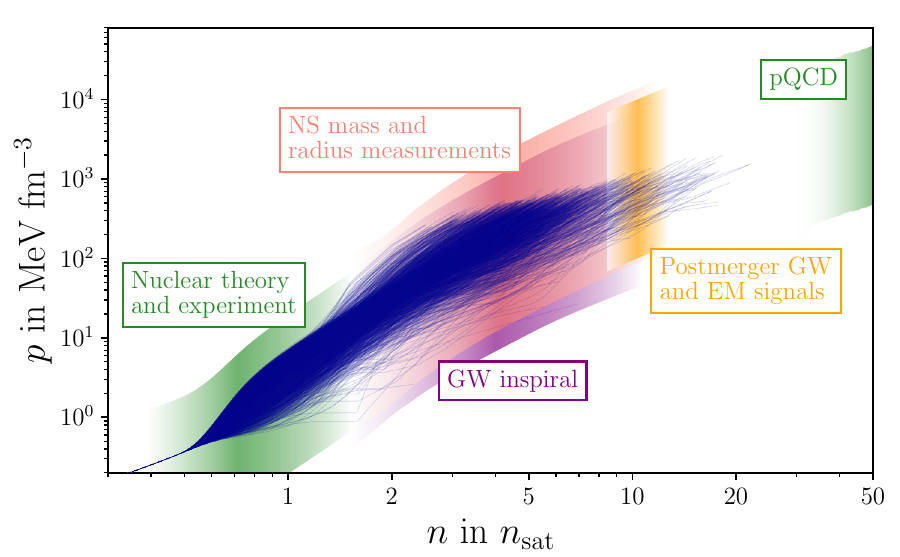}
    \caption{Left: internal stratified structure of a neutron star~\cite{MUSES:2023hyz}. Right: possible interpolations of dense matter equation of state between the nuclear and perturbative regimes, and a summary of the regions where different possible measurements can provide constraints on the equation of state, from Ref.~\cite{Koehn:2024set}.}
    \label{fig:NS}
\end{figure}

Since matter above saturation density cannot be created in 
the laboratory, the only available testbed for dense matter 
thermodynamics are neutron stars. These cold objects have typical 
radii of the order of $~10$~km, masses around $1-2 M_\odot$ 
($M_\odot$ is the Sun's mass), and thus are the second-densest  
known objects, after black holes. They are globally neutral 
objects with a stratified structure as sketched in 
Fig.~\ref{fig:NS} (left). The outer portion is 
mostly composed of heavy nuclei, and moving inward one finds more 
and more neutron-rich nuclei which do not exist in isolation. 
The inner layers comprise a rich structure, with possibly a 
neutron-pair superfluid~\cite{Haskell:2017lkl} and ``nuclear 
pasta" phases~\cite{Watanabe:2000rj}. The outer core contains mostly 
neutrons, with electrons, muons and protons contributing to 
$beta$-equilibrium. The inner core reaches densities up to a few 
times nuclear saturation density~\cite{Koehn:2024set}, and it is 
unclear what the effective degrees of freedom might be in this 
regime, as several different scenarios have been proposed. These 
include the existence of pions or kaon condensates, as well as nuclei 
including hyperons (strange baryons), which would be stable in a 
large density environment due to Pauli 
blocking~\cite{Blaschke:2020qrs}. Another possibility is provided
by \textit{quarkyonic} matter~\cite{McLerran:2007qj}, a conjectured 
phase where quarks populate the Fermi sea, and baryons live on the 
Fermi surface. 
Since baryons only populate a thin outer layer of the Fermi sphere, 
as the chemical potential (and thus the Fermi momentum) is increased, 
baryons can reach high momenta, thus increasing the pressure, without 
correspondingly increasing the energy density. This provides a 
mechanism for a very stiff equation of state~\cite{McLerran:2018hbz}. 
Originally based on large-$N_c$ arguments, it is not clear whether it 
does exist in QCD with $N_c=3$. Finally, densities might 
also be large enough that quarks become deconfined~\cite{Lattimer:2004pg}.

Regardless of the effective degrees of freedom, neutron stars  
owe their existence to the delicate equilibrium between the 
inward gravitational force and the outward force due to the 
pressure. 
Measurements of neutron stars masses and radii can provide 
significant constraints on the equation of state of dense matter, 
as the latter must be such to support star masses at least up to 
the observed ones, which currently lie at $m_{\rm NS} = 2.01 - 2.35 M_\odot $~\cite{Antoniadis:2013pzd,NANOGrav:2019jur,Romani:2021xmb,Romani:2022jhd}.
For non-rotating isolated neutron stars, the 
Tolman-Oppenheimer-Volkoff (TOV) equations~\cite{Tolman:1939jz,Oppenheimer:1939ne} 
relate the core density of the star to its radius and mass. A given 
equation of state is then one-to-one mapped onto a mass-radius curve 
$M(R)$~\cite{Lindblom:1992ApJ}. Also when not in isolation, 
and/or rotating, different solutions of Einstein's equations can 
relate the equation of state to the mass and radius, but also the 
moment of inertia and tidal deformability of the star, which can 
be constrained through gravitational wave 
detection~\cite{Hinderer:2009ca}. Through gravitational wave and 
electromagnetic astronomy, mergers of neutron stars and 
neutron star-black hole mergers can be observed, and numerical relativity 
simulations showed that temperatures up to $T \sim 80 -100$~MeV can 
be reached in these 
events~\cite{Sekiguchi:2011mc,Bernuzzi:2015opx,Perego:2019adq}. In 
particular, the gravitational wave signal from the inspiral stage 
can probe the regime $\sim 2-3 n_{\rm sat}$. On the other hand, 
post-merger signals would provide much more stringent constraints -- as 
their morphology is more sensitive to the equation of state than its 
pre-merger counterpart -- though their observation with current 
detectors is unlikely~\cite{Chatziioannou:2024tjq}.

From the theoretical side, the cold matter equation of state is 
constrained at the lower density end by chiral effective theory, 
and in the opposite regime by perturbation theory. While 
astrophysical observations can provide insights into the nuclear 
equation of state in the density range covered by neutron stars 
cores, above 10 times saturation densities they have little 
constraining power, which leaves a gap stretching from a few to a 
few tens of times saturation density.
Since thermodynamic relations (e.g. causality bounds on the speed 
of sound) cause perturbative QCD constraints to propagate to 
lower energy scales, modeling of the intermediate regions have 
been constructed to produce a posterior distribution for the 
pressure-energy density $p(\epsilon)$ 
relation~\cite{Annala:2017llu,Annala:2019puf,Komoltsev:2021jzg,Mroczek:2023zxo,Koehn:2024set}, see Fig.~\ref{fig:NS} (right), possibly 
hinting at the existence of deconfined matter in neutron stars 
cores~\cite{Annala:2019puf}. How much the equation of state in 
this regime in constrained by perturbative QCD results is however 
still unclear~\cite{Finch:2025bao}.


\subsection{Color superconductivity\label{sec:cfl}} 

The expectation, that asymptotic freedom also applies to superdense matter,
not just high temperatures or momentum scales, promising an analytically tractable
region for QCD, has a long history \cite{Collins:1974ky}. The dense phase, however,
is less trivial than its high temperature counterpart. Around the Fermi surface
of dense quark matter, even the smallest attractive interactions can introduce
instabilities \cite{Cooper:1956zz}. Unlike conventional superconductivity, here
a diquark pair is formed, and instead of phonons
a gluon exchange is sufficient to induce the bound state. We know that gluon exchange
is attractive in some channels, e.g. in those that bind baryons. Such pairs, being bosonic,
can form a condensate, the phase with such a condensate is called color
superconductivity \cite{Barrois:1977xd}. In presence of a diquark condensate
perturbation theory is still
possible, though around an improved ground state \cite{Pisarski:1999tv}. 
This state is characterized by broken chiral and color symmetries. Most strikingly, the originally
massless gluons and quarks become massive. See
Refs.~\cite{Rajagopal:2000wf,Alford:2007xm} for in-depth reviews.

The pattern of symmetry breaking depends on the channel of attractive interaction.
At asymptotic densities of the three-flavor theory the most symmetric option is
the color-flavor locked phase \cite{Alford:1998mk} with the condensate form
\begin{equation}
    \langle \psi^a_i C\gamma_5 \psi^b\rangle \sim
    \Delta_{\rm CFL} \epsilon^{abc}\epsilon_{ijc} +   \Delta_{\rm CFL} \kappa \left(
    \delta^a_i \delta^b_j+
    \delta^a_j \delta^b_i\right)\,.
\end{equation}
Here $a,b$ are color indices and $i,j$ are flavor indices, both running from 1 to 3. $C$
is the Dirac charge conjugation matrix, and $\Delta_{\rm CFL}$ is the dimensionful
gap parameter. The pattern of symmetry breaking is then
\begin{equation}
\left[SU(3)_c\right]\times U(1)_B \times SU(3)_L \times SU(3)_R \to SU(3)_{c+L+R}\times {Z}_2
\end{equation}
The attractive interaction is through the first term, which is antisymmetric both in color
and flavor. From the Dirac structure it is a parity-even spin-singlet.
Unlike the QCD vacuum, the chiral symmetry breaking is not caused by the pairing of left-handed
quarks with right-handed antiquarks, here the pairs are formed by quarks of equal chirality.
The breaking occurs due to the locking of the flavor-space and color-space rotation, and is manifest
in an order parameter of $\langle\bar\psi\bar\psi\psi\psi\rangle$ type. Even the $U(1)_B$ is broken
and reduced to $Z_2$, the quantum numbers of the associated order parameter matches the di-Lambda
condensate $\langle\Lambda\Lambda\rangle$ \cite{Schafer:1998ef}. While the gluon fields produce an octet of massive
spin 1 bosons through the Meissner effect and fermionic excitations have a gap
\cite{Son:1999cm,Son:2000tu}, there will also be Goldstone bosons associated
with the broken symmetries. Those associated with chiral symmetry breaking will be lifted
by the small quark masses, but the scalar singlet coming from the $U(1)_B$ stays massless
(baryon number superfluidity). 

There are several effective models that can address superconducting phases.
For example in Ref.~\cite{Casalbuoni:1999wu} the Goldstone bosons
are explicitly represented. The NJL model was also extended with the diquark channel
to describe superconductivity \cite{Gastineau:2001zke,Buballa:2003qv}.
The order of magnitude estimates for the superconducting gap
hint for the extension of these phases to finite temperature, as high as 50~MeV \cite{Alford:1997zt,Rapp:1997zu}. The emerging picture is shown in the high density side
of Fig.~\ref{fig:pdstephanov}, where color-flavor-locking is labeled as CFL. In this
figure, other possibilities at lower densities, such as the locking of the up and down
quarks with two of the colors in a 2SC phase is omitted, as its significance is questionable
for the most important inhabitants of the intermediate density part of the diagram, the charge neutral
neutron stars \cite{Alford:1997zt}. There are many other phases suggested for the high density
region of the phase diagram (see \cite{Buballa:2003qv} for a review). Presently, these are far
from the range of applicability of first principles non-perturbative approaches, such as lattice QCD.


\section{Conclusions}
\label{sec:conclusions}

The study of the QCD phase diagram over the past few decades has gathered an enormous
amount of information, from the theory community as well as from experiment. As a 
result, we can draw a quantitative picture of the phase structure and thermodynamic 
features of strongly interacting matter under extreme conditions.

The existence of the QGP has been established thanks to huge experimental efforts
at several generation of accelerator facilities, such as SPS, RHIC and LHC. 
Moreover, the experimental description of the QGP and of the conditions needed to 
produce it have developed to a precision science. Thanks mostly 
to lattice QCD simulations, the low net-density region of the phase diagram is 
quantitatively well established, with precise determinations of the QCD transition 
line and the equation of state. At the same time, other diagrammatic techniques
emerged in the form of the functional renormalization group and Dyson-Schwinger
equations that are in quantitative agreement with lattice simulations.

The search for the conjectured critical end-point has dominated the recent years of research in the community, with RHIC's beam energy scan program ultimately producing hints for its existence, though not unambiguous evidence. Various models, functional methods and indirect evidence from lattice
simulations suggest a possible location in the chemical potential range ($\mu_B \simeq 550 - 650 $~MeV) available to 
fixed-target heavy ion experiments. Current lattice methods do not extend to this range but, at the very least, the critical point's existence could be excluded up to $\mu_B = 450$~MeV. 

The structures on the phase diagram with physical quark masses can be viewed as a 
point in a quark-mass-dependent diagram (Columbia plot). 
How the physical critical point is connected with other critical lines at zero
chemical potential and smaller quark masses is known only in some effective
models. In view of the latest lattice results it is not clear whether such a
critical surface exists at all. If not the case, the existence of the critical point 
would be less likely. These findings call for further investigations,
which remain challenging not only because of the finite density, but the proximity 
to the chiral limit. 

In fact, although experiment and theory alike seem to be closing in on the QCD critical 
point, it should be stressed that it is still not settled whether it exists or not. 
In the coming years, more data from fixed-target experiments at RHIC and CBM 
will help clarify whether the hints we see now are indeed signatures of critical 
behavior. At the same time, theoretical predictions will become more accurate, 
refining the region where the critical point is more likely expected to be.
As the reach of lattice QCD continues to grow to larger chemical potentials,
thanks to new methods and more computing power, precise quantitative knowledge of
QCD phases keeps extending. 

The larger phase diagram of QCD, including finite isospin density or magnetic fields,
has also been thoroughly explored in recent years via lattice methods, and can now
provide a comprehensive picture on the thermodynamic properties of the theory in
more exotic scenarios, which are nonetheless relevant for experimental and 
astrophysical conditions. As a welcome by-product, the equation of state of QCD is 
now known over a very broad range of conditions, which is crucial for charting QCD 
phases, as well as for implementations in hydrodynamic simulations of different 
systems. 

Since the first observation of gravitational waves from a neutron star merger, 
stronger constraints have been placed on the equation of state of dense matter, 
which needs to be stiff enough to support very dense stars. Whether the 
inner cores of neutron stars are formed by hadrons, quark matter or other exotic 
phases of high-density QCD, and whether these phases are separated by a first order 
transition, remain crucial open questions.

\begin{ack}[Acknowledgments]%
We thank Anton Andronic, Roberta Arnaldi, Andrea Beraudo, Livio Bianchi, Gergely Endr\H odi, Tetyana Galatyuk, Tam\'as G. Kov\'acs, Jan Pawlowski,
P\'eter Petreczky and Enrico Scomparin for their feedback and 
enlightening discussions while preparing this manuscript.
\end{ack}


\bibliographystyle{Numbered-Style} 
\bibliography{thermo}

\begin{thebibliography*}{100}
\providecommand{\bibtype}[1]{}
\providecommand{\url}[1]{{\tt #1}}
\providecommand{\urlprefix}{URL }
\expandafter\ifx\csname urlstyle\endcsname\relax
  \providecommand{\doi}[1]{doi:\discretionary{}{}{}#1}\else
  \providecommand{\doi}{doi:\discretionary{}{}{}\begingroup
  \urlstyle{rm}\Url}\fi
\providecommand{\bibinfo}[2]{#2}
\providecommand{\eprint}[2][]{\url{#2}}
\makeatletter\def\@biblabel#1{\bibinfo{label}{[#1]}}\makeatother

\bibtype{Article}%
\bibitem{LIGOScientific:2017vwq}
\bibinfo{author}{B.~P. Abbott}, et al. (\bibinfo{collaboration}{LIGO
  Scientific, Virgo}), \bibinfo{title}{{GW170817: Observation of Gravitational
  Waves from a Binary Neutron Star Inspiral}}, \bibinfo{journal}{Phys. Rev.
  Lett.} \bibinfo{volume}{119} (\bibinfo{number}{16}) (\bibinfo{year}{2017})
  \bibinfo{pages}{161101}, \bibinfo{doi}{\doi{10.1103/PhysRevLett.119.161101}},
  \eprint{1710.05832}.

\bibtype{Article}%
\bibitem{LIGOScientific:2017ync}
\bibinfo{author}{B.~P. Abbott}, et al. (\bibinfo{collaboration}{LIGO
  Scientific, Virgo, Fermi GBM, INTEGRAL, IceCube, AstroSat Cadmium Zinc
  Telluride Imager Team, IPN, Insight-Hxmt, ANTARES, Swift, AGILE Team, 1M2H
  Team, Dark Energy Camera GW-EM, DES, DLT40, GRAWITA, Fermi-LAT, ATCA, ASKAP,
  Las Cumbres Observatory Group, OzGrav, DWF (Deeper Wider Faster Program),
  AST3, CAASTRO, VINROUGE, MASTER, J-GEM, GROWTH, JAGWAR, CaltechNRAO,
  TTU-NRAO, NuSTAR, Pan-STARRS, MAXI Team, TZAC Consortium, KU, Nordic Optical
  Telescope, ePESSTO, GROND, Texas Tech University, SALT Group, TOROS, BOOTES,
  MWA, CALET, IKI-GW Follow-up, H.E.S.S., LOFAR, LWA, HAWC, Pierre Auger, ALMA,
  Euro VLBI Team, Pi of Sky, Chandra Team at McGill University, DFN, ATLAS
  Telescopes, High Time Resolution Universe Survey, RIMAS, RATIR, SKA South
  Africa/MeerKAT}), \bibinfo{title}{{Multi-messenger Observations of a Binary
  Neutron Star Merger}}, \bibinfo{journal}{Astrophys. J. Lett.}
  \bibinfo{volume}{848} (\bibinfo{number}{2}) (\bibinfo{year}{2017})
  \bibinfo{pages}{L12}, \bibinfo{doi}{\doi{10.3847/2041-8213/aa91c9}},
  \eprint{1710.05833}.

\bibtype{Article}%
\bibitem{Aoki:2006we}
\bibinfo{author}{Y. Aoki}, \bibinfo{author}{G. Endrodi}, \bibinfo{author}{Z.
  Fodor}, \bibinfo{author}{S.D. Katz}, \bibinfo{author}{K.K. Szabo},
  \bibinfo{title}{{The Order of the quantum chromodynamics transition predicted
  by the standard model of particle physics}}, \bibinfo{journal}{Nature}
  \bibinfo{volume}{443} (\bibinfo{year}{2006}) \bibinfo{pages}{675--678},
  \bibinfo{doi}{\doi{10.1038/nature05120}}, \eprint{hep-lat/0611014}.

\bibtype{Article}%
\bibitem{Borsanyi:2010bp}
\bibinfo{author}{Szabolcs Borsanyi}, et al.
  (\bibinfo{collaboration}{Wuppertal-Budapest Collaboration}),
  \bibinfo{title}{{Is there still any $T_c$ mystery in lattice QCD? Results
  with physical masses in the continuum limit III}}, \bibinfo{journal}{JHEP}
  \bibinfo{volume}{1009} (\bibinfo{year}{2010}) \bibinfo{pages}{073},
  \bibinfo{doi}{\doi{10.1007/JHEP09(2010)073}}, \eprint{1005.3508}.

\bibtype{Article}%
\bibitem{Bazavov:2011nk}
\bibinfo{author}{A. Bazavov}, \bibinfo{author}{T. Bhattacharya},
  \bibinfo{author}{M. Cheng}, \bibinfo{author}{C. DeTar}, \bibinfo{author}{H.T.
  Ding}, et al., \bibinfo{title}{{The chiral and deconfinement aspects of the
  QCD transition}}, \bibinfo{journal}{Phys.Rev.} \bibinfo{volume}{D85}
  (\bibinfo{year}{2012}) \bibinfo{pages}{054503},
  \bibinfo{doi}{\doi{10.1103/PhysRevD.85.054503}}, \eprint{1111.1710}.

\bibtype{Article}%
\bibitem{Borsanyi:2020fev}
\bibinfo{author}{Szabolcs Borsanyi}, \bibinfo{author}{Zoltan Fodor},
  \bibinfo{author}{Jana~N. Guenther}, \bibinfo{author}{Ruben Kara},
  \bibinfo{author}{Sandor~D. Katz}, \bibinfo{author}{Paolo Parotto},
  \bibinfo{author}{Attila Pasztor}, \bibinfo{author}{Claudia Ratti},
  \bibinfo{author}{Kalman~K. Szabo}, \bibinfo{title}{{The QCD crossover at
  finite chemical potential from lattice simulations}}, \bibinfo{journal}{Phys.
  Rev. Lett.} \bibinfo{volume}{125} (\bibinfo{year}{2020})
  \bibinfo{pages}{052001}, \eprint{2002.02821}.

\bibtype{Article}%
\bibitem{Cleymans:1998fq}
\bibinfo{author}{J. Cleymans}, \bibinfo{author}{K. Redlich},
  \bibinfo{title}{{Unified description of freezeout parameters in relativistic
  heavy ion collisions}}, \bibinfo{journal}{Phys. Rev. Lett.}
  \bibinfo{volume}{81} (\bibinfo{year}{1998}) \bibinfo{pages}{5284--5286},
  \bibinfo{doi}{\doi{10.1103/PhysRevLett.81.5284}}, \eprint{nucl-th/9808030}.

\bibtype{Article}%
\bibitem{Vovchenko:2015idt}
\bibinfo{author}{V. Vovchenko}, \bibinfo{author}{V.~V. Begun},
  \bibinfo{author}{M.~I. Gorenstein}, \bibinfo{title}{{Hadron multiplicities
  and chemical freeze-out conditions in proton-proton and nucleus-nucleus
  collisions}}, \bibinfo{journal}{Phys. Rev.} \bibinfo{volume}{C93}
  (\bibinfo{number}{6}) (\bibinfo{year}{2016}) \bibinfo{pages}{064906},
  \bibinfo{doi}{\doi{10.1103/PhysRevC.93.064906}}, \eprint{1512.08025}.

\bibtype{Article}%
\bibitem{Becattini:2016xct}
\bibinfo{author}{F. Becattini}, \bibinfo{author}{J. Steinheimer},
  \bibinfo{author}{R. Stock}, \bibinfo{author}{M. Bleicher},
  \bibinfo{title}{{Hadronization conditions in relativistic nuclear collisions
  and the QCD pseudo-critical line}}, \bibinfo{journal}{Phys. Lett. B}
  \bibinfo{volume}{764} (\bibinfo{year}{2017}) \bibinfo{pages}{241--246},
  \bibinfo{doi}{\doi{10.1016/j.physletb.2016.11.033}}, \eprint{1605.09694}.

\bibtype{Article}%
\bibitem{Vovchenko:2018fmh}
\bibinfo{author}{Volodymyr Vovchenko}, \bibinfo{author}{Mark~I. Gorenstein},
  \bibinfo{author}{Horst Stoecker}, \bibinfo{title}{{Finite resonance widths
  influence the thermal-model description of hadron yields}},
  \bibinfo{journal}{Phys. Rev. C} \bibinfo{volume}{98} (\bibinfo{number}{3})
  (\bibinfo{year}{2018}) \bibinfo{pages}{034906},
  \bibinfo{doi}{\doi{10.1103/PhysRevC.98.034906}}, \eprint{1807.02079}.

\bibtype{Article}%
\bibitem{STAR:2017sal}
\bibinfo{author}{L. Adamczyk}, et al. (\bibinfo{collaboration}{STAR}),
  \bibinfo{title}{{Bulk Properties of the Medium Produced in Relativistic
  Heavy-Ion Collisions from the Beam Energy Scan Program}},
  \bibinfo{journal}{Phys. Rev. C} \bibinfo{volume}{96} (\bibinfo{number}{4})
  (\bibinfo{year}{2017}) \bibinfo{pages}{044904},
  \bibinfo{doi}{\doi{10.1103/PhysRevC.96.044904}}, \eprint{1701.07065}.

\bibtype{Article}%
\bibitem{Andronic:2017pug}
\bibinfo{author}{Anton Andronic}, \bibinfo{author}{Peter Braun-Munzinger},
  \bibinfo{author}{Krzysztof Redlich}, \bibinfo{author}{Johanna Stachel},
  \bibinfo{title}{{Decoding the phase structure of QCD via particle production
  at high energy}}, \bibinfo{journal}{Nature} \bibinfo{volume}{561}
  (\bibinfo{number}{7723}) (\bibinfo{year}{2018}) \bibinfo{pages}{321--330},
  \bibinfo{doi}{\doi{10.1038/s41586-018-0491-6}}, \eprint{1710.09425}.

\bibtype{Article}%
\bibitem{Lysenko:2024hqp}
\bibinfo{author}{Artemiy Lysenko}, \bibinfo{author}{Mark~I. Gorenstein},
  \bibinfo{author}{Roman Poberezhniuk}, \bibinfo{author}{Volodymyr Vovchenko},
  \bibinfo{title}{{Chemical freeze-out curve in heavy-ion collisions and the
  QCD critical point}}, \bibinfo{journal}{Phys. Rev. C} \bibinfo{volume}{111}
  (\bibinfo{number}{5}) (\bibinfo{year}{2025}) \bibinfo{pages}{054903},
  \bibinfo{doi}{\doi{10.1103/PhysRevC.111.054903}}, \eprint{2408.06473}.

\bibtype{Article}%
\bibitem{Elliott:2013pna}
\bibinfo{author}{J.~B. Elliott}, \bibinfo{author}{P.~T. Lake},
  \bibinfo{author}{L.~G. Moretto}, \bibinfo{author}{L. Phair},
  \bibinfo{title}{{Determination of the coexistence curve, critical
  temperature, density, and pressure of bulk nuclear matter from fragment
  emission data}}, \bibinfo{journal}{Phys. Rev. C} \bibinfo{volume}{87}
  (\bibinfo{number}{5}) (\bibinfo{year}{2013}) \bibinfo{pages}{054622},
  \bibinfo{doi}{\doi{10.1103/PhysRevC.87.054622}}.

\bibtype{Article}%
\bibitem{MUSES:2023hyz}
\bibinfo{author}{Rajesh Kumar}, et al. (\bibinfo{collaboration}{MUSES}),
  \bibinfo{title}{{Theoretical and experimental constraints for the equation of
  state of dense and hot matter}}, \bibinfo{journal}{Living Rev. Rel.}
  \bibinfo{volume}{27} (\bibinfo{number}{1}) (\bibinfo{year}{2024})
  \bibinfo{pages}{3}, \bibinfo{doi}{\doi{10.1007/s41114-024-00049-6}},
  \eprint{2303.17021}.

\bibtype{Article}%
\bibitem{Gross:2022hyw}
\bibinfo{author}{Franz Gross}, et al., \bibinfo{title}{{50 Years of Quantum
  Chromodynamics}}, \bibinfo{journal}{Eur. Phys. J. C} \bibinfo{volume}{83}
  (\bibinfo{year}{2023}) \bibinfo{pages}{1125},
  \bibinfo{doi}{\doi{10.1140/epjc/s10052-023-11949-2}}, \eprint{2212.11107}.

\bibtype{Article}%
\bibitem{Durr:2008zz}
\bibinfo{author}{S. Durr}, \bibinfo{author}{Z. Fodor}, \bibinfo{author}{J.
  Frison}, \bibinfo{author}{C. Hoelbling}, \bibinfo{author}{R. Hoffmann}, et
  al., \bibinfo{title}{{Ab-Initio Determination of Light Hadron Masses}},
  \bibinfo{journal}{Science} \bibinfo{volume}{322} (\bibinfo{year}{2008})
  \bibinfo{pages}{1224--1227}, \bibinfo{doi}{\doi{10.1126/science.1163233}},
  \eprint{0906.3599}.

\bibtype{Article}%
\bibitem{BMW:2014pzb}
\bibinfo{author}{Sz. Borsanyi}, \bibinfo{author}{S. Durr}, \bibinfo{author}{Z.
  Fodor}, \bibinfo{author}{C. Hoelbling}, \bibinfo{author}{S.D. Katz}, et al.,
  \bibinfo{title}{{Ab initio calculation of the neutron-proton mass
  difference}}, \bibinfo{journal}{Science} \bibinfo{volume}{347}
  (\bibinfo{year}{2015}) \bibinfo{pages}{1452--1455},
  \bibinfo{doi}{\doi{10.1126/science.1257050}}, \eprint{1406.4088}.

\bibtype{Article}%
\bibitem{Lahde:2013uqa}
\bibinfo{author}{Timo~A. L\"ahde}, \bibinfo{author}{Evgeny Epelbaum},
  \bibinfo{author}{Hermann Krebs}, \bibinfo{author}{Dean Lee},
  \bibinfo{author}{Ulf-G. Mei\ss{}ner}, \bibinfo{author}{Gautam Rupak},
  \bibinfo{title}{{Lattice Effective Field Theory for Medium-Mass Nuclei}},
  \bibinfo{journal}{Phys. Lett. B} \bibinfo{volume}{732} (\bibinfo{year}{2014})
  \bibinfo{pages}{110--115},
  \bibinfo{doi}{\doi{10.1016/j.physletb.2014.03.023}}, \eprint{1311.0477}.

\bibtype{Article}%
\bibitem{Ding:2019prx}
\bibinfo{author}{H.~T. Ding}, et al., \bibinfo{title}{{Chiral phase transition
  temperature in (2+1)-flavor QCD}}, \bibinfo{journal}{Phys. Rev. Lett.}
  \bibinfo{volume}{123} (\bibinfo{year}{2019}) \bibinfo{pages}{062002},
  \bibinfo{doi}{\doi{10.1103/PhysRevLett.123.062002}}, \eprint{1903.04801}.

\bibtype{Article}%
\bibitem{Stephanov:2006dn}
\bibinfo{author}{M.~A. Stephanov}, \bibinfo{title}{{QCD critical point and
  complex chemical potential singularities}}, \bibinfo{journal}{Phys. Rev.}
  \bibinfo{volume}{D73} (\bibinfo{year}{2006}) \bibinfo{pages}{094508},
  \bibinfo{doi}{\doi{10.1103/PhysRevD.73.094508}}, \eprint{hep-lat/0603014}.

\bibtype{Article}%
\bibitem{Bonati:2015bha}
\bibinfo{author}{Claudio Bonati}, \bibinfo{author}{Massimo D'Elia},
  \bibinfo{author}{Marco Mariti}, \bibinfo{author}{Michele Mesiti},
  \bibinfo{author}{Francesco Negro}, \bibinfo{author}{Francesco Sanfilippo},
  \bibinfo{title}{{Curvature of the chiral pseudocritical line in QCD:
  Continuum extrapolated results}}, \bibinfo{journal}{Phys. Rev.}
  \bibinfo{volume}{D92} (\bibinfo{number}{5}) (\bibinfo{year}{2015})
  \bibinfo{pages}{054503}, \bibinfo{doi}{\doi{10.1103/PhysRevD.92.054503}},
  \eprint{1507.03571}.

\bibtype{Article}%
\bibitem{Bellwied:2015rza}
\bibinfo{author}{R. Bellwied}, \bibinfo{author}{S. Borsanyi},
  \bibinfo{author}{Z. Fodor}, \bibinfo{author}{J. G{\"u}nther},
  \bibinfo{author}{S.~D. Katz}, \bibinfo{author}{C. Ratti},
  \bibinfo{author}{K.~K. Szabo}, \bibinfo{title}{{The QCD phase diagram from
  analytic continuation}}, \bibinfo{journal}{Phys. Lett.}
  \bibinfo{volume}{B751} (\bibinfo{year}{2015}) \bibinfo{pages}{559--564},
  \bibinfo{doi}{\doi{10.1016/j.physletb.2015.11.011}}, \eprint{1507.07510}.

\bibtype{Article}%
\bibitem{Bazavov:2018mes}
\bibinfo{author}{A. Bazavov}, et al., \bibinfo{title}{{Chiral crossover in QCD
  at zero and non-zero chemical potentials}}, \bibinfo{journal}{Physics Letters
  B} \bibinfo{volume}{795} (\bibinfo{year}{2019}) \bibinfo{pages}{15--21},
  \eprint{1812.08235}.

\bibtype{Article}%
\bibitem{Borsanyi:2025dyp}
\bibinfo{author}{Szabolcs Borsanyi}, \bibinfo{author}{Zoltan Fodor},
  \bibinfo{author}{Jana~N. Guenther}, \bibinfo{author}{Paolo Parotto},
  \bibinfo{author}{Attila Pasztor}, \bibinfo{author}{Claudia Ratti},
  \bibinfo{author}{Volodymyr Vovchenko}, \bibinfo{author}{Chik~Him Wong},
  \bibinfo{title}{{Lattice QCD constraints on the critical point from an
  improved precision equation of state}}  (\bibinfo{year}{2025}),
  \eprint{2502.10267}.

\bibtype{Article}%
\bibitem{Stephanov:1999zu}
\bibinfo{author}{Misha~A. Stephanov}, \bibinfo{author}{K. Rajagopal},
  \bibinfo{author}{Edward~V. Shuryak}, \bibinfo{title}{{Event-by-event
  fluctuations in heavy ion collisions and the QCD critical point}},
  \bibinfo{journal}{Phys.Rev.} \bibinfo{volume}{D60} (\bibinfo{year}{1999})
  \bibinfo{pages}{114028}, \bibinfo{doi}{\doi{10.1103/PhysRevD.60.114028}},
  \eprint{hep-ph/9903292}.

\bibtype{Article}%
\bibitem{Stephanov:2008qz}
\bibinfo{author}{M.A. Stephanov}, \bibinfo{title}{{Non-Gaussian fluctuations
  near the QCD critical point}}, \bibinfo{journal}{Phys.Rev.Lett.}
  \bibinfo{volume}{102} (\bibinfo{year}{2009}) \bibinfo{pages}{032301},
  \bibinfo{doi}{\doi{10.1103/PhysRevLett.102.032301}}, \eprint{0809.3450}.

\bibtype{Article}%
\bibitem{Stephanov:2011pb}
\bibinfo{author}{M.A. Stephanov}, \bibinfo{title}{{On the sign of kurtosis near
  the QCD critical point}}, \bibinfo{journal}{Phys.Rev.Lett.}
  \bibinfo{volume}{107} (\bibinfo{year}{2011}) \bibinfo{pages}{052301},
  \bibinfo{doi}{\doi{10.1103/PhysRevLett.107.052301}}, \eprint{1104.1627}.

\bibtype{Article}%
\bibitem{Adamczyk:2013dal}
\bibinfo{author}{L. Adamczyk}, et al. (\bibinfo{collaboration}{STAR
  Collaboration}), \bibinfo{title}{{Energy Dependence of Moments of Net-proton
  Multiplicity Distributions at RHIC}}, \bibinfo{journal}{Phys.Rev.Lett.}
  \bibinfo{volume}{112} (\bibinfo{year}{2014}) \bibinfo{pages}{032302},
  \bibinfo{doi}{\doi{10.1103/PhysRevLett.112.032302}}, \eprint{1309.5681}.

\bibtype{Article}%
\bibitem{STAR:2021rls}
\bibinfo{author}{Mohamed Abdallah}, et al. (\bibinfo{collaboration}{STAR}),
  \bibinfo{title}{{Measurement of the Sixth-Order Cumulant of Net-Proton
  Multiplicity Distributions in Au+Au Collisions at $\sqrt{s_{NN}}=$ 27, 54.4,
  and 200 GeV at RHIC}}, \bibinfo{journal}{Phys. Rev. Lett.}
  \bibinfo{volume}{127} (\bibinfo{number}{26}) (\bibinfo{year}{2021})
  \bibinfo{pages}{262301}, \bibinfo{doi}{\doi{10.1103/PhysRevLett.127.262301}},
  \eprint{2105.14698}.

\bibtype{Article}%
\bibitem{STAR:2021iop}
\bibinfo{author}{Mohamed Abdallah}, et al. (\bibinfo{collaboration}{STAR}),
  \bibinfo{title}{{Cumulants and correlation functions of net-proton, proton,
  and antiproton multiplicity distributions in Au+Au collisions at energies
  available at the BNL Relativistic Heavy Ion Collider}},
  \bibinfo{journal}{Phys. Rev. C} \bibinfo{volume}{104} (\bibinfo{number}{2})
  (\bibinfo{year}{2021}) \bibinfo{pages}{024902},
  \bibinfo{doi}{\doi{10.1103/PhysRevC.104.024902}}, \bibinfo{note}{[Erratum:
  Phys.Rev.C 111, 029902 (2025)]}, \eprint{2101.12413}.

\bibtype{Article}%
\bibitem{STAR:2025zdq}
\bibinfo{author}{B.~E. Aboona}, et al. (\bibinfo{collaboration}{STAR}),
  \bibinfo{title}{{Precision Measurement of Net-Proton-Number Fluctuations in
  Au+Au Collisions at RHIC}}, \bibinfo{journal}{Phys. Rev. Lett.}
  \bibinfo{volume}{135} (\bibinfo{number}{14}) (\bibinfo{year}{2025})
  \bibinfo{pages}{142301}, \bibinfo{doi}{\doi{10.1103/9l69-2d7p}},
  \eprint{2504.00817}.

\bibtype{Article}%
\bibitem{Lu:2019nbg}
\bibinfo{author}{Bing-Nan Lu}, \bibinfo{author}{Ning Li},
  \bibinfo{author}{Serdar Elhatisari}, \bibinfo{author}{Dean Lee},
  \bibinfo{author}{Joaqu{\'\i}n~E. Drut}, \bibinfo{author}{Timo~A. L{\"a}hde},
  \bibinfo{author}{Evgeny Epelbaum}, \bibinfo{author}{Ulf-G. Mei{\ss}ner},
  \bibinfo{title}{{$Ab Initio$ Nuclear Thermodynamics}},
  \bibinfo{journal}{Phys. Rev. Lett.} \bibinfo{volume}{125}
  (\bibinfo{number}{19}) (\bibinfo{year}{2020}) \bibinfo{pages}{192502},
  \bibinfo{doi}{\doi{10.1103/PhysRevLett.125.192502}}, \eprint{1912.05105}.

\bibtype{Article}%
\bibitem{Fukushima:2023wnl}
\bibinfo{author}{Kenji Fukushima}, \bibinfo{author}{Jan Horak},
  \bibinfo{author}{Jan~M. Pawlowski}, \bibinfo{author}{Nicolas Wink},
  \bibinfo{author}{Carl~Philipp Zelle}, \bibinfo{title}{{Nuclear liquid-gas
  transition in QCD}}, \bibinfo{journal}{Phys. Rev. D} \bibinfo{volume}{110}
  (\bibinfo{number}{7}) (\bibinfo{year}{2024}) \bibinfo{pages}{076022},
  \bibinfo{doi}{\doi{10.1103/PhysRevD.110.076022}}, \eprint{2308.16594}.

\bibtype{Article}%
\bibitem{Du:2024wjm}
\bibinfo{author}{Lipei Du}, \bibinfo{author}{Agnieszka Sorensen},
  \bibinfo{author}{Mikhail Stephanov}, \bibinfo{title}{{The QCD phase diagram
  and Beam Energy Scan physics: a theory overview}}, \bibinfo{journal}{Int. J.
  Mod. Phys. E} \bibinfo{volume}{33} (\bibinfo{number}{07})
  (\bibinfo{year}{2024}) \bibinfo{pages}{2430008},
  \bibinfo{doi}{\doi{10.1142/S021830132430008X}}, \eprint{2402.10183}.

\bibtype{Article}%
\bibitem{Harris:2023tti}
\bibinfo{author}{John~W. Harris}, \bibinfo{author}{Berndt M\"uller},
  \bibinfo{title}{{''QGP Signatures'' Revisited}}, \bibinfo{journal}{Eur. Phys.
  J. C} \bibinfo{volume}{84} (\bibinfo{number}{3}) (\bibinfo{year}{2024})
  \bibinfo{pages}{247}, \bibinfo{doi}{\doi{10.1140/epjc/s10052-024-12533-y}},
  \eprint{2308.05743}.

\bibtype{Article}%
\bibitem{Plumberg:2024leb}
\bibinfo{author}{Christopher Plumberg}, et al., \bibinfo{title}{{Conservation
  of B, S, and Q charges in relativistic viscous hydrodynamics solved with
  smoothed particle hydrodynamics}}, \bibinfo{journal}{Phys. Rev. C}
  \bibinfo{volume}{111} (\bibinfo{number}{4}) (\bibinfo{year}{2025})
  \bibinfo{pages}{044905}, \bibinfo{doi}{\doi{10.1103/PhysRevC.111.044905}},
  \eprint{2405.09648}.

\bibtype{Article}%
\bibitem{Gale:2013da}
\bibinfo{author}{Charles Gale}, \bibinfo{author}{Sangyong Jeon},
  \bibinfo{author}{Bjoern Schenke}, \bibinfo{title}{{Hydrodynamic Modeling of
  Heavy-Ion Collisions}}, \bibinfo{journal}{Int. J. Mod. Phys.}
  \bibinfo{volume}{A28} (\bibinfo{year}{2013}) \bibinfo{pages}{1340011},
  \bibinfo{doi}{\doi{10.1142/S0217751X13400113}}, \eprint{1301.5893}.

\bibtype{Article}%
\bibitem{An:2021wof}
\bibinfo{author}{Xin An}, et al., \bibinfo{title}{{The BEST framework for the
  search for the QCD critical point and the chiral magnetic effect}},
  \bibinfo{journal}{Nucl. Phys. A} \bibinfo{volume}{1017}
  (\bibinfo{year}{2022}) \bibinfo{pages}{122343},
  \bibinfo{doi}{\doi{10.1016/j.nuclphysa.2021.122343}}, \eprint{2108.13867}.

\bibtype{Article}%
\bibitem{Lisa:2016buz}
\bibinfo{author}{Mike Lisa}, \bibinfo{title}{{Timescales in heavy ion
  collisions}}, \bibinfo{journal}{Acta Phys. Polon. B} \bibinfo{volume}{47}
  (\bibinfo{year}{2016}) \bibinfo{pages}{1847},
  \bibinfo{doi}{\doi{10.5506/APhysPolB.47.1847}}, \eprint{1607.06188}.

\bibtype{Article}%
\bibitem{ALICE:2022wpn}
\bibinfo{author}{Shreyasi Acharya}, et al. (\bibinfo{collaboration}{ALICE}),
  \bibinfo{title}{{The ALICE experiment: a journey through QCD}},
  \bibinfo{journal}{Eur. Phys. J. C} \bibinfo{volume}{84} (\bibinfo{number}{8})
  (\bibinfo{year}{2024}) \bibinfo{pages}{813},
  \bibinfo{doi}{\doi{10.1140/epjc/s10052-024-12935-y}}, \eprint{2211.04384}.

\bibtype{Article}%
\bibitem{CMS:2017uuv}
\bibinfo{author}{Albert~M. Sirunyan}, et al. (\bibinfo{collaboration}{CMS}),
  \bibinfo{title}{{Measurement of prompt and nonprompt charmonium suppression
  in $\text {PbPb}$ collisions at 5.02 $\,\text {Te}\text {V}$}},
  \bibinfo{journal}{Eur. Phys. J. C} \bibinfo{volume}{78} (\bibinfo{number}{6})
  (\bibinfo{year}{2018}) \bibinfo{pages}{509},
  \bibinfo{doi}{\doi{10.1140/epjc/s10052-018-5950-6}}, \bibinfo{note}{[Erratum:
  Eur.Phys.J.C 83, 145 (2023)]}, \eprint{1712.08959}.

\bibtype{Article}%
\bibitem{ALICE:2018rtz}
\bibinfo{author}{S. Acharya}, et al. (\bibinfo{collaboration}{ALICE}),
  \bibinfo{title}{{Energy dependence and fluctuations of anisotropic flow in
  Pb-Pb collisions at $ \sqrt{s_{\mathrm{NN}}}=5.02 $ and 2.76 TeV}},
  \bibinfo{journal}{JHEP} \bibinfo{volume}{07} (\bibinfo{year}{2018})
  \bibinfo{pages}{103}, \bibinfo{doi}{\doi{10.1007/JHEP07(2018)103}},
  \eprint{1804.02944}.

\bibtype{Article}%
\bibitem{Harris:1996zx}
\bibinfo{author}{John~W. Harris}, \bibinfo{author}{Berndt Muller},
  \bibinfo{title}{{The Search for the quark - gluon plasma}},
  \bibinfo{journal}{Ann. Rev. Nucl. Part. Sci.} \bibinfo{volume}{46}
  (\bibinfo{year}{1996}) \bibinfo{pages}{71--107},
  \bibinfo{doi}{\doi{10.1146/annurev.nucl.46.1.71}}, \eprint{hep-ph/9602235}.

\bibtype{Article}%
\bibitem{Schenke:2020mbo}
\bibinfo{author}{Bjoern Schenke}, \bibinfo{author}{Chun Shen},
  \bibinfo{author}{Prithwish Tribedy}, \bibinfo{title}{{Running the gamut of
  high energy nuclear collisions}}, \bibinfo{journal}{Phys. Rev. C}
  \bibinfo{volume}{102} (\bibinfo{number}{4}) (\bibinfo{year}{2020})
  \bibinfo{pages}{044905}, \bibinfo{doi}{\doi{10.1103/PhysRevC.102.044905}},
  \eprint{2005.14682}.

\bibtype{Techreport}%
\bibitem{Ahdida:2932302}
\bibinfo{author}{C Ahdida}, \bibinfo{author}{G Alocco}, \bibinfo{author}{M
  Arba}, \bibinfo{author}{R Arnaldi}, \bibinfo{author}{S Beole},
  \bibinfo{author}{J Bernhard}, \bibinfo{author}{L Bianchi}, \bibinfo{author}{I
  Bilinskyi}, \bibinfo{author}{E Borisova}, \bibinfo{author}{S Bressler},
  \bibinfo{author}{S Bufalino}, \bibinfo{author}{R Cerri}, \bibinfo{author}{M
  Ciacco}, \bibinfo{author}{C Cicalo}, \bibinfo{author}{S Coli},
  \bibinfo{author}{P Cortese}, \bibinfo{author}{A Dainese}, \bibinfo{author}{H
  Danielsson}, \bibinfo{author}{J Datta}, \bibinfo{author}{A De~Falco},
  \bibinfo{author}{A Drees}, \bibinfo{author}{L Epshteyn}, \bibinfo{author}{A
  Ferretti}, \bibinfo{author}{F Fionda}, \bibinfo{author}{M Gagliardi},
  \bibinfo{author}{G~M Galimberti}, \bibinfo{author}{S Girod},
  \bibinfo{author}{T Gunji}, \bibinfo{author}{G Jin}, \bibinfo{author}{F
  Geurts}, \bibinfo{author}{L Hu}, \bibinfo{author}{L Levinson},
  \bibinfo{author}{F Li}, \bibinfo{author}{W Li}, \bibinfo{author}{Z Liu},
  \bibinfo{author}{D Marras}, \bibinfo{author}{M Masera}, \bibinfo{author}{A
  Masoni}, \bibinfo{author}{F Mazzaschi}, \bibinfo{author}{P Mereu},
  \bibinfo{author}{J Metselaar}, \bibinfo{author}{L Micheletti},
  \bibinfo{author}{A Milov}, \bibinfo{author}{L Mirasola}, \bibinfo{author}{A
  Mulliri}, \bibinfo{author}{L Musa}, \bibinfo{author}{E Nowak},
  \bibinfo{author}{C Oppedisano}, \bibinfo{author}{N Pacifico},
  \bibinfo{author}{M Ploskon}, \bibinfo{author}{T Prebibaj}, \bibinfo{author}{F
  Prino}, \bibinfo{author}{M Puccio}, \bibinfo{author}{C Puggioni},
  \bibinfo{author}{A Rossi}, \bibinfo{author}{V Sarritzu}, \bibinfo{author}{B
  Schmidt}, \bibinfo{author}{E Scomparin}, \bibinfo{author}{D Sekihata},
  \bibinfo{author}{Q Shou}, \bibinfo{author}{R Shahoyan}, \bibinfo{author}{M
  Shoa}, \bibinfo{author}{S Siddhanta}, \bibinfo{author}{X Su},
  \bibinfo{author}{Z Tang}, \bibinfo{author}{S Trogolo}, \bibinfo{author}{M
  Tuveri}, \bibinfo{author}{A Uras}, \bibinfo{author}{G Usai},
  \bibinfo{author}{M Van~Dijk}, \bibinfo{author}{E Vercellin},
  \bibinfo{author}{I Vorobyev}, \bibinfo{author}{B Yankovsky},
  \bibinfo{author}{D Zavazieva} (\bibinfo{collaboration}{NA60+/DiCE}),
  \bibinfo{title}{{NA60+/DiCE: study of rare probes of the Quark-Gluon Plasma
  at SPS energies}}, \bibinfo{type}{\bibinfo{comment}{tech. rep.}},
  \bibinfo{institution}{CERN}, \bibinfo{address}{Geneva} \bibinfo{year}{2025},
  \bibinfo{url}{\urlprefix\url{https://cds.cern.ch/record/2932302}}.

\bibtype{Article}%
\bibitem{Matsui:1986dk}
\bibinfo{author}{T. Matsui}, \bibinfo{author}{H. Satz},
  \bibinfo{title}{{$J/\psi$ Suppression by Quark-Gluon Plasma Formation}},
  \bibinfo{journal}{Phys.Lett.} \bibinfo{volume}{B178} (\bibinfo{year}{1986})
  \bibinfo{pages}{416}, \bibinfo{doi}{\doi{10.1016/0370-2693(86)91404-8}}.

\bibtype{Article}%
\bibitem{Andronic:2025jbp}
\bibinfo{author}{Anton Andronic}, \bibinfo{author}{Roberta Arnaldi},
  \bibinfo{title}{{Quarkonia and Deconfined Quark{\textendash}Gluon Matter in
  Heavy-Ion Collisions}}, \bibinfo{journal}{Ann. Rev. Nucl. Part. Sci.}
  \bibinfo{volume}{75} (\bibinfo{number}{1}) (\bibinfo{year}{2025})
  \bibinfo{pages}{351--375},
  \bibinfo{doi}{\doi{10.1146/annurev-nucl-121423-101041}}, \eprint{2501.08290}.

\bibtype{Article}%
\bibitem{Bazavov:2023dci}
\bibinfo{author}{Alexei Bazavov}, \bibinfo{author}{Daniel Hoying},
  \bibinfo{author}{Rasmus~N. Larsen}, \bibinfo{author}{Swagato Mukherjee},
  \bibinfo{author}{Peter Petreczky}, \bibinfo{author}{Alexander Rothkopf},
  \bibinfo{author}{Johannes~Heinrich Weber} (\bibinfo{collaboration}{HotQCD}),
  \bibinfo{title}{{Unscreened forces in the quark-gluon plasma?}},
  \bibinfo{journal}{Phys. Rev. D} \bibinfo{volume}{109} (\bibinfo{number}{7})
  (\bibinfo{year}{2024}) \bibinfo{pages}{074504},
  \bibinfo{doi}{\doi{10.1103/PhysRevD.109.074504}}, \eprint{2308.16587}.

\bibtype{Article}%
\bibitem{Laine:2006ns}
\bibinfo{author}{M. Laine}, \bibinfo{author}{O. Philipsen}, \bibinfo{author}{P.
  Romatschke}, \bibinfo{author}{M. Tassler}, \bibinfo{title}{{Real-time static
  potential in hot QCD}}, \bibinfo{journal}{JHEP} \bibinfo{volume}{03}
  (\bibinfo{year}{2007}) \bibinfo{pages}{054},
  \bibinfo{doi}{\doi{10.1088/1126-6708/2007/03/054}}, \eprint{hep-ph/0611300}.

\bibtype{Article}%
\bibitem{Beraudo:2007ky}
\bibinfo{author}{A. Beraudo}, \bibinfo{author}{J.~P. Blaizot},
  \bibinfo{author}{C. Ratti}, \bibinfo{title}{{Real and imaginary-time Q anti-Q
  correlators in a thermal medium}}, \bibinfo{journal}{Nucl. Phys.}
  \bibinfo{volume}{A806} (\bibinfo{year}{2008}) \bibinfo{pages}{312--338},
  \bibinfo{doi}{\doi{10.1016/j.nuclphysa.2008.03.001}}, \eprint{0712.4394}.

\bibtype{Article}%
\bibitem{Akamatsu:2014qsa}
\bibinfo{author}{Yukinao Akamatsu}, \bibinfo{title}{{Heavy quark master
  equations in the Lindblad form at high temperatures}},
  \bibinfo{journal}{Phys. Rev. D} \bibinfo{volume}{91} (\bibinfo{number}{5})
  (\bibinfo{year}{2015}) \bibinfo{pages}{056002},
  \bibinfo{doi}{\doi{10.1103/PhysRevD.91.056002}}, \eprint{1403.5783}.

\bibtype{Article}%
\bibitem{Brambilla:2016wgg}
\bibinfo{author}{Nora Brambilla}, \bibinfo{author}{Miguel~A. Escobedo},
  \bibinfo{author}{Joan Soto}, \bibinfo{author}{Antonio Vairo},
  \bibinfo{title}{{Quarkonium suppression in heavy-ion collisions: an open
  quantum system approach}}, \bibinfo{journal}{Phys. Rev. D}
  \bibinfo{volume}{96} (\bibinfo{number}{3}) (\bibinfo{year}{2017})
  \bibinfo{pages}{034021}, \bibinfo{doi}{\doi{10.1103/PhysRevD.96.034021}},
  \eprint{1612.07248}.

\bibtype{Article}%
\bibitem{Akamatsu:2020ypb}
\bibinfo{author}{Yukinao Akamatsu}, \bibinfo{title}{{Quarkonium in
  quark{\textendash}gluon plasma: Open quantum system approaches re-examined}},
  \bibinfo{journal}{Prog. Part. Nucl. Phys.} \bibinfo{volume}{123}
  (\bibinfo{year}{2022}) \bibinfo{pages}{103932},
  \bibinfo{doi}{\doi{10.1016/j.ppnp.2021.103932}}, \eprint{2009.10559}.

\bibtype{Article}%
\bibitem{Brambilla:2023hkw}
\bibinfo{author}{Nora Brambilla}, \bibinfo{author}{Miguel~{\'A}ngel Escobedo},
  \bibinfo{author}{Ajaharul Islam}, \bibinfo{author}{Michael Strickland},
  \bibinfo{author}{Anurag Tiwari}, \bibinfo{author}{Antonio Vairo},
  \bibinfo{author}{Peter Vander~Griend}, \bibinfo{title}{{Regeneration of
  bottomonia in an open quantum systems approach}}, \bibinfo{journal}{Phys.
  Rev. D} \bibinfo{volume}{108} (\bibinfo{number}{1}) (\bibinfo{year}{2023})
  \bibinfo{pages}{L011502}, \bibinfo{doi}{\doi{10.1103/PhysRevD.108.L011502}},
  \eprint{2302.11826}.

\bibtype{Article}%
\bibitem{ATLAS:2018ezv}
\bibinfo{author}{Morad Aaboud}, et al. (\bibinfo{collaboration}{ATLAS}),
  \bibinfo{title}{{Measurement of the azimuthal anisotropy of charged particles
  produced in $\sqrt{s_{_\text {NN}}}$ = 5.02 TeV Pb+Pb collisions with the
  ATLAS detector}}, \bibinfo{journal}{Eur. Phys. J. C} \bibinfo{volume}{78}
  (\bibinfo{number}{12}) (\bibinfo{year}{2018}) \bibinfo{pages}{997},
  \bibinfo{doi}{\doi{10.1140/epjc/s10052-018-6468-7}}, \eprint{1808.03951}.

\bibtype{Article}%
\bibitem{Shuryak:1978ij}
\bibinfo{author}{Edward~V. Shuryak}, \bibinfo{title}{{Quark-Gluon Plasma and
  Hadronic Production of Leptons, Photons and Psions}}, \bibinfo{journal}{Phys.
  Lett. B} \bibinfo{volume}{78} (\bibinfo{year}{1978}) \bibinfo{pages}{150},
  \bibinfo{doi}{\doi{10.1016/0370-2693(78)90370-2}}.

\bibtype{Article}%
\bibitem{PhysRevLett.97.102301}
\bibinfo{author}{Hendrik van Hees}, \bibinfo{author}{Ralf Rapp},
  \bibinfo{title}{Comprehensive Interpretation of Thermal Dileptons Measured at
  the CERN Super Proton Synchrotron}, \bibinfo{journal}{Phys. Rev. Lett.}
  \bibinfo{volume}{97} (\bibinfo{year}{2006}) \bibinfo{pages}{102301},
  \bibinfo{doi}{\doi{10.1103/PhysRevLett.97.102301}},
  \bibinfo{url}{\urlprefix\url{https://link.aps.org/doi/10.1103/PhysRevLett.97.102301}}.

\bibtype{Article}%
\bibitem{Rapp:2014hha}
\bibinfo{author}{Ralf Rapp}, \bibinfo{author}{Hendrik van Hees},
  \bibinfo{title}{{Thermal Dileptons as Fireball Thermometer and Chronometer}},
  \bibinfo{journal}{Phys. Lett. B} \bibinfo{volume}{753} (\bibinfo{year}{2016})
  \bibinfo{pages}{586--590},
  \bibinfo{doi}{\doi{10.1016/j.physletb.2015.12.065}}, \eprint{1411.4612}.

\bibtype{Article}%
\bibitem{Seck:2020qbx}
\bibinfo{author}{Florian Seck}, \bibinfo{author}{Tetyana Galatyuk},
  \bibinfo{author}{Ayon Mukherjee}, \bibinfo{author}{Ralf Rapp},
  \bibinfo{author}{Jan Steinheimer}, \bibinfo{author}{Joachim Stroth},
  \bibinfo{author}{Maximilian Wiest}, \bibinfo{title}{{Dilepton signature of a
  first-order phase transition}}, \bibinfo{journal}{Phys. Rev. C}
  \bibinfo{volume}{106} (\bibinfo{number}{1}) (\bibinfo{year}{2022})
  \bibinfo{pages}{014904}, \bibinfo{doi}{\doi{10.1103/PhysRevC.106.014904}},
  \eprint{2010.04614}.

\bibtype{Article}%
\bibitem{Salabura:2020tou}
\bibinfo{author}{Piotr Salabura}, \bibinfo{author}{Joachim Stroth},
  \bibinfo{title}{{Dilepton radiation from strongly interacting systems}},
  \bibinfo{journal}{Prog. Part. Nucl. Phys.} \bibinfo{volume}{120}
  (\bibinfo{year}{2021}) \bibinfo{pages}{103869},
  \bibinfo{doi}{\doi{10.1016/j.ppnp.2021.103869}}, \eprint{2005.14589}.

\bibtype{Article}%
\bibitem{Geurts:2022xmk}
\bibinfo{author}{Frank Geurts}, \bibinfo{author}{Ralf-Arno Tripolt},
  \bibinfo{title}{{Electromagnetic probes: Theory and experiment}},
  \bibinfo{journal}{Prog. Part. Nucl. Phys.} \bibinfo{volume}{128}
  (\bibinfo{year}{2023}) \bibinfo{pages}{104004},
  \bibinfo{doi}{\doi{10.1016/j.ppnp.2022.104004}}, \eprint{2210.01622}.

\bibtype{Inbook}%
\bibitem{Xu:2022mqn}
\bibinfo{author}{N. Xu}, et al., \bibinfo{title}{{Nuclear Matter at High
  Density and Equation of State}} \bibinfo{year}{2022}
  \bibinfo{doi}{\doi{10.1007/978-981-19-4441-3_4}}.

\bibtype{Article}%
\bibitem{Savchuk:2022aev}
\bibinfo{author}{Oleh Savchuk}, \bibinfo{author}{Anton Motornenko},
  \bibinfo{author}{Jan Steinheimer}, \bibinfo{author}{Volodymyr Vovchenko},
  \bibinfo{author}{Marcus Bleicher}, \bibinfo{author}{Mark Gorenstein},
  \bibinfo{author}{Tetyana Galatyuk}, \bibinfo{title}{{Enhanced dilepton
  emission from a phase transition in dense matter}}, \bibinfo{journal}{J.
  Phys. G} \bibinfo{volume}{50} (\bibinfo{number}{12}) (\bibinfo{year}{2023})
  \bibinfo{pages}{125104}, \bibinfo{doi}{\doi{10.1088/1361-6471/acfccf}},
  \eprint{2209.05267}.

\bibtype{Article}%
\bibitem{NA60:2008dcb}
\bibinfo{author}{R Arnaldi}, et al. (\bibinfo{collaboration}{NA60}),
  \bibinfo{title}{{Evidence for the production of thermal-like muon pairs with
  masses above 1-GeV/c**2 in 158-A-GeV Indium-Indium Collisions}},
  \bibinfo{journal}{Eur. Phys. J. C} \bibinfo{volume}{59}
  (\bibinfo{year}{2009}) \bibinfo{pages}{607--623},
  \bibinfo{doi}{\doi{10.1140/epjc/s10052-008-0857-2}}, \eprint{0810.3204}.

\bibtype{Article}%
\bibitem{Specht:2010xu}
\bibinfo{author}{Hans~J. Specht} (\bibinfo{collaboration}{NA60}),
  \bibinfo{title}{{Thermal Dileptons from Hot and Dense Strongly Interacting
  Matter}}, \bibinfo{journal}{AIP Conf. Proc.} \bibinfo{volume}{1322}
  (\bibinfo{number}{1}) (\bibinfo{year}{2010}) \bibinfo{pages}{1--10},
  \bibinfo{doi}{\doi{10.1063/1.3541982}}, \eprint{1011.0615}.

\bibtype{Article}%
\bibitem{STAR:2015zal}
\bibinfo{author}{L. Adamczyk}, et al. (\bibinfo{collaboration}{STAR}),
  \bibinfo{title}{{Energy dependence of acceptance-corrected dielectron excess
  mass spectrum at mid-rapidity in Au$+$Au collisions at $\sqrt{s_{NN}} =$ 19.6
  and 200 GeV}}, \bibinfo{journal}{Phys. Lett. B} \bibinfo{volume}{750}
  (\bibinfo{year}{2015}) \bibinfo{pages}{64--71},
  \bibinfo{doi}{\doi{10.1016/j.physletb.2015.08.044}}, \eprint{1501.05341}.

\bibtype{Article}%
\bibitem{PhysRevC.107.L061901}
\bibinfo{author}{M.~I. Abdulhamid}, \bibinfo{author}{B.~E. Aboona},
  \bibinfo{author}{J. Adam}, \bibinfo{author}{L. Adamczyk},
  \bibinfo{author}{J.~R. Adams}, \bibinfo{author}{I. Aggarwal},
  \bibinfo{author}{M.~M. Aggarwal}, \bibinfo{author}{Z. Ahammed},
  \bibinfo{author}{D.~M. Anderson}, \bibinfo{author}{E.~C. Aschenauer},
  \bibinfo{author}{S. Aslam}, \bibinfo{author}{J. Atchison},
  \bibinfo{author}{V. Bairathi}, \bibinfo{author}{W. Baker},
  \bibinfo{author}{J.~G. Ball~Cap}, \bibinfo{author}{K. Barish},
  \bibinfo{author}{R. Bellwied}, \bibinfo{author}{P. Bhagat},
  \bibinfo{author}{A. Bhasin}, \bibinfo{author}{S. Bhatta}, \bibinfo{author}{J.
  Bielcik}, \bibinfo{author}{J. Bielcikova}, \bibinfo{author}{J.~D.
  Brandenburg}, \bibinfo{author}{J. Butterworth}, \bibinfo{author}{X.~Z. Cai},
  \bibinfo{author}{H. Caines}, \bibinfo{author}{M. Calder\'on de~la
  Barca~S\'anchez}, \bibinfo{author}{D. Cebra}, \bibinfo{author}{J. Ceska},
  \bibinfo{author}{I. Chakaberia}, \bibinfo{author}{P. Chaloupka},
  \bibinfo{author}{B.~K. Chan}, \bibinfo{author}{Z. Chang}, \bibinfo{author}{A.
  Chatterjee}, \bibinfo{author}{D. Chen}, \bibinfo{author}{J. Chen},
  \bibinfo{author}{J.~H. Chen}, \bibinfo{author}{Z. Chen}, \bibinfo{author}{J.
  Cheng}, \bibinfo{author}{Y. Cheng}, \bibinfo{author}{S. Choudhury},
  \bibinfo{author}{W. Christie}, \bibinfo{author}{X. Chu},
  \bibinfo{author}{H.~J. Crawford}, \bibinfo{author}{M. Csan\'ad},
  \bibinfo{author}{G. Dale-Gau}, \bibinfo{author}{A. Das}, \bibinfo{author}{M.
  Daugherity}, \bibinfo{author}{I.~M. Deppner}, \bibinfo{author}{A. Dhamija},
  \bibinfo{author}{L. Di~Carlo}, \bibinfo{author}{L. Didenko},
  \bibinfo{author}{P. Dixit}, \bibinfo{author}{X. Dong}, \bibinfo{author}{J.~L.
  Drachenberg}, \bibinfo{author}{E. Duckworth}, \bibinfo{author}{J.~C. Dunlop},
  \bibinfo{author}{J. Engelage}, \bibinfo{author}{G. Eppley},
  \bibinfo{author}{S. Esumi}, \bibinfo{author}{O. Evdokimov},
  \bibinfo{author}{A. Ewigleben}, \bibinfo{author}{O. Eyser},
  \bibinfo{author}{R. Fatemi}, \bibinfo{author}{S. Fazio},
  \bibinfo{author}{C.~J. Feng}, \bibinfo{author}{Y. Feng}, \bibinfo{author}{E.
  Finch}, \bibinfo{author}{Y. Fisyak}, \bibinfo{author}{F.~A. Flor},
  \bibinfo{author}{C. Fu}, \bibinfo{author}{C.~A. Gagliardi},
  \bibinfo{author}{T. Galatyuk}, \bibinfo{author}{F. Geurts},
  \bibinfo{author}{N. Ghimire}, \bibinfo{author}{A. Gibson},
  \bibinfo{author}{K. Gopal}, \bibinfo{author}{X. Gou}, \bibinfo{author}{D.
  Grosnick}, \bibinfo{author}{Y. Guo}, \bibinfo{author}{A. Gupta},
  \bibinfo{author}{W. Guryn}, \bibinfo{author}{A. Hamed}, \bibinfo{author}{Y.
  Han}, \bibinfo{author}{S. Harabasz}, \bibinfo{author}{M.~D. Harasty},
  \bibinfo{author}{J.~W. Harris}, \bibinfo{author}{H. Harrison-Smith},
  \bibinfo{author}{W. He}, \bibinfo{author}{X.~H. He}, \bibinfo{author}{Y. He},
  \bibinfo{author}{N. Herrmann}, \bibinfo{author}{L. Holub},
  \bibinfo{author}{C. Hu}, \bibinfo{author}{Q. Hu}, \bibinfo{author}{Y. Hu},
  \bibinfo{author}{B. Huang}, \bibinfo{author}{H. Huang},
  \bibinfo{author}{H.~Z. Huang}, \bibinfo{author}{S.~L. Huang},
  \bibinfo{author}{T. Huang}, \bibinfo{author}{X. Huang}, \bibinfo{author}{Y.
  Huang}, \bibinfo{author}{Y. Huang}, \bibinfo{author}{P. Huck},
  \bibinfo{author}{T.~J. Humanic}, \bibinfo{author}{D. Isenhower},
  \bibinfo{author}{M. Isshiki}, \bibinfo{author}{W.~W. Jacobs},
  \bibinfo{author}{A. Jalotra}, \bibinfo{author}{C. Jena}, \bibinfo{author}{A.
  Jentsch}, \bibinfo{author}{Y. Ji}, \bibinfo{author}{J. Jia},
  \bibinfo{author}{C. Jin}, \bibinfo{author}{X. Ju}, \bibinfo{author}{E.~G.
  Judd}, \bibinfo{author}{S. Kabana}, \bibinfo{author}{M.~L. Kabir},
  \bibinfo{author}{S. Kagamaster}, \bibinfo{author}{D. Kalinkin},
  \bibinfo{author}{K. Kang}, \bibinfo{author}{D. Kapukchyan},
  \bibinfo{author}{K. Kauder}, \bibinfo{author}{H.~W. Ke}, \bibinfo{author}{D.
  Keane}, \bibinfo{author}{M. Kelsey}, \bibinfo{author}{Y.~V. Khyzhniak},
  \bibinfo{author}{D.~P. Kiko\l{}a}, \bibinfo{author}{B. Kimelman},
  \bibinfo{author}{D. Kincses}, \bibinfo{author}{I. Kisel}, \bibinfo{author}{A.
  Kiselev}, \bibinfo{author}{A.~G. Knospe}, \bibinfo{author}{H.~S. Ko},
  \bibinfo{author}{L.~K. Kosarzewski}, \bibinfo{author}{L. Kramarik},
  \bibinfo{author}{L. Kumar}, \bibinfo{author}{S. Kumar}, \bibinfo{author}{R.
  Kunnawalkam~Elayavalli}, \bibinfo{author}{R. Lacey}, \bibinfo{author}{J.~M.
  Landgraf}, \bibinfo{author}{J. Lauret}, \bibinfo{author}{A. Lebedev},
  \bibinfo{author}{J.~H. Lee}, \bibinfo{author}{Y.~H. Leung},
  \bibinfo{author}{N. Lewis}, \bibinfo{author}{C. Li}, \bibinfo{author}{W. Li},
  \bibinfo{author}{X. Li}, \bibinfo{author}{Y. Li}, \bibinfo{author}{Y. Li},
  \bibinfo{author}{Z. Li}, \bibinfo{author}{X. Liang}, \bibinfo{author}{Y.
  Liang}, \bibinfo{author}{R. Licenik}, \bibinfo{author}{T. Lin},
  \bibinfo{author}{M.~A. Lisa}, \bibinfo{author}{C. Liu}, \bibinfo{author}{F.
  Liu}, \bibinfo{author}{G. Liu}, \bibinfo{author}{H. Liu}, \bibinfo{author}{H.
  Liu}, \bibinfo{author}{L. Liu}, \bibinfo{author}{T. Liu}, \bibinfo{author}{X.
  Liu}, \bibinfo{author}{Y. Liu}, \bibinfo{author}{Z. Liu}, \bibinfo{author}{T.
  Ljubicic}, \bibinfo{author}{W.~J. Llope}, \bibinfo{author}{O. Lomicky},
  \bibinfo{author}{R.~S. Longacre}, \bibinfo{author}{E.~M. Loyd},
  \bibinfo{author}{T. Lu}, \bibinfo{author}{N.~S. Lukow},
  \bibinfo{author}{X.~F. Luo}, \bibinfo{author}{L. Ma}, \bibinfo{author}{R.
  Ma}, \bibinfo{author}{Y.~G. Ma}, \bibinfo{author}{N. Magdy},
  \bibinfo{author}{D. Mallick}, \bibinfo{author}{S. Margetis},
  \bibinfo{author}{C. Markert}, \bibinfo{author}{H.~S. Matis},
  \bibinfo{author}{J.~A. Mazer}, \bibinfo{author}{G. McNamara},
  \bibinfo{author}{K. Mi}, \bibinfo{author}{S. Mioduszewski},
  \bibinfo{author}{B. Mohanty}, \bibinfo{author}{M.~M. Mondal},
  \bibinfo{author}{I. Mooney}, \bibinfo{author}{A. Mukherjee},
  \bibinfo{author}{M.~I. Nagy}, \bibinfo{author}{A.~S. Nain},
  \bibinfo{author}{J.~D. Nam}, \bibinfo{author}{Md. Nasim}, \bibinfo{author}{D.
  Neff}, \bibinfo{author}{J.~M. Nelson}, \bibinfo{author}{D.~B. Nemes},
  \bibinfo{author}{M. Nie}, \bibinfo{author}{T. Niida}, \bibinfo{author}{R.
  Nishitani}, \bibinfo{author}{T. Nonaka}, \bibinfo{author}{G. Odyniec},
  \bibinfo{author}{A. Ogawa}, \bibinfo{author}{S. Oh}, \bibinfo{author}{K.
  Okubo}, \bibinfo{author}{B.~S. Page}, \bibinfo{author}{R. Pak},
  \bibinfo{author}{J. Pan}, \bibinfo{author}{A. Pandav}, \bibinfo{author}{A.~K.
  Pandey}, \bibinfo{author}{T. Pani}, \bibinfo{author}{A. Paul},
  \bibinfo{author}{B. Pawlik}, \bibinfo{author}{D. Pawlowska},
  \bibinfo{author}{C. Perkins}, \bibinfo{author}{J. Pluta},
  \bibinfo{author}{B.~R. Pokhrel}, \bibinfo{author}{M. Posik},
  \bibinfo{author}{T. Protzman}, \bibinfo{author}{V. Prozorova},
  \bibinfo{author}{N.~K. Pruthi}, \bibinfo{author}{M. Przybycien},
  \bibinfo{author}{J. Putschke}, \bibinfo{author}{Z. Qin}, \bibinfo{author}{H.
  Qiu}, \bibinfo{author}{A. Quintero}, \bibinfo{author}{C. Racz},
  \bibinfo{author}{S.~K. Radhakrishnan}, \bibinfo{author}{N. Raha},
  \bibinfo{author}{R.~L. Ray}, \bibinfo{author}{R. Reed},
  \bibinfo{author}{H.~G. Ritter}, \bibinfo{author}{C.~W. Robertson},
  \bibinfo{author}{M. Robotkova}, \bibinfo{author}{M.~A. Rosales~Aguilar},
  \bibinfo{author}{D. Roy}, \bibinfo{author}{P. Roy~Chowdhury},
  \bibinfo{author}{L. Ruan}, \bibinfo{author}{A.~K. Sahoo},
  \bibinfo{author}{N.~R. Sahoo}, \bibinfo{author}{H. Sako}, \bibinfo{author}{S.
  Salur}, \bibinfo{author}{S. Sato}, \bibinfo{author}{W.~B. Schmidke},
  \bibinfo{author}{N. Schmitz}, \bibinfo{author}{F-J. Seck},
  \bibinfo{author}{J. Seger}, \bibinfo{author}{R. Seto}, \bibinfo{author}{P.
  Seyboth}, \bibinfo{author}{N. Shah}, \bibinfo{author}{P.~V. Shanmuganathan},
  \bibinfo{author}{T. Shao}, \bibinfo{author}{M. Sharma}, \bibinfo{author}{N.
  Sharma}, \bibinfo{author}{R. Sharma}, \bibinfo{author}{S.~R. Sharma},
  \bibinfo{author}{A.~I. Sheikh}, \bibinfo{author}{D.~Y. Shen},
  \bibinfo{author}{K. Shen}, \bibinfo{author}{S.~S. Shi}, \bibinfo{author}{Y.
  Shi}, \bibinfo{author}{Q.~Y. Shou}, \bibinfo{author}{F. Si},
  \bibinfo{author}{J. Singh}, \bibinfo{author}{S. Singha}, \bibinfo{author}{P.
  Sinha}, \bibinfo{author}{M.~J. Skoby}, \bibinfo{author}{N. Smirnov},
  \bibinfo{author}{Y. S\"ohngen}, \bibinfo{author}{Y. Song},
  \bibinfo{author}{B. Srivastava}, \bibinfo{author}{T.~D.~S. Stanislaus},
  \bibinfo{author}{M. Stefaniak}, \bibinfo{author}{D.~J. Stewart},
  \bibinfo{author}{B. Stringfellow}, \bibinfo{author}{Y. Su},
  \bibinfo{author}{A.~A.~P. Suaide}, \bibinfo{author}{M. Sumbera},
  \bibinfo{author}{C. Sun}, \bibinfo{author}{X. Sun}, \bibinfo{author}{Y. Sun},
  \bibinfo{author}{Y. Sun}, \bibinfo{author}{B. Surrow}, \bibinfo{author}{Z.~W.
  Sweger}, \bibinfo{author}{P. Szymanski}, \bibinfo{author}{A. Tamis},
  \bibinfo{author}{A.~H. Tang}, \bibinfo{author}{Z. Tang}, \bibinfo{author}{T.
  Tarnowsky}, \bibinfo{author}{J.~H. Thomas}, \bibinfo{author}{A.~R. Timmins},
  \bibinfo{author}{D. Tlusty}, \bibinfo{author}{T. Todoroki},
  \bibinfo{author}{C.~A. Tomkiel}, \bibinfo{author}{S. Trentalange},
  \bibinfo{author}{R.~E. Tribble}, \bibinfo{author}{P. Tribedy},
  \bibinfo{author}{T. Truhlar}, \bibinfo{author}{B.~A. Trzeciak},
  \bibinfo{author}{O.~D. Tsai}, \bibinfo{author}{C.~Y. Tsang},
  \bibinfo{author}{Z. Tu}, \bibinfo{author}{T. Ullrich}, \bibinfo{author}{D.~G.
  Underwood}, \bibinfo{author}{I. Upsal}, \bibinfo{author}{G. Van~Buren},
  \bibinfo{author}{J. Vanek}, \bibinfo{author}{I. Vassiliev},
  \bibinfo{author}{V. Verkest}, \bibinfo{author}{F. Videb\ae{}k},
  \bibinfo{author}{S.~A. Voloshin}, \bibinfo{author}{F. Wang},
  \bibinfo{author}{G. Wang}, \bibinfo{author}{J.~S. Wang}, \bibinfo{author}{X.
  Wang}, \bibinfo{author}{Y. Wang}, \bibinfo{author}{Y. Wang},
  \bibinfo{author}{Y. Wang}, \bibinfo{author}{Z. Wang}, \bibinfo{author}{J.~C.
  Webb}, \bibinfo{author}{P.~C. Weidenkaff}, \bibinfo{author}{G.~D. Westfall},
  \bibinfo{author}{D. Wielanek}, \bibinfo{author}{H. Wieman},
  \bibinfo{author}{G. Wilks}, \bibinfo{author}{S.~W. Wissink},
  \bibinfo{author}{R. Witt}, \bibinfo{author}{J. Wu}, \bibinfo{author}{J. Wu},
  \bibinfo{author}{X. Wu}, \bibinfo{author}{Y. Wu}, \bibinfo{author}{B. Xi},
  \bibinfo{author}{Z.~G. Xiao}, \bibinfo{author}{G. Xie}, \bibinfo{author}{W.
  Xie}, \bibinfo{author}{H. Xu}, \bibinfo{author}{N. Xu},
  \bibinfo{author}{Q.~H. Xu}, \bibinfo{author}{Y. Xu}, \bibinfo{author}{Y. Xu},
  \bibinfo{author}{Z. Xu}, \bibinfo{author}{Z. Xu}, \bibinfo{author}{G. Yan},
  \bibinfo{author}{Z. Yan}, \bibinfo{author}{C. Yang}, \bibinfo{author}{Q.
  Yang}, \bibinfo{author}{S. Yang}, \bibinfo{author}{Y. Yang},
  \bibinfo{author}{Z. Ye}, \bibinfo{author}{Z. Ye}, \bibinfo{author}{L. Yi},
  \bibinfo{author}{K. Yip}, \bibinfo{author}{Y. Yu}, \bibinfo{author}{H.
  Zbroszczyk}, \bibinfo{author}{W. Zha}, \bibinfo{author}{C. Zhang},
  \bibinfo{author}{D. Zhang}, \bibinfo{author}{J. Zhang}, \bibinfo{author}{S.
  Zhang}, \bibinfo{author}{W. Zhang}, \bibinfo{author}{X. Zhang},
  \bibinfo{author}{Y. Zhang}, \bibinfo{author}{Y. Zhang}, \bibinfo{author}{Y.
  Zhang}, \bibinfo{author}{Z.~J. Zhang}, \bibinfo{author}{Z. Zhang},
  \bibinfo{author}{Z. Zhang}, \bibinfo{author}{F. Zhao}, \bibinfo{author}{J.
  Zhao}, \bibinfo{author}{M. Zhao}, \bibinfo{author}{C. Zhou},
  \bibinfo{author}{J. Zhou}, \bibinfo{author}{S. Zhou}, \bibinfo{author}{Y.
  Zhou}, \bibinfo{author}{X. Zhu}, \bibinfo{author}{M. Zurek},
  \bibinfo{author}{M. Zyzak} (\bibinfo{collaboration}{STAR Collaboration}),
  \bibinfo{title}{Measurements of dielectron production in
  $\mathrm{Au}+\mathrm{Au}$ collisions at $\sqrt{{s}_{NN}}=27$, 39, and 62.4
  GeV from the STAR experiment}, \bibinfo{journal}{Phys. Rev. C}
  \bibinfo{volume}{107} (\bibinfo{year}{2023}) \bibinfo{pages}{L061901},
  \bibinfo{doi}{\doi{10.1103/PhysRevC.107.L061901}},
  \bibinfo{url}{\urlprefix\url{https://link.aps.org/doi/10.1103/PhysRevC.107.L061901}}.

\bibtype{Article}%
\bibitem{STAR:2024bpc}
\bibinfo{author}{B.~E. Aboona}, et al. (\bibinfo{collaboration}{STAR}),
  \bibinfo{title}{{Temperature measurement of Quark-Gluon plasma at different
  stages}}, \bibinfo{journal}{Nature Commun.} \bibinfo{volume}{16}
  (\bibinfo{number}{1}) (\bibinfo{year}{2025}) \bibinfo{pages}{9098},
  \bibinfo{doi}{\doi{10.1038/s41467-025-63216-5}}, \eprint{2402.01998}.

\bibtype{Article}%
\bibitem{HADES:2019auv}
\bibinfo{author}{J. Adamczewski-Musch}, et al.
  (\bibinfo{collaboration}{HADES}), \bibinfo{title}{{Probing dense baryon-rich
  matter with virtual photons}}, \bibinfo{journal}{Nature Phys.}
  \bibinfo{volume}{15} (\bibinfo{number}{10}) (\bibinfo{year}{2019})
  \bibinfo{pages}{1040--1045}, \bibinfo{doi}{\doi{10.1038/s41567-019-0583-8}}.

\bibtype{Article}%
\bibitem{Alba:2014eba}
\bibinfo{author}{Paolo Alba}, \bibinfo{author}{Wanda Alberico},
  \bibinfo{author}{Rene Bellwied}, \bibinfo{author}{Marcus Bluhm},
  \bibinfo{author}{Valentina Mantovani~Sarti}, et al.,
  \bibinfo{title}{{Freeze-out conditions from net-proton and net-charge
  fluctuations at RHIC}}, \bibinfo{journal}{Phys.Lett.} \bibinfo{volume}{B738}
  (\bibinfo{year}{2014}) \bibinfo{pages}{305--310},
  \bibinfo{doi}{\doi{10.1016/j.physletb.2014.09.052}}, \eprint{1403.4903}.

\bibtype{Article}%
\bibitem{Bellwied:2018tkc}
\bibinfo{author}{Rene Bellwied}, \bibinfo{author}{Jacquelyn Noronha-Hostler},
  \bibinfo{author}{Paolo Parotto}, \bibinfo{author}{Israel Portillo~Vazquez},
  \bibinfo{author}{Claudia Ratti}, \bibinfo{author}{Jamie~M. Stafford},
  \bibinfo{title}{{Freeze-out temperature from net-kaon fluctuations at
  energies available at the BNL Relativistic Heavy Ion Collider}},
  \bibinfo{journal}{Phys. Rev. C} \bibinfo{volume}{99} (\bibinfo{number}{3})
  (\bibinfo{year}{2019}) \bibinfo{pages}{034912},
  \bibinfo{doi}{\doi{10.1103/PhysRevC.99.034912}}, \eprint{1805.00088}.

\bibtype{Article}%
\bibitem{Bellwied:2019pxh}
\bibinfo{author}{Rene Bellwied}, \bibinfo{author}{Szabolcs Borsanyi},
  \bibinfo{author}{Zoltan Fodor}, \bibinfo{author}{Jana~N. Guenther},
  \bibinfo{author}{Jacquelyn Noronha-Hostler}, \bibinfo{author}{Paolo Parotto},
  \bibinfo{author}{Attila Pasztor}, \bibinfo{author}{Claudia Ratti},
  \bibinfo{author}{Jamie~M. Stafford}, \bibinfo{title}{{Off-diagonal
  correlators of conserved charges from lattice QCD and experiment}},
  \bibinfo{journal}{Physical Review D} \bibinfo{volume}{101}
  (\bibinfo{number}{3}) (\bibinfo{year}{2020}), \eprint{1910.14592}.

\bibtype{Article}%
\bibitem{Bluhm:2018aei}
\bibinfo{author}{Marcus Bluhm}, \bibinfo{author}{Marlene Nahrgang},
  \bibinfo{title}{{Freeze-out conditions from strangeness observables at
  RHIC}}, \bibinfo{journal}{Eur. Phys. J. C} \bibinfo{volume}{79}
  (\bibinfo{number}{2}) (\bibinfo{year}{2019}) \bibinfo{pages}{155},
  \bibinfo{doi}{\doi{10.1140/epjc/s10052-019-6661-3}}, \eprint{1806.04499}.

\bibtype{Article}%
\bibitem{Flor:2020fdw}
\bibinfo{author}{Fernando~Antonio Flor}, \bibinfo{author}{Gabrielle Olinger},
  \bibinfo{author}{Rene Bellwied}, \bibinfo{title}{{Flavour and Energy
  Dependence of Chemical Freeze-out Temperatures in Relativistic Heavy Ion
  Collisions from RHIC-BES to LHC Energies}}, \bibinfo{journal}{Phys. Lett. B}
  \bibinfo{volume}{814} (\bibinfo{year}{2021}) \bibinfo{pages}{136098},
  \bibinfo{doi}{\doi{10.1016/j.physletb.2021.136098}}, \eprint{2009.14781}.

\bibtype{Article}%
\bibitem{Wilson:1974sk}
\bibinfo{author}{Kenneth~G. Wilson}, \bibinfo{title}{{Confinement of Quarks}},
  \bibinfo{journal}{Phys.Rev.} \bibinfo{volume}{D10} (\bibinfo{year}{1974})
  \bibinfo{pages}{2445--2459}, \bibinfo{doi}{\doi{10.1103/PhysRevD.10.2445}}.

\bibtype{Article}%
\bibitem{Allton:2003vx}
\bibinfo{author}{C.~R. Allton}, \bibinfo{author}{S. Ejiri},
  \bibinfo{author}{S.~J. Hands}, \bibinfo{author}{O. Kaczmarek},
  \bibinfo{author}{F. Karsch}, \bibinfo{author}{E. Laermann},
  \bibinfo{author}{C. Schmidt}, \bibinfo{title}{{The Equation of state for two
  flavor QCD at nonzero chemical potential}}, \bibinfo{journal}{Phys. Rev.}
  \bibinfo{volume}{D68} (\bibinfo{year}{2003}) \bibinfo{pages}{014507},
  \bibinfo{doi}{\doi{10.1103/PhysRevD.68.014507}}, \eprint{hep-lat/0305007}.

\bibtype{Article}%
\bibitem{Allton:2005gk}
\bibinfo{author}{C.R. Allton}, \bibinfo{author}{M. Doring}, \bibinfo{author}{S.
  Ejiri}, \bibinfo{author}{S.J. Hands}, \bibinfo{author}{O. Kaczmarek}, et al.,
  \bibinfo{title}{{Thermodynamics of two flavor QCD to sixth order in quark
  chemical potential}}, \bibinfo{journal}{Phys.Rev.} \bibinfo{volume}{D71}
  (\bibinfo{year}{2005}) \bibinfo{pages}{054508},
  \bibinfo{doi}{\doi{10.1103/PhysRevD.71.054508}}, \eprint{hep-lat/0501030}.

\bibtype{Article}%
\bibitem{Kaczmarek:2011zz}
\bibinfo{author}{O. Kaczmarek}, \bibinfo{author}{F. Karsch},
  \bibinfo{author}{E. Laermann}, \bibinfo{author}{C. Miao}, \bibinfo{author}{S.
  Mukherjee}, et al., \bibinfo{title}{{Phase boundary for the chiral transition
  in (2+1) -flavor QCD at small values of the chemical potential}},
  \bibinfo{journal}{Phys.Rev.} \bibinfo{volume}{D83} (\bibinfo{year}{2011})
  \bibinfo{pages}{014504}, \bibinfo{doi}{\doi{10.1103/PhysRevD.83.014504}},
  \eprint{1011.3130}.

\bibtype{Article}%
\bibitem{Endrodi:2011gv}
\bibinfo{author}{G. Endrodi}, \bibinfo{author}{Z. Fodor}, \bibinfo{author}{S.D.
  Katz}, \bibinfo{author}{K.K. Szabo}, \bibinfo{title}{{The QCD phase diagram
  at nonzero quark density}}, \bibinfo{journal}{JHEP} \bibinfo{volume}{1104}
  (\bibinfo{year}{2011}) \bibinfo{pages}{001},
  \bibinfo{doi}{\doi{10.1007/JHEP04(2011)001}}, \eprint{1102.1356}.

\bibtype{Article}%
\bibitem{Borsanyi:2012cr}
\bibinfo{author}{Sz. Borsanyi}, \bibinfo{author}{G. Endrodi},
  \bibinfo{author}{Z. Fodor}, \bibinfo{author}{S.D. Katz}, \bibinfo{author}{S.
  Krieg}, et al., \bibinfo{title}{{QCD equation of state at nonzero chemical
  potential: continuum results with physical quark masses at order $mu^2$}},
  \bibinfo{journal}{JHEP} \bibinfo{volume}{1208} (\bibinfo{year}{2012})
  \bibinfo{pages}{053}, \bibinfo{doi}{\doi{10.1007/JHEP08(2012)053}},
  \eprint{1204.6710}.

\bibtype{Article}%
\bibitem{Bazavov:2017dus}
\bibinfo{author}{A. Bazavov}, et al., \bibinfo{title}{{The QCD Equation of
  State to $\mathcal{O}(\mu_B^6)$ from Lattice QCD}}, \bibinfo{journal}{Phys.
  Rev.} \bibinfo{volume}{D95} (\bibinfo{number}{5}) (\bibinfo{year}{2017})
  \bibinfo{pages}{054504}, \bibinfo{doi}{\doi{10.1103/PhysRevD.95.054504}},
  \eprint{1701.04325}.

\bibtype{Article}%
\bibitem{Bonati:2018nut}
\bibinfo{author}{Claudio Bonati}, \bibinfo{author}{Massimo D'Elia},
  \bibinfo{author}{Francesco Negro}, \bibinfo{author}{Francesco Sanfilippo},
  \bibinfo{author}{Kevin Zambello}, \bibinfo{title}{{Curvature of the
  pseudocritical line in QCD: Taylor expansion matches analytic continuation}},
  \bibinfo{journal}{Phys. Rev.} \bibinfo{volume}{D98} (\bibinfo{number}{5})
  (\bibinfo{year}{2018}) \bibinfo{pages}{054510},
  \bibinfo{doi}{\doi{10.1103/PhysRevD.98.054510}}, \eprint{1805.02960}.

\bibtype{Article}%
\bibitem{HotQCD:2018pds}
\bibinfo{author}{A. Bazavov}, et al. (\bibinfo{collaboration}{HotQCD}),
  \bibinfo{title}{{Chiral crossover in QCD at zero and non-zero chemical
  potentials}}, \bibinfo{journal}{Phys. Lett. B} \bibinfo{volume}{795}
  (\bibinfo{year}{2019}) \bibinfo{pages}{15--21},
  \bibinfo{doi}{\doi{10.1016/j.physletb.2019.05.013}}, \eprint{1812.08235}.

\bibtype{Article}%
\bibitem{Bollweg:2022rps}
\bibinfo{author}{D. Bollweg}, \bibinfo{author}{J. Goswami}, \bibinfo{author}{O.
  Kaczmarek}, \bibinfo{author}{F. Karsch}, \bibinfo{author}{Swagato Mukherjee},
  \bibinfo{author}{P. Petreczky}, \bibinfo{author}{C. Schmidt},
  \bibinfo{author}{P. Scior} (\bibinfo{collaboration}{HotQCD}),
  \bibinfo{title}{{Taylor expansions and Pad\'e approximants for cumulants of
  conserved charge fluctuations at nonvanishing chemical potentials}},
  \bibinfo{journal}{Phys. Rev. D} \bibinfo{volume}{105} (\bibinfo{number}{7})
  (\bibinfo{year}{2022}) \bibinfo{pages}{074511},
  \bibinfo{doi}{\doi{10.1103/PhysRevD.105.074511}}, \eprint{2202.09184}.

\bibtype{Article}%
\bibitem{DElia:2002tig}
\bibinfo{author}{Massimo D'Elia}, \bibinfo{author}{Maria-Paola Lombardo},
  \bibinfo{title}{{Finite density QCD via imaginary chemical potential}},
  \bibinfo{journal}{Phys. Rev.} \bibinfo{volume}{D67} (\bibinfo{year}{2003})
  \bibinfo{pages}{014505}, \bibinfo{doi}{\doi{10.1103/PhysRevD.67.014505}},
  \eprint{hep-lat/0209146}.

\bibtype{Inproceedings}%
\bibitem{deForcrand:2003bz}
\bibinfo{author}{P. de Forcrand}, \bibinfo{author}{O. Philipsen},
  \bibinfo{title}{{QCD phase diagram at small densities from simulations with
  imaginary $\mu$}}, in: \bibinfo{booktitle}{{5th Internationa Conference on
  Strong and Electroweak Matter}} \bibinfo{year}{2003}, pp.
  \bibinfo{pages}{271--275}, \bibinfo{doi}{\doi{10.1142/9789812704498_0027}},
  \eprint{hep-ph/0301209}.

\bibtype{Article}%
\bibitem{DElia:2007bkz}
\bibinfo{author}{Massimo D'Elia}, \bibinfo{author}{Francesco Di~Renzo},
  \bibinfo{author}{Maria~Paola Lombardo}, \bibinfo{title}{{The Strongly
  interacting quark gluon plasma, and the critical behaviour of QCD at
  imaginary mu}}, \bibinfo{journal}{Phys. Rev. D} \bibinfo{volume}{76}
  (\bibinfo{year}{2007}) \bibinfo{pages}{114509},
  \bibinfo{doi}{\doi{10.1103/PhysRevD.76.114509}}, \eprint{0705.3814}.

\bibtype{Article}%
\bibitem{Cea:2009ba}
\bibinfo{author}{Paolo Cea}, \bibinfo{author}{Leonardo Cosmai},
  \bibinfo{author}{Massimo D'Elia}, \bibinfo{author}{Chiara Manneschi},
  \bibinfo{author}{Alessandro Papa}, \bibinfo{title}{{Analytic continuation of
  the critical line: Suggestions for QCD}}, \bibinfo{journal}{Phys.Rev.}
  \bibinfo{volume}{D80} (\bibinfo{year}{2009}) \bibinfo{pages}{034501},
  \bibinfo{doi}{\doi{10.1103/PhysRevD.80.034501}}, \eprint{0905.1292}.

\bibtype{Article}%
\bibitem{Cea:2015cya}
\bibinfo{author}{Paolo Cea}, \bibinfo{author}{Leonardo Cosmai},
  \bibinfo{author}{Alessandro Papa}, \bibinfo{title}{{Critical line of 2+1
  flavor QCD: Toward the continuum limit}}, \bibinfo{journal}{Phys. Rev.}
  \bibinfo{volume}{D93} (\bibinfo{number}{1}) (\bibinfo{year}{2016})
  \bibinfo{pages}{014507}, \bibinfo{doi}{\doi{10.1103/PhysRevD.93.014507}},
  \eprint{1508.07599}.

\bibtype{Article}%
\bibitem{Vovchenko:2017xad}
\bibinfo{author}{Volodymyr Vovchenko}, \bibinfo{author}{Attila Pasztor},
  \bibinfo{author}{Zoltan Fodor}, \bibinfo{author}{Sandor~D. Katz},
  \bibinfo{author}{Horst Stoecker}, \bibinfo{title}{{Repulsive baryonic
  interactions and lattice QCD observables at imaginary chemical potential}},
  \bibinfo{journal}{Phys. Lett.} \bibinfo{volume}{B775} (\bibinfo{year}{2017})
  \bibinfo{pages}{71--78}, \bibinfo{doi}{\doi{10.1016/j.physletb.2017.10.042}},
  \eprint{1708.02852}.

\bibtype{Article}%
\bibitem{Vovchenko:2017gkg}
\bibinfo{author}{Volodymyr Vovchenko}, \bibinfo{author}{Jan Steinheimer},
  \bibinfo{author}{Owe Philipsen}, \bibinfo{author}{Horst Stoecker},
  \bibinfo{title}{{Cluster Expansion Model for QCD Baryon Number Fluctuations:
  No Phase Transition at $\mu_B / T < \pi$}}, \bibinfo{journal}{Phys. Rev.}
  \bibinfo{volume}{D97} (\bibinfo{number}{11}) (\bibinfo{year}{2018})
  \bibinfo{pages}{114030}, \bibinfo{doi}{\doi{10.1103/PhysRevD.97.114030}},
  \eprint{1711.01261}.

\bibtype{Article}%
\bibitem{Borsanyi:2021sxv}
\bibinfo{author}{S. Bors\'anyi}, \bibinfo{author}{Z. Fodor},
  \bibinfo{author}{J.~N. Guenther}, \bibinfo{author}{R. Kara},
  \bibinfo{author}{S.~D. Katz}, \bibinfo{author}{P. Parotto},
  \bibinfo{author}{A. P\'asztor}, \bibinfo{author}{C. Ratti},
  \bibinfo{author}{K.~K. Szab\'o}, \bibinfo{title}{{Lattice QCD equation of
  state at finite chemical potential from an alternative expansion scheme}},
  \bibinfo{journal}{Phys. Rev. Lett.} \bibinfo{volume}{126}
  (\bibinfo{number}{23}) (\bibinfo{year}{2021}) \bibinfo{pages}{232001},
  \bibinfo{doi}{\doi{10.1103/PhysRevLett.126.232001}}, \eprint{2102.06660}.

\bibtype{Article}%
\bibitem{Borsanyi:2022qlh}
\bibinfo{author}{Szabolcs Borsanyi}, \bibinfo{author}{Zoltan Fodor},
  \bibinfo{author}{Jana~N. Guenther}, \bibinfo{author}{Ruben Kara},
  \bibinfo{author}{Paolo Parotto}, \bibinfo{author}{Attila Pasztor},
  \bibinfo{author}{Claudia Ratti}, \bibinfo{author}{Kalman~K. Szabo},
  \bibinfo{title}{{Resummed lattice QCD equation of state at finite baryon
  density: Strangeness neutrality and beyond}}, \bibinfo{journal}{Phys. Rev. D}
  \bibinfo{volume}{105} (\bibinfo{number}{11}) (\bibinfo{year}{2022})
  \bibinfo{pages}{114504}, \bibinfo{doi}{\doi{10.1103/PhysRevD.105.114504}},
  \eprint{2202.05574}.

\bibtype{Article}%
\bibitem{Barbour:1997ej}
\bibinfo{author}{Ian~M. Barbour}, \bibinfo{author}{Susan~E. Morrison},
  \bibinfo{author}{Elyakum~G. Klepfish}, \bibinfo{author}{John~B. Kogut},
  \bibinfo{author}{Maria-Paola Lombardo}, \bibinfo{title}{{Results on finite
  density QCD}}, \bibinfo{journal}{Nucl. Phys. Proc. Suppl.}
  \bibinfo{volume}{60A} (\bibinfo{year}{1998}) \bibinfo{pages}{220--234},
  \bibinfo{doi}{\doi{10.1016/S0920-5632(97)00484-2}},
  \bibinfo{note}{[,220(1997)]}, \eprint{hep-lat/9705042}.

\bibtype{Article}%
\bibitem{Fodor:2001au}
\bibinfo{author}{Z. Fodor}, \bibinfo{author}{S.D. Katz}, \bibinfo{title}{{A New
  method to study lattice QCD at finite temperature and chemical potential}},
  \bibinfo{journal}{Phys.Lett.} \bibinfo{volume}{B534} (\bibinfo{year}{2002})
  \bibinfo{pages}{87--92}, \bibinfo{doi}{\doi{10.1016/S0370-2693(02)01583-6}},
  \eprint{hep-lat/0104001}.

\bibtype{Article}%
\bibitem{Fodor:2001pe}
\bibinfo{author}{Z. Fodor}, \bibinfo{author}{S.D. Katz},
  \bibinfo{title}{{Lattice determination of the critical point of QCD at finite
  T and mu}}, \bibinfo{journal}{JHEP} \bibinfo{volume}{0203}
  (\bibinfo{year}{2002}) \bibinfo{pages}{014},
  \bibinfo{doi}{\doi{10.1088/1126-6708/2002/03/014}}, \eprint{hep-lat/0106002}.

\bibtype{Article}%
\bibitem{Fodor:2004nz}
\bibinfo{author}{Z. Fodor}, \bibinfo{author}{S.D. Katz},
  \bibinfo{title}{{Critical point of QCD at finite T and mu, lattice results
  for physical quark masses}}, \bibinfo{journal}{JHEP} \bibinfo{volume}{0404}
  (\bibinfo{year}{2004}) \bibinfo{pages}{050},
  \bibinfo{doi}{\doi{10.1088/1126-6708/2004/04/050}}, \eprint{hep-lat/0402006}.

\bibtype{Article}%
\bibitem{deForcrand:2002pa}
\bibinfo{author}{P. de Forcrand}, \bibinfo{author}{S. Kim}, \bibinfo{author}{T.
  Takaishi}, \bibinfo{title}{{QCD simulations at small chemical potential}},
  \bibinfo{journal}{Nucl. Phys. B Proc. Suppl.} \bibinfo{volume}{119}
  (\bibinfo{year}{2003}) \bibinfo{pages}{541--543},
  \bibinfo{doi}{\doi{10.1016/S0920-5632(03)80451-6}}, \eprint{hep-lat/0209126}.

\bibtype{Article}%
\bibitem{Alexandru:2005ix}
\bibinfo{author}{Andrei Alexandru}, \bibinfo{author}{Manfried Faber},
  \bibinfo{author}{Ivan Horvath}, \bibinfo{author}{Keh-Fei Liu},
  \bibinfo{title}{{Lattice QCD at finite density via a new canonical
  approach}}, \bibinfo{journal}{Phys. Rev.} \bibinfo{volume}{D72}
  (\bibinfo{year}{2005}) \bibinfo{pages}{114513},
  \bibinfo{doi}{\doi{10.1103/PhysRevD.72.114513}}, \eprint{hep-lat/0507020}.

\bibtype{Article}%
\bibitem{Fodor:2007vv}
\bibinfo{author}{Zoltan Fodor}, \bibinfo{author}{Sandor~D. Katz},
  \bibinfo{author}{Christian Schmidt}, \bibinfo{title}{{The Density of states
  method at non-zero chemical potential}}, \bibinfo{journal}{JHEP}
  \bibinfo{volume}{0703} (\bibinfo{year}{2007}) \bibinfo{pages}{121},
  \bibinfo{doi}{\doi{10.1088/1126-6708/2007/03/121}}, \eprint{hep-lat/0701022}.

\bibtype{Article}%
\bibitem{Endrodi:2018zda}
\bibinfo{author}{G. Endrodi}, \bibinfo{author}{Z. Fodor},
  \bibinfo{author}{S.~D. Katz}, \bibinfo{author}{D. Sexty},
  \bibinfo{author}{K.~K. Szabo}, \bibinfo{author}{Cs. Torok},
  \bibinfo{title}{{Applying constrained simulations for low temperature lattice
  QCD at finite baryon chemical potential}}, \bibinfo{journal}{Phys. Rev.}
  \bibinfo{volume}{D98} (\bibinfo{number}{7}) (\bibinfo{year}{2018})
  \bibinfo{pages}{074508}, \bibinfo{doi}{\doi{10.1103/PhysRevD.98.074508}},
  \eprint{1807.08326}.

\bibtype{Article}%
\bibitem{Giordano:2020uvk}
\bibinfo{author}{Matteo Giordano}, \bibinfo{author}{Kornel Kapas},
  \bibinfo{author}{Sandor~D. Katz}, \bibinfo{author}{Daniel Nogradi},
  \bibinfo{author}{Attila Pasztor}, \bibinfo{title}{{Effect of stout smearing
  on the phase diagram from multiparameter reweighting in lattice QCD}},
  \bibinfo{journal}{Phys. Rev. D} \bibinfo{volume}{102} (\bibinfo{number}{3})
  (\bibinfo{year}{2020}) \bibinfo{pages}{034503},
  \bibinfo{doi}{\doi{10.1103/PhysRevD.102.034503}}, \eprint{2003.04355}.

\bibtype{Article}%
\bibitem{Giordano:2020roi}
\bibinfo{author}{Matteo Giordano}, \bibinfo{author}{Kornel Kapas},
  \bibinfo{author}{Sandor~D. Katz}, \bibinfo{author}{Daniel Nogradi},
  \bibinfo{author}{Attila Pasztor}, \bibinfo{title}{{New approach to lattice
  QCD at finite density; results for the critical end point on coarse
  lattices}}, \bibinfo{journal}{JHEP} \bibinfo{volume}{05}
  (\bibinfo{year}{2020}) \bibinfo{pages}{088},
  \bibinfo{doi}{\doi{10.1007/JHEP05(2020)088}}, \eprint{2004.10800}.

\bibtype{Article}%
\bibitem{Borsanyi:2021hbk}
\bibinfo{author}{Szabolcs Borsanyi}, \bibinfo{author}{Zoltan Fodor},
  \bibinfo{author}{Matteo Giordano}, \bibinfo{author}{Sandor~D. Katz},
  \bibinfo{author}{Daniel Nogradi}, \bibinfo{author}{Attila Pasztor},
  \bibinfo{author}{Chik~Him Wong}, \bibinfo{title}{{Lattice simulations of the
  QCD chiral transition at real baryon density}}, \bibinfo{journal}{Phys. Rev.
  D} \bibinfo{volume}{105} (\bibinfo{number}{5}) (\bibinfo{year}{2022})
  \bibinfo{pages}{L051506}, \bibinfo{doi}{\doi{10.1103/PhysRevD.105.L051506}},
  \eprint{2108.09213}.

\bibtype{Article}%
\bibitem{Borsanyi:2022soo}
\bibinfo{author}{Szabolcs Borsanyi}, \bibinfo{author}{Zoltan Fodor},
  \bibinfo{author}{Matteo Giordano}, \bibinfo{author}{Jana~N. Guenther},
  \bibinfo{author}{Sandor~D. Katz}, \bibinfo{author}{Attila Pasztor},
  \bibinfo{author}{Chik~Him Wong}, \bibinfo{title}{{Equation of state of a
  hot-and-dense quark gluon plasma: Lattice simulations at real
  \ensuremath{\mu}B vs extrapolations}}, \bibinfo{journal}{Phys. Rev. D}
  \bibinfo{volume}{107} (\bibinfo{number}{9}) (\bibinfo{year}{2023})
  \bibinfo{pages}{L091503}, \bibinfo{doi}{\doi{10.1103/PhysRevD.107.L091503}},
  \eprint{2208.05398}.

\bibtype{Article}%
\bibitem{Hagedorn:1965st}
\bibinfo{author}{R. Hagedorn}, \bibinfo{title}{{Statistical thermodynamics of
  strong interactions at high-energies}}, \bibinfo{journal}{Nuovo Cim.Suppl.}
  \bibinfo{volume}{3} (\bibinfo{year}{1965}) \bibinfo{pages}{147--186}.

\bibtype{Article}%
\bibitem{Cabibbo:1975ig}
\bibinfo{author}{N. Cabibbo}, \bibinfo{author}{G. Parisi},
  \bibinfo{title}{{Exponential Hadronic Spectrum and Quark Liberation}},
  \bibinfo{journal}{Phys.Lett.} \bibinfo{volume}{B59} (\bibinfo{year}{1975})
  \bibinfo{pages}{67--69}, \bibinfo{doi}{\doi{10.1016/0370-2693(75)90158-6}}.

\bibtype{Article}%
\bibitem{Brown:1988qe}
\bibinfo{author}{F.~R. Brown}, \bibinfo{author}{N.~H. Christ},
  \bibinfo{author}{Y.~F. Deng}, \bibinfo{author}{M.~S. Gao},
  \bibinfo{author}{T.~J. Woch}, \bibinfo{title}{{Nature of the Deconfining
  Phase Transition in SU(3) Lattice Gauge Theory}}, \bibinfo{journal}{Phys.
  Rev. Lett.} \bibinfo{volume}{61} (\bibinfo{year}{1988})
  \bibinfo{pages}{2058}, \bibinfo{doi}{\doi{10.1103/PhysRevLett.61.2058}}.

\bibtype{Article}%
\bibitem{Fukugita:1989yb}
\bibinfo{author}{M. Fukugita}, \bibinfo{author}{M. Okawa}, \bibinfo{author}{A.
  Ukawa}, \bibinfo{title}{{Order of the Deconfining Phase Transition in SU(3)
  Lattice Gauge Theory}}, \bibinfo{journal}{Phys.Rev.Lett.}
  \bibinfo{volume}{63} (\bibinfo{year}{1989}) \bibinfo{pages}{1768},
  \bibinfo{doi}{\doi{10.1103/PhysRevLett.63.1768}}.

\bibtype{Article}%
\bibitem{Aoki:2006br}
\bibinfo{author}{Y. Aoki}, \bibinfo{author}{Z. Fodor}, \bibinfo{author}{S.D.
  Katz}, \bibinfo{author}{K.K. Szabo}, \bibinfo{title}{{The QCD transition
  temperature: Results with physical masses in the continuum limit}},
  \bibinfo{journal}{Phys.Lett.} \bibinfo{volume}{B643} (\bibinfo{year}{2006})
  \bibinfo{pages}{46--54}, \bibinfo{doi}{\doi{10.1016/j.physletb.2006.10.021}},
  \eprint{hep-lat/0609068}.

\bibtype{Article}%
\bibitem{Aoki:2009sc}
\bibinfo{author}{Y. Aoki}, \bibinfo{author}{Szabolcs Borsanyi},
  \bibinfo{author}{Stephan Durr}, \bibinfo{author}{Zoltan Fodor},
  \bibinfo{author}{Sandor~D. Katz}, et al., \bibinfo{title}{{The QCD transition
  temperature: results with physical masses in the continuum limit II.}},
  \bibinfo{journal}{JHEP} \bibinfo{volume}{0906} (\bibinfo{year}{2009})
  \bibinfo{pages}{088}, \bibinfo{doi}{\doi{10.1088/1126-6708/2009/06/088}},
  \eprint{0903.4155}.

\bibtype{Article}%
\bibitem{Ding:2024sux}
\bibinfo{author}{H.~T. Ding}, \bibinfo{author}{O. Kaczmarek},
  \bibinfo{author}{F. Karsch}, \bibinfo{author}{P. Petreczky},
  \bibinfo{author}{Mugdha Sarkar}, \bibinfo{author}{C. Schmidt},
  \bibinfo{author}{Sipaz Sharma}, \bibinfo{title}{{Curvature of the chiral
  phase transition line from the magnetic equation of state of (2+1)-flavor
  QCD}}, \bibinfo{journal}{Phys. Rev. D} \bibinfo{volume}{109}
  (\bibinfo{number}{11}) (\bibinfo{year}{2024}) \bibinfo{pages}{114516},
  \bibinfo{doi}{\doi{10.1103/PhysRevD.109.114516}}, \eprint{2403.09390}.

\bibtype{Article}%
\bibitem{Borsanyi:2024xrx}
\bibinfo{author}{Szabolcs Borsanyi}, \bibinfo{author}{Zoltan Fodor},
  \bibinfo{author}{Jana~N. Guenther}, \bibinfo{author}{Paolo Parotto},
  \bibinfo{author}{Attila Pasztor}, \bibinfo{author}{Ludovica Pirelli},
  \bibinfo{author}{Kalman~K. Szabo}, \bibinfo{author}{Chik~Him Wong},
  \bibinfo{title}{{QCD deconfinement transition line up to
  \ensuremath{\mu}B=400\,\,MeV from finite volume lattice simulations}},
  \bibinfo{journal}{Phys. Rev. D} \bibinfo{volume}{110} (\bibinfo{number}{11})
  (\bibinfo{year}{2024}) \bibinfo{pages}{114507},
  \bibinfo{doi}{\doi{10.1103/PhysRevD.110.114507}}, \eprint{2410.06216}.

\bibtype{Article}%
\bibitem{Bazavov:2016uvm}
\bibinfo{author}{A. Bazavov}, \bibinfo{author}{N. Brambilla},
  \bibinfo{author}{H.~T. Ding}, \bibinfo{author}{P. Petreczky},
  \bibinfo{author}{H.~P. Schadler}, \bibinfo{author}{A. Vairo},
  \bibinfo{author}{J.~H. Weber}, \bibinfo{title}{{Polyakov loop in 2+1 flavor
  QCD from low to high temperatures}}, \bibinfo{journal}{Phys. Rev.}
  \bibinfo{volume}{D93} (\bibinfo{number}{11}) (\bibinfo{year}{2016})
  \bibinfo{pages}{114502}, \bibinfo{doi}{\doi{10.1103/PhysRevD.93.114502}},
  \eprint{1603.06637}.

\bibtype{Article}%
\bibitem{Borsanyi:2025lim}
\bibinfo{author}{Szabolcs Bors\'anyi}, \bibinfo{author}{Zolt\'an Fodor},
  \bibinfo{author}{Jana~N. Guenther}, \bibinfo{author}{Ruben Kara},
  \bibinfo{author}{Paolo Parotto}, \bibinfo{author}{Attila P\'asztor},
  \bibinfo{author}{Ludovica Pirelli}, \bibinfo{author}{Chik~Him Wong},
  \bibinfo{title}{{Chiral versus deconfinement properties of the QCD crossover:
  Differences in the volume and chemical potential dependence from the
  lattice}}, \bibinfo{journal}{Phys. Rev. D} \bibinfo{volume}{111}
  (\bibinfo{number}{1}) (\bibinfo{year}{2025}) \bibinfo{pages}{014506},
  \bibinfo{doi}{\doi{10.1103/PhysRevD.111.014506}}.

\bibtype{Article}%
\bibitem{Bazavov:2014pvz}
\bibinfo{author}{A. Bazavov}, et al. (\bibinfo{collaboration}{HotQCD
  Collaboration}), \bibinfo{title}{{Equation of state in ( 2+1 )-flavor QCD}},
  \bibinfo{journal}{Phys.Rev.} \bibinfo{volume}{D90} (\bibinfo{number}{9})
  (\bibinfo{year}{2014}) \bibinfo{pages}{094503},
  \bibinfo{doi}{\doi{10.1103/PhysRevD.90.094503}}, \eprint{1407.6387}.

\bibtype{Article}%
\bibitem{Bollweg:2022fqq}
\bibinfo{author}{D. Bollweg}, \bibinfo{author}{D.~A. Clarke},
  \bibinfo{author}{J. Goswami}, \bibinfo{author}{O. Kaczmarek},
  \bibinfo{author}{F. Karsch}, \bibinfo{author}{Swagato Mukherjee},
  \bibinfo{author}{P. Petreczky}, \bibinfo{author}{C. Schmidt},
  \bibinfo{author}{Sipaz Sharma} (\bibinfo{collaboration}{HotQCD}),
  \bibinfo{title}{{Equation of state and speed of sound of (2+1)-flavor QCD in
  strangeness-neutral matter at nonvanishing net baryon-number density}},
  \bibinfo{journal}{Phys. Rev. D} \bibinfo{volume}{108} (\bibinfo{number}{1})
  (\bibinfo{year}{2023}) \bibinfo{pages}{014510},
  \bibinfo{doi}{\doi{10.1103/PhysRevD.108.014510}}, \eprint{2212.09043}.

\bibtype{Article}%
\bibitem{Engels:1990vr}
\bibinfo{author}{J. Engels}, \bibinfo{author}{J. Fingberg}, \bibinfo{author}{F.
  Karsch}, \bibinfo{author}{D. Miller}, \bibinfo{author}{M. Weber},
  \bibinfo{title}{{Nonperturbative thermodynamics of SU(N) gauge theories}},
  \bibinfo{journal}{Phys.Lett.} \bibinfo{volume}{B252} (\bibinfo{year}{1990})
  \bibinfo{pages}{625--630}, \bibinfo{doi}{\doi{10.1016/0370-2693(90)90496-S}}.

\bibtype{Article}%
\bibitem{Borsanyi:2013bia}
\bibinfo{author}{Szabolcs Borsanyi}, \bibinfo{author}{Zoltan Fodor},
  \bibinfo{author}{Christian Hoelbling}, \bibinfo{author}{Sandor~D. Katz},
  \bibinfo{author}{Stefan Krieg}, et al., \bibinfo{title}{{Full result for the
  QCD equation of state with 2+1 flavors}}, \bibinfo{journal}{Phys.Lett.}
  \bibinfo{volume}{B730} (\bibinfo{year}{2014}) \bibinfo{pages}{99--104},
  \bibinfo{doi}{\doi{10.1016/j.physletb.2014.01.007}}, \eprint{1309.5258}.

\bibtype{Article}%
\bibitem{Allton:2002zi}
\bibinfo{author}{C.R. Allton}, \bibinfo{author}{S. Ejiri},
  \bibinfo{author}{S.J. Hands}, \bibinfo{author}{O. Kaczmarek},
  \bibinfo{author}{F. Karsch}, et al., \bibinfo{title}{{The QCD thermal phase
  transition in the presence of a small chemical potential}},
  \bibinfo{journal}{Phys.Rev.} \bibinfo{volume}{D66} (\bibinfo{year}{2002})
  \bibinfo{pages}{074507}, \bibinfo{doi}{\doi{10.1103/PhysRevD.66.074507}},
  \eprint{hep-lat/0204010}.

\bibtype{Article}%
\bibitem{Borsanyi:2023wno}
\bibinfo{author}{Szabolcs Borsanyi}, \bibinfo{author}{Zoltan Fodor},
  \bibinfo{author}{Jana~N. Guenther}, \bibinfo{author}{Sandor~D. Katz},
  \bibinfo{author}{Paolo Parotto}, \bibinfo{author}{Attila Pasztor},
  \bibinfo{author}{David Pesznyak}, \bibinfo{author}{Kalman~K. Szabo},
  \bibinfo{author}{Chik~Him Wong}, \bibinfo{title}{{Continuum-extrapolated
  high-order baryon fluctuations}}, \bibinfo{journal}{Phys. Rev. D}
  \bibinfo{volume}{110} (\bibinfo{number}{1}) (\bibinfo{year}{2024})
  \bibinfo{pages}{L011501}, \bibinfo{doi}{\doi{10.1103/PhysRevD.110.L011501}},
  \eprint{2312.07528}.

\bibtype{Article}%
\bibitem{Monnai:2019hkn}
\bibinfo{author}{Akihiko Monnai}, \bibinfo{author}{Bj{\"o}rn Schenke},
  \bibinfo{author}{Chun Shen}, \bibinfo{title}{{Equation of state at finite
  densities for QCD matter in nuclear collisions}}, \bibinfo{journal}{Phys.
  Rev. C} \bibinfo{volume}{100} (\bibinfo{number}{2}) (\bibinfo{year}{2019})
  \bibinfo{pages}{024907}, \bibinfo{doi}{\doi{10.1103/PhysRevC.100.024907}},
  \eprint{1902.05095}.

\bibtype{Article}%
\bibitem{Noronha-Hostler:2019ayj}
\bibinfo{author}{J. Noronha-Hostler}, \bibinfo{author}{P. Parotto},
  \bibinfo{author}{C. Ratti}, \bibinfo{author}{J.~M. Stafford},
  \bibinfo{title}{{Lattice-based equation of state at finite baryon number,
  electric charge and strangeness chemical potentials}},
  \bibinfo{journal}{Phys. Rev. C} \bibinfo{volume}{100} (\bibinfo{number}{6})
  (\bibinfo{year}{2019}) \bibinfo{pages}{064910},
  \bibinfo{doi}{\doi{10.1103/PhysRevC.100.064910}}, \eprint{1902.06723}.

\bibtype{Article}%
\bibitem{DElia:2016jqh}
\bibinfo{author}{Massimo D'Elia}, \bibinfo{author}{Giuseppe Gagliardi},
  \bibinfo{author}{Francesco Sanfilippo}, \bibinfo{title}{{Higher order quark
  number fluctuations via imaginary chemical potentials in $N_f=2+1$ QCD}},
  \bibinfo{journal}{Phys. Rev.} \bibinfo{volume}{D95} (\bibinfo{number}{9})
  (\bibinfo{year}{2017}) \bibinfo{pages}{094503},
  \bibinfo{doi}{\doi{10.1103/PhysRevD.95.094503}}, \eprint{1611.08285}.

\bibtype{Article}%
\bibitem{Borsanyi:2018grb}
\bibinfo{author}{Szabolcs Borsanyi}, \bibinfo{author}{Zoltan Fodor},
  \bibinfo{author}{Jana~N. Guenther}, \bibinfo{author}{Sandor~K. Katz},
  \bibinfo{author}{Kalman~K. Szabo}, \bibinfo{author}{Attila Pasztor},
  \bibinfo{author}{Israel Portillo}, \bibinfo{author}{Claudia Ratti},
  \bibinfo{title}{{Higher order fluctuations and correlations of conserved
  charges from lattice QCD}}, \bibinfo{journal}{JHEP} \bibinfo{volume}{10}
  (\bibinfo{year}{2018}) \bibinfo{pages}{205},
  \bibinfo{doi}{\doi{10.1007/JHEP10(2018)205}}, \eprint{1805.04445}.

\bibtype{Article}%
\bibitem{Abuali:2025tbd}
\bibinfo{author}{Ahmed Abuali}, \bibinfo{author}{Szabolcs Bors{\'a}nyi},
  \bibinfo{author}{Zolt{\'a}n Fodor}, \bibinfo{author}{Johannes Jahan},
  \bibinfo{author}{Micheal Kahangirwe}, \bibinfo{author}{Paolo Parotto},
  \bibinfo{author}{Attila P{\'a}sztor}, \bibinfo{author}{Claudia Ratti},
  \bibinfo{author}{Hitansh Shah}, \bibinfo{author}{Seth~A. Trabulsi},
  \bibinfo{title}{{New 4D lattice QCD equation of state: Extended density
  coverage from a generalized T' expansion}}, \bibinfo{journal}{Phys. Rev. D}
  \bibinfo{volume}{112} (\bibinfo{number}{5}) (\bibinfo{year}{2025})
  \bibinfo{pages}{054502}, \bibinfo{doi}{\doi{10.1103/2dmh-26yh}},
  \eprint{2504.01881}.

\bibtype{Article}%
\bibitem{Kajantie:2002wa}
\bibinfo{author}{K. Kajantie}, \bibinfo{author}{M. Laine}, \bibinfo{author}{K.
  Rummukainen}, \bibinfo{author}{Y. Schroder}, \bibinfo{title}{{The Pressure of
  hot QCD up to g6 ln(1/g)}}, \bibinfo{journal}{Phys.Rev.}
  \bibinfo{volume}{D67} (\bibinfo{year}{2003}) \bibinfo{pages}{105008},
  \bibinfo{doi}{\doi{10.1103/PhysRevD.67.105008}}, \eprint{hep-ph/0211321}.

\bibtype{Article}%
\bibitem{Linde:1980ts}
\bibinfo{author}{Andrei~D. Linde}, \bibinfo{title}{{Infrared Problem in
  Thermodynamics of the Yang-Mills Gas}}, \bibinfo{journal}{Phys. Lett. B}
  \bibinfo{volume}{96} (\bibinfo{year}{1980}) \bibinfo{pages}{289--292},
  \bibinfo{doi}{\doi{10.1016/0370-2693(80)90769-8}}.

\bibtype{Article}%
\bibitem{Gross:1980br}
\bibinfo{author}{David~J. Gross}, \bibinfo{author}{Robert~D. Pisarski},
  \bibinfo{author}{Laurence~G. Yaffe}, \bibinfo{title}{{QCD and Instantons at
  Finite Temperature}}, \bibinfo{journal}{Rev. Mod. Phys.} \bibinfo{volume}{53}
  (\bibinfo{year}{1981}) \bibinfo{pages}{43},
  \bibinfo{doi}{\doi{10.1103/RevModPhys.53.43}}.

\bibtype{Article}%
\bibitem{Braaten:1991gm}
\bibinfo{author}{Eric Braaten}, \bibinfo{author}{Robert~D. Pisarski},
  \bibinfo{title}{{Simple effective Lagrangian for hard thermal loops}},
  \bibinfo{journal}{Phys.Rev.} \bibinfo{volume}{D45} (\bibinfo{year}{1992})
  \bibinfo{pages}{R1827--1830}, \bibinfo{doi}{\doi{10.1103/PhysRevD.45.R1827}}.

\bibtype{Article}%
\bibitem{Andersen:2002ey}
\bibinfo{author}{Jens~O. Andersen}, \bibinfo{author}{Eric Braaten},
  \bibinfo{author}{Emmanuel Petitgirard}, \bibinfo{author}{Michael Strickland},
  \bibinfo{title}{{HTL perturbation theory to two loops}},
  \bibinfo{journal}{Phys.Rev.} \bibinfo{volume}{D66} (\bibinfo{year}{2002})
  \bibinfo{pages}{085016}, \bibinfo{doi}{\doi{10.1103/PhysRevD.66.085016}},
  \eprint{hep-ph/0205085}.

\bibtype{Article}%
\bibitem{Andersen:2011sf}
\bibinfo{author}{Jens~O. Andersen}, \bibinfo{author}{Lars~E. Leganger},
  \bibinfo{author}{Michael Strickland}, \bibinfo{author}{Nan Su},
  \bibinfo{title}{{Three-loop HTL QCD thermodynamics}}, \bibinfo{journal}{JHEP}
  \bibinfo{volume}{1108} (\bibinfo{year}{2011}) \bibinfo{pages}{053},
  \bibinfo{doi}{\doi{10.1007/JHEP08(2011)053}}, \eprint{1103.2528}.

\bibtype{Article}%
\bibitem{Braaten:1995cm}
\bibinfo{author}{Eric Braaten}, \bibinfo{author}{Agustin Nieto},
  \bibinfo{title}{{Effective field theory approach to high temperature
  thermodynamics}}, \bibinfo{journal}{Phys.Rev.} \bibinfo{volume}{D51}
  (\bibinfo{year}{1995}) \bibinfo{pages}{6990--7006},
  \bibinfo{doi}{\doi{10.1103/PhysRevD.51.6990}}, \eprint{hep-ph/9501375}.

\bibtype{Article}%
\bibitem{Borsanyi:2016ksw}
\bibinfo{author}{Sz. Borsanyi}, et al., \bibinfo{title}{{Calculation of the
  axion mass based on high-temperature lattice quantum chromodynamics}},
  \bibinfo{journal}{Nature} \bibinfo{volume}{539} (\bibinfo{number}{7627})
  (\bibinfo{year}{2016}) \bibinfo{pages}{69--71},
  \bibinfo{doi}{\doi{10.1038/nature20115}}, \eprint{1606.07494}.

\bibtype{Article}%
\bibitem{Ghiglieri:2020dpq}
\bibinfo{author}{Jacopo Ghiglieri}, \bibinfo{author}{Aleksi Kurkela},
  \bibinfo{author}{Michael Strickland}, \bibinfo{author}{Aleksi Vuorinen},
  \bibinfo{title}{{Perturbative Thermal QCD: Formalism and Applications}},
  \bibinfo{journal}{Phys. Rept.} \bibinfo{volume}{880} (\bibinfo{year}{2020})
  \bibinfo{pages}{1--73}, \bibinfo{doi}{\doi{10.1016/j.physrep.2020.07.004}},
  \eprint{2002.10188}.

\bibtype{Article}%
\bibitem{Blaizot:2001vr}
\bibinfo{author}{J.P. Blaizot}, \bibinfo{author}{Edmond Iancu},
  \bibinfo{author}{A. Rebhan}, \bibinfo{title}{{Quark number susceptibilities
  from HTL resummed thermodynamics}}, \bibinfo{journal}{Phys.Lett.}
  \bibinfo{volume}{B523} (\bibinfo{year}{2001}) \bibinfo{pages}{143--150},
  \bibinfo{doi}{\doi{10.1016/S0370-2693(01)01316-8}}, \eprint{hep-ph/0110369}.

\bibtype{Article}%
\bibitem{Haque:2013sja}
\bibinfo{author}{Najmul Haque}, \bibinfo{author}{Jens~O. Andersen},
  \bibinfo{author}{Munshi~G. Mustafa}, \bibinfo{author}{Michael Strickland},
  \bibinfo{author}{Nan Su}, \bibinfo{title}{{Three-loop HTLpt Pressure and
  Susceptibilities at Finite Temperature and Density}},
  \bibinfo{journal}{Phys.Rev.} \bibinfo{volume}{D89} (\bibinfo{year}{2014})
  \bibinfo{pages}{061701}, \bibinfo{doi}{\doi{10.1103/PhysRevD.89.061701}},
  \eprint{1309.3968}.

\bibtype{Article}%
\bibitem{Andersen:2012wr}
\bibinfo{author}{Jens~O. Andersen}, \bibinfo{author}{Sylvain Mogliacci},
  \bibinfo{author}{Nan Su}, \bibinfo{author}{Aleksi Vuorinen},
  \bibinfo{title}{{Quark number susceptibilities from resummed perturbation
  theory}}, \bibinfo{journal}{Phys.Rev.} \bibinfo{volume}{D87}
  (\bibinfo{year}{2013}) \bibinfo{pages}{074003},
  \bibinfo{doi}{\doi{10.1103/PhysRevD.87.074003}}, \eprint{1210.0912}.

\bibtype{Article}%
\bibitem{Haque:2013qta}
\bibinfo{author}{Najmul Haque}, \bibinfo{author}{Munshi~G. Mustafa},
  \bibinfo{author}{Michael Strickland}, \bibinfo{title}{{Quark Number
  Susceptibilities from Two-Loop Hard Thermal Loop Perturbation Theory}},
  \bibinfo{journal}{JHEP} \bibinfo{volume}{1307} (\bibinfo{year}{2013})
  \bibinfo{pages}{184}, \bibinfo{doi}{\doi{10.1007/JHEP07(2013)184}},
  \eprint{1302.3228}.

\bibtype{Article}%
\bibitem{Andersen:2015eoa}
\bibinfo{author}{Jens~O. Andersen}, \bibinfo{author}{Najmul Haque},
  \bibinfo{author}{Munshi~G. Mustafa}, \bibinfo{author}{Michael Strickland},
  \bibinfo{title}{{Three-loop HTLpt thermodynamics at finite temperature and
  isospin chemical potential}}, \bibinfo{journal}{Phys. Rev.}
  \bibinfo{volume}{D93} (\bibinfo{year}{2016}) \bibinfo{pages}{054045},
  \bibinfo{doi}{\doi{10.1103/PhysRevD.93.054045}}, \eprint{1511.04660}.

\bibtype{Article}%
\bibitem{Bellwied:2015lba}
\bibinfo{author}{R. Bellwied}, \bibinfo{author}{S. Borsanyi},
  \bibinfo{author}{Z. Fodor}, \bibinfo{author}{S.~D. Katz}, \bibinfo{author}{A.
  Pasztor}, \bibinfo{author}{C. Ratti}, \bibinfo{author}{K.~K. Szabo},
  \bibinfo{title}{{Fluctuations and correlations in high temperature QCD}},
  \bibinfo{journal}{Phys. Rev.} \bibinfo{volume}{D92} (\bibinfo{number}{11})
  (\bibinfo{year}{2015}) \bibinfo{pages}{114505},
  \bibinfo{doi}{\doi{10.1103/PhysRevD.92.114505}}, \eprint{1507.04627}.

\bibtype{Article}%
\bibitem{Ding:2015fca}
\bibinfo{author}{H.~T. Ding}, \bibinfo{author}{Swagato Mukherjee},
  \bibinfo{author}{H. Ohno}, \bibinfo{author}{P. Petreczky},
  \bibinfo{author}{H.~P. Schadler}, \bibinfo{title}{{Diagonal and off-diagonal
  quark number susceptibilities at high temperatures}}, \bibinfo{journal}{Phys.
  Rev.} \bibinfo{volume}{D92} (\bibinfo{number}{7}) (\bibinfo{year}{2015})
  \bibinfo{pages}{074043}, \bibinfo{doi}{\doi{10.1103/PhysRevD.92.074043}},
  \eprint{1507.06637}.

\bibtype{Article}%
\bibitem{Haque:2014rua}
\bibinfo{author}{Najmul Haque}, \bibinfo{author}{Aritra Bandyopadhyay},
  \bibinfo{author}{Jens~O. Andersen}, \bibinfo{author}{Munshi~G. Mustafa},
  \bibinfo{author}{Michael Strickland}, et al., \bibinfo{title}{{Three-loop
  HTLpt thermodynamics at finite temperature and chemical potential}},
  \bibinfo{journal}{JHEP} \bibinfo{volume}{1405} (\bibinfo{year}{2014})
  \bibinfo{pages}{027}, \bibinfo{doi}{\doi{10.1007/JHEP05(2014)027}},
  \eprint{1402.6907}.

\bibtype{Article}%
\bibitem{Mogliacci:2013mca}
\bibinfo{author}{Sylvain Mogliacci}, \bibinfo{author}{Jens~O. Andersen},
  \bibinfo{author}{Michael Strickland}, \bibinfo{author}{Nan Su},
  \bibinfo{author}{Aleksi Vuorinen}, \bibinfo{title}{{Equation of State of hot
  and dense QCD: Resummed perturbation theory confronts lattice data}},
  \bibinfo{journal}{JHEP} \bibinfo{volume}{1312} (\bibinfo{year}{2013})
  \bibinfo{pages}{055}, \bibinfo{doi}{\doi{10.1007/JHEP12(2013)055}},
  \eprint{1307.8098}.

\bibtype{Article}%
\bibitem{Bazavov:2013uja}
\bibinfo{author}{A. Bazavov}, \bibinfo{author}{H.-T. Ding}, \bibinfo{author}{P.
  Hegde}, \bibinfo{author}{F. Karsch}, \bibinfo{author}{C. Miao}, et al.,
  \bibinfo{title}{{Quark number susceptibilities at high temperatures}},
  \bibinfo{journal}{Phys.Rev.} \bibinfo{volume}{D88} (\bibinfo{number}{9})
  (\bibinfo{year}{2013}) \bibinfo{pages}{094021},
  \bibinfo{doi}{\doi{10.1103/PhysRevD.88.094021}}, \eprint{1309.2317}.

\bibtype{Article}%
\bibitem{Appelquist:1981vg}
\bibinfo{author}{Thomas Appelquist}, \bibinfo{author}{Robert~D. Pisarski},
  \bibinfo{title}{{High-Temperature Yang-Mills Theories and Three-Dimensional
  Quantum Chromodynamics}}, \bibinfo{journal}{Phys. Rev. D}
  \bibinfo{volume}{23} (\bibinfo{year}{1981}) \bibinfo{pages}{2305},
  \bibinfo{doi}{\doi{10.1103/PhysRevD.23.2305}}.

\bibtype{Article}%
\bibitem{Kajantie:1995dw}
\bibinfo{author}{K. Kajantie}, \bibinfo{author}{M. Laine}, \bibinfo{author}{K.
  Rummukainen}, \bibinfo{author}{Mikhail~E. Shaposhnikov},
  \bibinfo{title}{{Generic rules for high temperature dimensional reduction and
  their application to the standard model}}, \bibinfo{journal}{Nucl. Phys.}
  \bibinfo{volume}{B458} (\bibinfo{year}{1996}) \bibinfo{pages}{90--136},
  \bibinfo{doi}{\doi{10.1016/0550-3213(95)00549-8}}, \eprint{hep-ph/9508379}.

\bibtype{Article}%
\bibitem{Braaten:1995jr}
\bibinfo{author}{Eric Braaten}, \bibinfo{author}{Agustin Nieto},
  \bibinfo{title}{{Free energy of QCD at high temperature}},
  \bibinfo{journal}{Phys. Rev. D} \bibinfo{volume}{53} (\bibinfo{year}{1996})
  \bibinfo{pages}{3421--3437}, \bibinfo{doi}{\doi{10.1103/PhysRevD.53.3421}},
  \eprint{hep-ph/9510408}.

\bibtype{Article}%
\bibitem{Hietanen:2008tv}
\bibinfo{author}{A. Hietanen}, \bibinfo{author}{K. Kajantie},
  \bibinfo{author}{M. Laine}, \bibinfo{author}{K. Rummukainen},
  \bibinfo{author}{Y. Schroder}, \bibinfo{title}{{Three-dimensional physics and
  the pressure of hot QCD}}, \bibinfo{journal}{Phys.Rev.} \bibinfo{volume}{D79}
  (\bibinfo{year}{2009}) \bibinfo{pages}{045018},
  \bibinfo{doi}{\doi{10.1103/PhysRevD.79.045018}}, \eprint{0811.4664}.

\bibtype{Article}%
\bibitem{Hietanen:2008xb}
\bibinfo{author}{Ari Hietanen}, \bibinfo{author}{Kari Rummukainen},
  \bibinfo{title}{{The Diagonal and off-diagonal quark number susceptibility of
  high temperature and finite density QCD}}, \bibinfo{journal}{JHEP}
  \bibinfo{volume}{04} (\bibinfo{year}{2008}) \bibinfo{pages}{078},
  \bibinfo{doi}{\doi{10.1088/1126-6708/2008/04/078}}, \eprint{0802.3979}.

\bibtype{Article}%
\bibitem{Rummukainen:2021qyl}
\bibinfo{author}{Kari Rummukainen}, \bibinfo{author}{Niels Schlusser},
  \bibinfo{title}{{High order quark number susceptibilities in hot QCD from
  lattice electrostatic QCD}}, \bibinfo{journal}{Phys. Rev. D}
  \bibinfo{volume}{106} (\bibinfo{number}{5}) (\bibinfo{year}{2022})
  \bibinfo{pages}{054507}, \bibinfo{doi}{\doi{10.1103/PhysRevD.106.054507}},
  \eprint{2108.13236}.

\bibtype{Article}%
\bibitem{Laine:2006cp}
\bibinfo{author}{Mikko Laine}, \bibinfo{author}{York Schroder},
  \bibinfo{title}{{Quark mass thresholds in QCD thermodynamics}},
  \bibinfo{journal}{Phys.Rev.} \bibinfo{volume}{D73} (\bibinfo{year}{2006})
  \bibinfo{pages}{085009}, \bibinfo{doi}{\doi{10.1103/PhysRevD.73.085009}},
  \eprint{hep-ph/0603048}.

\bibtype{Article}%
\bibitem{Wetterich:1992yh}
\bibinfo{author}{Christof Wetterich}, \bibinfo{title}{{Exact evolution equation
  for the effective potential}}, \bibinfo{journal}{Phys. Lett. B}
  \bibinfo{volume}{301} (\bibinfo{year}{1993}) \bibinfo{pages}{90--94},
  \bibinfo{doi}{\doi{10.1016/0370-2693(93)90726-X}}, \eprint{1710.05815}.

\bibtype{Article}%
\bibitem{Ellwanger:1993mw}
\bibinfo{author}{Ulrich Ellwanger}, \bibinfo{title}{{FLow equations for N point
  functions and bound states}}, \bibinfo{journal}{Z. Phys. C}
  \bibinfo{volume}{62} (\bibinfo{year}{1994}) \bibinfo{pages}{503--510},
  \bibinfo{doi}{\doi{10.1007/BF01555911}}, \eprint{hep-ph/9308260}.

\bibtype{Article}%
\bibitem{Morris:1993qb}
\bibinfo{author}{Tim~R. Morris}, \bibinfo{title}{{The Exact renormalization
  group and approximate solutions}}, \bibinfo{journal}{Int. J. Mod. Phys.}
  \bibinfo{volume}{A9} (\bibinfo{year}{1994}) \bibinfo{pages}{2411--2450},
  \bibinfo{doi}{\doi{10.1142/S0217751X94000972}}, \eprint{hep-ph/9308265}.

\bibtype{Article}%
\bibitem{Dupuis:2020fhh}
\bibinfo{author}{N. Dupuis}, \bibinfo{author}{L. Canet}, \bibinfo{author}{A.
  Eichhorn}, \bibinfo{author}{W. Metzner}, \bibinfo{author}{J.~M. Pawlowski},
  \bibinfo{author}{M. Tissier}, \bibinfo{author}{N. Wschebor},
  \bibinfo{title}{{The nonperturbative functional renormalization group and its
  applications}}, \bibinfo{journal}{Phys. Rept.} \bibinfo{volume}{910}
  (\bibinfo{year}{2021}) \bibinfo{pages}{1--114},
  \bibinfo{doi}{\doi{10.1016/j.physrep.2021.01.001}}, \eprint{2006.04853}.

\bibtype{Article}%
\bibitem{Ihssen:2024miv}
\bibinfo{author}{Friederike Ihssen}, \bibinfo{author}{Jan~M. Pawlowski},
  \bibinfo{author}{Franz~R. Sattler}, \bibinfo{author}{Nicolas Wink},
  \bibinfo{title}{{Towards quantitative precision in functional QCD I}}
  (\bibinfo{year}{2024}), \eprint{2408.08413}.

\bibtype{Article}%
\bibitem{Fischer:2008uz}
\bibinfo{author}{Christian~S. Fischer}, \bibinfo{author}{Axel Maas},
  \bibinfo{author}{Jan~M. Pawlowski}, \bibinfo{title}{{On the infrared behavior
  of Landau gauge Yang-Mills theory}}, \bibinfo{journal}{Annals Phys.}
  \bibinfo{volume}{324} (\bibinfo{year}{2009}) \bibinfo{pages}{2408--2437},
  \bibinfo{doi}{\doi{10.1016/j.aop.2009.07.009}}, \eprint{0810.1987}.

\bibtype{Article}%
\bibitem{Cyrol:2016tym}
\bibinfo{author}{Anton~K. Cyrol}, \bibinfo{author}{Leonard Fister},
  \bibinfo{author}{Mario Mitter}, \bibinfo{author}{Jan~M. Pawlowski},
  \bibinfo{author}{Nils Strodthoff}, \bibinfo{title}{{Landau gauge Yang-Mills
  correlation functions}}, \bibinfo{journal}{Phys. Rev. D} \bibinfo{volume}{94}
  (\bibinfo{number}{5}) (\bibinfo{year}{2016}) \bibinfo{pages}{054005},
  \bibinfo{doi}{\doi{10.1103/PhysRevD.94.054005}}, \eprint{1605.01856}.

\bibtype{Article}%
\bibitem{Huber:2020keu}
\bibinfo{author}{Markus~Q. Huber}, \bibinfo{title}{{Correlation functions of
  Landau gauge Yang-Mills theory}}, \bibinfo{journal}{Phys. Rev. D}
  \bibinfo{volume}{101} (\bibinfo{year}{2020}) \bibinfo{pages}{114009},
  \bibinfo{doi}{\doi{10.1103/PhysRevD.101.114009}}, \eprint{2003.13703}.

\bibtype{Article}%
\bibitem{Braun:2007bx}
\bibinfo{author}{Jens Braun}, \bibinfo{author}{Holger Gies},
  \bibinfo{author}{Jan~M. Pawlowski}, \bibinfo{title}{{Quark Confinement from
  Color Confinement}}, \bibinfo{journal}{Phys.Lett.} \bibinfo{volume}{B684}
  (\bibinfo{year}{2010}) \bibinfo{pages}{262--267},
  \bibinfo{doi}{\doi{10.1016/j.physletb.2010.01.009}}, \eprint{0708.2413}.

\bibtype{Article}%
\bibitem{Gies:2001nw}
\bibinfo{author}{Holger Gies}, \bibinfo{author}{Christof Wetterich},
  \bibinfo{title}{{Renormalization flow of bound states}},
  \bibinfo{journal}{Phys. Rev. D} \bibinfo{volume}{65} (\bibinfo{year}{2002})
  \bibinfo{pages}{065001}, \bibinfo{doi}{\doi{10.1103/PhysRevD.65.065001}},
  \eprint{hep-th/0107221}.

\bibtype{Article}%
\bibitem{Gies:2002hq}
\bibinfo{author}{Holger Gies}, \bibinfo{author}{Christof Wetterich},
  \bibinfo{title}{{Universality of spontaneous chiral symmetry breaking in
  gauge theories}}, \bibinfo{journal}{Phys. Rev. D} \bibinfo{volume}{69}
  (\bibinfo{year}{2004}) \bibinfo{pages}{025001},
  \bibinfo{doi}{\doi{10.1103/PhysRevD.69.025001}}, \eprint{hep-th/0209183}.

\bibtype{Article}%
\bibitem{Pawlowski:2005xe}
\bibinfo{author}{Jan~M. Pawlowski}, \bibinfo{title}{{Aspects of the functional
  renormalisation group}}, \bibinfo{journal}{Annals Phys.}
  \bibinfo{volume}{322} (\bibinfo{year}{2007}) \bibinfo{pages}{2831--2915},
  \bibinfo{doi}{\doi{10.1016/j.aop.2007.01.007}}, \eprint{hep-th/0512261}.

\bibtype{Article}%
\bibitem{Fukushima:2021ctq}
\bibinfo{author}{Kenji Fukushima}, \bibinfo{author}{Jan~M. Pawlowski},
  \bibinfo{author}{Nils Strodthoff}, \bibinfo{title}{{Emergent hadrons and
  diquarks}}, \bibinfo{journal}{Annals Phys.} \bibinfo{volume}{446}
  (\bibinfo{year}{2022}) \bibinfo{pages}{169106},
  \bibinfo{doi}{\doi{10.1016/j.aop.2022.169106}}, \eprint{2103.01129}.

\bibtype{Article}%
\bibitem{Braun:2014ata}
\bibinfo{author}{Jens Braun}, \bibinfo{author}{Leonard Fister},
  \bibinfo{author}{Jan~M. Pawlowski}, \bibinfo{author}{Fabian Rennecke},
  \bibinfo{title}{{From Quarks and Gluons to Hadrons: Chiral Symmetry Breaking
  in Dynamical QCD}}, \bibinfo{journal}{Phys. Rev. D} \bibinfo{volume}{94}
  (\bibinfo{number}{3}) (\bibinfo{year}{2016}) \bibinfo{pages}{034016},
  \bibinfo{doi}{\doi{10.1103/PhysRevD.94.034016}}, \eprint{1412.1045}.

\bibtype{Article}%
\bibitem{Fu:2019hdw}
\bibinfo{author}{Wei-jie Fu}, \bibinfo{author}{Jan~M. Pawlowski},
  \bibinfo{author}{Fabian Rennecke}, \bibinfo{title}{{QCD phase structure at
  finite temperature and density}}, \bibinfo{journal}{Phys. Rev. D}
  \bibinfo{volume}{101} (\bibinfo{number}{5}) (\bibinfo{year}{2020})
  \bibinfo{pages}{054032}, \bibinfo{doi}{\doi{10.1103/PhysRevD.101.054032}},
  \eprint{1909.02991}.

\bibtype{Article}%
\bibitem{Braun:2006jd}
\bibinfo{author}{Jens Braun}, \bibinfo{author}{Holger Gies},
  \bibinfo{title}{{Chiral phase boundary of QCD at finite temperature}},
  \bibinfo{journal}{JHEP} \bibinfo{volume}{06} (\bibinfo{year}{2006})
  \bibinfo{pages}{024}, \bibinfo{doi}{\doi{10.1088/1126-6708/2006/06/024}},
  \eprint{hep-ph/0602226}.

\bibtype{Article}%
\bibitem{Mitter:2014wpa}
\bibinfo{author}{Mario Mitter}, \bibinfo{author}{Jan~M. Pawlowski},
  \bibinfo{author}{Nils Strodthoff}, \bibinfo{title}{{Chiral symmetry breaking
  in continuum QCD}}, \bibinfo{journal}{Phys. Rev. D} \bibinfo{volume}{91}
  (\bibinfo{year}{2015}) \bibinfo{pages}{054035},
  \bibinfo{doi}{\doi{10.1103/PhysRevD.91.054035}}, \eprint{1411.7978}.

\bibtype{Article}%
\bibitem{Braun:2009gm}
\bibinfo{author}{Jens Braun}, \bibinfo{author}{Lisa~M. Haas},
  \bibinfo{author}{Florian Marhauser}, \bibinfo{author}{Jan~M. Pawlowski},
  \bibinfo{title}{{Phase Structure of Two-Flavor QCD at Finite Chemical
  Potential}}, \bibinfo{journal}{Phys. Rev. Lett.} \bibinfo{volume}{106}
  (\bibinfo{year}{2011}) \bibinfo{pages}{022002},
  \bibinfo{doi}{\doi{10.1103/PhysRevLett.106.022002}}, \eprint{0908.0008}.

\bibtype{Article}%
\bibitem{Fischer:2018sdj}
\bibinfo{author}{Christian~S. Fischer}, \bibinfo{title}{{QCD at finite
  temperature and chemical potential from DysonSchwinger equations}},
  \bibinfo{journal}{Prog. Part. Nucl. Phys.} \bibinfo{volume}{105}
  (\bibinfo{year}{2019}) \bibinfo{pages}{1--60},
  \bibinfo{doi}{\doi{10.1016/j.ppnp.2019.01.002}}, \eprint{1810.12938}.

\bibtype{Article}%
\bibitem{Eichmann:2016yit}
\bibinfo{author}{Gernot Eichmann}, \bibinfo{author}{Helios Sanchis-Alepuz},
  \bibinfo{author}{Richard Williams}, \bibinfo{author}{Reinhard Alkofer},
  \bibinfo{author}{Christian~S. Fischer}, \bibinfo{title}{{Baryons as
  relativistic three-quark bound states}}, \bibinfo{journal}{Prog. Part. Nucl.
  Phys.} \bibinfo{volume}{91} (\bibinfo{year}{2016}) \bibinfo{pages}{1--100},
  \bibinfo{doi}{\doi{10.1016/j.ppnp.2016.07.001}}, \eprint{1606.09602}.

\bibtype{Article}%
\bibitem{Fischer:2010fx}
\bibinfo{author}{Christian~S. Fischer}, \bibinfo{author}{Axel Maas},
  \bibinfo{author}{Jens~A. Muller}, \bibinfo{title}{{Chiral and deconfinement
  transition from correlation functions: SU(2) vs. SU(3)}},
  \bibinfo{journal}{Eur. Phys. J. C} \bibinfo{volume}{68}
  (\bibinfo{year}{2010}) \bibinfo{pages}{165--181},
  \bibinfo{doi}{\doi{10.1140/epjc/s10052-010-1343-1}}, \eprint{1003.1960}.

\bibtype{Article}%
\bibitem{Fischer:2014vxa}
\bibinfo{author}{Christian~S. Fischer}, \bibinfo{author}{Jan Luecker},
  \bibinfo{author}{Jan~M. Pawlowski}, \bibinfo{title}{{Phase structure of QCD
  for heavy quarks}}, \bibinfo{journal}{Phys. Rev.} \bibinfo{volume}{D91}
  (\bibinfo{number}{1}) (\bibinfo{year}{2015}) \bibinfo{pages}{014024},
  \bibinfo{doi}{\doi{10.1103/PhysRevD.91.014024}}, \eprint{1409.8462}.

\bibtype{Article}%
\bibitem{Roberts:2000aa}
\bibinfo{author}{Craig~D. Roberts}, \bibinfo{author}{Sebastian~M. Schmidt},
  \bibinfo{title}{{Dyson-Schwinger equations: Density, temperature and
  continuum strong QCD}}, \bibinfo{journal}{Prog. Part. Nucl. Phys.}
  \bibinfo{volume}{45} (\bibinfo{year}{2000}) \bibinfo{pages}{S1--S103},
  \bibinfo{doi}{\doi{10.1016/S0146-6410(00)90011-5}}, \eprint{nucl-th/0005064}.

\bibtype{Article}%
\bibitem{Fischer:2011pk}
\bibinfo{author}{Christian~S. Fischer}, \bibinfo{author}{Jens~A. Mueller},
  \bibinfo{title}{{On critical scaling at the QCD $N_f$ = 2 chiral phase
  transition}}, \bibinfo{journal}{Phys. Rev. D} \bibinfo{volume}{84}
  (\bibinfo{year}{2011}) \bibinfo{pages}{054013},
  \bibinfo{doi}{\doi{10.1103/PhysRevD.84.054013}}, \eprint{1106.2700}.

\bibtype{Article}%
\bibitem{Fischer:2012vc}
\bibinfo{author}{Christian~S. Fischer}, \bibinfo{author}{Jan Luecker},
  \bibinfo{title}{{Propagators and phase structure of Nf=2 and Nf=2+1 QCD}},
  \bibinfo{journal}{Phys. Lett. B} \bibinfo{volume}{718} (\bibinfo{year}{2013})
  \bibinfo{pages}{1036--1043},
  \bibinfo{doi}{\doi{10.1016/j.physletb.2012.11.054}}, \eprint{1206.5191}.

\bibtype{Article}%
\bibitem{Fischer:2013eca}
\bibinfo{author}{Christian~S. Fischer}, \bibinfo{author}{Leonard Fister},
  \bibinfo{author}{Jan Luecker}, \bibinfo{author}{Jan~M. Pawlowski},
  \bibinfo{title}{{Polyakov loop potential at finite density}},
  \bibinfo{journal}{Phys.Lett.} \bibinfo{volume}{B732} (\bibinfo{year}{2014})
  \bibinfo{pages}{273--277},
  \bibinfo{doi}{\doi{10.1016/j.physletb.2014.03.057}}, \eprint{1306.6022}.

\bibtype{Article}%
\bibitem{Fischer:2014ata}
\bibinfo{author}{Christian~S. Fischer}, \bibinfo{author}{Jan Luecker},
  \bibinfo{author}{Christian~A. Welzbacher}, \bibinfo{title}{{Phase structure
  of three and four flavor QCD}}, \bibinfo{journal}{Phys.Rev.}
  \bibinfo{volume}{D90} (\bibinfo{number}{3}) (\bibinfo{year}{2014})
  \bibinfo{pages}{034022}, \bibinfo{doi}{\doi{10.1103/PhysRevD.90.034022}},
  \eprint{1405.4762}.

\bibtype{Article}%
\bibitem{Eichmann:2015kfa}
\bibinfo{author}{Gernot Eichmann}, \bibinfo{author}{Christian~S. Fischer},
  \bibinfo{author}{Christian~A. Welzbacher}, \bibinfo{title}{{Baryon effects on
  the location of QCD's critical end point}}, \bibinfo{journal}{Phys. Rev. D}
  \bibinfo{volume}{93} (\bibinfo{number}{3}) (\bibinfo{year}{2016})
  \bibinfo{pages}{034013}, \bibinfo{doi}{\doi{10.1103/PhysRevD.93.034013}},
  \eprint{1509.02082}.

\bibtype{Article}%
\bibitem{Gunkel:2021oya}
\bibinfo{author}{Pascal~J. Gunkel}, \bibinfo{author}{Christian~S. Fischer},
  \bibinfo{title}{{Locating the critical endpoint of QCD: Mesonic backcoupling
  effects}}, \bibinfo{journal}{Phys. Rev. D} \bibinfo{volume}{104}
  (\bibinfo{number}{5}) (\bibinfo{year}{2021}) \bibinfo{pages}{054022},
  \bibinfo{doi}{\doi{10.1103/PhysRevD.104.054022}}, \eprint{2106.08356}.

\bibtype{Article}%
\bibitem{Gao:2015kea}
\bibinfo{author}{Fei Gao}, \bibinfo{author}{Jing Chen}, \bibinfo{author}{Yu-Xin
  Liu}, \bibinfo{author}{Si-Xue Qin}, \bibinfo{author}{Craig~D. Roberts},
  \bibinfo{author}{Sebastian~M. Schmidt}, \bibinfo{title}{{Phase diagram and
  thermal properties of strong-interaction matter}}, \bibinfo{journal}{Phys.
  Rev. D} \bibinfo{volume}{93} (\bibinfo{number}{9}) (\bibinfo{year}{2016})
  \bibinfo{pages}{094019}, \bibinfo{doi}{\doi{10.1103/PhysRevD.93.094019}},
  \eprint{1507.00875}.

\bibtype{Article}%
\bibitem{Isserstedt:2019pgx}
\bibinfo{author}{Philipp Isserstedt}, \bibinfo{author}{Michael Buballa},
  \bibinfo{author}{Christian~S. Fischer}, \bibinfo{author}{Pascal~J. Gunkel},
  \bibinfo{title}{{Baryon number fluctuations in the QCD phase diagram from
  Dyson-Schwinger equations}}, \bibinfo{journal}{Phys. Rev.}
  \bibinfo{volume}{D100} (\bibinfo{number}{7}) (\bibinfo{year}{2019})
  \bibinfo{pages}{074011}, \bibinfo{doi}{\doi{10.1103/PhysRevD.100.074011}},
  \eprint{1906.11644}.

\bibtype{Article}%
\bibitem{Cyrol:2017ewj}
\bibinfo{author}{Anton~K. Cyrol}, \bibinfo{author}{Mario Mitter},
  \bibinfo{author}{Jan~M. Pawlowski}, \bibinfo{author}{Nils Strodthoff},
  \bibinfo{title}{{Nonperturbative quark, gluon, and meson correlators of
  unquenched QCD}}, \bibinfo{journal}{Phys. Rev. D} \bibinfo{volume}{97}
  (\bibinfo{number}{5}) (\bibinfo{year}{2018}) \bibinfo{pages}{054006},
  \bibinfo{doi}{\doi{10.1103/PhysRevD.97.054006}}, \eprint{1706.06326}.

\bibtype{Article}%
\bibitem{Zafeiropoulos:2019flq}
\bibinfo{author}{S. Zafeiropoulos}, \bibinfo{author}{P. Boucaud},
  \bibinfo{author}{F. De~Soto}, \bibinfo{author}{J. Rodr{\'\i}guez-Quintero},
  \bibinfo{author}{J. Segovia}, \bibinfo{title}{{Strong Running Coupling from
  the Gauge Sector of Domain Wall Lattice QCD with Physical Quark Masses}},
  \bibinfo{journal}{Phys. Rev. Lett.} \bibinfo{volume}{122}
  (\bibinfo{number}{16}) (\bibinfo{year}{2019}) \bibinfo{pages}{162002},
  \bibinfo{doi}{\doi{10.1103/PhysRevLett.122.162002}}, \eprint{1902.08148}.

\bibtype{Article}%
\bibitem{Gao:2021wun}
\bibinfo{author}{Fei Gao}, \bibinfo{author}{Joannis Papavassiliou},
  \bibinfo{author}{Jan~M. Pawlowski}, \bibinfo{title}{{Fully coupled functional
  equations for the quark sector of QCD}}, \bibinfo{journal}{Phys. Rev. D}
  \bibinfo{volume}{103} (\bibinfo{number}{9}) (\bibinfo{year}{2021})
  \bibinfo{pages}{094013}, \bibinfo{doi}{\doi{10.1103/PhysRevD.103.094013}},
  \eprint{2102.13053}.

\bibtype{Article}%
\bibitem{Lu:2023mkn}
\bibinfo{author}{Yi Lu}, \bibinfo{author}{Fei Gao}, \bibinfo{author}{Yu-Xin
  Liu}, \bibinfo{author}{Jan~M. Pawlowski}, \bibinfo{title}{{QCD equation of
  state and thermodynamic observables from computationally minimal
  Dyson-Schwinger equations}}, \bibinfo{journal}{Phys. Rev. D}
  \bibinfo{volume}{110} (\bibinfo{number}{1}) (\bibinfo{year}{2024})
  \bibinfo{pages}{014036}, \bibinfo{doi}{\doi{10.1103/PhysRevD.110.014036}},
  \eprint{2310.18383}.

\bibtype{Article}%
\bibitem{Lu:2025cls}
\bibinfo{author}{Yi Lu}, \bibinfo{author}{Fei Gao}, \bibinfo{author}{Yu-xin
  Liu}, \bibinfo{author}{Jan~M. Pawlowski}, \bibinfo{title}{{Finite density
  signatures of confining and chiral dynamics in QCD thermodynamics and
  fluctuations of conserved charges}}  (\bibinfo{year}{2025}),
  \eprint{2504.05099}.

\bibtype{Article}%
\bibitem{Gao:2020fbl}
\bibinfo{author}{Fei Gao}, \bibinfo{author}{Jan~M. Pawlowski},
  \bibinfo{title}{{Chiral phase structure and critical end point in QCD}},
  \bibinfo{journal}{Phys. Lett. B} \bibinfo{volume}{820} (\bibinfo{year}{2021})
  \bibinfo{pages}{136584}, \bibinfo{doi}{\doi{10.1016/j.physletb.2021.136584}},
  \eprint{2010.13705}.

\bibtype{Article}%
\bibitem{Fu:2021oaw}
\bibinfo{author}{Wei-jie Fu}, \bibinfo{author}{Xiaofeng Luo},
  \bibinfo{author}{Jan~M. Pawlowski}, \bibinfo{author}{Fabian Rennecke},
  \bibinfo{author}{Rui Wen}, \bibinfo{author}{Shi Yin},
  \bibinfo{title}{{Hyper-order baryon number fluctuations at finite temperature
  and density}}, \bibinfo{journal}{Phys. Rev. D} \bibinfo{volume}{104}
  (\bibinfo{number}{9}) (\bibinfo{year}{2021}) \bibinfo{pages}{094047},
  \bibinfo{doi}{\doi{10.1103/PhysRevD.104.094047}}, \eprint{2101.06035}.

\bibtype{Article}%
\bibitem{Fu:2023lcm}
\bibinfo{author}{Wei-jie Fu}, \bibinfo{author}{Xiaofeng Luo},
  \bibinfo{author}{Jan~M. Pawlowski}, \bibinfo{author}{Fabian Rennecke},
  \bibinfo{author}{Shi Yin}, \bibinfo{title}{{Ripples of the QCD critical
  point}}, \bibinfo{journal}{Phys. Rev. D} \bibinfo{volume}{111}
  (\bibinfo{number}{3}) (\bibinfo{year}{2025}) \bibinfo{pages}{L031502},
  \bibinfo{doi}{\doi{10.1103/PhysRevD.111.L031502}}, \eprint{2308.15508}.

\bibtype{Article}%
\bibitem{Hansen:1990yg}
\bibinfo{author}{F.~C. Hansen}, \bibinfo{author}{H. Leutwyler},
  \bibinfo{title}{{Charge correlations and topological susceptibility in QCD}},
  \bibinfo{journal}{Nucl. Phys. B} \bibinfo{volume}{350} (\bibinfo{year}{1991})
  \bibinfo{pages}{201--227}, \bibinfo{doi}{\doi{10.1016/0550-3213(91)90259-Z}}.

\bibtype{Article}%
\bibitem{Beutler:2010cha}
\bibinfo{author}{F. Beutler}, \bibinfo{author}{A. Andronic},
  \bibinfo{author}{P. Braun-Munzinger}, \bibinfo{author}{K. Redlich},
  \bibinfo{author}{J. Stachel}, \bibinfo{title}{{The Canonical partition
  function for relativistic hadron gases}}, \bibinfo{journal}{Eur. Phys. J. C}
  \bibinfo{volume}{67} (\bibinfo{year}{2010}) \bibinfo{pages}{439--444},
  \bibinfo{doi}{\doi{10.1140/epjc/s10052-010-1309-3}}, \eprint{0910.1697}.

\bibtype{Article}%
\bibitem{Vovchenko:2019kes}
\bibinfo{author}{Volodymyr Vovchenko}, \bibinfo{author}{Benjamin D{\"o}nigus},
  \bibinfo{author}{Horst Stoecker}, \bibinfo{title}{{Canonical statistical
  model analysis of p-p , p -Pb, and Pb-Pb collisions at energies available at
  the CERN Large Hadron Collider}}, \bibinfo{journal}{Phys. Rev. C}
  \bibinfo{volume}{100} (\bibinfo{number}{5}) (\bibinfo{year}{2019})
  \bibinfo{pages}{054906}, \bibinfo{doi}{\doi{10.1103/PhysRevC.100.054906}},
  \eprint{1906.03145}.

\bibtype{Article}%
\bibitem{Borsanyi:2011sw}
\bibinfo{author}{Szabolcs Borsanyi}, \bibinfo{author}{Zoltan Fodor},
  \bibinfo{author}{Sandor~D. Katz}, \bibinfo{author}{Stefan Krieg},
  \bibinfo{author}{Claudia Ratti}, et al., \bibinfo{title}{{Fluctuations of
  conserved charges at finite temperature from lattice QCD}},
  \bibinfo{journal}{JHEP} \bibinfo{volume}{1201} (\bibinfo{year}{2012})
  \bibinfo{pages}{138}, \bibinfo{doi}{\doi{10.1007/JHEP01(2012)138}},
  \eprint{1112.4416}.

\bibtype{Article}%
\bibitem{Bazavov:2012jq}
\bibinfo{author}{A. Bazavov}, et al. (\bibinfo{collaboration}{HotQCD
  Collaboration}), \bibinfo{title}{{Fluctuations and Correlations of net baryon
  number, electric charge, and strangeness: A comparison of lattice QCD results
  with the hadron resonance gas model}}, \bibinfo{journal}{Phys.Rev.}
  \bibinfo{volume}{D86} (\bibinfo{year}{2012}) \bibinfo{pages}{034509},
  \bibinfo{doi}{\doi{10.1103/PhysRevD.86.034509}}, \eprint{1203.0784}.

\bibtype{Article}%
\bibitem{HotQCD:2014kol}
\bibinfo{author}{A. Bazavov}, et al. (\bibinfo{collaboration}{HotQCD}),
  \bibinfo{title}{{Equation of state in ( 2+1 )-flavor QCD}},
  \bibinfo{journal}{Phys. Rev. D} \bibinfo{volume}{90} (\bibinfo{year}{2014})
  \bibinfo{pages}{094503}, \bibinfo{doi}{\doi{10.1103/PhysRevD.90.094503}},
  \eprint{1407.6387}.

\bibtype{Article}%
\bibitem{Vovchenko:2020lju}
\bibinfo{author}{Volodymyr Vovchenko}, \bibinfo{title}{{Hadron resonance gas
  with van der Waals interactions}}, \bibinfo{journal}{Int. J. Mod. Phys. E}
  \bibinfo{volume}{29} (\bibinfo{number}{05}) (\bibinfo{year}{2020})
  \bibinfo{pages}{2040002}, \bibinfo{doi}{\doi{10.1142/S0218301320400029}},
  \eprint{2004.06331}.

\bibtype{Article}%
\bibitem{Huovinen:2017ogf}
\bibinfo{author}{Pasi Huovinen}, \bibinfo{author}{Peter Petreczky},
  \bibinfo{title}{{Hadron Resonance Gas with Repulsive Interactions and
  Fluctuations of Conserved Charges}}, \bibinfo{journal}{Phys. Lett.}
  \bibinfo{volume}{B777} (\bibinfo{year}{2018}) \bibinfo{pages}{125--130},
  \bibinfo{doi}{\doi{10.1016/j.physletb.2017.12.001}}, \eprint{1708.00879}.

\bibtype{Article}%
\bibitem{Dashen:1969ep}
\bibinfo{author}{Roger Dashen}, \bibinfo{author}{Shang-Keng Ma},
  \bibinfo{author}{Herbert~J. Bernstein}, \bibinfo{title}{{S Matrix formulation
  of statistical mechanics}}, \bibinfo{journal}{Phys.Rev.}
  \bibinfo{volume}{187} (\bibinfo{year}{1969}) \bibinfo{pages}{345--370},
  \bibinfo{doi}{\doi{10.1103/PhysRev.187.345}}.

\bibtype{Article}%
\bibitem{Venugopalan:1992hy}
\bibinfo{author}{R. Venugopalan}, \bibinfo{author}{M. Prakash},
  \bibinfo{title}{{Thermal properties of interacting hadrons}},
  \bibinfo{journal}{Nucl.Phys.} \bibinfo{volume}{A546} (\bibinfo{year}{1992})
  \bibinfo{pages}{718--760}, \bibinfo{doi}{\doi{10.1016/0375-9474(92)90005-5}}.

\bibtype{Article}%
\bibitem{Fernandez-Ramirez:2018vzu}
\bibinfo{author}{C\'esar Fern\'andez-Ram\'\i{}rez}, \bibinfo{author}{Pok~Man
  Lo}, \bibinfo{author}{Peter Petreczky}, \bibinfo{title}{{Thermodynamics of
  the strange baryon system from a coupled-channels analysis and missing
  states}}, \bibinfo{journal}{Phys. Rev. C} \bibinfo{volume}{98}
  (\bibinfo{number}{4}) (\bibinfo{year}{2018}) \bibinfo{pages}{044910},
  \bibinfo{doi}{\doi{10.1103/PhysRevC.98.044910}}, \eprint{1806.02177}.

\bibtype{Article}%
\bibitem{Andronic:2018qqt}
\bibinfo{author}{Anton Andronic}, \bibinfo{author}{Peter Braun-Munzinger},
  \bibinfo{author}{Bengt Friman}, \bibinfo{author}{Pok~Man Lo},
  \bibinfo{author}{Krzysztof Redlich}, \bibinfo{author}{Johanna Stachel},
  \bibinfo{title}{{The thermal proton yield anomaly in Pb-Pb collisions at the
  LHC and its resolution}}, \bibinfo{journal}{Phys. Lett.}
  \bibinfo{volume}{B792} (\bibinfo{year}{2019}) \bibinfo{pages}{304--309},
  \bibinfo{doi}{\doi{10.1016/j.physletb.2019.03.052}}, \eprint{1808.03102}.

\bibtype{Article}%
\bibitem{Bazavov:2014xya}
\bibinfo{author}{A. Bazavov}, \bibinfo{author}{H.~T. Ding}, \bibinfo{author}{P.
  Hegde}, \bibinfo{author}{O. Kaczmarek}, \bibinfo{author}{F. Karsch}, et al.,
  \bibinfo{title}{{Additional Strange Hadrons from QCD Thermodynamics and
  Strangeness Freeze-out in Heavy Ion Collisions}},
  \bibinfo{journal}{Phys.Rev.Lett.} \bibinfo{volume}{113}
  (\bibinfo{year}{2014}) \bibinfo{pages}{072001},
  \bibinfo{doi}{\doi{10.1103/PhysRevLett.113.072001}}, \eprint{1404.6511}.

\bibtype{Article}%
\bibitem{Alba:2017mqu}
\bibinfo{author}{Paolo Alba}, et al., \bibinfo{title}{{Constraining the
  hadronic spectrum through QCD thermodynamics on the lattice}},
  \bibinfo{journal}{Phys. Rev.} \bibinfo{volume}{D96} (\bibinfo{number}{3})
  (\bibinfo{year}{2017}) \bibinfo{pages}{034517},
  \bibinfo{doi}{\doi{10.1103/PhysRevD.96.034517}}, \eprint{1702.01113}.

\bibtype{Article}%
\bibitem{Alba:2020jir}
\bibinfo{author}{P. Alba}, \bibinfo{author}{V.~Mantovani Sarti},
  \bibinfo{author}{J. Noronha-Hostler}, \bibinfo{author}{P. Parotto},
  \bibinfo{author}{I. Portillo-Vazquez}, \bibinfo{author}{C. Ratti},
  \bibinfo{author}{J.~M. Stafford}, \bibinfo{title}{{Influence of hadronic
  resonances on the chemical freeze-out in heavy-ion collisions}},
  \bibinfo{journal}{Phys. Rev. C} \bibinfo{volume}{101} (\bibinfo{number}{5})
  (\bibinfo{year}{2020}) \bibinfo{pages}{054905},
  \bibinfo{doi}{\doi{10.1103/PhysRevC.101.054905}}, \eprint{2002.12395}.

\bibtype{Article}%
\bibitem{Bollweg:2021vqf}
\bibinfo{author}{D. Bollweg}, \bibinfo{author}{J. Goswami}, \bibinfo{author}{O.
  Kaczmarek}, \bibinfo{author}{F. Karsch}, \bibinfo{author}{Swagato Mukherjee},
  \bibinfo{author}{P. Petreczky}, \bibinfo{author}{C. Schmidt},
  \bibinfo{author}{P. Scior} (\bibinfo{collaboration}{HotQCD}),
  \bibinfo{title}{{Second order cumulants of conserved charge fluctuations
  revisited: Vanishing chemical potentials}}, \bibinfo{journal}{Phys. Rev. D}
  \bibinfo{volume}{104} (\bibinfo{number}{7}) (\bibinfo{year}{2021}),
  \bibinfo{doi}{\doi{10.1103/PhysRevD.104.074512}}, \eprint{2107.10011}.

\bibtype{Article}%
\bibitem{Capstick:1986bm}
\bibinfo{author}{Simon Capstick}, \bibinfo{author}{Nathan Isgur},
  \bibinfo{title}{{Baryons in a Relativized Quark Model with Chromodynamics}},
  \bibinfo{journal}{Phys. Rev.} \bibinfo{volume}{D34} (\bibinfo{year}{1986})
  \bibinfo{pages}{2809}, \bibinfo{doi}{\doi{10.1103/PhysRevD.34.2809}}.

\bibtype{Article}%
\bibitem{Ebert:2009ub}
\bibinfo{author}{D. Ebert}, \bibinfo{author}{R.~N. Faustov},
  \bibinfo{author}{V.~O. Galkin}, \bibinfo{title}{{Mass spectra and Regge
  trajectories of light mesons in the relativistic quark model}},
  \bibinfo{journal}{Phys. Rev.} \bibinfo{volume}{D79} (\bibinfo{year}{2009})
  \bibinfo{pages}{114029}, \bibinfo{doi}{\doi{10.1103/PhysRevD.79.114029}},
  \eprint{0903.5183}.

\bibtype{Article}%
\bibitem{Ferraris:1995ui}
\bibinfo{author}{M. Ferraris}, \bibinfo{author}{M.~M. Giannini},
  \bibinfo{author}{M. Pizzo}, \bibinfo{author}{E. Santopinto},
  \bibinfo{author}{L. Tiator}, \bibinfo{title}{{A Three body force model for
  the baryon spectrum}}, \bibinfo{journal}{Phys. Lett.} \bibinfo{volume}{B364}
  (\bibinfo{year}{1995}) \bibinfo{pages}{231--238},
  \bibinfo{doi}{\doi{10.1016/0370-2693(95)01091-2}}.

\bibtype{Article}%
\bibitem{Pal:2021qav}
\bibinfo{author}{Somenath Pal}, \bibinfo{author}{Guruprasad Kadam},
  \bibinfo{author}{Abhijit Bhattacharyya}, \bibinfo{title}{{Hadron resonance
  gas model with repulsive mean-field interactions: Specific heat, isothermal
  compressibility and speed of sound}}, \bibinfo{journal}{Nucl. Phys. A}
  \bibinfo{volume}{1023} (\bibinfo{year}{2022}) \bibinfo{pages}{122464},
  \bibinfo{doi}{\doi{10.1016/j.nuclphysa.2022.122464}}, \eprint{2104.08531}.

\bibtype{Article}%
\bibitem{Biswas:2024xxh}
\bibinfo{author}{Deeptak Biswas}, \bibinfo{author}{Peter Petreczky},
  \bibinfo{author}{Sayantan Sharma}, \bibinfo{title}{{Chiral condensate and the
  equation~of state at nonzero baryon density from the hadron resonance gas
  model with a repulsive mean field}}, \bibinfo{journal}{Phys. Rev. C}
  \bibinfo{volume}{109} (\bibinfo{number}{5}) (\bibinfo{year}{2024})
  \bibinfo{pages}{055206}, \bibinfo{doi}{\doi{10.1103/PhysRevC.109.055206}},
  \eprint{2401.02874}.

\bibtype{Article}%
\bibitem{Gell-Mann:1960mvl}
\bibinfo{author}{Murray Gell-Mann}, \bibinfo{author}{M Levy},
  \bibinfo{title}{{The axial vector current in beta decay}},
  \bibinfo{journal}{Nuovo Cim.} \bibinfo{volume}{16} (\bibinfo{year}{1960})
  \bibinfo{pages}{705}, \bibinfo{doi}{\doi{10.1007/BF02859738}}.

\bibtype{Article}%
\bibitem{Meissner:2024ona}
\bibinfo{author}{Ulf-G. Mei{\ss}ner}, \bibinfo{title}{{Chiral perturbation
  theory}}  (\bibinfo{year}{2024}), \eprint{2410.21912}.

\bibtype{Article}%
\bibitem{Gasser:1983yg}
\bibinfo{author}{J. Gasser}, \bibinfo{author}{H. Leutwyler},
  \bibinfo{title}{{Chiral Perturbation Theory to One Loop}},
  \bibinfo{journal}{Annals Phys.} \bibinfo{volume}{158} (\bibinfo{year}{1984})
  \bibinfo{pages}{142}, \bibinfo{doi}{\doi{10.1016/0003-4916(84)90242-2}}.

\bibtype{Article}%
\bibitem{Gerber:1988tt}
\bibinfo{author}{P. Gerber}, \bibinfo{author}{H. Leutwyler},
  \bibinfo{title}{{Hadrons Below the Chiral Phase Transition}},
  \bibinfo{journal}{Nucl.Phys.} \bibinfo{volume}{B321} (\bibinfo{year}{1989})
  \bibinfo{pages}{387}, \bibinfo{doi}{\doi{10.1016/0550-3213(89)90349-0}}.

\bibtype{Article}%
\bibitem{Nambu:1961tp}
\bibinfo{author}{Yoichiro Nambu}, \bibinfo{author}{G. Jona-Lasinio},
  \bibinfo{title}{{Dynamical Model of Elementary Particles Based on an Analogy
  with Superconductivity. 1.}}, \bibinfo{journal}{Phys. Rev.}
  \bibinfo{volume}{122} (\bibinfo{year}{1961}) \bibinfo{pages}{345--358},
  \bibinfo{doi}{\doi{10.1103/PhysRev.122.345}}.

\bibtype{Article}%
\bibitem{Nambu:1961fr}
\bibinfo{author}{Yoichiro Nambu}, \bibinfo{author}{G. Jona-Lasinio},
  \bibinfo{title}{{Dynamical model of elementary particles based on an analogy
  with superconductivity. II.}}, \bibinfo{journal}{Phys. Rev.}
  \bibinfo{volume}{124} (\bibinfo{year}{1961}) \bibinfo{pages}{246--254},
  \bibinfo{doi}{\doi{10.1103/PhysRev.124.246}}.

\bibtype{Article}%
\bibitem{Klevansky:1992qe}
\bibinfo{author}{S.~P. Klevansky}, \bibinfo{title}{{The Nambu-Jona-Lasinio
  model of quantum chromodynamics}}, \bibinfo{journal}{Rev. Mod. Phys.}
  \bibinfo{volume}{64} (\bibinfo{year}{1992}) \bibinfo{pages}{649--708},
  \bibinfo{doi}{\doi{10.1103/RevModPhys.64.649}}.

\bibtype{Article}%
\bibitem{Asakawa:1989bq}
\bibinfo{author}{M. Asakawa}, \bibinfo{author}{K. Yazaki},
  \bibinfo{title}{{Chiral Restoration at Finite Density and Temperature}},
  \bibinfo{journal}{Nucl. Phys. A} \bibinfo{volume}{504} (\bibinfo{year}{1989})
  \bibinfo{pages}{668--684}, \bibinfo{doi}{\doi{10.1016/0375-9474(89)90002-X}}.

\bibtype{Article}%
\bibitem{Buballa:2003qv}
\bibinfo{author}{Michael Buballa}, \bibinfo{title}{{NJL model analysis of quark
  matter at large density}}, \bibinfo{journal}{Phys. Rept.}
  \bibinfo{volume}{407} (\bibinfo{year}{2005}) \bibinfo{pages}{205--376},
  \bibinfo{doi}{\doi{10.1016/j.physrep.2004.11.004}}, \eprint{hep-ph/0402234}.

\bibtype{Article}%
\bibitem{Schaefer:2004en}
\bibinfo{author}{Bernd-Jochen Schaefer}, \bibinfo{author}{Jochen Wambach},
  \bibinfo{title}{{The Phase diagram of the quark meson model}},
  \bibinfo{journal}{Nucl.Phys.} \bibinfo{volume}{A757} (\bibinfo{year}{2005})
  \bibinfo{pages}{479--492},
  \bibinfo{doi}{\doi{10.1016/j.nuclphysa.2005.04.012}},
  \eprint{nucl-th/0403039}.

\bibtype{Article}%
\bibitem{Fukushima:2008wg}
\bibinfo{author}{Kenji Fukushima}, \bibinfo{title}{{Phase diagrams in the
  three-flavor Nambu-Jona-Lasinio model with the Polyakov loop}},
  \bibinfo{journal}{Phys. Rev. D} \bibinfo{volume}{77} (\bibinfo{year}{2008})
  \bibinfo{pages}{114028}, \bibinfo{doi}{\doi{10.1103/PhysRevD.77.114028}},
  \bibinfo{note}{[Erratum: Phys.Rev.D 78, 039902 (2008)]}, \eprint{0803.3318}.

\bibtype{Article}%
\bibitem{Ratti:2006wg}
\bibinfo{author}{C. Ratti}, \bibinfo{author}{Simon Roessner},
  \bibinfo{author}{M.~A. Thaler}, \bibinfo{author}{W. Weise},
  \bibinfo{title}{{Thermodynamics of the PNJL model}}, \bibinfo{journal}{Eur.
  Phys. J. C} \bibinfo{volume}{49} (\bibinfo{year}{2007})
  \bibinfo{pages}{213--217},
  \bibinfo{doi}{\doi{10.1140/epjc/s10052-006-0065-x}}, \eprint{hep-ph/0609218}.

\bibtype{Article}%
\bibitem{Ratti:2006gh}
\bibinfo{author}{Claudia Ratti}, \bibinfo{author}{Michael~A. Thaler},
  \bibinfo{author}{Wolfram Weise}, \bibinfo{title}{{Phase diagram and
  thermodynamics of the PNJL model}}  (\bibinfo{year}{2006}),
  \eprint{nucl-th/0604025}.

\bibtype{Article}%
\bibitem{Schaefer:2007pw}
\bibinfo{author}{Bernd-Jochen Schaefer}, \bibinfo{author}{Jan~M. Pawlowski},
  \bibinfo{author}{Jochen Wambach}, \bibinfo{title}{{The Phase Structure of the
  Polyakov--Quark-Meson Model}}, \bibinfo{journal}{Phys. Rev. D}
  \bibinfo{volume}{76} (\bibinfo{year}{2007}) \bibinfo{pages}{074023},
  \bibinfo{doi}{\doi{10.1103/PhysRevD.76.074023}}, \eprint{0704.3234}.

\bibtype{Article}%
\bibitem{Haas:2013qwp}
\bibinfo{author}{Lisa~M. Haas}, \bibinfo{author}{Rainer Stiele},
  \bibinfo{author}{Jens Braun}, \bibinfo{author}{Jan~M. Pawlowski},
  \bibinfo{author}{J{\"u}rgen Schaffner-Bielich}, \bibinfo{title}{{Improved
  Polyakov-loop potential for effective models from functional calculations}},
  \bibinfo{journal}{Phys. Rev. D} \bibinfo{volume}{87} (\bibinfo{number}{7})
  (\bibinfo{year}{2013}) \bibinfo{pages}{076004},
  \bibinfo{doi}{\doi{10.1103/PhysRevD.87.076004}}, \eprint{1302.1993}.

\bibtype{Article}%
\bibitem{Skokov:2012kw}
\bibinfo{author}{V. Skokov}, \bibinfo{title}{{Fluctuations of conserved charges
  in the Polyakov loop extended quark-meson model at finite baryon density}},
  \bibinfo{journal}{Acta Phys.Polon.Supp.} \bibinfo{volume}{5}
  (\bibinfo{year}{2012}) \bibinfo{pages}{877--886},
  \bibinfo{doi}{\doi{10.5506/APhysPolBSupp.5.877}}.

\bibtype{Article}%
\bibitem{Alford:1999pb}
\bibinfo{author}{Mark~G. Alford}, \bibinfo{author}{Juergen Berges},
  \bibinfo{author}{Krishna Rajagopal}, \bibinfo{title}{{Magnetic fields within
  color superconducting neutron star cores}}, \bibinfo{journal}{Nucl. Phys. B}
  \bibinfo{volume}{571} (\bibinfo{year}{2000}) \bibinfo{pages}{269--284},
  \bibinfo{doi}{\doi{10.1016/S0550-3213(99)00830-5}}, \eprint{hep-ph/9910254}.

\bibtype{Article}%
\bibitem{Alford:2007xm}
\bibinfo{author}{Mark~G. Alford}, \bibinfo{author}{Andreas Schmitt},
  \bibinfo{author}{Krishna Rajagopal}, \bibinfo{author}{Thomas Sch{\"a}fer},
  \bibinfo{title}{{Color superconductivity in dense quark matter}},
  \bibinfo{journal}{Rev.Mod.Phys.} \bibinfo{volume}{80} (\bibinfo{year}{2008})
  \bibinfo{pages}{1455--1515}, \bibinfo{doi}{\doi{10.1103/RevModPhys.80.1455}},
  \eprint{0709.4635}.

\bibtype{Article}%
\bibitem{Herbst:2013ail}
\bibinfo{author}{Tina~K. Herbst}, \bibinfo{author}{Jan~M. Pawlowski},
  \bibinfo{author}{Bernd-Jochen Schaefer}, \bibinfo{title}{{Phase structure and
  thermodynamics of QCD}}, \bibinfo{journal}{Phys. Rev.} \bibinfo{volume}{D88}
  (\bibinfo{number}{1}) (\bibinfo{year}{2013}) \bibinfo{pages}{014007},
  \bibinfo{doi}{\doi{10.1103/PhysRevD.88.014007}}, \eprint{1302.1426}.

\bibtype{Article}%
\bibitem{Herbst:2013ufa}
\bibinfo{author}{Tina~Katharina Herbst}, \bibinfo{author}{Mario Mitter},
  \bibinfo{author}{Jan~M. Pawlowski}, \bibinfo{author}{Bernd-Jochen Schaefer},
  \bibinfo{author}{Rainer Stiele}, \bibinfo{title}{{Thermodynamics of QCD at
  vanishing density}}, \bibinfo{journal}{Phys. Lett.} \bibinfo{volume}{B731}
  (\bibinfo{year}{2014}) \bibinfo{pages}{248--256},
  \bibinfo{doi}{\doi{10.1016/j.physletb.2014.02.045}}, \eprint{1308.3621}.

\bibtype{Article}%
\bibitem{Gubser:2008yx}
\bibinfo{author}{Steven~S. Gubser}, \bibinfo{author}{Abhinav Nellore},
  \bibinfo{author}{Silviu~S. Pufu}, \bibinfo{author}{Fabio~D. Rocha},
  \bibinfo{title}{{Thermodynamics and bulk viscosity of approximate black hole
  duals to finite temperature quantum chromodynamics}},
  \bibinfo{journal}{Phys.Rev.Lett.} \bibinfo{volume}{101}
  (\bibinfo{year}{2008}) \bibinfo{pages}{131601},
  \bibinfo{doi}{\doi{10.1103/PhysRevLett.101.131601}}, \eprint{0804.1950}.

\bibtype{Article}%
\bibitem{Gubser:2008ny}
\bibinfo{author}{Steven~S. Gubser}, \bibinfo{author}{Abhinav Nellore},
  \bibinfo{title}{{Mimicking the QCD equation of state with a dual black
  hole}}, \bibinfo{journal}{Phys. Rev. D} \bibinfo{volume}{78}
  (\bibinfo{year}{2008}) \bibinfo{pages}{086007},
  \bibinfo{doi}{\doi{10.1103/PhysRevD.78.086007}}, \eprint{0804.0434}.

\bibtype{Article}%
\bibitem{Rougemont:2023gfz}
\bibinfo{author}{Romulo Rougemont}, \bibinfo{author}{Joaquin Grefa},
  \bibinfo{author}{Mauricio Hippert}, \bibinfo{author}{Jorge Noronha},
  \bibinfo{author}{Jacquelyn Noronha-Hostler}, \bibinfo{author}{Israel
  Portillo}, \bibinfo{author}{Claudia Ratti}, \bibinfo{title}{{Hot QCD phase
  diagram from holographic Einstein{\textendash}Maxwell{\textendash}Dilaton
  models}}, \bibinfo{journal}{Prog. Part. Nucl. Phys.} \bibinfo{volume}{135}
  (\bibinfo{year}{2024}) \bibinfo{pages}{104093},
  \bibinfo{doi}{\doi{10.1016/j.ppnp.2023.104093}}, \eprint{2307.03885}.

\bibtype{Article}%
\bibitem{DeWolfe:2010he}
\bibinfo{author}{Oliver DeWolfe}, \bibinfo{author}{Steven~S. Gubser},
  \bibinfo{author}{Christopher Rosen}, \bibinfo{title}{{A holographic critical
  point}}, \bibinfo{journal}{Phys. Rev. D} \bibinfo{volume}{83}
  (\bibinfo{year}{2011}) \bibinfo{pages}{086005},
  \bibinfo{doi}{\doi{10.1103/PhysRevD.83.086005}}, \eprint{1012.1864}.

\bibtype{Article}%
\bibitem{DeWolfe:2011ts}
\bibinfo{author}{Oliver DeWolfe}, \bibinfo{author}{Steven~S. Gubser},
  \bibinfo{author}{Christopher Rosen}, \bibinfo{title}{{Dynamic critical
  phenomena at a holographic critical point}}, \bibinfo{journal}{Phys. Rev. D}
  \bibinfo{volume}{84} (\bibinfo{year}{2011}) \bibinfo{pages}{126014},
  \bibinfo{doi}{\doi{10.1103/PhysRevD.84.126014}}, \eprint{1108.2029}.

\bibtype{Article}%
\bibitem{Gubser:2008sz}
\bibinfo{author}{Steven~S. Gubser}, \bibinfo{author}{Silviu~S. Pufu},
  \bibinfo{author}{Fabio~D. Rocha}, \bibinfo{title}{{Bulk viscosity of strongly
  coupled plasmas with holographic duals}}, \bibinfo{journal}{JHEP}
  \bibinfo{volume}{08} (\bibinfo{year}{2008}) \bibinfo{pages}{085},
  \bibinfo{doi}{\doi{10.1088/1126-6708/2008/08/085}}, \eprint{0806.0407}.

\bibtype{Article}%
\bibitem{Finazzo:2014cna}
\bibinfo{author}{Stefano~I. Finazzo}, \bibinfo{author}{Romulo Rougemont},
  \bibinfo{author}{Hugo Marrochio}, \bibinfo{author}{Jorge Noronha},
  \bibinfo{title}{{Hydrodynamic transport coefficients for the non-conformal
  quark-gluon plasma from holography}}, \bibinfo{journal}{JHEP}
  \bibinfo{volume}{02} (\bibinfo{year}{2015}) \bibinfo{pages}{051},
  \bibinfo{doi}{\doi{10.1007/JHEP02(2015)051}}, \eprint{1412.2968}.

\bibtype{Article}%
\bibitem{Rougemont:2015ona}
\bibinfo{author}{Romulo Rougemont}, \bibinfo{author}{Jorge Noronha},
  \bibinfo{author}{Jacquelyn Noronha-Hostler}, \bibinfo{title}{{Suppression of
  baryon diffusion and transport in a baryon rich strongly coupled quark-gluon
  plasma}}, \bibinfo{journal}{Phys. Rev. Lett.} \bibinfo{volume}{115}
  (\bibinfo{number}{20}) (\bibinfo{year}{2015}) \bibinfo{pages}{202301},
  \bibinfo{doi}{\doi{10.1103/PhysRevLett.115.202301}}, \eprint{1507.06972}.

\bibtype{Article}%
\bibitem{Grefa:2022sav}
\bibinfo{author}{Joaquin Grefa}, \bibinfo{author}{Mauricio Hippert},
  \bibinfo{author}{Jorge Noronha}, \bibinfo{author}{Jacquelyn Noronha-Hostler},
  \bibinfo{author}{Israel Portillo}, \bibinfo{author}{Claudia Ratti},
  \bibinfo{author}{Romulo Rougemont}, \bibinfo{title}{{Transport coefficients
  of the quark-gluon plasma at the critical point and across the first-order
  line}}, \bibinfo{journal}{Phys. Rev. D} \bibinfo{volume}{106}
  (\bibinfo{number}{3}) (\bibinfo{year}{2022}) \bibinfo{pages}{034024},
  \bibinfo{doi}{\doi{10.1103/PhysRevD.106.034024}}, \eprint{2203.00139}.

\bibtype{Article}%
\bibitem{Critelli:2017oub}
\bibinfo{author}{Renato Critelli}, \bibinfo{author}{Jorge Noronha},
  \bibinfo{author}{Jacquelyn Noronha-Hostler}, \bibinfo{author}{Israel
  Portillo}, \bibinfo{author}{Claudia Ratti}, \bibinfo{author}{Romulo
  Rougemont}, \bibinfo{title}{{Critical point in the phase diagram of
  primordial quark-gluon matter from black hole physics}},
  \bibinfo{journal}{Phys. Rev.} \bibinfo{volume}{D96} (\bibinfo{number}{9})
  (\bibinfo{year}{2017}) \bibinfo{pages}{096026},
  \bibinfo{doi}{\doi{10.1103/PhysRevD.96.096026}}, \eprint{1706.00455}.

\bibtype{Article}%
\bibitem{Hippert:2023bel}
\bibinfo{author}{Mauricio Hippert}, \bibinfo{author}{Joaquin Grefa},
  \bibinfo{author}{T.~Andrew Manning}, \bibinfo{author}{Jorge Noronha},
  \bibinfo{author}{Jacquelyn Noronha-Hostler}, \bibinfo{author}{Israel
  Portillo~Vazquez}, \bibinfo{author}{Claudia Ratti}, \bibinfo{author}{Romulo
  Rougemont}, \bibinfo{author}{Michael Trujillo}, \bibinfo{title}{{Bayesian
  location of the QCD critical point from a holographic perspective}},
  \bibinfo{journal}{Phys. Rev. D} \bibinfo{volume}{110} (\bibinfo{number}{9})
  (\bibinfo{year}{2024}) \bibinfo{pages}{094006},
  \bibinfo{doi}{\doi{10.1103/PhysRevD.110.094006}}, \eprint{2309.00579}.

\bibtype{Article}%
\bibitem{Grefa:2021qvt}
\bibinfo{author}{Joaquin Grefa}, \bibinfo{author}{Jorge Noronha},
  \bibinfo{author}{Jacquelyn Noronha-Hostler}, \bibinfo{author}{Israel
  Portillo}, \bibinfo{author}{Claudia Ratti}, \bibinfo{author}{Romulo
  Rougemont}, \bibinfo{title}{{Hot and dense quark-gluon plasma thermodynamics
  from holographic black holes}}, \bibinfo{journal}{Phys. Rev. D}
  \bibinfo{volume}{104} (\bibinfo{number}{3}) (\bibinfo{year}{2021})
  \bibinfo{pages}{034002}, \bibinfo{doi}{\doi{10.1103/PhysRevD.104.034002}},
  \eprint{2102.12042}.

\bibtype{Article}%
\bibitem{Stephanov:2004wx}
\bibinfo{author}{Mikhail~A. Stephanov}, \bibinfo{title}{{QCD phase diagram and
  the critical point}}, \bibinfo{journal}{Prog. Theor. Phys. Suppl.}
  \bibinfo{volume}{153} (\bibinfo{year}{2004}) \bibinfo{pages}{139--156},
  \bibinfo{doi}{\doi{10.1142/S0217751X05027965}}, \bibinfo{note}{[Int. J. Mod.
  Phys.A20,4387(2005)]}, \eprint{hep-ph/0402115}.

\bibtype{Article}%
\bibitem{Brown:1990ev}
\bibinfo{author}{Frank~R. Brown}, \bibinfo{author}{Frank~P. Butler},
  \bibinfo{author}{Hong Chen}, \bibinfo{author}{Norman~H. Christ},
  \bibinfo{author}{Zhi-hua Dong}, et al., \bibinfo{title}{{On the existence of
  a phase transition for QCD with three light quarks}},
  \bibinfo{journal}{Phys.Rev.Lett.} \bibinfo{volume}{65} (\bibinfo{year}{1990})
  \bibinfo{pages}{2491--2494},
  \bibinfo{doi}{\doi{10.1103/PhysRevLett.65.2491}}.

\bibtype{Article}%
\bibitem{Pisarski:1983ms}
\bibinfo{author}{Robert~D. Pisarski}, \bibinfo{author}{Frank Wilczek},
  \bibinfo{title}{{Remarks on the Chiral Phase Transition in Chromodynamics}},
  \bibinfo{journal}{Phys.Rev.} \bibinfo{volume}{D29} (\bibinfo{year}{1984})
  \bibinfo{pages}{338--341}, \bibinfo{doi}{\doi{10.1103/PhysRevD.29.338}}.

\bibtype{Article}%
\bibitem{Rajagopal:1995bc}
\bibinfo{author}{Krishna Rajagopal}, \bibinfo{title}{{The Chiral phase
  transition in QCD: Critical phenomena and long wavelength pion oscillations}}
   (\bibinfo{year}{1995}) \bibinfo{pages}{484--554},
  \bibinfo{doi}{\doi{10.1142/9789812830661_0009}}, \eprint{hep-ph/9504310}.

\bibtype{Article}%
\bibitem{Cuteri:2021ikv}
\bibinfo{author}{Francesca Cuteri}, \bibinfo{author}{Owe Philipsen},
  \bibinfo{author}{Alessandro Sciarra}, \bibinfo{title}{{On the order of the
  QCD chiral phase transition for different numbers of quark flavours}},
  \bibinfo{journal}{JHEP} \bibinfo{volume}{11} (\bibinfo{year}{2021})
  \bibinfo{pages}{141}, \bibinfo{doi}{\doi{10.1007/JHEP11(2021)141}},
  \eprint{2107.12739}.

\bibtype{Article}%
\bibitem{Giacosa:2024orp}
\bibinfo{author}{Francesco Giacosa}, \bibinfo{author}{Gy\H{o}z\H{o} Kov\'acs},
  \bibinfo{author}{P\'eter Kov\'acs}, \bibinfo{author}{Robert~D. Pisarski},
  \bibinfo{author}{Fabian Rennecke}, \bibinfo{title}{{Anomalous U(1)A couplings
  and the Columbia plot}}, \bibinfo{journal}{Phys. Rev. D}
  \bibinfo{volume}{111} (\bibinfo{number}{1}) (\bibinfo{year}{2025})
  \bibinfo{pages}{016014}, \bibinfo{doi}{\doi{10.1103/PhysRevD.111.016014}},
  \eprint{2410.08185}.

\bibtype{Article}%
\bibitem{McLerran:1981pb}
\bibinfo{author}{Larry~D. McLerran}, \bibinfo{author}{Benjamin Svetitsky},
  \bibinfo{title}{{Quark Liberation at High Temperature: A Monte Carlo Study of
  SU(2) Gauge Theory}}, \bibinfo{journal}{Phys. Rev. D} \bibinfo{volume}{24}
  (\bibinfo{year}{1981}) \bibinfo{pages}{450},
  \bibinfo{doi}{\doi{10.1103/PhysRevD.24.450}}.

\bibtype{Article}%
\bibitem{McLerran:1980pk}
\bibinfo{author}{Larry~D. McLerran}, \bibinfo{author}{Benjamin Svetitsky},
  \bibinfo{title}{{A Monte Carlo Study of SU(2) Yang-Mills Theory at Finite
  Temperature}}, \bibinfo{journal}{Phys.Lett.} \bibinfo{volume}{B98}
  (\bibinfo{year}{1981}) \bibinfo{pages}{195},
  \bibinfo{doi}{\doi{10.1016/0370-2693(81)90986-2}}.

\bibtype{Article}%
\bibitem{Kuti:1980gh}
\bibinfo{author}{J. Kuti}, \bibinfo{author}{J. Polonyi}, \bibinfo{author}{K.
  Szlachanyi}, \bibinfo{title}{{Monte Carlo Study of SU(2) Gauge Theory at
  Finite Temperature}}, \bibinfo{journal}{Phys.Lett.} \bibinfo{volume}{B98}
  (\bibinfo{year}{1981}) \bibinfo{pages}{199},
  \bibinfo{doi}{\doi{10.1016/0370-2693(81)90987-4}}.

\bibtype{Article}%
\bibitem{Engels:1989fz}
\bibinfo{author}{J. Engels}, \bibinfo{author}{J. Fingberg}, \bibinfo{author}{M.
  Weber}, \bibinfo{title}{{Finite Size Scaling Analysis of SU(2) Lattice Gauge
  Theory in (3+1)-dimensions}}, \bibinfo{journal}{Nucl.Phys.}
  \bibinfo{volume}{B332} (\bibinfo{year}{1990}) \bibinfo{pages}{737},
  \bibinfo{doi}{\doi{10.1016/0550-3213(90)90010-B}}.

\bibtype{Article}%
\bibitem{Lucini:2002ku}
\bibinfo{author}{Biagio Lucini}, \bibinfo{author}{Michael Teper},
  \bibinfo{author}{Urs Wenger}, \bibinfo{title}{{The Deconfinement transition
  in SU(N) gauge theories}}, \bibinfo{journal}{Phys. Lett. B}
  \bibinfo{volume}{545} (\bibinfo{year}{2002}) \bibinfo{pages}{197--206},
  \bibinfo{doi}{\doi{10.1016/S0370-2693(02)02556-X}}, \eprint{hep-lat/0206029}.

\bibtype{Article}%
\bibitem{Yaffe:1982qf}
\bibinfo{author}{L.G. Yaffe}, \bibinfo{author}{B. Svetitsky},
  \bibinfo{title}{{First Order Phase Transition in the SU(3) Gauge Theory at
  Finite Temperature}}, \bibinfo{journal}{Phys.Rev.} \bibinfo{volume}{D26}
  (\bibinfo{year}{1982}) \bibinfo{pages}{963},
  \bibinfo{doi}{\doi{10.1103/PhysRevD.26.963}}.

\bibtype{Article}%
\bibitem{Lucini:2005vg}
\bibinfo{author}{Biagio Lucini}, \bibinfo{author}{Michael Teper},
  \bibinfo{author}{Urs Wenger}, \bibinfo{title}{{Properties of the deconfining
  phase transition in SU(N) gauge theories}}, \bibinfo{journal}{JHEP}
  \bibinfo{volume}{0502} (\bibinfo{year}{2005}) \bibinfo{pages}{033},
  \bibinfo{doi}{\doi{10.1088/1126-6708/2005/02/033}}, \eprint{hep-lat/0502003}.

\bibtype{Article}%
\bibitem{Borsanyi:2022xml}
\bibinfo{author}{S. Borsanyi}, \bibinfo{author}{Kara R.}, \bibinfo{author}{Z.
  Fodor}, \bibinfo{author}{D.~A. Godzieba}, \bibinfo{author}{P. Parotto},
  \bibinfo{author}{D. Sexty}, \bibinfo{title}{{Precision study of the continuum
  SU(3) Yang-Mills theory: How to use parallel tempering to improve on
  supercritical slowing down for first order phase transitions}},
  \bibinfo{journal}{Phys. Rev. D} \bibinfo{volume}{105} (\bibinfo{number}{7})
  (\bibinfo{year}{2022}) \bibinfo{pages}{074513},
  \bibinfo{doi}{\doi{10.1103/PhysRevD.105.074513}}, \eprint{2202.05234}.

\bibtype{Article}%
\bibitem{Shirogane:2020muc}
\bibinfo{author}{Mizuki Shirogane}, \bibinfo{author}{Shinji Ejiri},
  \bibinfo{author}{Ryo Iwami}, \bibinfo{author}{Kazuyuki Kanaya},
  \bibinfo{author}{Masakiyo Kitazawa}, \bibinfo{author}{Hiroshi Suzuki},
  \bibinfo{author}{Yusuke Taniguchi}, \bibinfo{author}{Takashi Umeda}
  (\bibinfo{collaboration}{WHOT-QCD}), \bibinfo{title}{{Latent heat and
  pressure gap at the first-order deconfining phase transition of SU(3)
  Yang-Mills theory using the small flow-time expansion method}},
  \bibinfo{journal}{PTEP} \bibinfo{volume}{2021} (\bibinfo{number}{1})
  (\bibinfo{year}{2021}) \bibinfo{pages}{013B08},
  \bibinfo{doi}{\doi{10.1093/ptep/ptaa184}}, \eprint{2011.10292}.

\bibtype{Article}%
\bibitem{Giusti:2025fxu}
\bibinfo{author}{Leonardo Giusti}, \bibinfo{author}{Mitsuaki Hirasawa},
  \bibinfo{author}{Michele Pepe}, \bibinfo{author}{Luca Virz{\`\i}},
  \bibinfo{title}{{A precise study of the thermodynamic properties of the SU(3)
  Yang-Mills theory across the deconfinement transition}},
  \bibinfo{journal}{Phys. Lett. B} \bibinfo{volume}{868} (\bibinfo{year}{2025})
  \bibinfo{pages}{139775}, \bibinfo{doi}{\doi{10.1016/j.physletb.2025.139775}},
  \eprint{2501.10284}.

\bibtype{Article}%
\bibitem{Boyd:1996bx}
\bibinfo{author}{G. Boyd}, \bibinfo{author}{J. Engels}, \bibinfo{author}{F.
  Karsch}, \bibinfo{author}{E. Laermann}, \bibinfo{author}{C. Legeland}, et
  al., \bibinfo{title}{{Thermodynamics of SU(3) lattice gauge theory}},
  \bibinfo{journal}{Nucl.Phys.} \bibinfo{volume}{B469} (\bibinfo{year}{1996})
  \bibinfo{pages}{419--444}, \bibinfo{doi}{\doi{10.1016/0550-3213(96)00170-8}},
  \eprint{hep-lat/9602007}.

\bibtype{Article}%
\bibitem{Borsanyi:2012ve}
\bibinfo{author}{Sz. Borsanyi}, \bibinfo{author}{G. Endrodi},
  \bibinfo{author}{Z. Fodor}, \bibinfo{author}{S.D. Katz},
  \bibinfo{author}{K.K. Szabo}, \bibinfo{title}{{Precision SU(3) lattice
  thermodynamics for a large temperature range}}, \bibinfo{journal}{JHEP}
  \bibinfo{volume}{1207} (\bibinfo{year}{2012}) \bibinfo{pages}{056},
  \bibinfo{doi}{\doi{10.1007/JHEP07(2012)056}}, \eprint{1204.6184}.

\bibtype{Article}%
\bibitem{Caselle:2018kap}
\bibinfo{author}{Michele Caselle}, \bibinfo{author}{Alessandro Nada},
  \bibinfo{author}{Marco Panero}, \bibinfo{title}{{QCD thermodynamics from
  lattice calculations with nonequilibrium methods: The SU(3) equation of
  state}}, \bibinfo{journal}{Phys. Rev. D} \bibinfo{volume}{98}
  (\bibinfo{number}{5}) (\bibinfo{year}{2018}) \bibinfo{pages}{054513},
  \bibinfo{doi}{\doi{10.1103/PhysRevD.98.054513}}, \eprint{1801.03110}.

\bibtype{Article}%
\bibitem{Kiyohara:2021smr}
\bibinfo{author}{Atsushi Kiyohara}, \bibinfo{author}{Masakiyo Kitazawa},
  \bibinfo{author}{Shinji Ejiri}, \bibinfo{author}{Kazuyuki Kanaya},
  \bibinfo{title}{{Finite-size scaling around the critical point in the heavy
  quark region of QCD}}, \bibinfo{journal}{Phys. Rev. D} \bibinfo{volume}{104}
  (\bibinfo{number}{11}) (\bibinfo{year}{2021}) \bibinfo{pages}{114509},
  \bibinfo{doi}{\doi{10.1103/PhysRevD.104.114509}}, \eprint{2108.00118}.

\bibtype{Article}%
\bibitem{Karsch:2001ya}
\bibinfo{author}{F. Karsch}, \bibinfo{author}{C. Schmidt}, \bibinfo{author}{S.
  Stickan}, \bibinfo{title}{{Common features of deconfining and chiral critical
  points in QCD and the three state Potts model in an external field}},
  \bibinfo{journal}{Comput. Phys. Commun.} \bibinfo{volume}{147}
  (\bibinfo{year}{2002}) \bibinfo{pages}{451--454},
  \bibinfo{doi}{\doi{10.1016/S0010-4655(02)00327-2}}, \eprint{hep-lat/0111059}.

\bibtype{Article}%
\bibitem{Saito:2011fs}
\bibinfo{author}{H. Saito}, et al. (\bibinfo{collaboration}{WHOT-QCD
  Collaboration}), \bibinfo{title}{{Phase structure of finite temperature QCD
  in the heavy quark region}}, \bibinfo{journal}{Phys.Rev.}
  \bibinfo{volume}{D84} (\bibinfo{year}{2011}) \bibinfo{pages}{054502},
  \bibinfo{doi}{\doi{10.1103/PhysRevD.85.079902, 10.1103/PhysRevD.84.054502}},
  \eprint{1106.0974}.

\bibtype{Article}%
\bibitem{Cuteri:2020yke}
\bibinfo{author}{Francesca Cuteri}, \bibinfo{author}{Owe Philipsen},
  \bibinfo{author}{Alena Sch\"on}, \bibinfo{author}{Alessandro Sciarra},
  \bibinfo{title}{{Deconfinement critical point of lattice QCD with $N_f$=2
  Wilson fermions}}, \bibinfo{journal}{Phys. Rev. D} \bibinfo{volume}{103}
  (\bibinfo{number}{1}) (\bibinfo{year}{2021}) \bibinfo{pages}{014513},
  \bibinfo{doi}{\doi{10.1103/PhysRevD.103.014513}}, \eprint{2009.14033}.

\bibtype{Article}%
\bibitem{Kashiwa:2012wa}
\bibinfo{author}{Kouji Kashiwa}, \bibinfo{author}{Robert~D. Pisarski},
  \bibinfo{author}{Vladimir~V. Skokov}, \bibinfo{title}{{Critical endpoint for
  deconfinement in matrix and other effective models}}, \bibinfo{journal}{Phys.
  Rev. D} \bibinfo{volume}{85} (\bibinfo{year}{2012}) \bibinfo{pages}{114029},
  \bibinfo{doi}{\doi{10.1103/PhysRevD.85.114029}}, \eprint{1205.0545}.

\bibtype{Article}%
\bibitem{Clarke:2020htu}
\bibinfo{author}{David~Anthony Clarke}, \bibinfo{author}{Olaf Kaczmarek},
  \bibinfo{author}{Frithjof Karsch}, \bibinfo{author}{Anirban Lahiri},
  \bibinfo{author}{Mugdha Sarkar}, \bibinfo{title}{{Sensitivity of the Polyakov
  loop and related observables to chiral symmetry restoration}},
  \bibinfo{journal}{Phys. Rev. D} \bibinfo{volume}{103} (\bibinfo{number}{1})
  (\bibinfo{year}{2021}) \bibinfo{pages}{L011501},
  \bibinfo{doi}{\doi{10.1103/PhysRevD.103.L011501}}, \eprint{2008.11678}.

\bibtype{Article}%
\bibitem{Vafa:1983tf}
\bibinfo{author}{C. Vafa}, \bibinfo{author}{Edward Witten},
  \bibinfo{title}{{Restrictions on Symmetry Breaking in Vector-Like Gauge
  Theories}}, \bibinfo{journal}{Nucl. Phys.} \bibinfo{volume}{B234}
  (\bibinfo{year}{1984}) \bibinfo{pages}{173--188},
  \bibinfo{doi}{\doi{10.1016/0550-3213(84)90230-X}}.

\bibtype{Article}%
\bibitem{Weingarten:1983uj}
\bibinfo{author}{Don Weingarten}, \bibinfo{title}{{Mass Inequalities for QCD}},
  \bibinfo{journal}{Phys. Rev. Lett.} \bibinfo{volume}{51}
  (\bibinfo{year}{1983}) \bibinfo{pages}{1830},
  \bibinfo{doi}{\doi{10.1103/PhysRevLett.51.1830}}.

\bibtype{Book}%
\bibitem{Shuryak:2021vnj}
\bibinfo{author}{Edward Shuryak}, \bibinfo{title}{{Nonperturbative Topological
  Phenomena in QCD and Related Theories}}, \bibinfo{series}{Lecture Notes in
  Physics}, \bibinfo{comment}{vol.} \bibinfo{volume}{977} \bibinfo{year}{2021},
  ISBN \bibinfo{isbn}{978-3-030-62989-2, 978-3-030-62990-8},
  \bibinfo{doi}{\doi{10.1007/978-3-030-62990-8}}.

\bibtype{Article}%
\bibitem{Pelissetto:2013hqa}
\bibinfo{author}{Andrea Pelissetto}, \bibinfo{author}{Ettore Vicari},
  \bibinfo{title}{{Relevance of the axial anomaly at the finite-temperature
  chiral transition in QCD}}, \bibinfo{journal}{Phys.Rev.}
  \bibinfo{volume}{D88} (\bibinfo{number}{10}) (\bibinfo{year}{2013})
  \bibinfo{pages}{105018}, \bibinfo{doi}{\doi{10.1103/PhysRevD.88.105018}},
  \eprint{1309.5446}.

\bibtype{Article}%
\bibitem{Azcoiti:2021gst}
\bibinfo{author}{Vicente Azcoiti}, \bibinfo{title}{{Axial $U_A(1)$ Anomaly: A
  New Mechanism to Generate Massless Bosons}}, \bibinfo{journal}{Symmetry}
  \bibinfo{volume}{13} (\bibinfo{number}{2}) (\bibinfo{year}{2021})
  \bibinfo{pages}{209}, \bibinfo{doi}{\doi{10.3390/sym13020209}},
  \eprint{2101.06439}.

\bibtype{Article}%
\bibitem{HotQCD:2019xnw}
\bibinfo{author}{H.~T. Ding}, et al. (\bibinfo{collaboration}{HotQCD}),
  \bibinfo{title}{{Chiral Phase Transition Temperature in ( 2+1 )-Flavor QCD}},
  \bibinfo{journal}{Phys. Rev. Lett.} \bibinfo{volume}{123}
  (\bibinfo{number}{6}) (\bibinfo{year}{2019}) \bibinfo{pages}{062002},
  \bibinfo{doi}{\doi{10.1103/PhysRevLett.123.062002}}, \eprint{1903.04801}.

\bibtype{Article}%
\bibitem{Kotov:2021rah}
\bibinfo{author}{Andrey~Yu. Kotov}, \bibinfo{author}{Maria~Paola Lombardo},
  \bibinfo{author}{Anton Trunin}, \bibinfo{title}{{QCD transition at the
  physical point, and its scaling window from twisted mass Wilson fermions}},
  \bibinfo{journal}{Phys. Lett. B} \bibinfo{volume}{823} (\bibinfo{year}{2021})
  \bibinfo{pages}{136749}, \bibinfo{doi}{\doi{10.1016/j.physletb.2021.136749}},
  \eprint{2105.09842}.

\bibtype{Article}%
\bibitem{Braun:2020ada}
\bibinfo{author}{Jens Braun}, \bibinfo{author}{Wei-jie Fu},
  \bibinfo{author}{Jan~M. Pawlowski}, \bibinfo{author}{Fabian Rennecke},
  \bibinfo{author}{Daniel Rosenbl\"uh}, \bibinfo{author}{Shi Yin},
  \bibinfo{title}{{Chiral susceptibility in ( 2+1 )-flavor QCD}},
  \bibinfo{journal}{Phys. Rev. D} \bibinfo{volume}{102} (\bibinfo{number}{5})
  (\bibinfo{year}{2020}) \bibinfo{pages}{056010},
  \bibinfo{doi}{\doi{10.1103/PhysRevD.102.056010}}, \eprint{2003.13112}.

\bibtype{Article}%
\bibitem{Gao:2021vsf}
\bibinfo{author}{Fei Gao}, \bibinfo{author}{Jan~M. Pawlowski},
  \bibinfo{title}{{Phase structure of (2+1)-flavor QCD and the magnetic
  equation of state}}, \bibinfo{journal}{Phys. Rev. D} \bibinfo{volume}{105}
  (\bibinfo{number}{9}) (\bibinfo{year}{2022}) \bibinfo{pages}{094020},
  \bibinfo{doi}{\doi{10.1103/PhysRevD.105.094020}}, \eprint{2112.01395}.

\bibtype{Article}%
\bibitem{Bernhardt:2023hpr}
\bibinfo{author}{Julian Bernhardt}, \bibinfo{author}{Christian~S. Fischer},
  \bibinfo{title}{{QCD phase transitions in the light quark chiral limit}},
  \bibinfo{journal}{Phys. Rev. D} \bibinfo{volume}{108} (\bibinfo{number}{11})
  (\bibinfo{year}{2023}) \bibinfo{pages}{114018},
  \bibinfo{doi}{\doi{10.1103/PhysRevD.108.114018}}, \eprint{2309.06737}.

\bibtype{Article}%
\bibitem{Braun:2023qak}
\bibinfo{author}{Jens Braun}, et al., \bibinfo{title}{{Soft modes in hot QCD
  matter}}, \bibinfo{journal}{Phys. Rev. D} \bibinfo{volume}{111}
  (\bibinfo{number}{9}) (\bibinfo{year}{2025}) \bibinfo{pages}{094010},
  \bibinfo{doi}{\doi{10.1103/PhysRevD.111.094010}}, \eprint{2310.19853}.

\bibtype{Article}%
\bibitem{Petreczky:2016vrs}
\bibinfo{author}{Peter Petreczky}, \bibinfo{author}{Hans-Peter Schadler},
  \bibinfo{author}{Sayantan Sharma}, \bibinfo{title}{{The topological
  susceptibility in finite temperature QCD and axion cosmology}},
  \bibinfo{journal}{Phys. Lett.} \bibinfo{volume}{B762} (\bibinfo{year}{2016})
  \bibinfo{pages}{498--505},
  \bibinfo{doi}{\doi{10.1016/j.physletb.2016.09.063}}, \eprint{1606.03145}.

\bibtype{Article}%
\bibitem{Shuryak:1993ee}
\bibinfo{author}{Edward~V. Shuryak}, \bibinfo{title}{{Which chiral symmetry is
  restored in hot QCD?}}, \bibinfo{journal}{Comments Nucl. Part. Phys.}
  \bibinfo{volume}{21} (\bibinfo{number}{4}) (\bibinfo{year}{1994})
  \bibinfo{pages}{235--248}, \eprint{hep-ph/9310253}.

\bibtype{Article}%
\bibitem{Chiu:2011dz}
\bibinfo{author}{Ting~Wai Chiu}, \bibinfo{author}{Tung~Han Hsieh},
  \bibinfo{author}{Yao~Yuan Mao} (\bibinfo{collaboration}{TWQCD}),
  \bibinfo{title}{{Topological Susceptibility in Two Flavors Lattice QCD with
  the Optimal Domain-Wall Fermion}}, \bibinfo{journal}{Phys. Lett. B}
  \bibinfo{volume}{702} (\bibinfo{year}{2011}) \bibinfo{pages}{131--134},
  \bibinfo{doi}{\doi{10.1016/j.physletb.2011.06.070}}, \eprint{1105.4414}.

\bibtype{Article}%
\bibitem{Cossu:2013uua}
\bibinfo{author}{Guido Cossu}, \bibinfo{author}{Sinya Aoki},
  \bibinfo{author}{Hidenori Fukaya}, \bibinfo{author}{Shoji Hashimoto},
  \bibinfo{author}{Takashi Kaneko}, \bibinfo{author}{Hideo Matsufuru},
  \bibinfo{author}{Jun-Ichi Noaki}, \bibinfo{title}{{Finite temperature study
  of the axial U(1) symmetry on the lattice with overlap fermion formulation}},
  \bibinfo{journal}{Phys. Rev. D} \bibinfo{volume}{87} (\bibinfo{number}{11})
  (\bibinfo{year}{2013}) \bibinfo{pages}{114514},
  \bibinfo{doi}{\doi{10.1103/PhysRevD.87.114514}}, \bibinfo{note}{[Erratum:
  Phys.Rev.D 88, 019901 (2013)]}, \eprint{1304.6145}.

\bibtype{Article}%
\bibitem{Brandt:2016daq}
\bibinfo{author}{Bastian~B. Brandt}, \bibinfo{author}{Anthony Francis},
  \bibinfo{author}{Harvey~B. Meyer}, \bibinfo{author}{Owe Philipsen},
  \bibinfo{author}{Daniel Robaina}, \bibinfo{author}{Hartmut Wittig},
  \bibinfo{title}{{On the strength of the $U_A(1)$ anomaly at the chiral phase
  transition in $N_f=2$ QCD}}, \bibinfo{journal}{JHEP} \bibinfo{volume}{12}
  (\bibinfo{year}{2016}) \bibinfo{pages}{158},
  \bibinfo{doi}{\doi{10.1007/JHEP12(2016)158}}, \eprint{1608.06882}.

\bibtype{Article}%
\bibitem{Tomiya:2016jwr}
\bibinfo{author}{A. Tomiya}, \bibinfo{author}{G. Cossu}, \bibinfo{author}{S.
  Aoki}, \bibinfo{author}{H. Fukaya}, \bibinfo{author}{S. Hashimoto},
  \bibinfo{author}{T. Kaneko}, \bibinfo{author}{J. Noaki},
  \bibinfo{title}{{Evidence of effective axial U(1) symmetry restoration at
  high temperature QCD}}, \bibinfo{journal}{Phys. Rev. D} \bibinfo{volume}{96}
  (\bibinfo{number}{3}) (\bibinfo{year}{2017}) \bibinfo{pages}{034509},
  \bibinfo{doi}{\doi{10.1103/PhysRevD.96.034509}}, \bibinfo{note}{[Addendum:
  Phys.Rev.D 96, 079902 (2017)]}, \eprint{1612.01908}.

\bibtype{Article}%
\bibitem{Aoki:2020noz}
\bibinfo{author}{S. Aoki}, \bibinfo{author}{Y. Aoki}, \bibinfo{author}{G.
  Cossu}, \bibinfo{author}{H. Fukaya}, \bibinfo{author}{S. Hashimoto},
  \bibinfo{author}{T. Kaneko}, \bibinfo{author}{C. Rohrhofer},
  \bibinfo{author}{K. Suzuki} (\bibinfo{collaboration}{JLQCD}),
  \bibinfo{title}{{Study of the axial $U(1)$ anomaly at high temperature with
  lattice chiral fermions}}, \bibinfo{journal}{Phys. Rev. D}
  \bibinfo{volume}{103} (\bibinfo{number}{7}) (\bibinfo{year}{2021})
  \bibinfo{pages}{074506}, \bibinfo{doi}{\doi{10.1103/PhysRevD.103.074506}},
  \eprint{2011.01499}.

\bibtype{Article}%
\bibitem{Ding:2020xlj}
\bibinfo{author}{H.~T. Ding}, \bibinfo{author}{S.~T. Li},
  \bibinfo{author}{Swagato Mukherjee}, \bibinfo{author}{A. Tomiya},
  \bibinfo{author}{X.~D. Wang}, \bibinfo{author}{Y. Zhang},
  \bibinfo{title}{{Correlated Dirac Eigenvalues and Axial Anomaly in Chiral
  Symmetric QCD}}, \bibinfo{journal}{Phys. Rev. Lett.} \bibinfo{volume}{126}
  (\bibinfo{number}{8}) (\bibinfo{year}{2021}) \bibinfo{pages}{082001},
  \bibinfo{doi}{\doi{10.1103/PhysRevLett.126.082001}}, \eprint{2010.14836}.

\bibtype{Article}%
\bibitem{Gavai:2024mcj}
\bibinfo{author}{Rajiv~V. Gavai}, \bibinfo{author}{Mischa~E. Jaensch},
  \bibinfo{author}{Olaf Kaczmarek}, \bibinfo{author}{Frithjof Karsch},
  \bibinfo{author}{Mugdha Sarkar}, \bibinfo{author}{Ravi Shanker},
  \bibinfo{author}{Sayantan Sharma}, \bibinfo{author}{Sipaz Sharma},
  \bibinfo{author}{Tristan Ueding}, \bibinfo{title}{{Aspects of the chiral
  crossover transition in (2+1)-flavor QCD with M\"obius domain-wall
  fermions}}, \bibinfo{journal}{Phys. Rev. D} \bibinfo{volume}{111}
  (\bibinfo{number}{3}) (\bibinfo{year}{2025}) \bibinfo{pages}{034507},
  \bibinfo{doi}{\doi{10.1103/PhysRevD.111.034507}}, \eprint{2411.10217}.

\bibtype{Article}%
\bibitem{Kanazawa:2014cua}
\bibinfo{author}{Takuya Kanazawa}, \bibinfo{author}{Naoki Yamamoto},
  \bibinfo{title}{{Quasi-instantons in QCD with chiral symmetry restoration}},
  \bibinfo{journal}{Phys. Rev. D} \bibinfo{volume}{91} (\bibinfo{year}{2015})
  \bibinfo{pages}{105015}, \bibinfo{doi}{\doi{10.1103/PhysRevD.91.105015}},
  \eprint{1410.3614}.

\bibtype{Article}%
\bibitem{Vig:2021oyt}
\bibinfo{author}{Reka~A. Vig}, \bibinfo{author}{Tamas~G. Kovacs},
  \bibinfo{title}{{Ideal topological gas in the high temperature phase of SU(3)
  gauge theory}}, \bibinfo{journal}{Phys. Rev. D} \bibinfo{volume}{103}
  (\bibinfo{number}{11}) (\bibinfo{year}{2021}) \bibinfo{pages}{114510},
  \bibinfo{doi}{\doi{10.1103/PhysRevD.103.114510}}, \eprint{2101.01498}.

\bibtype{Article}%
\bibitem{Atiyah:1968mp}
\bibinfo{author}{M.~F. Atiyah}, \bibinfo{author}{I.~M. Singer},
  \bibinfo{title}{{The Index of elliptic operators. 1}},
  \bibinfo{journal}{Annals Math.} \bibinfo{volume}{87} (\bibinfo{year}{1968})
  \bibinfo{pages}{484--530}, \bibinfo{doi}{\doi{10.2307/1970715}}.

\bibtype{Article}%
\bibitem{Edwards:1999zm}
\bibinfo{author}{Robert~G. Edwards}, \bibinfo{author}{Urs~M. Heller},
  \bibinfo{author}{Joe~E. Kiskis}, \bibinfo{author}{Rajamani Narayanan},
  \bibinfo{title}{{Chiral condensate in the deconfined phase of quenched gauge
  theories}}, \bibinfo{journal}{Phys. Rev. D} \bibinfo{volume}{61}
  (\bibinfo{year}{2000}) \bibinfo{pages}{074504},
  \bibinfo{doi}{\doi{10.1103/PhysRevD.61.074504}}, \eprint{hep-lat/9910041}.

\bibtype{Article}%
\bibitem{Dick:2015twa}
\bibinfo{author}{Viktor Dick}, \bibinfo{author}{Frithjof Karsch},
  \bibinfo{author}{Edwin Laermann}, \bibinfo{author}{Swagato Mukherjee},
  \bibinfo{author}{Sayantan Sharma}, \bibinfo{title}{{Microscopic origin of
  $U_A(1)$ symmetry violation in the high temperature phase of QCD}},
  \bibinfo{journal}{Phys. Rev. D} \bibinfo{volume}{91} (\bibinfo{number}{9})
  (\bibinfo{year}{2015}) \bibinfo{pages}{094504},
  \bibinfo{doi}{\doi{10.1103/PhysRevD.91.094504}}, \eprint{1502.06190}.

\bibtype{Article}%
\bibitem{Azcoiti:2023xvu}
\bibinfo{author}{Vicente Azcoiti}, \bibinfo{title}{{Spectral density of the
  Dirac-Ginsparg-Wilson operator, chiral U(1)A anomaly, and analyticity in the
  high temperature phase of QCD}}, \bibinfo{journal}{Phys. Rev. D}
  \bibinfo{volume}{107} (\bibinfo{number}{11}) (\bibinfo{year}{2023})
  \bibinfo{pages}{114516}, \bibinfo{doi}{\doi{10.1103/PhysRevD.107.114516}},
  \eprint{2304.14725}.

\bibtype{Article}%
\bibitem{Kovacs:2023vzi}
\bibinfo{author}{Tamas~G. Kovacs}, \bibinfo{title}{{Fate of Chiral Symmetries
  in the Quark-Gluon Plasma from an Instanton-Based Random Matrix Model of
  QCD}}, \bibinfo{journal}{Phys. Rev. Lett.} \bibinfo{volume}{132}
  (\bibinfo{number}{13}) (\bibinfo{year}{2024}) \bibinfo{pages}{131902},
  \bibinfo{doi}{\doi{10.1103/PhysRevLett.132.131902}}, \eprint{2311.04208}.

\bibtype{Article}%
\bibitem{Banks:1979yr}
\bibinfo{author}{Tom Banks}, \bibinfo{author}{A. Casher},
  \bibinfo{title}{{Chiral Symmetry Breaking in Confining Theories}},
  \bibinfo{journal}{Nucl. Phys. B} \bibinfo{volume}{169} (\bibinfo{year}{1980})
  \bibinfo{pages}{103--125}, \bibinfo{doi}{\doi{10.1016/0550-3213(80)90255-2}}.

\bibtype{Article}%
\bibitem{Alexandru:2019gdm}
\bibinfo{author}{Andrei Alexandru}, \bibinfo{author}{Ivan Horv\'ath},
  \bibinfo{title}{{Possible New Phase of Thermal QCD}}, \bibinfo{journal}{Phys.
  Rev. D} \bibinfo{volume}{100} (\bibinfo{number}{9}) (\bibinfo{year}{2019})
  \bibinfo{pages}{094507}, \bibinfo{doi}{\doi{10.1103/PhysRevD.100.094507}},
  \eprint{1906.08047}.

\bibtype{Article}%
\bibitem{Ding:2011du}
\bibinfo{author}{H.-T. Ding}, \bibinfo{author}{A. Bazavov}, \bibinfo{author}{P.
  Hegde}, \bibinfo{author}{F. Karsch}, \bibinfo{author}{S. Mukherjee}, et al.,
  \bibinfo{title}{{Exploring phase diagram of $N_f=3$ QCD at $\mu=0$ with HISQ
  fermions}}, \bibinfo{journal}{PoS} \bibinfo{volume}{LATTICE2011}
  (\bibinfo{year}{2011}) \bibinfo{pages}{191}, \eprint{1111.0185}.

\bibtype{Article}%
\bibitem{Varnhorst:2015lea}
\bibinfo{author}{Lukas Varnhorst}, \bibinfo{title}{{The $N_f$=3 critical
  endpoint with smeared staggered quarks}}, \bibinfo{journal}{PoS}
  \bibinfo{volume}{LATTICE2014} (\bibinfo{year}{2015}) \bibinfo{pages}{193}.

\bibtype{Article}%
\bibitem{Fejos:2022mso}
\bibinfo{author}{G. Fejos}, \bibinfo{title}{{Second-order chiral phase
  transition in three-flavor quantum chromodynamics?}}, \bibinfo{journal}{Phys.
  Rev. D} \bibinfo{volume}{105} (\bibinfo{number}{7}) (\bibinfo{year}{2022})
  \bibinfo{pages}{L071506}, \bibinfo{doi}{\doi{10.1103/PhysRevD.105.L071506}},
  \eprint{2201.07909}.

\bibtype{Article}%
\bibitem{Fejos:2024bgl}
\bibinfo{author}{G. Fejos}, \bibinfo{author}{T. Hatsuda},
  \bibinfo{title}{{Order of the SU(Nf)\texttimes{}SU(Nf) chiral transition via
  the functional renormalization group}}, \bibinfo{journal}{Phys. Rev. D}
  \bibinfo{volume}{110} (\bibinfo{number}{1}) (\bibinfo{year}{2024})
  \bibinfo{pages}{016021}, \bibinfo{doi}{\doi{10.1103/PhysRevD.110.016021}},
  \eprint{2404.00554}.

\bibtype{Article}%
\bibitem{Resch:2017vjs}
\bibinfo{author}{Simon Resch}, \bibinfo{author}{Fabian Rennecke},
  \bibinfo{author}{Bernd-Jochen Schaefer}, \bibinfo{title}{{Mass sensitivity of
  the three-flavor chiral phase transition}}, \bibinfo{journal}{Phys. Rev. D}
  \bibinfo{volume}{99} (\bibinfo{number}{7}) (\bibinfo{year}{2019})
  \bibinfo{pages}{076005}, \bibinfo{doi}{\doi{10.1103/PhysRevD.99.076005}},
  \eprint{1712.07961}.

\bibtype{Article}%
\bibitem{Roberge:1986mm}
\bibinfo{author}{Andre Roberge}, \bibinfo{author}{Nathan Weiss},
  \bibinfo{title}{{Gauge Theories With Imaginary Chemical Potential and the
  Phases of {QCD}}}, \bibinfo{journal}{Nucl.Phys.} \bibinfo{volume}{B275}
  (\bibinfo{year}{1986}) \bibinfo{pages}{734},
  \bibinfo{doi}{\doi{10.1016/0550-3213(86)90582-1}}.

\bibtype{Article}%
\bibitem{Bonati:2016pwz}
\bibinfo{author}{Claudio Bonati}, \bibinfo{author}{Massimo D'Elia},
  \bibinfo{author}{Marco Mariti}, \bibinfo{author}{Michele Mesiti},
  \bibinfo{author}{Francesco Negro}, \bibinfo{author}{Francesco Sanfilippo},
  \bibinfo{title}{{Roberge-Weiss endpoint at the physical point of $N_f = 2+1$
  QCD}}, \bibinfo{journal}{Phys. Rev.} \bibinfo{volume}{D93}
  (\bibinfo{number}{7}) (\bibinfo{year}{2016}) \bibinfo{pages}{074504},
  \bibinfo{doi}{\doi{10.1103/PhysRevD.93.074504}}, \eprint{1602.01426}.

\bibtype{Book}%
\bibitem{kapusta:book}
\bibinfo{author}{Joseph~I. Kapusta}, \bibinfo{author}{Charles Gale},
  \bibinfo{title}{Finite-Temperature Field Theory}, \bibinfo{edition}{second}
  ed., \bibinfo{publisher}{Cambridge University Press} \bibinfo{year}{2006},
  ISBN \bibinfo{isbn}{9780511535130}, \bibinfo{note}{cambridge Books Online},
  \bibinfo{url}{\urlprefix\url{http://dx.doi.org/10.1017/CBO9780511535130}}.

\bibtype{Article}%
\bibitem{Philipsen:2014rpa}
\bibinfo{author}{Owe Philipsen}, \bibinfo{author}{Christopher Pinke},
  \bibinfo{title}{{The nature of the Roberge-Weiss transition in $N_f=2$ QCD
  with Wilson fermions}}, \bibinfo{journal}{Phys. Rev.} \bibinfo{volume}{D89}
  (\bibinfo{number}{9}) (\bibinfo{year}{2014}) \bibinfo{pages}{094504},
  \eprint{1402.0838}.

\bibtype{Article}%
\bibitem{Cuteri:2022vwk}
\bibinfo{author}{F. Cuteri}, \bibinfo{author}{J. Goswami}, \bibinfo{author}{F.
  Karsch}, \bibinfo{author}{Anirban Lahiri}, \bibinfo{author}{M. Neumann},
  \bibinfo{author}{O. Philipsen}, \bibinfo{author}{Christian Schmidt},
  \bibinfo{author}{A. Sciarra}, \bibinfo{title}{{Toward the chiral phase
  transition in the Roberge-Weiss plane}}, \bibinfo{journal}{Phys. Rev. D}
  \bibinfo{volume}{106} (\bibinfo{number}{1}) (\bibinfo{year}{2022})
  \bibinfo{pages}{014510}, \bibinfo{doi}{\doi{10.1103/PhysRevD.106.014510}},
  \eprint{2205.12707}.

\bibtype{Article}%
\bibitem{Bonati:2018fvg}
\bibinfo{author}{Claudio Bonati}, \bibinfo{author}{Enrico Calore},
  \bibinfo{author}{Massimo D'Elia}, \bibinfo{author}{Michele Mesiti},
  \bibinfo{author}{Francesco Negro}, \bibinfo{author}{Francesco Sanfilippo},
  \bibinfo{author}{Sebastiano~Fabio Schifano}, \bibinfo{author}{Giorgio Silvi},
  \bibinfo{author}{Raffaele Tripiccione}, \bibinfo{title}{{Roberge-Weiss
  endpoint and chiral symmetry restoration in $N_f = 2+1$ QCD}},
  \bibinfo{journal}{Phys. Rev.} \bibinfo{volume}{D99} (\bibinfo{year}{2019})
  \bibinfo{pages}{014502}, \eprint{1807.02106}.

\bibtype{Article}%
\bibitem{Yang:1952be}
\bibinfo{author}{Chen-Ning Yang}, \bibinfo{author}{T.~D. Lee},
  \bibinfo{title}{{Statistical theory of equations of state and phase
  transitions. 1. Theory of condensation}}, \bibinfo{journal}{Phys. Rev.}
  \bibinfo{volume}{87} (\bibinfo{year}{1952}) \bibinfo{pages}{404--409},
  \bibinfo{doi}{\doi{10.1103/PhysRev.87.404}}.

\bibtype{Article}%
\bibitem{Lee:1952ig}
\bibinfo{author}{T.~D. Lee}, \bibinfo{author}{Chen-Ning Yang},
  \bibinfo{title}{{Statistical theory of equations of state and phase
  transitions. 2. Lattice gas and Ising model}}, \bibinfo{journal}{Phys. Rev.}
  \bibinfo{volume}{87} (\bibinfo{year}{1952}) \bibinfo{pages}{410--419},
  \bibinfo{doi}{\doi{10.1103/PhysRev.87.410}}.

\bibtype{Article}%
\bibitem{Connelly:2020gwa}
\bibinfo{author}{Andrew Connelly}, \bibinfo{author}{Gregory Johnson},
  \bibinfo{author}{Fabian Rennecke}, \bibinfo{author}{Vladimir Skokov},
  \bibinfo{title}{{Universal Location of the Yang-Lee Edge Singularity in
  $O(N)$ Theories}}, \bibinfo{journal}{Phys. Rev. Lett.} \bibinfo{volume}{125}
  (\bibinfo{number}{19}) (\bibinfo{year}{2020}) \bibinfo{pages}{191602},
  \bibinfo{doi}{\doi{10.1103/PhysRevLett.125.191602}}, \eprint{2006.12541}.

\bibtype{Article}%
\bibitem{Johnson:2022cqv}
\bibinfo{author}{Gregory Johnson}, \bibinfo{author}{Fabian Rennecke},
  \bibinfo{author}{Vladimir~V. Skokov}, \bibinfo{title}{{Universal location of
  Yang-Lee edge singularity in classic O(N) universality classes}},
  \bibinfo{journal}{Phys. Rev. D} \bibinfo{volume}{107} (\bibinfo{number}{11})
  (\bibinfo{year}{2023}) \bibinfo{pages}{116013},
  \bibinfo{doi}{\doi{10.1103/PhysRevD.107.116013}}, \eprint{2211.00710}.

\bibtype{Article}%
\bibitem{Rennecke:2022ohx}
\bibinfo{author}{Fabian Rennecke}, \bibinfo{author}{Vladimir~V. Skokov},
  \bibinfo{title}{{Universal location of Yang\textendash{}Lee edge singularity
  for a one-component field theory in 1\ensuremath{\leq}d\ensuremath{\leq}4}},
  \bibinfo{journal}{Annals Phys.} \bibinfo{volume}{444} (\bibinfo{year}{2022})
  \bibinfo{pages}{169010}, \bibinfo{doi}{\doi{10.1016/j.aop.2022.169010}},
  \eprint{2203.16651}.

\bibtype{Article}%
\bibitem{Karsch:2023rfb}
\bibinfo{author}{Frithjof Karsch}, \bibinfo{author}{Christian Schmidt},
  \bibinfo{author}{Simran Singh}, \bibinfo{title}{{Lee-Yang and Langer edge
  singularities from analytic continuation of scaling functions}},
  \bibinfo{journal}{Phys. Rev. D} \bibinfo{volume}{109} (\bibinfo{number}{1})
  (\bibinfo{year}{2024}) \bibinfo{pages}{014508},
  \bibinfo{doi}{\doi{10.1103/PhysRevD.109.014508}}, \eprint{2311.13530}.

\bibtype{Article}%
\bibitem{Wada:2024qsk}
\bibinfo{author}{Tatsuya Wada}, \bibinfo{author}{Masakiyo Kitazawa},
  \bibinfo{author}{Kazuyuki Kanaya}, \bibinfo{title}{{Locating Critical Points
  Using Ratios of Lee-Yang Zeros}}, \bibinfo{journal}{Phys. Rev. Lett.}
  \bibinfo{volume}{134} (\bibinfo{number}{16}) (\bibinfo{year}{2025})
  \bibinfo{pages}{162302}, \bibinfo{doi}{\doi{10.1103/PhysRevLett.134.162302}},
  \eprint{2410.19345}.

\bibtype{Article}%
\bibitem{Skokov:2024fac}
\bibinfo{author}{Vladimir~V. Skokov}, \bibinfo{title}{{Two lectures on Yang-Lee
  edge singularity and analytic structure of QCD equation of state}},
  \bibinfo{journal}{SciPost Phys. Lect. Notes} \bibinfo{volume}{91}
  (\bibinfo{year}{2025}) \bibinfo{pages}{1},
  \bibinfo{doi}{\doi{10.21468/SciPostPhysLectNotes.91}}, \eprint{2411.02663}.

\bibtype{Article}%
\bibitem{Dimopoulos:2021vrk}
\bibinfo{author}{P. Dimopoulos}, \bibinfo{author}{L. Dini}, \bibinfo{author}{F.
  Di~Renzo}, \bibinfo{author}{J. Goswami}, \bibinfo{author}{G. Nicotra},
  \bibinfo{author}{C. Schmidt}, \bibinfo{author}{S. Singh}, \bibinfo{author}{K.
  Zambello}, \bibinfo{author}{F. Ziesch\'e}, \bibinfo{title}{{Contribution to
  understanding the phase structure of strong interaction matter: Lee-Yang edge
  singularities from lattice QCD}}, \bibinfo{journal}{Phys. Rev. D}
  \bibinfo{volume}{105} (\bibinfo{number}{3}) (\bibinfo{year}{2022})
  \bibinfo{pages}{034513}, \bibinfo{doi}{\doi{10.1103/PhysRevD.105.034513}},
  \eprint{2110.15933}.

\bibtype{Article}%
\bibitem{Clarke:2024ugt}
\bibinfo{author}{David~A. Clarke}, \bibinfo{author}{Petros Dimopoulos},
  \bibinfo{author}{Francesco Di~Renzo}, \bibinfo{author}{Jishnu Goswami},
  \bibinfo{author}{Christian Schmidt}, \bibinfo{author}{Simran Singh},
  \bibinfo{author}{Kevin Zambello}, \bibinfo{title}{{Searching for the QCD
  critical end point using multipoint Pad{\'e} approximations}},
  \bibinfo{journal}{Phys. Rev. D} \bibinfo{volume}{112} (\bibinfo{number}{9})
  (\bibinfo{year}{2025}) \bibinfo{pages}{L091504},
  \bibinfo{doi}{\doi{10.1103/y6kg-ry8x}}, \eprint{2405.10196}.

\bibtype{Article}%
\bibitem{Adam:2025phc}
\bibinfo{author}{Alexander Adam}, \bibinfo{author}{Szabolcs Bors{\'a}nyi},
  \bibinfo{author}{Zoltan Fodor}, \bibinfo{author}{Jana~N. Guenther},
  \bibinfo{author}{Piyush Kumar}, \bibinfo{author}{Paolo Parotto},
  \bibinfo{author}{Attila P{\'a}sztor}, \bibinfo{author}{Chik~Him Wong},
  \bibinfo{title}{{High-precision baryon number cumulants from lattice QCD in a
  finite box: cumulant ratios, Lee-Yang zeros and critical endpoint
  predictions}}  (\bibinfo{year}{2025}), \eprint{2507.13254}.

\bibtype{Article}%
\bibitem{Basar:2023nkp}
\bibinfo{author}{Gokce Basar}, \bibinfo{title}{{QCD critical point, Lee-Yang
  edge singularities, and Pad\'e resummations}}, \bibinfo{journal}{Phys. Rev.
  C} \bibinfo{volume}{110} (\bibinfo{number}{1}) (\bibinfo{year}{2024})
  \bibinfo{pages}{015203}, \bibinfo{doi}{\doi{10.1103/PhysRevC.110.015203}},
  \eprint{2312.06952}.

\bibtype{Article}%
\bibitem{Adam:2025pii}
\bibinfo{author}{Alexander Adam}, \bibinfo{author}{Szabolcs Bors{\'a}nyi},
  \bibinfo{author}{Zolt{\'a}n Fodor}, \bibinfo{author}{Jana~N. Guenther},
  \bibinfo{author}{Paolo Parotto}, \bibinfo{author}{Attila P{\'a}sztor},
  \bibinfo{author}{D{\'a}vid Peszny{\'a}k}, \bibinfo{author}{Ludovica Pirelli},
  \bibinfo{author}{Chik~Him Wong}, \bibinfo{title}{{Search for a Lee-Yang edge
  singularity in high-statistics Wuppertal-Budapest data}},
  \bibinfo{journal}{PoS} \bibinfo{volume}{LATTICE2024} (\bibinfo{year}{2025})
  \bibinfo{pages}{178}, \bibinfo{doi}{\doi{10.22323/1.466.0178}},
  \eprint{2502.03211}.

\bibtype{Article}%
\bibitem{Kharzeev:2007jp}
\bibinfo{author}{Dmitri~E. Kharzeev}, \bibinfo{author}{Larry~D. McLerran},
  \bibinfo{author}{Harmen~J. Warringa}, \bibinfo{title}{{The Effects of
  topological charge change in heavy ion collisions: 'Event by event P and CP
  violation'}}, \bibinfo{journal}{Nucl. Phys. A} \bibinfo{volume}{803}
  (\bibinfo{year}{2008}) \bibinfo{pages}{227--253},
  \bibinfo{doi}{\doi{10.1016/j.nuclphysa.2008.02.298}}, \eprint{0711.0950}.

\bibtype{Article}%
\bibitem{Andersen:2014xxa}
\bibinfo{author}{Jens~O. Andersen}, \bibinfo{author}{William~R. Naylor},
  \bibinfo{author}{Anders Tranberg}, \bibinfo{title}{{Phase diagram of QCD in a
  magnetic field: A review}}, \bibinfo{journal}{Rev. Mod. Phys.}
  \bibinfo{volume}{88} (\bibinfo{year}{2016}) \bibinfo{pages}{025001},
  \bibinfo{doi}{\doi{10.1103/RevModPhys.88.025001}}, \eprint{1411.7176}.

\bibtype{Article}%
\bibitem{Endrodi:2024cqn}
\bibinfo{author}{Gergely Endrodi}, \bibinfo{title}{{QCD with background
  electromagnetic fields on the lattice: A review}}, \bibinfo{journal}{Prog.
  Part. Nucl. Phys.} \bibinfo{volume}{141} (\bibinfo{year}{2025})
  \bibinfo{pages}{104153}, \bibinfo{doi}{\doi{10.1016/j.ppnp.2024.104153}},
  \eprint{2406.19780}.

\bibtype{Article}%
\bibitem{Shovkovy:2012zn}
\bibinfo{author}{Igor~A. Shovkovy}, \bibinfo{title}{{Magnetic Catalysis: A
  Review}}, \bibinfo{journal}{Lect. Notes Phys.} \bibinfo{volume}{871}
  (\bibinfo{year}{2013}) \bibinfo{pages}{13--49},
  \bibinfo{doi}{\doi{10.1007/978-3-642-37305-3}}, \eprint{1207.5081}.

\bibtype{Article}%
\bibitem{DElia:2011koc}
\bibinfo{author}{Massimo D'Elia}, \bibinfo{author}{Francesco Negro},
  \bibinfo{title}{{Chiral Properties of Strong Interactions in a Magnetic
  Background}}, \bibinfo{journal}{Phys. Rev. D} \bibinfo{volume}{83}
  (\bibinfo{year}{2011}) \bibinfo{pages}{114028},
  \bibinfo{doi}{\doi{10.1103/PhysRevD.83.114028}}, \eprint{1103.2080}.

\bibtype{Article}%
\bibitem{Bali:2012zg}
\bibinfo{author}{G.S. Bali}, \bibinfo{author}{F. Bruckmann},
  \bibinfo{author}{G. Endrodi}, \bibinfo{author}{Z. Fodor},
  \bibinfo{author}{S.D. Katz}, et al., \bibinfo{title}{{QCD quark condensate in
  external magnetic fields}}, \bibinfo{journal}{Phys.Rev.}
  \bibinfo{volume}{D86} (\bibinfo{year}{2012}) \bibinfo{pages}{071502},
  \bibinfo{doi}{\doi{10.1103/PhysRevD.86.071502}}, \eprint{1206.4205}.

\bibtype{Article}%
\bibitem{Brandt:2023dir}
\bibinfo{author}{B.~B. Brandt}, \bibinfo{author}{F. Cuteri},
  \bibinfo{author}{G. Endr\H{o}di}, \bibinfo{author}{G. Mark\'o},
  \bibinfo{author}{L. Sandbote}, \bibinfo{author}{A.~D.~M. Valois},
  \bibinfo{title}{{Thermal QCD in a non-uniform magnetic background}},
  \bibinfo{journal}{JHEP} \bibinfo{volume}{11} (\bibinfo{year}{2023})
  \bibinfo{pages}{229}, \bibinfo{doi}{\doi{10.1007/JHEP11(2023)229}},
  \eprint{2305.19029}.

\bibtype{Article}%
\bibitem{Bruckmann:2013oba}
\bibinfo{author}{Falk Bruckmann}, \bibinfo{author}{Gergely Endrodi},
  \bibinfo{author}{Tamas~G. Kovacs}, \bibinfo{title}{{Inverse magnetic
  catalysis and the Polyakov loop}}, \bibinfo{journal}{JHEP}
  \bibinfo{volume}{1304} (\bibinfo{year}{2013}) \bibinfo{pages}{112},
  \bibinfo{doi}{\doi{10.1007/JHEP04(2013)112}}, \eprint{1303.3972}.

\bibtype{Article}%
\bibitem{Bali:2011qj}
\bibinfo{author}{G.S. Bali}, \bibinfo{author}{F. Bruckmann},
  \bibinfo{author}{G. Endrodi}, \bibinfo{author}{Z. Fodor},
  \bibinfo{author}{S.D. Katz}, et al., \bibinfo{title}{{The QCD phase diagram
  for external magnetic fields}}, \bibinfo{journal}{JHEP}
  \bibinfo{volume}{1202} (\bibinfo{year}{2012}) \bibinfo{pages}{044},
  \bibinfo{doi}{\doi{10.1007/JHEP02(2012)044}}, \eprint{1111.4956}.

\bibtype{Article}%
\bibitem{Endrodi:2015oba}
\bibinfo{author}{Gergely Endrodi}, \bibinfo{title}{{Critical point in the QCD
  phase diagram for extremely strong background magnetic fields}},
  \bibinfo{journal}{JHEP} \bibinfo{volume}{07} (\bibinfo{year}{2015})
  \bibinfo{pages}{173}, \bibinfo{doi}{\doi{10.1007/JHEP07(2015)173}},
  \eprint{1504.08280}.

\bibtype{Article}%
\bibitem{Endrodi:2014vza}
\bibinfo{author}{Gergely Endr{\"o}di}, \bibinfo{title}{{QCD in magnetic fields:
  from Hofstadter's butterfly to the phase diagram}}, \bibinfo{journal}{PoS}
  \bibinfo{volume}{LATTICE2014} (\bibinfo{year}{2014}) \bibinfo{pages}{018},
  \eprint{1410.8028}.

\bibtype{Article}%
\bibitem{Cohen:2013zja}
\bibinfo{author}{Thomas~D. Cohen}, \bibinfo{author}{Naoki Yamamoto},
  \bibinfo{title}{{New critical point for QCD in a magnetic field}},
  \bibinfo{journal}{Phys. Rev. D} \bibinfo{volume}{89} (\bibinfo{number}{5})
  (\bibinfo{year}{2014}) \bibinfo{pages}{054029},
  \bibinfo{doi}{\doi{10.1103/PhysRevD.89.054029}}, \eprint{1310.2234}.

\bibtype{Article}%
\bibitem{DElia:2021yvk}
\bibinfo{author}{Massimo D'Elia}, \bibinfo{author}{Lorenzo Maio},
  \bibinfo{author}{Francesco Sanfilippo}, \bibinfo{author}{Alfredo Stanzione},
  \bibinfo{title}{{Phase diagram of QCD in a magnetic background}},
  \bibinfo{journal}{Phys. Rev. D} \bibinfo{volume}{105} (\bibinfo{number}{3})
  (\bibinfo{year}{2022}) \bibinfo{pages}{034511},
  \bibinfo{doi}{\doi{10.1103/PhysRevD.105.034511}}, \eprint{2111.11237}.

\bibtype{Article}%
\bibitem{DElia:2025ybj}
\bibinfo{author}{Massimo D'Elia}, \bibinfo{author}{Lorenzo Maio},
  \bibinfo{author}{Kevin Zambello}, \bibinfo{author}{Giuseppe Zanichelli},
  \bibinfo{title}{{Roberge-Weiss transition for QCD in a magnetic background}},
  \bibinfo{journal}{Phys. Rev. D} \bibinfo{volume}{111} (\bibinfo{number}{9})
  (\bibinfo{year}{2025}) \bibinfo{pages}{094509},
  \bibinfo{doi}{\doi{10.1103/PhysRevD.111.094509}}, \eprint{2502.19294}.

\bibtype{Article}%
\bibitem{Abbott:2023coj}
\bibinfo{author}{Ryan Abbott}, \bibinfo{author}{William Detmold},
  \bibinfo{author}{Fernando Romero-L\'opez}, \bibinfo{author}{Zohreh Davoudi},
  \bibinfo{author}{Marc Illa}, \bibinfo{author}{Assumpta Parre\~no},
  \bibinfo{author}{Robert~J. Perry}, \bibinfo{author}{Phiala~E. Shanahan},
  \bibinfo{author}{Michael~L. Wagman} (\bibinfo{collaboration}{NPLQCD}),
  \bibinfo{title}{{Lattice quantum chromodynamics at large isospin density}},
  \bibinfo{journal}{Phys. Rev. D} \bibinfo{volume}{108} (\bibinfo{number}{11})
  (\bibinfo{year}{2023}) \bibinfo{pages}{114506},
  \bibinfo{doi}{\doi{10.1103/PhysRevD.108.114506}}, \eprint{2307.15014}.

\bibtype{Article}%
\bibitem{Abbott:2024vhj}
\bibinfo{author}{Ryan Abbott}, \bibinfo{author}{William Detmold},
  \bibinfo{author}{Marc Illa}, \bibinfo{author}{Assumpta Parre\~no},
  \bibinfo{author}{Robert~J. Perry}, \bibinfo{author}{Fernando Romero-L\'opez},
  \bibinfo{author}{Phiala~E. Shanahan}, \bibinfo{author}{Michael~L. Wagman}
  (\bibinfo{collaboration}{NPLQCD}), \bibinfo{title}{{QCD Constraints on
  Isospin-Dense Matter and the Nuclear Equation of State}},
  \bibinfo{journal}{Phys. Rev. Lett.} \bibinfo{volume}{134}
  (\bibinfo{number}{1}) (\bibinfo{year}{2025}) \bibinfo{pages}{011903},
  \bibinfo{doi}{\doi{10.1103/PhysRevLett.134.011903}}, \eprint{2406.09273}.

\bibtype{Article}%
\bibitem{Son:2000xc}
\bibinfo{author}{D.~T. Son}, \bibinfo{author}{Misha~A. Stephanov},
  \bibinfo{title}{{QCD at finite isospin density}}, \bibinfo{journal}{Phys.
  Rev. Lett.} \bibinfo{volume}{86} (\bibinfo{year}{2001})
  \bibinfo{pages}{592--595}, \bibinfo{doi}{\doi{10.1103/PhysRevLett.86.592}},
  \eprint{hep-ph/0005225}.

\bibtype{Article}%
\bibitem{Brandt:2018omg}
\bibinfo{author}{Bastian~B. Brandt}, \bibinfo{author}{Gergely Endrodi},
  \bibinfo{title}{{Reliability of Taylor expansions in QCD}},
  \bibinfo{journal}{Phys. Rev. D} \bibinfo{volume}{99} (\bibinfo{number}{1})
  (\bibinfo{year}{2019}) \bibinfo{pages}{014518},
  \bibinfo{doi}{\doi{10.1103/PhysRevD.99.014518}}, \eprint{1810.11045}.

\bibtype{Article}%
\bibitem{Borsanyi:2023tdp}
\bibinfo{author}{Szabolcs Borsanyi}, \bibinfo{author}{Zoltan Fodor},
  \bibinfo{author}{Matteo Giordano}, \bibinfo{author}{Jana~N. Guenther},
  \bibinfo{author}{Sandor~D. Katz}, \bibinfo{author}{Attila Pasztor},
  \bibinfo{author}{Chik~Him Wong}, \bibinfo{title}{{Can rooted staggered
  fermions describe nonzero baryon density at low temperatures?}},
  \bibinfo{journal}{Phys. Rev. D} \bibinfo{volume}{109} (\bibinfo{number}{5})
  (\bibinfo{year}{2024}) \bibinfo{pages}{054509},
  \bibinfo{doi}{\doi{10.1103/PhysRevD.109.054509}}, \eprint{2308.06105}.

\bibtype{Article}%
\bibitem{Brandt:2017oyy}
\bibinfo{author}{B.~B. Brandt}, \bibinfo{author}{G. Endrodi},
  \bibinfo{author}{S. Schmalzbauer}, \bibinfo{title}{{QCD phase diagram for
  nonzero isospin-asymmetry}}, \bibinfo{journal}{Phys. Rev. D}
  \bibinfo{volume}{97} (\bibinfo{number}{5}) (\bibinfo{year}{2018})
  \bibinfo{pages}{054514}, \bibinfo{doi}{\doi{10.1103/PhysRevD.97.054514}},
  \eprint{1712.08190}.

\bibtype{Article}%
\bibitem{Koehn:2024set}
\bibinfo{author}{Hauke Koehn}, et al., \bibinfo{title}{{From existing and new
  nuclear and astrophysical constraints to stringent limits on the equation of
  state of neutron-rich dense matter}}, \bibinfo{journal}{Phys. Rev. X}
  \bibinfo{volume}{15} (\bibinfo{number}{2}) (\bibinfo{year}{2025})
  \bibinfo{pages}{021014}, \bibinfo{doi}{\doi{10.1103/PhysRevX.15.021014}},
  \eprint{2402.04172}.

\bibtype{Article}%
\bibitem{Haskell:2017lkl}
\bibinfo{author}{Brynmor Haskell}, \bibinfo{author}{Armen Sedrakian},
  \bibinfo{title}{{Superfluidity and Superconductivity in Neutron Stars}},
  \bibinfo{journal}{Astrophys. Space Sci. Libr.} \bibinfo{volume}{457}
  (\bibinfo{year}{2018}) \bibinfo{pages}{401--454},
  \bibinfo{doi}{\doi{10.1007/978-3-319-97616-7_8}}, \eprint{1709.10340}.

\bibtype{Article}%
\bibitem{Watanabe:2000rj}
\bibinfo{author}{Gentaro Watanabe}, \bibinfo{author}{Kei Iida},
  \bibinfo{author}{Katsuhiko Sato}, \bibinfo{title}{{Thermodynamic properties
  of nuclear 'pasta' in neutron star crusts}}, \bibinfo{journal}{Nucl. Phys. A}
  \bibinfo{volume}{676} (\bibinfo{year}{2000}) \bibinfo{pages}{455--473},
  \bibinfo{doi}{\doi{10.1016/S0375-9474(00)00197-4}}, \bibinfo{note}{[Erratum:
  Nucl.Phys.A 726, 357--365 (2003)]}, \eprint{astro-ph/0001273}.

\bibtype{Article}%
\bibitem{Blaschke:2020qrs}
\bibinfo{author}{David Blaschke}, \bibinfo{author}{Hovik Grigorian},
  \bibinfo{author}{Gerd R{\"o}pke}, \bibinfo{title}{{Chirally improved quark
  Pauli blocking in nuclear matter and applications to quark deconfinement in
  neutron stars}}, \bibinfo{journal}{Particles} \bibinfo{volume}{3}
  (\bibinfo{number}{2}) (\bibinfo{year}{2020}) \bibinfo{pages}{477--499},
  \bibinfo{doi}{\doi{10.3390/particles3020033}}, \eprint{2005.10218}.

\bibtype{Article}%
\bibitem{McLerran:2007qj}
\bibinfo{author}{Larry McLerran}, \bibinfo{author}{Robert~D. Pisarski},
  \bibinfo{title}{{Phases of cold, dense quarks at large N(c)}},
  \bibinfo{journal}{Nucl. Phys. A} \bibinfo{volume}{796} (\bibinfo{year}{2007})
  \bibinfo{pages}{83--100},
  \bibinfo{doi}{\doi{10.1016/j.nuclphysa.2007.08.013}}, \eprint{0706.2191}.

\bibtype{Article}%
\bibitem{McLerran:2018hbz}
\bibinfo{author}{Larry McLerran}, \bibinfo{author}{Sanjay Reddy},
  \bibinfo{title}{{Quarkyonic Matter and Neutron Stars}},
  \bibinfo{journal}{Phys. Rev. Lett.} \bibinfo{volume}{122}
  (\bibinfo{number}{12}) (\bibinfo{year}{2019}) \bibinfo{pages}{122701},
  \bibinfo{doi}{\doi{10.1103/PhysRevLett.122.122701}}, \eprint{1811.12503}.

\bibtype{Article}%
\bibitem{Lattimer:2004pg}
\bibinfo{author}{J.~M. Lattimer}, \bibinfo{author}{M. Prakash},
  \bibinfo{title}{{The physics of neutron stars}}, \bibinfo{journal}{Science}
  \bibinfo{volume}{304} (\bibinfo{year}{2004}) \bibinfo{pages}{536--542},
  \bibinfo{doi}{\doi{10.1126/science.1090720}}, \eprint{astro-ph/0405262}.

\bibtype{Article}%
\bibitem{Antoniadis:2013pzd}
\bibinfo{author}{John Antoniadis}, et al., \bibinfo{title}{{A Massive Pulsar in
  a Compact Relativistic Binary}}, \bibinfo{journal}{Science}
  \bibinfo{volume}{340} (\bibinfo{year}{2013}) \bibinfo{pages}{6131},
  \bibinfo{doi}{\doi{10.1126/science.1233232}}, \eprint{1304.6875}.

\bibtype{Article}%
\bibitem{NANOGrav:2019jur}
\bibinfo{author}{H.~T. Cromartie}, et al. (\bibinfo{collaboration}{NANOGrav}),
  \bibinfo{title}{{Relativistic Shapiro delay measurements of an extremely
  massive millisecond pulsar}}, \bibinfo{journal}{Nature Astron.}
  \bibinfo{volume}{4} (\bibinfo{number}{1}) (\bibinfo{year}{2019})
  \bibinfo{pages}{72--76}, \bibinfo{doi}{\doi{10.1038/s41550-019-0880-2}},
  \eprint{1904.06759}.

\bibtype{Article}%
\bibitem{Romani:2021xmb}
\bibinfo{author}{Roger~W. Romani}, \bibinfo{author}{D. Kandel},
  \bibinfo{author}{Alexei~V. Filippenko}, \bibinfo{author}{Thomas~G. Brink},
  \bibinfo{author}{WeiKang Zheng}, \bibinfo{title}{{PSR J1810+1744: Companion
  Darkening and a Precise High Neutron Star Mass}},
  \bibinfo{journal}{Astrophys. J. Lett.} \bibinfo{volume}{908}
  (\bibinfo{number}{2}) (\bibinfo{year}{2021}) \bibinfo{pages}{L46},
  \bibinfo{doi}{\doi{10.3847/2041-8213/abe2b4}}, \eprint{2101.09822}.

\bibtype{Article}%
\bibitem{Romani:2022jhd}
\bibinfo{author}{Roger~W. Romani}, \bibinfo{author}{D. Kandel},
  \bibinfo{author}{Alexei~V. Filippenko}, \bibinfo{author}{Thomas~G. Brink},
  \bibinfo{author}{WeiKang Zheng}, \bibinfo{title}{{PSR
  J0952{\ensuremath{-}}0607: The Fastest and Heaviest Known Galactic Neutron
  Star}}, \bibinfo{journal}{Astrophys. J. Lett.} \bibinfo{volume}{934}
  (\bibinfo{number}{2}) (\bibinfo{year}{2022}) \bibinfo{pages}{L17},
  \bibinfo{doi}{\doi{10.3847/2041-8213/ac8007}}, \eprint{2207.05124}.

\bibtype{Article}%
\bibitem{Tolman:1939jz}
\bibinfo{author}{Richard~C. Tolman}, \bibinfo{title}{{Static solutions of
  Einstein's field equations for spheres of fluid}}, \bibinfo{journal}{Phys.
  Rev.} \bibinfo{volume}{55} (\bibinfo{year}{1939}) \bibinfo{pages}{364--373},
  \bibinfo{doi}{\doi{10.1103/PhysRev.55.364}}.

\bibtype{Article}%
\bibitem{Oppenheimer:1939ne}
\bibinfo{author}{J.~R. Oppenheimer}, \bibinfo{author}{G.~M. Volkoff},
  \bibinfo{title}{{On massive neutron cores}}, \bibinfo{journal}{Phys. Rev.}
  \bibinfo{volume}{55} (\bibinfo{year}{1939}) \bibinfo{pages}{374--381},
  \bibinfo{doi}{\doi{10.1103/PhysRev.55.374}}.

\bibtype{Article}%
\bibitem{Lindblom:1992ApJ}
\bibinfo{author}{Lee {Lindblom}}, \bibinfo{title}{{Determining the Nuclear
  Equation of State from Neutron-Star Masses and Radii}},
  \bibinfo{journal}{ApJ} \bibinfo{volume}{398} (\bibinfo{year}{1992})
  \bibinfo{pages}{569}, \bibinfo{doi}{\doi{10.1086/171882}}.

\bibtype{Article}%
\bibitem{Hinderer:2009ca}
\bibinfo{author}{Tanja Hinderer}, \bibinfo{author}{Benjamin~D. Lackey},
  \bibinfo{author}{Ryan~N. Lang}, \bibinfo{author}{Jocelyn~S. Read},
  \bibinfo{title}{{Tidal deformability of neutron stars with realistic
  equations of state and their gravitational wave signatures in binary
  inspiral}}, \bibinfo{journal}{Phys. Rev. D} \bibinfo{volume}{81}
  (\bibinfo{year}{2010}) \bibinfo{pages}{123016},
  \bibinfo{doi}{\doi{10.1103/PhysRevD.81.123016}}, \eprint{0911.3535}.

\bibtype{Article}%
\bibitem{Sekiguchi:2011mc}
\bibinfo{author}{Yuichiro Sekiguchi}, \bibinfo{author}{Kenta Kiuchi},
  \bibinfo{author}{Koutarou Kyutoku}, \bibinfo{author}{Masaru Shibata},
  \bibinfo{title}{{Effects of hyperons in binary neutron star mergers}},
  \bibinfo{journal}{Phys. Rev. Lett.} \bibinfo{volume}{107}
  (\bibinfo{year}{2011}) \bibinfo{pages}{211101},
  \bibinfo{doi}{\doi{10.1103/PhysRevLett.107.211101}}, \eprint{1110.4442}.

\bibtype{Article}%
\bibitem{Bernuzzi:2015opx}
\bibinfo{author}{Sebastiano Bernuzzi}, \bibinfo{author}{David Radice},
  \bibinfo{author}{Christian~D. Ott}, \bibinfo{author}{Luke~F. Roberts},
  \bibinfo{author}{Philipp Moesta}, \bibinfo{author}{Filippo Galeazzi},
  \bibinfo{title}{{How loud are neutron star mergers?}},
  \bibinfo{journal}{Phys. Rev. D} \bibinfo{volume}{94} (\bibinfo{number}{2})
  (\bibinfo{year}{2016}) \bibinfo{pages}{024023},
  \bibinfo{doi}{\doi{10.1103/PhysRevD.94.024023}}, \eprint{1512.06397}.

\bibtype{Article}%
\bibitem{Perego:2019adq}
\bibinfo{author}{Albino Perego}, \bibinfo{author}{Sebastiano Bernuzzi},
  \bibinfo{author}{David Radice}, \bibinfo{title}{{Thermodynamics conditions of
  matter in neutron star mergers}}, \bibinfo{journal}{Eur. Phys. J. A}
  \bibinfo{volume}{55} (\bibinfo{number}{8}) (\bibinfo{year}{2019})
  \bibinfo{pages}{124}, \bibinfo{doi}{\doi{10.1140/epja/i2019-12810-7}},
  \eprint{1903.07898}.

\bibtype{Article}%
\bibitem{Chatziioannou:2024tjq}
\bibinfo{author}{Katerina Chatziioannou}, \bibinfo{author}{H.~Thankful
  Cromartie}, \bibinfo{author}{Stefano Gandolfi}, \bibinfo{author}{Ingo Tews},
  \bibinfo{author}{David Radice}, \bibinfo{author}{Andrew~W. Steiner},
  \bibinfo{author}{Anna~L. Watts}, \bibinfo{title}{{Neutron stars and the dense
  matter equation of state: from microscopic theory to macroscopic
  observations}}  (\bibinfo{year}{2024}), \eprint{2407.11153}.

\bibtype{Article}%
\bibitem{Annala:2017llu}
\bibinfo{author}{Eemeli Annala}, \bibinfo{author}{Tyler Gorda},
  \bibinfo{author}{Aleksi Kurkela}, \bibinfo{author}{Aleksi Vuorinen},
  \bibinfo{title}{{Gravitational-wave constraints on the neutron-star-matter
  Equation of State}}, \bibinfo{journal}{Phys. Rev. Lett.}
  \bibinfo{volume}{120} (\bibinfo{number}{17}) (\bibinfo{year}{2018})
  \bibinfo{pages}{172703}, \bibinfo{doi}{\doi{10.1103/PhysRevLett.120.172703}},
  \eprint{1711.02644}.

\bibtype{Article}%
\bibitem{Annala:2019puf}
\bibinfo{author}{Eemeli Annala}, \bibinfo{author}{Tyler Gorda},
  \bibinfo{author}{Aleksi Kurkela}, \bibinfo{author}{Joonas N{\"a}ttil{\"a}},
  \bibinfo{author}{Aleksi Vuorinen}, \bibinfo{title}{{Evidence for quark-matter
  cores in massive neutron stars}}, \bibinfo{journal}{Nature Phys.}
  \bibinfo{volume}{16} (\bibinfo{number}{9}) (\bibinfo{year}{2020})
  \bibinfo{pages}{907--910}, \bibinfo{doi}{\doi{10.1038/s41567-020-0914-9}},
  \eprint{1903.09121}.

\bibtype{Article}%
\bibitem{Komoltsev:2021jzg}
\bibinfo{author}{Oleg Komoltsev}, \bibinfo{author}{Aleksi Kurkela},
  \bibinfo{title}{{How Perturbative QCD Constrains the Equation of State at
  Neutron-Star Densities}}, \bibinfo{journal}{Phys. Rev. Lett.}
  \bibinfo{volume}{128} (\bibinfo{number}{20}) (\bibinfo{year}{2022})
  \bibinfo{pages}{202701}, \bibinfo{doi}{\doi{10.1103/PhysRevLett.128.202701}},
  \eprint{2111.05350}.

\bibtype{Article}%
\bibitem{Mroczek:2023zxo}
\bibinfo{author}{Debora Mroczek}, \bibinfo{author}{M.~Coleman Miller},
  \bibinfo{author}{Jacquelyn Noronha-Hostler}, \bibinfo{author}{Nicolas Yunes},
  \bibinfo{title}{{Nontrivial features in the speed of sound inside neutron
  stars}}, \bibinfo{journal}{Phys. Rev. D} \bibinfo{volume}{110}
  (\bibinfo{number}{12}) (\bibinfo{year}{2024}) \bibinfo{pages}{123009},
  \bibinfo{doi}{\doi{10.1103/PhysRevD.110.123009}}, \eprint{2309.02345}.

\bibtype{Article}%
\bibitem{Finch:2025bao}
\bibinfo{author}{Eliot Finch}, \bibinfo{author}{Isaac Legred},
  \bibinfo{author}{Katerina Chatziioannou}, \bibinfo{author}{Reed Essick},
  \bibinfo{author}{Sophia Han}, \bibinfo{author}{Philippe Landry},
  \bibinfo{title}{{Unified nonparametric equation-of-state inference from the
  neutron-star crust to perturbative-QCD densities}}, \bibinfo{journal}{Phys.
  Rev. D} \bibinfo{volume}{112} (\bibinfo{number}{10}) (\bibinfo{year}{2025})
  \bibinfo{pages}{103023}, \bibinfo{doi}{\doi{10.1103/krc7-kz2l}},
  \eprint{2505.13691}.

\bibtype{Article}%
\bibitem{Collins:1974ky}
\bibinfo{author}{John~C. Collins}, \bibinfo{author}{M.~J. Perry},
  \bibinfo{title}{{Superdense Matter: Neutrons Or Asymptotically Free
  Quarks?}}, \bibinfo{journal}{Phys. Rev. Lett.} \bibinfo{volume}{34}
  (\bibinfo{year}{1975}) \bibinfo{pages}{1353},
  \bibinfo{doi}{\doi{10.1103/PhysRevLett.34.1353}}.

\bibtype{Article}%
\bibitem{Cooper:1956zz}
\bibinfo{author}{Leon~N. Cooper}, \bibinfo{title}{{Bound electron pairs in a
  degenerate Fermi gas}}, \bibinfo{journal}{Phys. Rev.} \bibinfo{volume}{104}
  (\bibinfo{year}{1956}) \bibinfo{pages}{1189--1190},
  \bibinfo{doi}{\doi{10.1103/PhysRev.104.1189}}.

\bibtype{Article}%
\bibitem{Barrois:1977xd}
\bibinfo{author}{Bertrand~C. Barrois}, \bibinfo{title}{{Superconducting Quark
  Matter}}, \bibinfo{journal}{Nucl. Phys. B} \bibinfo{volume}{129}
  (\bibinfo{year}{1977}) \bibinfo{pages}{390--396},
  \bibinfo{doi}{\doi{10.1016/0550-3213(77)90123-7}}.

\bibtype{Article}%
\bibitem{Pisarski:1999tv}
\bibinfo{author}{Robert~D. Pisarski}, \bibinfo{author}{Dirk~H. Rischke},
  \bibinfo{title}{{Color superconductivity in weak coupling}},
  \bibinfo{journal}{Phys. Rev. D} \bibinfo{volume}{61} (\bibinfo{year}{2000})
  \bibinfo{pages}{074017}, \bibinfo{doi}{\doi{10.1103/PhysRevD.61.074017}},
  \eprint{nucl-th/9910056}.

\bibtype{Inbook}%
\bibitem{Rajagopal:2000wf}
\bibinfo{author}{Krishna Rajagopal}, \bibinfo{author}{Frank Wilczek},
  \bibinfo{title}{{The Condensed matter physics of QCD}} \bibinfo{year}{2000}
  pp. \bibinfo{pages}{2061--2151},
  \bibinfo{doi}{\doi{10.1142/9789812810458_0043}}, \eprint{hep-ph/0011333}.

\bibtype{Article}%
\bibitem{Alford:1998mk}
\bibinfo{author}{Mark~G. Alford}, \bibinfo{author}{Krishna Rajagopal},
  \bibinfo{author}{Frank Wilczek}, \bibinfo{title}{{Color flavor locking and
  chiral symmetry breaking in high density QCD}}, \bibinfo{journal}{Nucl.Phys.}
  \bibinfo{volume}{B537} (\bibinfo{year}{1999}) \bibinfo{pages}{443--458},
  \bibinfo{doi}{\doi{10.1016/S0550-3213(98)00668-3}}, \eprint{hep-ph/9804403}.

\bibtype{Article}%
\bibitem{Schafer:1998ef}
\bibinfo{author}{Thomas Sch\"afer}, \bibinfo{author}{Frank Wilczek},
  \bibinfo{title}{{Continuity of quark and hadron matter}},
  \bibinfo{journal}{Phys. Rev. Lett.} \bibinfo{volume}{82}
  (\bibinfo{year}{1999}) \bibinfo{pages}{3956--3959},
  \bibinfo{doi}{\doi{10.1103/PhysRevLett.82.3956}}, \eprint{hep-ph/9811473}.

\bibtype{Article}%
\bibitem{Son:1999cm}
\bibinfo{author}{D.~T. Son}, \bibinfo{author}{Misha~A. Stephanov},
  \bibinfo{title}{{Inverse meson mass ordering in color flavor locking phase of
  high density QCD}}, \bibinfo{journal}{Phys. Rev. D} \bibinfo{volume}{61}
  (\bibinfo{year}{2000}) \bibinfo{pages}{074012},
  \bibinfo{doi}{\doi{10.1103/PhysRevD.61.074012}}, \eprint{hep-ph/9910491}.

\bibtype{Article}%
\bibitem{Son:2000tu}
\bibinfo{author}{D.~T. Son}, \bibinfo{author}{Misha~A. Stephanov},
  \bibinfo{title}{{Inverse meson mass ordering in color flavor locking phase of
  high density QCD: Erratum}}, \bibinfo{journal}{Phys. Rev. D}
  \bibinfo{volume}{62} (\bibinfo{year}{2000}) \bibinfo{pages}{059902},
  \bibinfo{doi}{\doi{10.1103/PhysRevD.62.059902}}, \eprint{hep-ph/0004095}.

\bibtype{Article}%
\bibitem{Casalbuoni:1999wu}
\bibinfo{author}{R. Casalbuoni}, \bibinfo{author}{Raoul Gatto},
  \bibinfo{title}{{Effective theory for color flavor locking in high density
  QCD}}, \bibinfo{journal}{Phys. Lett. B} \bibinfo{volume}{464}
  (\bibinfo{year}{1999}) \bibinfo{pages}{111--116},
  \bibinfo{doi}{\doi{10.1016/S0370-2693(99)01032-1}}, \eprint{hep-ph/9908227}.

\bibtype{Article}%
\bibitem{Gastineau:2001zke}
\bibinfo{author}{F. Gastineau}, \bibinfo{author}{R. Nebauer},
  \bibinfo{author}{J. Aichelin}, \bibinfo{title}{{Thermodynamics of the three
  flavor NJL model: Chiral symmetry breaking and color superconductivity}},
  \bibinfo{journal}{Phys. Rev. C} \bibinfo{volume}{65} (\bibinfo{year}{2002})
  \bibinfo{pages}{045204}, \bibinfo{doi}{\doi{10.1103/PhysRevC.65.045204}},
  \eprint{hep-ph/0101289}.

\bibtype{Article}%
\bibitem{Alford:1997zt}
\bibinfo{author}{Mark~G. Alford}, \bibinfo{author}{Krishna Rajagopal},
  \bibinfo{author}{Frank Wilczek}, \bibinfo{title}{{QCD at finite baryon
  density: Nucleon droplets and color superconductivity}},
  \bibinfo{journal}{Phys.Lett.} \bibinfo{volume}{B422} (\bibinfo{year}{1998})
  \bibinfo{pages}{247--256},
  \bibinfo{doi}{\doi{10.1016/S0370-2693(98)00051-3}}, \eprint{hep-ph/9711395}.

\bibtype{Article}%
\bibitem{Rapp:1997zu}
\bibinfo{author}{R. Rapp}, \bibinfo{author}{Thomas Sch{\"a}fer},
  \bibinfo{author}{Edward~V. Shuryak}, \bibinfo{author}{M. Velkovsky},
  \bibinfo{title}{{Diquark Bose condensates in high density matter and
  instantons}}, \bibinfo{journal}{Phys. Rev. Lett.} \bibinfo{volume}{81}
  (\bibinfo{year}{1998}) \bibinfo{pages}{53--56},
  \bibinfo{doi}{\doi{10.1103/PhysRevLett.81.53}}, \eprint{hep-ph/9711396}.

\end{thebibliography*}

\end{document}